# STUDIES ON REGGE BEHAVIOUR AND SPIN-INDEPENDENT AND SPIN-DEPENDENT STRUCTURE FUNCTIONS

A thesis submitted in partial fulfillment of
the requirements for award of the degree of
Doctor of Philosophy

**Begum Umme Jamil**

Regn. No. : 053 of 1999

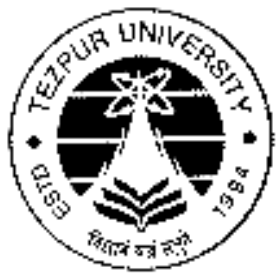

Department of Physics
School of Science and Technology
Tezpur University
Napaam, Tezpur – 784 028
Assam, India

August, 2008

*I would like to dedicate this thesis to*

*My*

***Abba** 'Md. Shirajul Haque'*

*And*

***Amma** 'Ms. Sarifa Begum'*

*Whose*

*Affection*

*Encouragement*

*Inspiration*

*Order*

*&*

*Understanding*

*Helped me to attain the position where I am today.*

# Abstract

Deep Inelastic Scattering (DIS) experiments have provided important information on the structure of hadrons and ultimately the structure of matter and on the nature of interactions between leptons and hadrons, since the discovery of partons. Various high energy deep inelastic interactions lead to different evolution equations from which we obtain various structure functions giving information about the partons i.e. quarks and gluons involved in different scattering processes. Actually structure function is a mathematical picture of the hadron structure in the high energy region.

Understanding the behaviour of the structure functions of the nucleon at low-x, where x is the Bjorken variable, is interesting both theoretically and phenomenologically. Structure functions are important inputs in many high energy processes and also important for examination of perturbative quantum chromodynamics (PQCD), the underlying dynamics of quarks and gluons. In PQCD, for high-$Q^2$, where $Q^2$ is the four momentum transfer in a DIS process, the $Q^2$-evolutions of these densities (at fixed-x) are given by mainly Dokshitzer-Gribov-Lipatov-Altarelli-Parisi (DGLAP) Evolution Equations.

DGLAP evolution equations can be solved either by numerical integration in steps or by taking the moments of the distributions. Among various solutions of these equations, most of the methods are numerical. Mellin moment space with subsequent inversion, Brute force method, Laguerre method, Matrix method etc. are different methods used to solve DGLAP evolution equations. The shortcomings common to all are the computer time required and decreasing accuracy for x → 0. More precise approach is the matrix approach to the solution of the DGLAP evolution equations, but it is also a numerical method. Thus, though numerical solutions are available in the literature, the explorations of the possibilities to obtain analytical solutions of DGLAP evolution equations are always interesting. Some approximated analytical solutions of DGLAP evolution equations suitable at low-x have been reported in recent years with considerable phenomenological success. Of these methods using Taylor expansion, applying Regge behaviour of structure functions, method of characteristics etc. are important. In this connection, general solutions of DGLAP evolution equations at high-x, medium-x and low-x in leading order and next-to-leading order have already been obtained by using Taylor expansion method. The structure functions thus calculated are expected to rise approximately with a power of x towards low-x which is supported by Regge theory. In this thesis, we solved both spin-independent and spin-dependent DGLAP evolution equations applying Regge behaviour of structure functions at low-x up-to next-next-to-leading order (NNLO) and have got the respective approximate analytical solutions of structure functions.

**In Chapter 1**, we have presented a brief introduction to the structure of matter, deep inelastic scattering, spin-independent DIS cross section and structure functions, spin-dependent structure functions, low-x physics, evolution equations and about some important research centres and experiments.

**In Chapter 2,** we have given the introductory discussion about the Regge theory including the complex angular momentum plane and Regge theory in DIS. Many models based on Regge theory are able to reproduce hadronic cross-sections. We have discussed extension of some of the simplest models to the DIS amplitudes and shown that Regge theory can be used to describe structure functions. In the subsequent Chapters we have considered Regge behaviour of structure functions at low-x and have solved both spin-independent and spin-dependent DGLAP evolution equations to get the both spin-independent and spin-dependent deuteron, proton, neutron and gluon structure functions at low-x.

**In Chapter 3,** we have presented our solutions of spin-independent DGLAP evolution equations for singlet, non-singlet and gluon structure functions at low-x in leading order (LO) and also the solution of the coupled equations for singlet and gluon structure functions. The t and x-evolutions of deuteron, proton and gluon structure functions thus obtained have been compared with NMC and E665 data sets and global MRST2001, MRST2004 and GRV1998LO gluon parameterizations respectively.

**In Chapter 4,** we have presented our solutions of spin-independent DGLAP evolution equations for singlet, non-singlet and gluon structure functions at low-x in next-to-leading order (NLO). We also solved the coupled equations for singlet and gluon structure functions. The t and x-evolutions of deuteron, proton and gluon structure functions thus obtained have been compared with NMC and E665 data sets and global MRST2001, MRST2004 and GRV1998LO and GRV1998NLO gluon parameterizations respectively. Along with the NLO results we also presented our LO results from Chapter 3.

**In Chapter 5,** we have presented our solutions of spin-independent DGLAP evolution equations for singlet and non-singlet structure functions at low-x in NNLO. The t-evolutions of deuteron and proton structure functions thus obtained from singlet and non-singlet structure functions have been compared with NMC and E665 data sets. Along with the NNLO results we have also presented our results of LO and NLO from Chapters 3 and 4.

**In Chapter 6,** we have presented our solutions of spin-dependent DGLAP evolution equations for singlet, non-singlet and gluon structure functions at low-x in LO. Here also we solved the coupled equations for singlet and gluon structure functions. The evolutions of deuteron, proton, neutron and gluon structure functions thus obtained have been compared with SLAC-E-154, SLAC-E-143, SMC collaborations data sets and the result obtained by numerical method.

**In Chapter 7,** we have presented our solutions of spin-dependent DGLAP evolution equations for singlet, non-singlet and gluon structure functions at low-x in NLO and also the solution of coupled equations for singlet and gluon structure functions. The evolutions of deuteron, proton, neutron and gluon structure functions thus obtained have been compared with SLAC-E-154, SLAC-E-143, SMC collaborations data sets and the result obtained by numerical method. Here we compared our LO and NLO results.

**In Chapter 8,** in the conclusion part, we have summarized the results drawn from our work. □

# DECLARATION

I hereby declare that the thesis entitled **'Studies on Regge Behaviour and Spin-independent and Spin-dependent Structure Functions'** being submitted to Tezpur University, Tezpur, Assam in partial fulfillment of the requirements for the award of the degree of Doctor of Philosophy, has previously not formed the basis for the award of any degree, diploma, associateship, fellowship or any other similar title or recognition.

Date     :                                                                                      ( Begum Umme Jamil )
Place    : Napaam, Tezpur                                                         Department of Physics
                                                                                                       Tezpur University
                                                                                                   Tezpur-784 028 (Assam)

# CERTIFICATE

**Dr. Jayanta Kumar Sarma**
Reader
Department of Physics
Tezpur University
Napaam, Tezpur- 784 028
Assam, India

This is to certify that **Begum Umme Jamil** has worked under my supervision and the thesis entitled '**Studies on Regge Behaviour and Spin-independent and Spin-dependent Structure Functions**' which is being submitted to Tezpur University in partial fulfillment of the requirements for the degree of Doctor of Philosophy, is a record of original bonafide research work carried out by her. She has fulfilled all the requirements under the Ph.D. rules and regulations of Tezpur University.  Also, to the best of my knowledge, the results contained in the thesis have not been submitted in part or full to any other university or institute for award of any degree or diploma.

Date    :                                                                         **(Dr. Jayanta Kumar Sarma)**
Place   : Napaam, Tezpur                                          Supervisor

# *Acknowledgements*

*There are many people whom I should acknowledge for the pain they took for helping me in some or the other way during my research period.*

*I find it difficult to write something in short to acknowledge my research supervisor Dr. Jayanta Kumar Sarma whose inspiration and invaluable guidance helped me to follow proper track in the field of High Energy Physics. I take this opportunity to express my intense reverence towards him for the extensive scientific discussions and for giving me the freedom in research.*

*I want to convey my sincere gratitude to Prof. A. Choudhury, Dr. A. Kumar, Dr. N. S. Bhattacharyya, Dr. N. Das, Dr. G. A. Ahmed, Dr. D. Mohanta, Dr. P. Deb and Dr. K. Barua of Dept. of Physics, Tezpur University for their encouragement, criticism and inspiration to carry out this work.*

*I am indebted to my aunty Prof. N. S. Islam for her affection, encouragement, help and time-to-time command to follow right directions during my research period and otherwise.*

*I am indebted to my elder brothers Bobby and Loni, elder sister Julie, sister in law Pinky, uncle Dr. Matiur Rahman. They remained in my heart and boosted me for whatever I did towards my carrier and life. I offer my love and sincere regards to little sweet sister Arshiya, niece Lollypop and nephew Babu whose melodious tune keeps me fresh and give immense happiness.*

*I would like to thank all my seniors, juniors colleagues and friends in Tezpur University of whom special thanks goes to Abuda, Anjanda, Ghanada, Digantada, Ranjitda, Navada, Panku, Upamanyu, Sovan, Ankur, Debashish, Bobby-baideu, Rasnaba, Jutiba, Swapnaliba, Nandini, Mithu, Nabanita, Maumita, Swati, Smriti and Mayuri for their company, help and goodwill. Also I take the opportunity to thank Pathakda and Narayanda, office staff, Dept. of Physics, Tezpur University.*

*Especially I am thankful to Sanjeevda as he remained near me and supported whenever needed from almost all directions.*

*I take the opportunity to thank Tezpur University for providing me the research facility and the University community for helping me in carrying my research work. I acknowledge the help extended by the Central Library staff of Tezpur University.*

*Finally I would like to acknowledge University Grants Commission for financial support that I received in some part of my research period that helped to carry out this work.*

*Date:*  (Begum Umme Jamil)

# STUDIES ON REGGE BEHAVIOUR AND SPIN-INDEPENDENT AND SPIN-DEPENDENT STRUCTURE FUNCTIONS

## Contents

<div style="text-align: right">**Chapter 1**</div>

# INTRODUCTION

## 1.1 Structure of Matter

Matter is composed of - what? Matter is composed of atoms or molecules what was suggested by John Dalton in 1805 in his Atomic Theory and according to this theory the atoms are the smallest indivisible particle. But with the discovery of some of the subatomic particles like electrons, protons, neutron etc. which are responsible for the more rich and complex structure of the atom, the Atomic Theory was discarded. Extensive researches, since the start of nineteenth century, have been carried out by the scientists to conclude about the ultimate representatives of the matter i.e. the basic building blocks called the elementary particles or sub-atomic particles [1]. By the end of the nineteenth century, in 1897, J. J. Thomson discovered the electron. In 1932, James Chadwick identified neutron and Werner Heisenberg suggested that atomic nuclei consist of neutrons and protons [2-6]. Thus atomic picture becomes somewhat clear with electron, neutron, proton and photon as the basic building blocks. Photon has been added as a field particle for electromagnetic force such as exists between the nucleus and electrons in the atom, i.e., it is a quantum unit of radiation. In the same year, Carl David Anderson found the positive electron or the positron while studying cosmic ray showers. The discovery of this particle, being the antiparticle of electron, predicted the existence of antimatter.

Then the concept of quark comes as they are the basic constituent of the elementary particles, such as the proton, neutron and pion. The quark concept [7, 8] was independently proposed in 1964 by the American physicists Murray Gell-Mann and George Zweig. Quarks were first believed to be of three kinds:

up(u), down(d), and strange(s) and in 1974 the existence of the fourth quark, named charm(c), was experimentally confirmed [9, 10]. Thereafter a fifth and sixth quarks-called bottom(b) and top(t), respectively – were proposed for theoretical reasons of symmetry. Experimental evidence for the existence of the bottom quark [9-10] was obtained in 1977. Again in 1994 physicists at Fermi National Accelerator Laboratory (Fermilab) announced the experimental evidence for the existence of top quark. Quarks have the extraordinary property of carrying electric charges that are fractions of the charge of the electron, previously believed to be the fundamental unit of charge. Quarks are also termed as flavor. Each kind of quark or flavor has its antiparticle. The carrier of the force between quarks is a particle called gluon [7-10]. Evidence for gluons came out in 1978 from an electron – positron machine, called PETRA [9-10], at Hamburg in Germany which is able to observe collisions up to 30 GeV.

A table showing the flavor and charges of quarks and anti-quarks is given below:

| Flavour | Charge | Flavour | Charge |
|---|---|---|---|
| ● Up | +2/3 | ● Anti-Up | -2/3 |
| ● Down | -1/3 | ● Anti-Down | +1/3 |
| ● Charm | +2/3 | ● Anti-Charm | -2/3 |
| ● Strange | -1/3 | ● Anti-Strange | +1/3 |
| ● Top | +2/3 | ● Anti-Top | -2/3 |
| ● Bottom | -1/3 | ● Anti-Bottom | +1/3 |

Quark structure of the positive Pion, Proton, and Neutron are shown below:

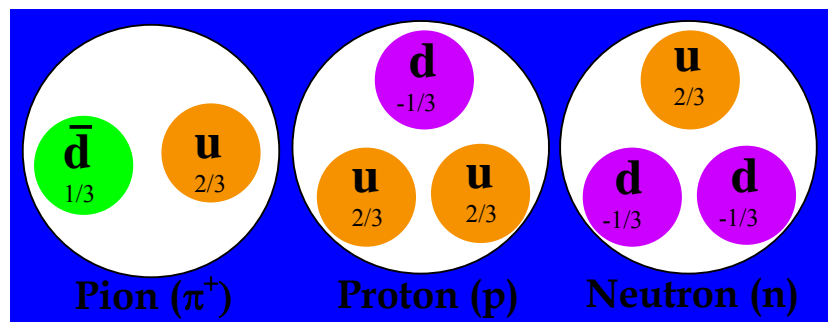

**Figure 1.1:** Quark Structure of the $\pi^+$, p and n

Quarks cannot be separated from each other, for this would require far more energy than even the most powerful particle accelerator [2, 9-10] can provide. They are observed bound together in pairs, forming particles called mesons, or in threes, forming particles called baryons. Mesons and baryons together are called hadrons. The positive pion consists of one up quark and one anti-down quark. The proton consists of two up quarks and one down quark, while the neutron consists of two down quarks and one up quark as depicted in Figure 1.1. Quantum chromodynamics (QCD) [11], physical theory of strong interaction, attempts to account for the behaviour of the elementary particles. Mathematically, QCD is quite similar to quantum electrodynamics (QED), the theory of electromagnetic interactions; it seeks to provide an equivalent basis for the strong nuclear force that binds particles into atomic nuclei. According to QCD each quark appears in three colours [7-10] – red (R), blue (B) and green (G). Antiquarks carry anticolours, anti-red (Cyan), anti-blue (Yellow) and anti-green (magenta), i.e, $(\bar{R}, \bar{B}, \bar{G})$. Colour has of course no relation with the traditional colours.

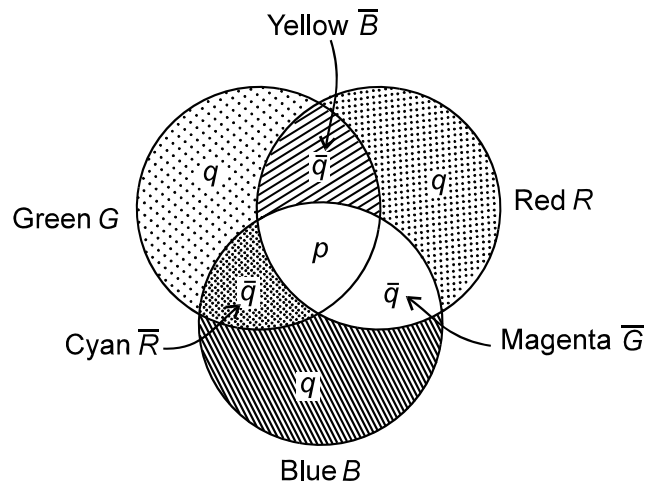

**Figure 1.2:** Colour composition of hadrons.

Equal mixture of Red, Green, Blue (R, G, B) or Cyan, Yellow and Magenta $(\bar{R}, \bar{B}, \bar{G})$ i. e. equal mixture of colour or anti colours, or colour-anti colours also ($R\bar{R}, B\bar{B}, G\bar{G}$) are white or colourless. This explains why observed particle states – baryon and mesons and their antiparticles in nature are colourless or white which means unchanged by rotation in colour space. It is easy to visualize the colour quantum number by associating the three possible colours of a quark with the three spots of primary red, green and blue light focused on a screen, as shown in figure 1.2.

Particles from massive one to tiny chunks experience four different types of interactions with different magnitude of strength and ranges. The basic forces [2-10] and some of their field properties are given below:

| Force | Experienced by | Exchange particle | Range | Relative Strength | Exchange particle | | |
|---|---|---|---|---|---|---|---|
| | | | | | Rest mass (GeV/c²) | Spin | Electric charge |
| Gravitational | All particles | graviton g* | Long, i.e. $F \propto 1/r^2$ | $10^{-41}$ | 0 | 2 | 0 |
| Weak nuclear | All particles except g* | W and Z bosons W⁺ W⁻ Z⁰ | $10^{-18}$ m | $10^{-16}$ | 81 81 92 | 1 1 1 | +1 -1 0 |
| Electro-magnetic | Particles with electric charge | Photons γ | Long, i.e. $F \propto 1/r^2$ | 1/137 | 0 | 1 | 0 |
| Strong nuclear | Quarks and gluons | Gluons g | 2×10⁻¹⁵ m | 1 | 0 | 1 | 0 |

## 1.2 Deep Inelastic Scattering

Deep Inelastic Scattering (DIS) experiments have provided important information on the structure of hadrons and ultimately the structure of matters and on the nature of interactions between leptons and hadrons.

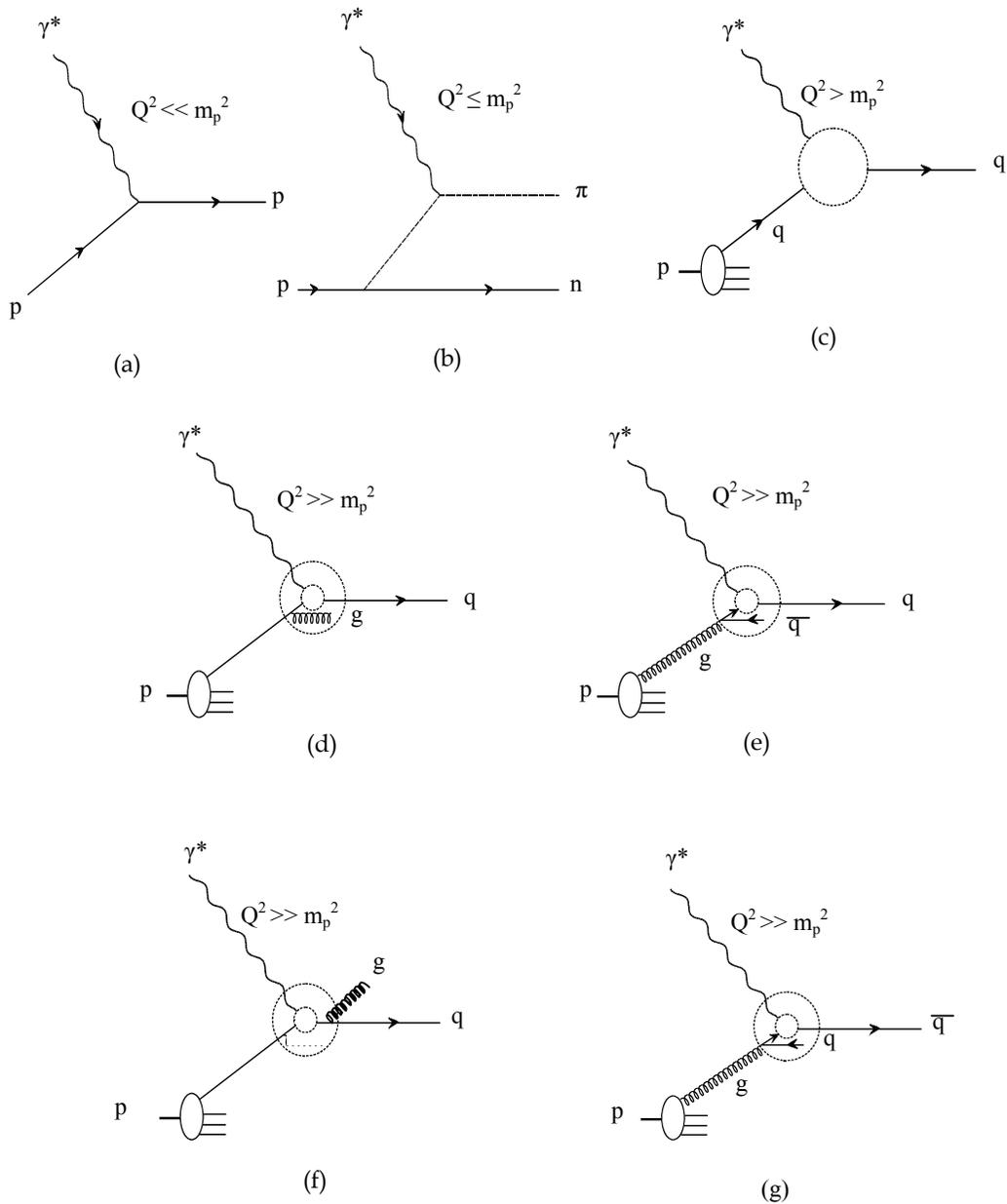

**Figure1.3:** The hadron as seen by a 'microscope' ≡ virtual photon: as $Q^2$ increases, a quark may be resolved into a quark and bremsstrahlung gluon or into a quark - antiquark pair.

When a very low mass virtual photon ($Q^2 = -q^2 \ll 1 GeV^2$) scatters off a hadron, the photon 'sees' only the total charge and magnetic moment of the hadron and the scattering appears point-like (Figure 1.3(a) ) [7, 12]. A higher-mass photon of a few hundred MeV² is able to resolve the individual constituents of the hadron's virtual pion cloud, as shown in Figure 1.3(b) [7, 12], and the hadron appears as a composite extended object. At high momentum transfers the photon probes the fine structure of the hadron's charge distribution and sees its elementary constituents (Figure 1.3(c)) [7, 12]. If quarks were non-interacting, no further structure would appear for increasing $Q^2$ and exact scaling would set in. However, in any renormalizable quantum field theory, we have to introduce a Bose-field (gluon) which mediates the interaction in order to form bound states of quarks, i.e. the observed hadrons. In such a picture, the quark is then always accompanied by a gluon cloud which will be probed as the momentum transfer is increased. The effect of gluons is then two-fold as illustrated in Figure 1.3(d-g) [7, 12].

In DIS, a lepton with four-momentum k scatters off a nucleon say, a proton with four momentum p as depicted in Figure 1.4.

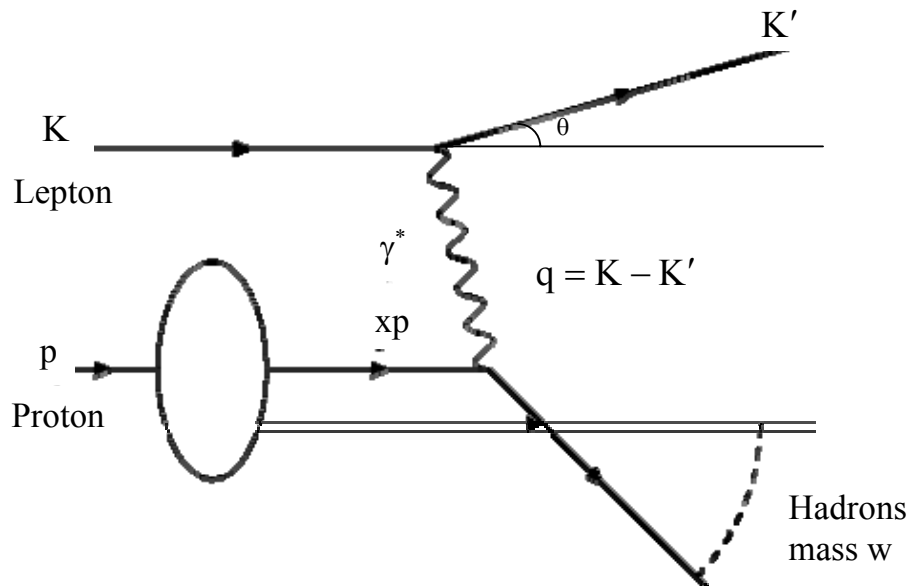

**Figure 1.4:** DIS reaction: the photon interacts with a quark inside the proton.

The final state of this reaction consists of the scattered lepton with four-momentum k′ and the hadronic fragmentation products xp. The exchanged virtual vector boson $\gamma^*$ carries a four-momentum q=k−k´. The first component of q is the energy transfer, ν = E − E´ where E is the energy of the incoming lepton and E´ is the energy of the scattered lepton. To describe the kinematics of the above process in the laboratory reference frame, the following variables are introduced [13-17].

- $Q^2 = -q^2$, the negative of the exchanged four-momentum squared.

- $x = Q^2/2p \cdot q = Q^2/2M\nu$, the Bjorken scaling variable, which describes the fraction of the nucleon momentum carried by the struck quark.

- $W^2 = (p + q)^2$, the invariant mass squared of the virtual-photon nucleon system.

- $y = p.q/p.k = \nu/E$, the fraction of the initial lepton energy transferred to the boson.

Neglecting the mass of the electron, the expressions for $Q^2$ and $W^2$ can be transformed into $Q^2 = 4EE' \sin^2(\theta/2)$ and $W^2 = M^2 + 2M(E − E') − Q^2$ where M is the mass of the nucleon and θ is the scattering angle in the laboratory reference frame. At large values of $Q^2$, i.e. at small scale distance DIS probes the constituents of the hadron (i.e. quarks) not the hadron as a whole. At small distance scales, the quarks act as almost free particles and because the interactions are relatively weak at those scales, perturbative QCD (PQCD) techniques can be used for DIS. A typical lower $Q^2$ limit for which PQCD is applicable, is 1 GeV². Similarly, to avoid contributions from the baryonic resonance region a minimum invariant mass W of 2 GeV is usually imposed on the data. In DIS, three types of events are distinguished: (i) inclusive events, where only the scattered lepton is detected; (ii) semi-inclusive events, where apart from the lepton also a hadron is detected; and (iii) exclusive events, where all reaction products are identified.

## 1.3 Spin-independent structure functions

The spin-independent DIS [7] cross section can be expressed as

$$\frac{d^2\sigma}{dxdQ^2} = \frac{4\pi\alpha_S^2}{Q^4}\left[F_1(x,Q^2)\cdot y^2 + \frac{F_2(x,Q^2)}{x}\left(1-y-\frac{Mxy}{2E}\right)\right],$$

where $\alpha_S(Q^2)$ is the running coupling constant which describes how the effective charge depends on the separation of two charged particles. $F_1$ and $F_2$ are two dimensionless structure functions. Actually structure function is a mathematical picture of the hadron structure at high-energy region [1, 7, 12]. Because quarks have spin 1/2, the two structure functions $F_1$ and $F_2$ are related by the Callan-Gross relation, $2xF_1(x, Q^2) = F_2(x, Q^2)$. In the quark parton model (QPM), the structure functions are independent of $Q^2$ for point-like quarks, and are only functions of the scaling variable x. Experimental data show sizeable deviations from the assumed $Q^2$-independence, which are known as scaling violations. The deviations are due to gluon radiation and the creation of quark-antiquark pairs. These deviations are only prominent at low values of x (< $10^{-2}$), and are well described by PQCD calculations, in which the quark and gluon distributions are used as free parameters. In this framework, the structure function $F_1$ can be interpreted as the parton density distribution which is given by the incoherent sum of the parton momentum distributions $q_f(x)$ for each quark flavor f,

$$F_1 = \frac{1}{2}\sum_f e_f^2 q_f(x),$$

here $e_f$ is the fractional electric charge of each of the quark flavors. Similarly, $F_2$ is the sum weighted by x, which is the momentum fraction carried by the parton,

$$F_2 = \sum_f xe_f^2 q_f(x).$$

For spin-independent beams and targets, the parton momentum distribution is defined as

$$q_f(x) = \vec{q}_f(x) + \vec{\bar{q}}_f(x),$$

where $\vec{q}_f(x)$ and $\vec{\bar{q}}_f(x)$ are the probability of finding a parton of type f with its spin aligned parallel or anti-parallel to the nucleon spin, respectively. The discussion here is limited to the longitudinal spin, i.e. parallel to the direction of motion of the proton. Therefore, the spins mentioned in the text actually correspond to helicities. Most experimental results on structure functions are obtained by inclusive measurements.

## 1.4 Spin-dependent structure functions

In analogy to the spin-independent structure functions $F_1$ and $F_2$, the spin-dependent structure functions $g_1$ and $g_2$ contain information on the helicity dependent contribution to the DIS cross section [18-20]. To access these structure functions, a polarized target and a polarized beam are needed. Results are obtained by measuring the difference in cross section for a parallel ($\rightarrow \Rightarrow$) or anti-parallel ($\rightarrow \Leftarrow$) orientation of the spins of the struck nucleon and the lepton. A measure for the helicity dependent contributions to the cross section is obtained by evaluating the asymmetry $\left(\sigma^{\vec{\Rightarrow}}_{-} - \sigma^{\vec{\Leftarrow}}\right) / \left(\sigma^{\vec{\Rightarrow}}_{+} + \sigma^{\vec{\Leftarrow}}\right)$. This is called a double spin asymmetry. Similarly, in case of a single spin asymmetry either the target or the beam is polarized, while the other is unpolarized.

The nucleon is a spin 1/2 particle and has a total spin that is given by the sum of the angular momentum components of its constituents. The total longitudinal spin of the nucleon is given by

$$\frac{1}{2}\Delta\Sigma_q + \Delta G + L_z = \frac{1}{2},$$

where $\Delta\Sigma_q$ is the contribution of the quark spins, $\Delta G$ is the gluon polarization, and $L_z$ is the (possible) contribution coming from the orbital angular momentum of the quarks and gluons. The longitudinal quark spin contribution $\Delta\Sigma_q$ is given by the sum over all flavors of the quark helicity distributions $\Delta q_f$

$$\Delta\Sigma_q = \sum_f \int_0^1 \Delta q_f(x)dx \text{ with } \Delta q_f(x) = \vec{q}_f(x) - \vec{\bar{q}}_f(x).$$

The distribution $\Delta q_f(x)$ can be interpreted as the probability of finding a quark with flavor f in the same helicity state as the nucleon. In 1988, the EMC experiment [21] found that only a small fraction of the nucleon spin seems carried by the quarks, which is about 20% in contrast to the naive quark-parton model. This lead to the so called "spin-crisis". This result is confirmed by a series of DIS experiments at CERN [22, 23] and DESY [24].

The inclusive scattering cross section gives access to longitudinally polarized structure function $g_1$, which is the sum of helicity distributions for different quark flavors weighted by the electric charge $e_f$ squared,

$$g_1 = \frac{1}{2}\sum_f e_f^2 \Delta q_f.$$

By determining the double spin asymmetry in semi-inclusive DIS for hadrons with a different quark composition, the helicity distributions of the individual quark flavors can be determined. Whereas transversely polarized structure function $g_2$ has contributions from quark-gluon correlations and other higher twist terms which cannot be described perturbatively. The contribution of the gluon spin ΔG to the nucleon spin can be determined from events created in the photon-gluon fusion process where the virtual photon interacts with a gluon from the nucleon by splitting into a quark-antiquark pair. The remaining contribution to the nucleon spin, almost 80% comes from the orbital angular momentum of the quarks and gluons ($L_z$). As the sum of the present best values for $\frac{1}{2}\Delta\Sigma_q$ and ΔG cannot make up for the spin of the nucleon, it is likely that quarks and gluons also carry a non-zero amount of orbital angular momentum.

## 1.5 Low-x physics

Low-x physics is always being the exciting field of DIS. The behaviour of the parton distributions of the hadrons in low-x region is of considerable importance both theoretically and phenomenologicaly. In the low-x region novel effects are expected to emerge. At very low-x region (less than $10^{-3}$ or $10^{-4}$), quarks and gluons radiate 'soft' gluons and thus the number of partons i.e. quarks and gluons increases rapidly. As the gluon density becomes higher several effects, like recombination of gluon to form higher-x gluons, shading of gluons by each other, collective effects like condensation or super fluidity or formation of local region (known as hot spots) etc. can occur. These may have dominant effect on non-perturbative physics at low-x. According to QCD, at low-x and at large-$Q^2$, a nucleon consists predominantly of gluons and sea quarks. Their densities grow rapidly in the limit x→0 leading to possible spatial overlap and to interactions between the partons. i.e. at low-x, the structure function is proportional to the sea quark density. Several DIS experiments have been performed on nuclear targets and various nuclear effects have shown up at low-x.

The low-x region of DIS obeys the Regge limit of PQCD [8, 25-28]. DIS corresponds to the region where both ν and $Q^2$ are large and x is finite. The low-x limit of DIS corresponds to the case when $2M\nu \gg Q^2$, yet $Q^2$ is still large, that is at least a couple of $GeV^2$. The limit $2M\nu \gg Q^2$ is equivalent to $S \gg Q^2$, that is to the limit when the center of mass energy squared S is large and much greater than $Q^2$. Since $Q^2$ is large it allows to use PQCD. The structure functions are expected to have Regge behaviour corresponding to Regge particle exchange i.e. structure function is proportional to $x^{-\lambda}$, where λ is the Regge intercept. Thus low-x physics represents an interesting area in DIS structure function of hadrons.

## 1.6 Evolution Equations

**1. DGLAP (Dokshitzer-Gribov-Lipatov-Altarelli-Parisi) Evolution Equations**

Keeping only leading powers of $\ln Q^2$ (i.e. $\alpha_s^n \ln^n Q^2$) terms in the perturbative expansion, the DGLAP evolution equation comes in the Leading Logarithmic $Q^2$ (LLQ$^2$) approximation. The DGLAP evolution equations for quark and gluon in Leading Order (LO) are respectively

$$\frac{\partial q_i(x,Q^2)}{\partial t} = \frac{\alpha_s(Q^2)}{2\pi} \int_x^1 \frac{dy}{y} \left[ P_{qq}(x/y) q_i(y,Q^2) + P_{qg}(x/y) G(y,Q^2) \right] \quad (1.1)$$

and

$$\frac{\partial G(x,Q^2)}{\partial t} = \frac{\alpha_s(Q^2)}{2\pi} \int_x^1 \frac{dy}{y} \left[ \sum_i P_{gq}(x/y) q_i(y,Q^2) + P_{gg}(x/y) G(y,Q^2) \right], \quad (1.2)$$

where $t = \ln(Q^2/\Lambda^2)$ and $P_{qq}$, $P_{qg}$, $P_{gg}$, $P_{gq}$ denoting the splitting functions. The sum $i = 1\ldots\ldots 2n_f$, $n_f$ being the number of flavours, runs over quarks and antiquarks of all flavors. $P_{gq}$ does not depend on the index i if the quark masses are neglected [7].

For example, in the equation (1.1), the first term mathematically expresses the fact that a quark with momentum fraction x [q(x, Q$^2$) on the left hand side] could have come form a parent quark with a larger momentum fraction y [q(y, Q$^2$) on the right-hand side] which has radiated a gluon. The probability is proportional to $\alpha_s P_{qq}(x/y)$. The second term considers the possibility that a quark with momentum fraction x is the result of $q\bar{q}$ pair creation by a parent gluon with momentum fraction y (>x). The probability is proportional to $\alpha_s P_{qg}(x/y)$. The integral in the equation is the sum over all possible momentum fractions y (>x) of the parent [7]. For gluon we can give a symbolic representation of the gluon evolution equation (1.2) as in Figure 1.5.

$$\frac{d}{d\log Q^2}\left(g(x,Q^2)\right) = \sum_i \frac{q_i(y,Q^2) \quad g(x,Q^2)}{P_{gq}(x/y)} + \frac{g(y,Q^2) \quad g(x,Q^2)}{P_{gg}(x/y)}$$

**Figure 1.5:** Symbolic representation of the gluon evolution equation in LO.

Figure 1.6 gives a schematic ladder diagram of quark and gluon exchange in LLQ² approximation of DIS.

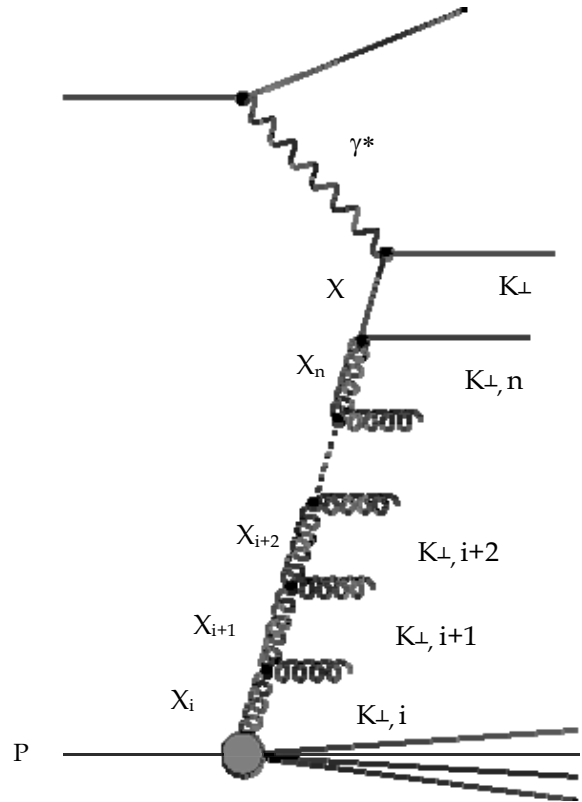

**Figure 1.6**: Ladder diagram for the DIS in LLQ²

When the appropriate gauge is chosen, the diagrams which contribute in the DGLAP approximation are the ladder diagrams with gluon and quark exchange as depicted in Figure 1.6. In ladder diagrams, the longitudinal momenta ~$X_i$ are ordered along the chain $(x_i \geq x_{i+1})$ and the transverse momenta are strongly ordered, that is, $k^2_{\perp,i} << k^2_{\perp,i+1}$. It is this strong ordering of transverse momenta towards $Q^2$ which gives the maximal power of $\ln(Q^2)$, since the integration over transverse momentum in each cell is logarithmic.

2. **BKFL (Balitsky-Kuraev-Fadin-Lipatov) Evolution Equation**

Keeping only leading powers of LL (1/x) terms in the perturbative expansion, the BKFL evolution equation comes in the Leading Logarithmic 1/x (LL(1/x)) approximation[29-31]. The BKFL evolution equation is

$$f(x,k^2) = f^0(x,k^2) + \frac{3\alpha_s(k^2)}{\pi} k^2 \int_x^1 \frac{dx'}{x'} \int_{k_0^2}^\infty \frac{dk'^2}{k'^2} \left\{ \frac{f(x',k'^2) - f(x',k^2)}{|k'^2 - k^2|} + \frac{f(x',k^2)}{\sqrt{4k'^4 + k^4}} \right\}, \quad (1.3)$$

where the function $f(x, k^2)$ is the nonintegrated gluon distribution, that is $f(x,k^2) = \partial xG(x,k^2)/\partial \ln k^2$, $f^0(x,k^2)$ is a suitably defined inhomogeneous term; $k^2$, $k'^2$ are the transverse momenta squared of the gluon in the final and initial states respectively, and $k_0^2$ is the lower limit cut-off. The important point here is that, unlike the case of the LLQ² approximation, the transverse momenta are no longer ordered along the chain.

3. **GLR (Gribov-Levin-Ryskin) Evolution Equation**

The GLR evolution equation is obtained in the double logarithmic approximation (DLA). DLA is the approximation where both leading power of lnQ² and ln(1/x) are kept. The compact forms of GLR equations are shown in the recent literature [32-34]. Further approximation is that the coupling of n ≥ 2 ladder to the hadrons is proportional to the n-th power of a single ladder and the probability of finding two gluons (at low momentum $Q_0^2$) with momentum fraction $x_1$ and $x_2$ is proportional to $g(x_1,Q_0^2).g(x_2,Q_0^2)$. It leads to a non-linear integro-differential equation for structure function which gives the GLR equation as

$$\frac{\partial \Phi(x,Q^2)}{\partial \ln(1/x)} = \int \hat{k}(q^2,q'^2)\Phi(x,q'^2).\frac{4N_f\alpha_s(q'^2)}{4\pi} - \frac{1}{4\pi R^2}\left(\frac{\alpha_s(q'^2)}{4\pi}\right)^2 .V.\Phi^2(x,Q^2), \quad (1.4)$$

where $\Phi = \partial F(x, Q^2)/\partial Q^2$, R denotes the transverse radius of the hadron and V stands for the triple ladder vertex.

4. **CCFM (Ciafaloni-Catani-Fiorani-Marchesini) Evolution Equation**

The CCFM [35-38] evolution equation with respect to the scale $\bar{q}_i^2$ can be written in a differential form [30]

$$\bar{q}_i \frac{d}{d\bar{q}^2} \frac{xA(x,k_\perp^2,\bar{q}^2)}{\Delta_S(\bar{q}^2,Q_0^2)} = \int dz \frac{d\phi}{2\pi} \frac{\tilde{P}(z,(\bar{q}/z)^2,k_\perp^2)}{\Delta_S(\bar{q}^2,Q_0^2)} x' A(x',k_\perp'^2,(\bar{q}/z)^2), \quad (1.5)$$

where $A(x,k_\perp^2,\bar{q}^2)$ is the unintegrated gluon density which depends on longitudinal momentum fraction x, transverse momentum $k_\perp^2$ and the evolution variable $\mu^2$ (factorization scale)$=\bar{q}^2$. The splitting variables are $z=x/x'$ and $\vec{k}_\perp' = (1-z)/z\vec{q} + \vec{k}_\perp$ where the vector $\vec{q}$ is at an azimuthal angle $\phi$. $\Delta_s$ is the Sudakov form factor and is given as

$$\Delta_S(\bar{q}^2,Q_0^2) = \exp\left(-\int_{Q_0^2}^{\bar{q}^2} \frac{dq^2}{q^2} \int_0^{1-Q_0/q} dz \frac{\bar{\alpha}_s(q^2(1-z)^2)}{1-z}\right), \text{ where } \bar{\alpha}_s = 3\alpha_s/\pi.$$

And the splitting function $\tilde{P}$ for branching i is given by

$$\tilde{P}_g(z_i,q_i^2,k_{\perp i}^2) = \frac{\alpha_s(q_i^2(1-z_i)^2)}{1-z_i} + \frac{\bar{\alpha}_s(k_{\perp i}^2)}{z_i}\Delta_{ns}(z_i,q_i^2,k_{\perp i}^2),$$

where $\Delta_{ns}$ is the non-Sudakov form factor defined as

$$\log \Delta_{ns}\left(z_i, q_i^2, k_{\perp i}^2\right) = -\overline{\alpha}_s \int_{z_i}^{1} \frac{dz'}{z'} \int \frac{dq^2}{q^2} \Theta(k_{\perp i} - q)\Theta(q - z'q_i) \cdot$$

## 1.7: Some Important Research Centres and Experiments

### 1. CERN (Conseil Europeen pour la Recherche Nucleaire)

CERN is the EUROPEAN ORGANIZATION FOR NUCLEAR RESEARCH which is the international scientific organization for collaborative research in sub-nuclear physics (high-energy, or particle physics). Head office of CERN is in Geneva, Swizerland. The activation of a 600-mega volt synchrocyclotron in 1957 enabled CERN physicists to observe the decay of a pion, into an electron and a neutrino. The event was instrumental in the development of the theory of weak interaction. The laboratory grew steadily, activating the particle accelerator known as the Proton Synchrotron (1959), which used 'strong focusing' of particle beams; the Intersecting Storage Rings (ISR; 1971), enabling head-on collisions between protons; and the Super Proton Synchrotron (SPS; 1976), with a 7-kilometre circumference. With the addition of an Antiproton Accumulator Ring, the SPS was converted into a proton-antiproton collider in 1981 and provided experimenters with the discovery of the W and Z particles in 1983 by Carlo Rubbia and Simon van der Meer. In November 2000 the Large Electron-Positron Collider (LEP), a particle accelerator installed at CERN is an underground tunnel 27 km in circumference, closed down after 11 years service. LEP was used to counter-rotate accelerated electrons and positrons in a narrow evacuated tube at velocities close to that of light, making a complete circuit about 11,000 times per second. Their paths crossed at four points around the ring. DELPHI, one of the four LEP detectors, was a horizontal cylinder about 10 m in diameter, 10 m long and weighing about 3,000 tones. It was made of concentric sub-detectors, each designed for a specialized recording task.

CERN also has LHC (Large Hadron Collider) which is a particle accelerator and hadron collider. The LHC has started its operation from May 2008. It is expected to become the world's largest and highest-energy particle accelerator. When activated, it is theorized that the collider will produce the elusive Higgs boson, the observation of which could confirm the predictions and 'missing links' in the Standard Model of physics and could explain how other elementary particles acquire properties such as mass. Six detectors are being constructed at the LHC. They are located underground, in large caverns excavated at the LHC's intersection points. Two of them, ATLAS and CMS are large particle detectors. ALICE is a large detector designed to search for a quark-gluon plasma in the very messy debris of heavy ion collisions. The other three (LHCb, TOTEM, and LHCf) are smaller and more specialized. A seventh experiment, FP420 (Forward Physics at 420 m), has been proposed which would add detectors to four available spaces located 420 m on either side of the ATLAS and CMS detectors.

Parton distribution functions (PDF) are vital for reliable predictions for new physics signals and their background cross sections at the LHC. Since QCD does not predict the parton content of the proton, the PDF parameters are determined by fit to data from experimental observables in various processes, using the DGLAP evolution equation. Recently PDF's also provide uncertainties which take into account experimental errors and their correlations. Since the LHC kinematic region is much broader than currently explored, we will have the unique opportunity to test QCD at very high and low-$x$, where predictions are extremely important for precise measurements and new physics searches at the LHC.

2. **FNAL (Fermi National Accelerator Laboratory)**

FNAL, also called FERMILAB, centre for particle-physics research is located at Batavia, Illions in USA named after the Italian-American physicist Enrico Fermi, who headed the team that first achieved a controlled nuclear reaction. The major components of Fermilab are two large particle accelerators called proton synchrotrons, configured in the form of a ring with a circumference of 6.3 km. The first, which went into operation in 1972, is capable of accelerating particles to 400 billion electron volts. The second, called the Tevatron, is installed below the first and incorporates more powerful superconducting magnets; it can accelerate particles to 1 trillion electron volts. The older instrument, operating at lower energy levels, now is used as an injector for the Tevatron. The high-energy beams of particles (notably muons and neutrinos) produced at the laboratory, have been used to study the structure of protons in terms of their most fundamental components, the quarks. In 1972 a team of scientists at Fermilab isolated the bottom quark and its associated antiquark. In 1977 a team led by Leon Lederman discovered the upsilon meson, which revealed the existence of the bottom quark and its accompanying antiquark. The existence of the top quark predicted by the standard model was established at Fermilab in March 1994.

3. **SLAC (Stanford Linear Accelerator Center)**

SLAC was established in 1962 at Stanford University in Menlo Park, California, USA. Its mission is to design, construct and operate electron accelerators and related experimental facilities for use in high-energy physics and synchrotron radiation research. It houses the longest linear accelerator (linac) in the world-a machine of 3.2 km long that accelerates electrons up to energies of 50 GeV. In 1966 a new machine, designed to reach 20 GeV was completed. In 1968 experiments at SLAC found the first direct evidence for further structure (i.e., quarks) inside protons and neutrons. In 1972, an electron-positron collider called SPEAR (Stanford Positron-Electron Asymmetric Rings) producing collisions at energies of 2.5 GeV per beam was constructed. In 1974 SPEAR was upgraded to reach 4.0 GeV per beam. A new type of quark, known as charm, and a new, heavy leptons relative of the electron, called the tau were discovered using SPEAR. SPEAR was followed by a larger, higher-energy colliding-beam machine, the PEP (Positron-Electron Project), which began operation in 1980 and took electron-positron collisions to a total energy of 36 GeV. The SLAC Linear Collider (SLC) was completed in 1987. SLC uses the original linac, upgraded to reach 50 GeV, to accelerate electrons and positrons before sending them in opposite directions around a 600-metre loop, where they collide at a total energy of 100 GeV. This is sufficient to produce the Z particle, the neutral carrier of the weak nuclear force that acts on fundamental particles.

4. **DESY (Deutsches Elektronen-Synchrotron)**

DESY, the largest centre for particle-physics research located in Hamburg, Germany was founded in 1959. The construction of an electron synchrotron to generate an energy level of 7.4 billion electron-volts was completed in 1964. Ten years later the Double Ring Storage Facility (DORIS) was completed which is capable of colliding beams of electrons and positrons at 3.5 GeV per beam. In 1978 its power was upgraded to 5 GeV per beam. DORIS is no longer used as a collider, but its electron beam provides synchrotron radiation (mainly at X-ray and ultraviolet wavelengths) for experiments on a variety of materials. A larger collider capable of reaching 19 GeV per beam, the Positron-Electron Tandem Ring Accelerator (PETRA), began operational in 1978. Experiments with PETRA in the following year gave the first direct evidence of the existence of gluons. The Hadron-Electron Ring Accelerator (HERA) capable of colliding electrons and protons was completed in 1992. HERA consists of two rings in a single tunnel with a circumference of 6.3 km, one ring accelerates electrons to 30 GeV and the other protons to 820 GeV. It is being used to continue the study of quarks.

5. **KEK (Koh-Ene-Ken)**

KEK is a NATIONAL LABORATORY FOR HIGH ENERGY PHYSICS located at Tsukuba, Ibaraki Prefecture, Japan. Both proton accelerators and electron/positron accelerators, including storage rings and colliders, are in operation in KEK to support various activities, ranging from particle physics to structure biology. High-intensity proton accelerators was also constructed in this laboratory in collaboration with Japan Atomic Energy Research Institute. KEK is associated with two research institutes, Institute of Particle and Nuclear Studies (IPNS) and Institute of Materials Structure Science (IMSS) and two laboratories, Accelerator Laboratory and Applied Research Laboratory. IPNS carries out research programs in particle physics and nuclear physics. IMSS offers three types of probes for research programs in material science. Its two major accelerators are the 12 GeV Proton Synchrotron and the KEKB electron-positron collider where the Belle experiment is currently running. The Belle collaboration at the KEKB factory was highlighted by its observation of the CP violation of B-mesons. The Applied Research Laboratory, which has four research centers (Radiation Science Center, Computing Research Center, Cryogenics Science Center and Mechanical Engineering Center), provide basic technical support for all KEK activities with their high-level technologies. KEK is also associated in the J-PARC proton accelerator under construction in Tokaimura.

6. **VECC (Variable Energy Cyclotron Centre)**

VECC is a research and development unit located in Kolkata, India. The variable energy cyclotron (VEC) set up is used for research in Accelerator Science & Technology, Nuclear Science (Theoretical and Experimental), Material Science, Computer Science & Technology and in other relevant areas. The Variable Energy Cyclotron (VEC) is the main accelerator, operational at the Centre since 1980. The Centre is also constructing Radioactive Ion Beam (RIB) accelerators – highly complex and sophisticated – for most modern nuclear physics and nuclear astrophysics experiments. High level scientific activity goes on at the Centre for International collaborations in the areas of high energy physics experiments at large accelerators in other parts of the world. The Centre has also developed frontline computational facilities to carry out research and development in the above mentioned areas. Exploration and recovery of helium gas from hot spring emanations and earthquake prediction utilizing related observations is another important area in which the Centre is actively engaged.

7. **BNL (Brookhaven National Laboratory)**

Brookhaven National Laboratory is located at Upton, New York. The setup of Relativistic Heavy Ion Collider (RHIC) is a heavy-ion collider used to collide ions at relativistic speeds. At present, RHIC is the most powerful heavy-ion collider in the world. The RHIC double storage ring is itself hexagonally shaped and its circumference is 3834 m with curved edges in which stored particles are deflected by 1,740 superconducting niobium titanium magnets. The six interaction points are at the middle of the six relatively straight sections, where the two rings cross, allowing the particles to collide. The interaction points are enumerated by clock positions, with the injection point at 6 o'clock. There are four detectors at RHIC: STAR (6 o'clock, and near the ATR), PHENIX (8 o'clock), PHOBOS (10 o'clock), and BRAHMS (2 o'clock). PHOBOS has the largest pseudorapidity coverage of all detectors, and tailored for bulk particle multiplicity measurement and it has completed its operation after 2005. BRAHMS is designed for momentum spectroscopy, in order to study low-x and saturation physics and it has completed its operation after 2006. STAR is aimed at the detection of hadrons with its system of time projection chambers covering a large solid angle and in a conventionally generated solenoidal magnetic field, while PHENIX is further specialized in detecting rare and electromagnetic particles, using a partial coverage detector system in a superconductively generated axial magnetic field. There is an additional experiment PP2PP, investigating spin dependence in p + p scattering.

Another collider eRHIC, also known as spin-dependent electron-hadron collider was designed based on the RHIC hadron rings and 10 to 20 GeV energy recovery electron linac. The designs of eRHIC, based on a high current super-conducting energy-recovery linac (ERL) with energy of electrons up to 20 GeV, have a number of specific requirements on the ERL optics. Two of the most attractive features of this scheme are full spin transparency of the ERL at all operational energies and the capability to support up to four interaction points. The main goal of the eRHIC is to explore the physics at low-x, and the physics of colour-glass condensate in electron-hadron collisions. □

# REGGE THEORY

Regge theory (also known as the S-matrix theory) since 1959 describes hadronic interactions starting with basic principles such as unitarity or analyticity where Regge introduced a theory of complex orbital momenta j that allows to constrain the energy dependence of high energy interactions.

## 2.1 S-matrix theory

Let in a typical scattering experiment the initial state is represented as $|i\rangle$ and after the interaction the final state is represented as $|f\rangle$. If 'S' is the scattering operator such that its matrix elements between the initial and final states, $\langle f|S|i\rangle$, gives the probability $P_{fi}$, that after the interaction the final state $|f\rangle$ comes from the initial state $|i\rangle$,

$$P_{fi} = \left|\langle f|S|i\rangle\right|^2,$$

then the scattering operator is known as the scattering matrix or S-matrix.

Postulates

The S-matrix theory starts with the basic assumptions,

1. Free particle states, containing any number of particles, satisfy the superposition principle [39], so that if $|A\rangle$ and $|B\rangle$ are different physical

states, $|C\rangle$ will be another physical state given by $|C\rangle = a|A\rangle + b|B\rangle$, where a and b are arbitrary complex numbers.

2. Strong interaction forces are of short range, i. e. we regard the particles as free and non-interacting except when they are very close together. So the asymptotic states, before and after an experiment, consists just of free particles, neglecting the long range forces.

3. S-matrix remains invariant under Lorentz transformation [39].

4. S-matrix is unitary [40].

5. Maximum analyticity of the first kind [41].

## Analyticity

The scattering amplitude A of S-matrix can be written as arbitrary functions of the four momenta of the particles involved in the scattering process and hence must be written as a function of Lorentz scalars. Thus A will be a Lorentz scalar. For the four-line process 1+2→3+4, the amplitude A ($P_1$, $P_2$; $P_3$, $P_4$) will be a function of Lorentz scalars such as $(P_1+P_2)^2$, $(P_3+P_4)^2$, $(P_1+P_2+P_3)^2$ etc. however not all these are independent quantities, since, for example $(P_1+P_2)^2 = (P_3+P_4)^2$ by four momentum conservation. For an n-line process in the 4-dimensional space, ultimately we are left with (3n-10) independent variables. So we denote these variables by the Lorentz invariants

$$S_{ijk\ldots} = (\pm P_i \pm P_j \pm P_k \pm \ldots\ldots\ldots)^2 .$$

The 5-th postulate of S-matrix: Maximum analyticity of the first kind is stated as: The scattering amplitudes are the real boundary values of analytic functions of the invariants $S_{ijk\ldots}$ which are regarded as complex variables with only such singularities as are demanded by the unitarity equations [41]. The most important type of singularity which can be identified in the unitarity equations is a simple pole which corresponds to the exchange of a physical particle. Another requirement for the S-matrix is that it should be TCP invariant, where T is the time reversal, C is the charge conjugation and P is the parity inversion.

Crossing

Crossing is an important result of the analyticity property which is a relation that implies between quite separate scattering processes.

As an example we can consider the 2→2 amplitude, i.e. the scattering process 1+2→3+4 (figure 2.1(a)). By crossing and TCP theorem all the six processes are

$$\left.\begin{array}{ll} 1+2 \rightarrow 3+4 \qquad \overline{3}+\overline{4} \rightarrow \overline{1}+\overline{2} & (s-\text{channel}), \\ 1+\overline{3} \rightarrow \overline{2}+4 \qquad 2+\overline{4} \rightarrow \overline{1}+3 & (t-\text{channel}), \\ 1+\overline{4} \rightarrow \overline{2}+3 \qquad 2+\overline{3} \rightarrow \overline{1}+4 & (u-\text{channel}), \end{array}\right\} \qquad (2.1)$$

where 1 and 2 are the incoming particles and 3 and 4 are the outgoing particles. $\overline{1},\overline{2},\overline{3}$ and $\overline{4}$ are the antiparticles of 1, 2, 3 and 4 respectively. The channels are named after their respective energy invariants. These processes will share the same scattering amplitude, but the pairs of channels s, t and u will occupy different regions of the variables [41]. As we know that the four line amplitude depends only on two independent variables (3×4-10=2), so there must be a relation between s, t and u. The relation can be found as

$$s+t+u = \Sigma = m_1^2 + m_2^2 + m_3^2 + m_4^2,$$

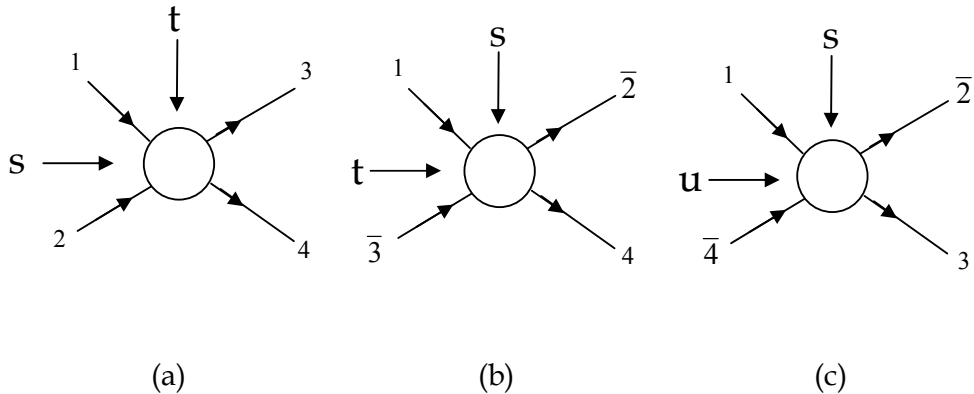

(a) (b) (c)

**Figure 2.1**: The scattering processes in the s, t and u channels

where $m_1, m_2, m_3$ and $m_4$ are the masses of the free particles and $\Sigma$ represents the sum of the squares of these masses. s and t are regarded as the independent variables. The physical region for the s-channel is given by

$$s \geq \max\left\{(m_1+m_2)^2, (m_3+m_4)^2\right\}$$

i.e. the threshold for the process and $-1 \leq \cos\theta_S \leq 1$, where $\theta_s$ is the scattering angle between the directions of motion of particles 1 and 3 in the s-channel centre of mass system. The various singularities may also be plotted on the Mandelstam diagram [41].

## 2.2 The complex angular momentum plane

The idea which Regge [42, 43] introduced into the scattering theory was the importance of analytically continuing scattering amplitudes in the complex angular momentum plane.

### Partial-wave amplitude

Throughout this discussion we will consider the 2→2 scattering amplitude and spinless particles, so that the total angular momentum of the initial state is just the relative orbital angular momentum of the two particles. Since the angular momentum is a conserved quantity, the orbital angular momentum of the final state must be the same as that of the initial state, so it is frequently convenient to consider the scattering amplitude for each individual angular momentum separately, i.e. the so-called 'Partial-wave amplitudes'. However, the initial state will not in general be an eigenstate of angular momentum, but a sum over many possible angular momentum eigenstates and hence the total scattering amplitude will be a sum over all these partial-wave amplitudes.

For spinless particles the angular dependence of the wave function describing a state of orbital angular momentum l in the s-channel is given by the Legendre polynomial of the first kind $P_l(Z_S)$, where $Z_S = \cos\theta_S$. $\cos\theta_S$ can be shown to be a function of t, s and u. At fixed s, the scattering angle is just given by t (or u), so $t = t(Z_s, s)$. The centre-of mass partial-wave scattering amplitude of angular momentum l in the s-channel is defined from the total scattering amplitude by

$$A_l(s) = \frac{1}{16\pi} \frac{1}{2} \int_{-1}^{1} dz_s P_l(Z_s) A(s, t(Z_s, s)), \qquad (2.2)$$

where l=0, 1, 2……….. and the factor $1/(16\pi)$ is purely a matter of convention in order to simplify the unitary equations. We can convert equation (2.2) to give its inverse as

$$A(s, t) = 16\pi \sum_{l=0}^{\infty} (2l+1) \, A_l(s) \, P_l(Z_s), \qquad (2.3)$$

which is called the partial-wave series for the total scattering amplitude A(s, t). We can obviously make an exactly similar partial-wave decomposition in the t-channel, defining

$$A_l(t) = \frac{1}{16\pi} \frac{1}{2} \int_{-1}^{1} dZ_t \, P_l(Z_t) \, A(s(Z_t, t), t), \qquad (2.4)$$

with inverse

$$A(s, t) = 16\pi \sum_{l=0}^{\infty} (2l+1) \, A_l(t) \, P_l(Z_t), \qquad (2.5)$$

Equation (2.5) provides a representation of the scattering amplitude which is satisfactory throughout the t-channel physical region. Since $A_l(t)$ contains the t-channel thresholds and resonance poles, the amplitude obtained from equation (2.5) has all the t singularities. But its s-dependence is completely contained in the Legendre polynomials which are entire functions of $Z_t$ and hence of s at fixed t. It is evident that this representation must break down if we continue it beyond the t-channel physical region ($-1 \leq Z_t \leq 1$) to the nearest singularity in s (or u) at $s=s_0$ say, where the series will diverge. For example the pole

$$\left(m^2 - s\right)^{-1} = m^{-2}\left(1 + \frac{s}{m^2} + \left(\frac{s}{m^2}\right)^2 + \ldots\ldots\ldots\ldots\right)$$ can be represented as a polynomial in s which diverges at $s=m^2$.

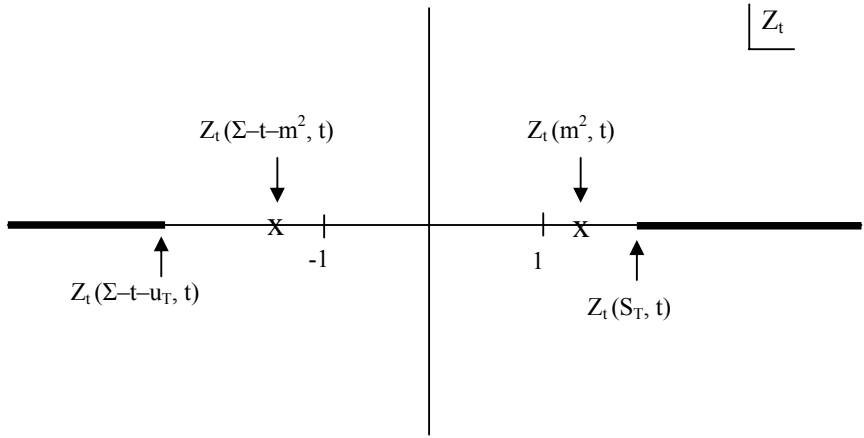

**Figure 2.2:** The singularities in $Z_t$ at fixed t ($>t_T$). Outside the physical region ($-1 \leq Z_t \leq 1$) these are the s-channel poles and threshold branch point for $Z_t>1$ and the u-channel singularities for $Z_t<-1$.

In figure 2.2 we have plotted the nearest s and u-channel poles and branch points in terms of the variable $Z_t$. They always occur outside the physical region of the t-channel but it is clear that the use of equation (2.5) is restricted to only a small region of the plot beyond physical region. Here $t_T$, $s_T$ and $u_T$ are the t, s and u-channel thresholds.

To obtain an expression for the partial-wave amplitudes which incorporates the s and u singularities and hence valid over the whole Mandelstam plane, the dispersion relation used is

$$A(s,t,u) = \frac{g_s(t)}{m^2 - s} + \frac{g_u(t)}{m^2 - u} + \frac{1}{\pi}\int_{s_T}^{\infty}\frac{D_s(s',t)}{s' - s}dS' + \frac{1}{\pi}\int_{u_T}^{\infty}\frac{D_u(u',t)}{u' - u}du',$$

where $g_s$ and $g_u$ are some functions of t, $D_s$ and $D_u$ are discontinuity functions. After some rigorous calculations we get the Froissart-Gribov projection [44, 45] as

$$A_l(t) = \frac{1}{16\pi}\frac{g_s(t)}{2q_{t13}q_{t24}}Q_l\big(Z_t(m^2,t)\big) + \frac{1}{16\pi}\frac{g_u(t)}{2q_{t13}q_{t24}}Q_l\big(Z_t(\Sigma - t - m^2,t)\big)$$

$$+ \frac{1}{16\pi^2}\int_{Z_t(s_T,t)}^{\infty} D_s(s',t)Q_l(Z_t')dZ_t' + \frac{1}{16\pi^2}\int_{Z_t(u_T,t)}^{\infty} D_u(u',t)Q_l(Z_t')dZ_t', \qquad (2.6)$$

where $q_{t13}$ and $q_{t24}$ are the three momentum, equal but opposite for the two particles and $Q_l$ is the Legendre polynomial of second kind. This Froissart-Gribov projection is completely equivalent to equation (2.4) provided the dispersion relation is valid. However equations (2.4) and (2.6) involve completely different regions of $Z_t$ and hence s. Since equation (2.4) requires integration only over a finite region, the partial-wave amplitudes can always be so defined, at least in the t-channel physical region, but equation (2.6) involves an infinite integration and can be used only if the integration converges.

Froissart bound

Froissart showed that, for amplitudes which satisfy the Mandelstam representation, s-channel unitarity limits the asymptotic behaviour of the scattering amplitude in the s-channel physical region, $t \leq 0$. Since Legendre polynomial of second kind we have,

$$Q_l(z) \underset{l\to\infty}{\approx} l^{-\frac{1}{2}} e^{-\left(l+\frac{1}{2}\right)\xi(Z)},$$

where $\xi(Z) \equiv \log\{Z + \sqrt{(Z^2 - 1)}\}$, the Froissart-Gribov projection (equation (2.6)) for S-channel partial waves gives

$$A_l(s) \to f(s)\, e^{-l\xi(Z_0)},$$
$$l, s \to \infty$$

where $Z_0$ is the lowest t-singularity of A(s, t) and f(s) is some function of s. This means that all the partial waves with $l \gg l_M \equiv \xi^{-1}(Z_0)$ will be very small. $l_M$ is some maximum value of l and the range of the force can be defined as $R\,q_s \equiv l_M$ and particle passing the target at impact parameter b>R effectively miss the target and are not scattered much. After some mathematics, equation (2.3) may be truncated as

$$A(s,t) \approx 16\pi \sum_{l=0}^{C(\sqrt{s})\log s}(2l+1)A_l(s)\,P_l(Z_s). \tag{2.7}$$

Then using the bound conditions

$0 \le |A_l^{ii}|^2 \le \mathrm{Im}\{A^{ii}\} \le 1$ and $|P_l(Z)| \le 1$ for $-1 \le Z \le 1$, we get the scattering amplitude as

$$|A(s,t)| \le 16\pi \sum_{l=0}^{C(\sqrt{s})\log s}(2l+1) \le C.s.\log^2 s, \qquad \text{for } s \to \infty,\ t \le 0. \tag{2.8}$$

where C is a constant. Using optical theorem (Appendix A), the total scattering cross-section take the form,

$$\sigma^{tot}(s) \le C\log^2 s, \qquad s \to \infty \tag{2.9}$$

Which is called the Froissart bound [46].

Analytic continuation in angular momentum

In the t-channel physical region we can obtain the signatured partial-wave amplitude as

$$A_l^{\tilde{\sigma}} = \frac{1}{16\pi^2}\int_{Z_t}^{\infty} D_s^{\tilde{\sigma}}(s,t)Q_l(Z_t)dZ_t. \tag{2.10}$$

Which is the Froissart-Gribov projection and it may be used to define $A_l^{\tilde{3}}$ for all values of l, not necessarily integer or even real. In fact it can be used for all l values such that Re{l}>N(t), where $D_s^{\tilde{3}} \sim Z^{N(t)}$ and where $N(t) \leq 1$ for $t \leq 0$. The main advantage of using equation (2.10) rather than equation (2.4) for l≠integer is that $Q_l$ has a better behaviour than $P_l$ as l→∞. The only singularities of $Q_l(Z)$ are poles at l=-1, -2, -3………. So equation (2.10) defines a function of l which is holomorphic for Re{l}>max(N(t), -1). It is not immediately apparent that there is much merit to this extended definition of the partial-wave amplitudes, because of course it is only positive integer values of l that have physical significance, and there is clearly an infinite number of different ways of interpolating between the integers. However $A_l^{\tilde{3}}$ defined by equation (2.10) vanishes as $|l| \to \infty$ and a theorem due to Carlson (Appendix B) [47] tells us that equation (2.10) must be the unique continuation with this property. Hence equation (2.10) defines $A_l^{\tilde{3}}(t)$ uniquely as a holomorphic function of l with convergent behaviour as $|l| \to \infty$, for all Re{l}>N(t). However we are prevented from continuing below Re{l}>N(t) by the divergent behaviour of $D_s^{\tilde{3}}(s, t)$ as $s \to \infty$.

In this point another crucial assumption of S-matrix has to be made: the scattering amplitude A is an analytic function of orbital angular momentum l throughout the complex angular momentum plane, with only isolated singularities. It will be just these isolated singularities which cause the divergence problems, and we can easily continue past them. For example suppose that $D_l^{\tilde{3}}(s,t)$ has a leading asymptotic power behaviour $D_l^{\tilde{3}}(s,t) \approx s^{\alpha(t)} +$ lower order terms, so N(t)=α(t). Applying properties of Legendre polynomial of second kind, the large s region of equation (2.10) (s>$s_1$ say) gives

$$A_l^{\Im} \approx \int_{S_1}^{\infty} s^{\alpha(t)} s^{-l-1} ds = -\frac{e^{(\alpha(t)-l)\log s_1}}{\alpha(t)-l}.$$
$$l > \alpha(t)$$

Hence $A_l(t)$ has a pole at $l=\alpha(t)$. This is, by hypothesis, the rightmost singularity in the complex angular momentum plane and is this singularity which is preventing continuation to the left of $\text{Re}\{l\}=\alpha(t)$. However, once we have isolated this pole we can continue round it to the left, until we reach the singularity due to the next term in the asymptotic expansion of $D_l^{\Im}(s,t)$.

There may be logarithmic terms like

$$D_l^{\Im}(s,t) \approx s^{\alpha(t)} (\log s)^{\beta(t)},$$

giving

$$A_l^{\Im} \approx \int_{S_1}^{\infty} s^{\alpha(t)} (\log s)^{\beta(t)} s^{-l-1} ds = -\frac{1}{(\alpha(t)-l)^{1+\beta(t)}} + \ldots\ldots\ldots, \quad \beta(t) \neq -1$$
$$l > \alpha(t)$$
$$= \log(\alpha(t)-l), \qquad \beta(t) = -1,$$

so $A_l^{\Im}(t)$ has a branch point at $l=\alpha(t)$, or a multiple pole if $\beta$ is a positive integer. The assumption that $A_l^{\Im}(t)$ has only isolated singularities in $l$, and so can be analytically continued throughout the complex angular momentum plane, is sometimes called the postulate of 'maximal analyticity of the second kind'. It is the basic assumption upon which the applicability of Regge theory to particle physics rests.

It is known that two-body scattering of hadrons is strongly dominated by small momentum transfer or equivalently by small scattering angles. According to Regge theory this scattering amplitude is successfully described by the exchange of a particle with appropriate quantum numbers and these are known as Regge poles. The Regge poles, like elementary particles, are characterized by quantum numbers like charge, isospin, strangeness, etc. Regge pole exchange is a generalization of a single particle exchange (Figure 2.3). There are two types of regge poles: 1. Reggeon and 2. Pomeron.

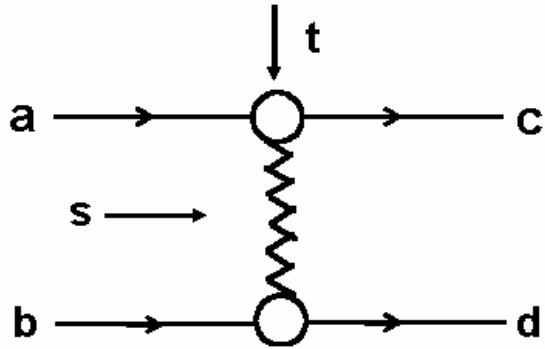

**Figure 2.3**: Regge pole exchange

A very simple expression for the behaviour of scattering amplitude A(s, t) is predicted by Regge after rigorous theoretical work [41] and which is given as

$$A(s,t) \approx s^{\alpha(t)}, \text{ for large s.}$$

The natural quantities to consider are the structure functions which are proportional to the total virtual photon-nucleon cross section and which are expected to have Regge behaviour corresponding to pomeron or reggeon exchange [26]. So the hadronic cross sections as well as structure functions will be dominated by two contributions: i) a pomeron, reproducing the rise of $F_2$, say, at low-x and ii) reggeons associated with meson trajectories.

It is useful to represent Regge pole exchange in terms of quarks and gluons (Figure 2.4).

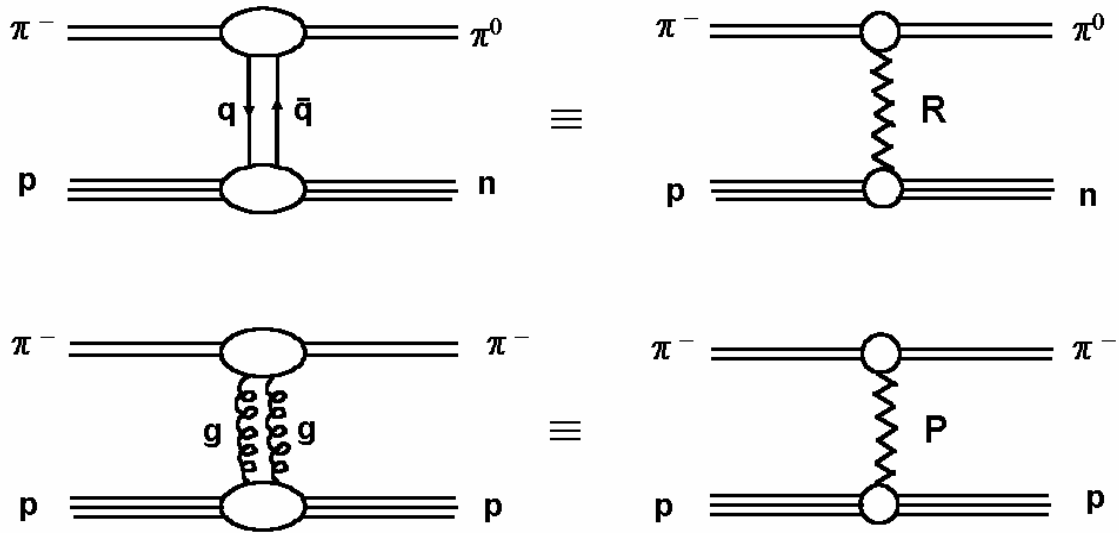

**Figure 2.4**: Reggeon and pomeron exchange.

## 2.3 Regge theory in DIS

Many models based on Regge theory are able to reproduce hadronic cross-sections. For discussion we will consider extension of some of the simplest models to the γ*p amplitudes and we will see how Regge theory can be used to describe structure functions. In DIS, Regge theory constrains the S behaviour but does not say anything about the $Q^2$ dependence. The Regge couplings are therefore functions of $Q^2$. DIS corresponds to the region where both ν and $Q^2$ are large. The low-x limit of DIS corresponds to the case when 2Mν>>$Q^2$, where x=$Q^2$/2Mν, yet $Q^2$ is still large. The limit 2Mν>>$Q^2$ is equivalent to S>>$Q^2$. The high energy limit, when the scattering energy is kept much greater than the external masses, is, by definition, the Regge limit. In DIS, $Q^2$ is, by definition, also kept large i.e. $Q^2$>>$\Lambda^2$. The limit of large ν and 2Mν>>$Q^2$ is therefore the Regge limit of DIS [26]. The fact that $Q^2$ is large allows to use PQCD and Regge theory is strictly applicable in the region of large s, i.e. in the region of low-x [27, 41].

## The pomeron term

For the pomeron contribution to $F_2$, we will give three different simple possibilities [27, 28]:

1. <u>A power behaviour</u>:
$$F_2(x,Q^2) = a(Q^2) x^{-\varepsilon},$$

where $a(Q^2)$ is a function of $Q^2$ and the exponent $\varepsilon$ is called intercept. This term, with $\varepsilon \approx 0.09$ is called the soft pomeron but is unable to describe the steeper rise of $\gamma^*p$ amplitudes. The solution is to add another contribution, called the hard pomeron, which leads to

$$F_2(x,Q^2) = a_s(Q^2) x^{-\varepsilon_s} + a_h(Q^2) x^{-\varepsilon_h}. \tag{2.11}$$

where $a_s(Q^2)$ and $a_h(Q^2)$ are functions of $Q^2$ and the exponents $\varepsilon_s$ and $\varepsilon_h$ are the intercepts for the soft and hard parts contributions to the structure function respectively. The hard pomeron has $\varepsilon_h \approx 0.4$. In the complex angular momentum plane, i.e. complex-j plane, this corresponds to two simple poles at $j = 1 + \varepsilon_h \approx 1.4$ and $j = 1 + \varepsilon_s \approx 1.1$:

$$F_2(j,Q^2) = \frac{a_s(Q^2)}{j-1-\varepsilon_s} + \frac{a_h(Q^2)}{j-1-\varepsilon_h}.$$

This is the Donnachie-Landshoff two-pomerons model.

2. <u>A logarithmic behaviour</u>:
$$F_2(\nu,Q^2) = A(Q^2) \log(2\nu) + B(Q^2),$$

where $A(Q^2)$ and $B(Q^2)$ are functions of $Q^2$. Here DIS variable $\nu$ is used instead of x. In the complex-j plane, this expression becomes

$$F_2(j,Q^2) = \frac{A(Q^2)}{(j-1)^2} + \frac{B(Q^2)}{j-1}.$$

And this behaviour is often called the double pole pomeron.

3. A squared-logarithmic behaviour:

$$F_2(\nu, Q^2) = A(Q^2)\log^2(2\nu) + B(Q^2)\log(2\nu) + C(Q^2)$$

$$= A(Q^2)\log^2\left[\frac{2\nu}{2\nu_0 Q^2}\right] + C(Q^2)'$$

where $C(Q^2)$ is also a function of $Q^2$. Here also DIS variable $\nu$ is used instead of x. In the complex-j plane, this expression becomes

$$F_2(j, Q^2) = \frac{2A(Q^2)}{(j-1)^3} + \frac{B(Q^2)}{(j-1)^2} + \frac{C(Q^2)}{j-1}.$$

And this behaviour is often called the triple pole pomeron.

From the above discussion, we found that one of the applications of the Regge behaviour is the Donnachie-Landshoff two-pomeron model where the rise of structure function is described by powers of $1/x$, which is given in equation(2.11), where the poles are given by $j = 1 + \varepsilon_h \approx 1.4$ and $j = 1 + \varepsilon_s \approx 1.1$. Now, in order to apply Regge theory to DGLAP evolution equations, let us take the functions of $Q^2$ to be the same as $T(Q^2)$. i.e. $A_h(Q^2) = A_S(Q^2) = T(Q^2)$. The contributions of the Regge poles solely determine the high energy behaviour of all QCD amplitudes in the multi-Regge kinematics given by namely Fadin, Fiore, Kozlov and Reznichenko [48].

So we can assume a simple form for Regge behaviour of spin-independent structure function to solve DGLAP evolution equation, as [49-56]

$$F_2(x, t) = T(t)x^{-\lambda}, \tag{2.12}$$

where $T(t)$ is a function of t and $\lambda$ is the Regge intercept for spin-independent structure function. This form of Regge behaviour is well supported by the work in this field carried out by namely Badelek [57], Soffer and Teryaev [58] and also Desgrolard, Lengyel and Martynov [59]. According to Regge theory, the high energy i. e. low-x behaviour of both gluons and sea quarks are controlled by the same singularity factor in the complex angular momentum plane [41]. And as the values of Regge intercepts for all the spin-independent singlet, non-singlet and gluon structure functions should be close to 0.5 in quite a broad range of low-x [49], we would also expect that our theoretical curves are best fitted to those of the experimental data and parameterization curves at $\lambda_S = \lambda_{NS} = \lambda_G \approx 0.5$, where at $\lambda_S$, $\lambda_{NS}$ and $\lambda_G$ are the Regge intercepts for singlet, non-singlet and gluon structure functions respectively.

The low-x behaviour of spin-dependent structure functions for fixed-$Q^2$ is the Regge limit of the polarized DIS, where the Regge pole exchange model should be applicable [60]. The Regge behaviour for polarized singlet, non-singlet and gluon structure functions has the general form $A_i(x,t) = T_i(t) x^{-\beta_i}$ [18, 41, 60], where $A_i(x, t)$ are the structure functions and $\beta_i$ are the respective Regge intercepts of the trajectory. Let us take $\beta_i$'s as $\beta_S$, $\beta_{NS}$ and $\beta_G$ for the spin-dependent singlet, non-singlet and gluon structure functions respectively. So we are in a state to solve the spin-dependent DGLAP evolution equations with this form of Regge behaviour.

# Part I

# Spin-independent DGLAP evolution equations at low-x

**Chapter 3**

# t and x- Evolutions of Spin-independent DGLAP Evolution Equations in Leading Order

Here in this chapter we have solved the spin-independent DGLAP evolution equations for singlet, non-singlet and gluon structure functions at low-x in leading order (LO) considering Regge behaviour of structure functions and also the coupled equations for singlet and gluon structure functions. The t and x-evolutions of deuteron, proton and gluon structure functions thus obtained have been compared with NMC and E665 collaborations data sets and global MRST 2001, MRST 2004 and GRV1998LO gluon parameterizations respectively.

## 3.1 Theory

The spin-independent DGLAP evolution equations for singlet, non-singlet and gluon structure functions in LO are given as [61, 62]

$$\frac{\partial F_2^S(x,t)}{\partial t} - \frac{\alpha_S(t)}{2\pi} I_1^S(x,t) = 0, \tag{3.1}$$

$$\frac{\partial F_2^{NS}(x,t)}{\partial t} - \frac{\alpha_S(t)}{2\pi} I_1^{NS}(x,t) = 0 \tag{3.2}$$

and

$$\frac{\partial G(x,t)}{\partial t} - \frac{\alpha_S(t)}{2\pi} I_1^G(x,t) = 0, \tag{3.3}$$

where

$$I_1^S(x,t) = \left[\frac{2}{3}\{3+4\ln(1-x)\} F_2^S(x,t)\right] + \frac{4}{3}\int_x^1 \frac{d\omega}{1-\omega}\left[(1+\omega^2) F_2^S\left(\frac{x}{\omega},t\right) - 2F_2^S(x,t)\right]$$

$$+ N_f \int_x^1 \{\omega^2 + (1-\omega)^2\} G\left(\frac{x}{\omega},t\right) d\omega,$$

$$I_1^{NS}(x,t) = \left[\frac{2}{3}\{3+4\ln(1-x)\} F_2^{NS}(x,t)\right] + \frac{4}{3}\int_x^1 \frac{d\omega}{1-\omega}\left[(1+\omega^2) F_2^{NS}\left(\frac{x}{\omega},t\right) - 2 F_2^{NS}(x,t)\right],$$

$$I_1^G(x,t) = \left\{6 \times \left(\frac{11}{12} - \frac{N_f}{18} + \ln(1-x)\right) G(x,t) + 6 \times I_g\right\}$$

and

$$I_g = \int_x^1 d\omega \left[\frac{\omega G\left(\frac{x}{\omega},t\right) - G(x,t)}{1-\omega} + \left(\omega(1-\omega) + \frac{1-\omega}{\omega}\right) G\left(\frac{x}{\omega},t\right) + \frac{2}{9}\left(\frac{1+(1-\omega)^2}{\omega}\right) F_2^s\left(\frac{x}{\omega},t\right)\right].$$

The strong coupling constant, $\alpha_S(Q^2)$ is related with the β-function as [63]

$$\beta(\alpha_S) = \frac{\partial \alpha_S(Q^2)}{\partial \log Q^2} = -\frac{\beta_0}{4\pi}\alpha_s^2 - \frac{\beta_1}{16\pi^2}\alpha_s^3 - \frac{\beta_2}{64\pi^3}\alpha_S^4 + \cdots,$$

where

$$\beta_0 = \frac{11}{3}N_C - \frac{4}{3}T_R = 11 - \frac{2}{3}N_f,$$

$$\beta_1 = \frac{34}{3}N_C^2 - \frac{10}{3}N_C N_f - 2C_F N_f = 102 - \frac{38}{3}N_f$$

and

$$\beta_2 = \frac{2857}{54}N_C^3 + 2C_F^2 T_R - \frac{205}{9}C_F N_C T_R + \frac{44}{9}C_F T_R^2 + \frac{158}{27}N_C T_R^2 = \frac{2857}{6} - \frac{6673}{18}N_f + \frac{325}{54}N_f^2$$

are the one loop, two loop and three loop corrections to the QCD β- function and $N_f$ being the number of flavour. $C_A$, $C_G$, $C_F$, and $T_R$ are constants associated with the colour SU(3) group where $C_A = C_G = N_C = 3$ and $T_R = 1/2$. $N_C$ is the number of colours. $C_F(\omega) = \dfrac{N_C^2 - 1}{2N_C} = \dfrac{4}{3}$. Running coupling constant in LO is $\alpha_S(t) = \dfrac{4\pi}{\beta_0 t}$.

Deuteron, proton and neutron spin-independent structure functions in terms of singlet and non-singlet spin-independent structure functions [62] can be written as

$$F_2^d(x,t) = \dfrac{5}{9} F_2^S(x,t), \tag{3.4}$$

$$F_2^p(x,t) = \dfrac{3}{18} F_2^{NS}(x,t) + \dfrac{5}{18} F_2^S(x,t) \tag{3.5}$$

and

$$F_2^n(x,t) = \dfrac{5}{18} F_2^S(x,t) - \dfrac{3}{18} F_2^{NS}(x,t). \tag{3.6}$$

Now let us consider the Regge behaviour of singlet, non-singlet and gluon structure functions [49-54] as discussed in Chapter 2:

$$F_2^S(x,t) = T_1(t) x^{-\lambda_S}, \tag{3.7}$$

$$F_2^{NS}(x,t) = T_2(t) x^{-\lambda_{NS}} \tag{3.8}$$

and

$$G(x,t) = T_3(t) x^{-\lambda_G}. \tag{3.9}$$

Therefore,

$$F_2^S\left(\dfrac{x}{\omega}, t\right) = T_1(t)\, \omega^{\lambda_S} x^{-\lambda_S} = F_2^S(x,t)\, \omega^{\lambda_S}, \tag{3.10}$$

$$F_2^{NS}\left(\dfrac{x}{\omega}, t\right) = T_2(t)\, \omega^{\lambda_{NS}} x^{-\lambda_{NS}} = F_2^{NS}(x,t)\, \omega^{\lambda_{NS}} \tag{3.11}$$

and

$$G\left(\dfrac{x}{\omega}, t\right) = T_3(t)\, \omega^{\lambda_G} x^{-\lambda_G} = G(x,t)\, \omega^{\lambda_G}, \tag{3.12}$$

where $T_1(t)$, $T_2(t)$ and $T_3(t)$ are functions of t, and $\lambda_S$, $\lambda_{NS}$ and $\lambda_G$ are the Regge intercepts for singlet, non-singlet and gluon structure functions respectively.

The DGLAP evolution equations of singlet, non-singlet and gluon structure functions are in the same forms of derivative with respect to t and also the input singlet and gluon parameterizations, taken for the global analysis to incorporate different high precision data, are also functions of x at fixed-$Q^2$. So the relation between singlet and gluon structure functions will come out in terms of x at fixed-$Q^2$, so, we can consider the ansatz [51, 52, 64-66]

$$G(x, t) = K(x) F_2^S(x, t) \qquad (3.13)$$

for simplicity, where $K(x)$ is a parameter to be determined from phenomenological analysis and we assume $K(x) = K$, $ax^b$ or $ce^{dx}$ where K, a, b, c and d are constants. Though we have assumed some simple standard functional forms of $K(x)$, yet we can not rule out the other possibilities.

Among the various methods to solve these equations, Taylor expansion [67] is the simple one to transform the integro-differential equations into partial differential equations and thus to solve them by standard methods [64, 65]. But when we consider Regge behaviour of structure functions, the use of Taylor expansion becomes limited. In this method, we introduce the variable $u = 1-\omega$ and we get

$$\frac{x}{\omega} = \frac{x}{1-u} = x \sum_{k=0}^{\infty} u^k.$$

Since $0 < u < (1-x)$, where $|u| < 1$, $\frac{x}{1-u} = x \sum_{k=0}^{\infty} u^k$ is convergent. Applying the Taylor expansion for the singlet structure function in equation (3.1), we get

$$F_2^S\left(\frac{x}{\omega}, t\right) = F_2^S\left(\frac{x}{1-u}, t\right) = F_2^S\left(x + x \sum_{k=1}^{\infty} u^k, t\right)$$

$$= F_2^S(x, t) + x \sum_{k=1}^{\infty} u^k \frac{\partial F_2^S(x, t)}{\partial x} + \frac{1}{2} x^2 \left(\sum_{k=1}^{\infty} u^k\right)^2 \frac{\partial^2 F_2^S(x, t)}{\partial x^2} + \ldots \ldots . \qquad (3.14)$$

When we apply Regge behaviour of structure functions, say, singlet structure function then

$$\frac{\partial F_2^S(x,t)}{\partial x} = (-1)\lambda_S x^{-1} F_2^S(x,t), \quad \frac{\partial^2 F_2^S(x,t)}{\partial x^2} = (-1)^2 \lambda_S(\lambda_S+1) x^{-2} F_2^S(x,t),$$

$$\frac{\partial^3 F_2^S(x,t)}{\partial x^3} = (-1)^3 \lambda_S(\lambda_S+1)(\lambda_S+2) x^{-3} F_2^S(x,t) \text{ and so on. So equation (3.14)}$$

becomes

$$F_2^S\left(\frac{x}{\omega},t\right) = F_2^S(x,t) + \left(\sum_{k=1}^{\infty} u^k\right)(-1)\lambda_S F_2^S(x,t) + \frac{1}{2}\left(\sum_{k=1}^{\infty} u^k\right)^2 (-1)^2 \lambda_S(\lambda_S+1) F_2^S(x,t) + \ldots\ldots .$$

So in the expansion series, we will get terms with alternate positive and negative signs and contribution from $\lambda_S$ to each term increases. So in this case, it is not possible to truncate this infinite series into finite number of terms by applying boundary condition such as low-x [68] and also this is not a convergent series [67]. So, in solving DGLAP evolution equations we can not apply Regge behaviour of singlet structure function and Taylor series expansion method simultaneously. Same is the case for non-singlet and gluon structure functions also.

Putting equations (3.7), (3.10) and (3.13) in equation (3.1) we arrive at

$$\frac{\partial F_2^S(x,t)}{\partial t} - \frac{F_2^S(x,t)}{t} H_1(x) = 0, \tag{3.15}$$

where

$$H_1(x) = A_f \left[ \{3 + 4\ln(1-x)\} + 2\int_x^1 \frac{d\omega}{1-\omega} \{(1+\omega^2)\omega^{\lambda_S} - 2\} + \frac{3}{2} N_f \int_x^1 \{\omega^2 + (1-\omega)^2\} K\left(\frac{x}{\omega}\right) \omega^{\lambda_S} d\omega \right].$$

Integrating equation (3.15) we get

$$F_2^S(x,t) = C t^{H_1(x)}, \tag{3.16}$$

where C is a constant of integration and $A_f = 4/(33-2N_f)$. At $t = t_0$, equation (3.16) gives

$$F_2^S(x,t_0) = C t_0^{H_1(x)} . \tag{3.17}$$

From equations (3.16) and (3.17) we get

$$F_2^S(x,t) = F_2^S(x,t_0)\left(\frac{t}{t_0}\right)^{H_1(x)}, \tag{3.18}$$

which gives the t-evolution of singlet structure function in LO. Again at $x = x_0$, equation (3.16) gives

$$F_2^S(x_0,t) = C t^{H_1(x_0)}. \tag{3.19}$$

From equations (3.16) and (3.19), we get

$$F_2^S(x,t) = F_2^S(x_0,t) t^{\{H_1(x) - H_1(x_0)\}}, \tag{3.20}$$

which gives the x-evolution of singlet structure function at LO.

Similarly, from the equations (3.2) and (3.3) we get the solutions of spin-independent DGLAP evolution equations for non-singlet and gluon structure functions in LO at low-x respectively

$$F_2^{NS}(x,t) = C t^{H_2(x)} \tag{3.21}$$

and

$$G(x,t) = C t^{H_3(x)}, \tag{3.22}$$

where

$$H_2(x) = A_f \left[ \{3 + 4\ln(1-x)\} + 2\int_x^1 \frac{d\omega}{1-\omega} \{(1+\omega^2)\omega^{\lambda_{NS}} - 2\} \right]$$

And

$$H_3(x) = 9A_f \left[ \left(\frac{11}{12} - \frac{N_f}{18} + \ln(1-x)\right) + \int_x^1 d\omega \left\{ \frac{(\omega^{\lambda_G+1} - 1)}{1-\omega} + \left(\omega(1-\omega) + \frac{1-\omega}{\omega}\right)\omega^{\lambda_G} \right.\right.$$

$$\left.\left. + \frac{2}{9}\left(\frac{1+(1-\omega)^2}{\omega}\right)\frac{\omega^{\lambda_G}}{K\left(\frac{x}{\omega}\right)} \right\} \right].$$

The t and x-evolutions of spin-independent non-singlet structure function from equation (3.21) are given as

$$F_2^{NS}(x,t) = F_2^{NS}(x,t_0)\left(\frac{t}{t_0}\right)^{H_2(x)} \quad (3.23)$$

and

$$F_2^{NS}(x,t) = F_2^{NS}(x_0,t) t^{\{H_2(x)-H_2(x_0)\}}. \quad (3.24)$$

The t and x-evolutions of deuteron and proton structure functions from equations (3.18), (3.20), (3.23), and (3.24) are respectively

$$F_2^d(x,t) = F_2^d(x,t_0)\left(\frac{t}{t_0}\right)^{H_1(x)}, \quad (3.25)$$

$$F_2^d(x,t) = F_2^d(x_0,t) t^{\{H_1(x)-H_1(x_0)\}}, \quad (3.26)$$

$$F_2^p(x,t) = F_2^p(x,t_0)\left(\frac{3t^{H_2(x)}+5t^{H_1(x)}}{3t_0^{H_2(x)}+5t_0^{H_1(x)}}\right) \quad (3.27)$$

and

$$F_2^p(x,t) = F_2^p(x_0,t)\left(\frac{3t^{H_2(x)}+5t^{H_1(x)}}{3t^{H_2(x_0)}+5t^{H_1(x_0)}}\right). \quad (3.28)$$

The t and x-evolutions of spin-independent gluon structure functions function from equation (3.22) are given as

$$G(x,t) = G(x,t_0)\left(\frac{t}{t_0}\right)^{H_3(x)} \quad (3.29)$$

and

$$G(x,t) = G(x_0,t) t^{\{H_3(x)-H_3(x_0)\}}. \quad (3.30)$$

Now ignoring the quark contribution to the gluon structure function we get from the DGLAP evolution equation (3.3)

$$\frac{\partial G(x,t)}{\partial t} - \frac{A_f}{t}\left\{\left(\frac{11}{12}-\frac{N_f}{18}+\ln(1-x)\right)G(x,t) + I_g'\right\} = 0, \quad (3.31)$$

where

$$I_g' = \int_x^1 d\omega \left[\frac{\omega G\left(\frac{x}{\omega},t\right)-G(x,t)}{1-\omega} + \left(\omega(1-\omega)+\frac{1-\omega}{\omega}\right)G\left(\frac{x}{\omega},t\right)\right].$$

By the same procedure as above, we get the t and x-evolution equations for the gluon structure function ignoring the quark contribution in LO at low-x as

$$G(x,t) = G(x,t_0)\left(\frac{t}{t_0}\right)^{B_1(x)} \tag{3.32}$$

and

$$G(x,t) = G(x_0,t)\, t^{\{B_1(x)-B_1(x_0)\}}, \tag{3.33}$$

where

$$B_1(x) = A_f\left[(A+\ln(1-x)) + \int_x^1 d\omega\left\{\frac{(\omega^{\lambda_G+1}-1)}{1-\omega} + \left(\omega(1-\omega) + \frac{1-\omega}{\omega}\right)\omega^{\lambda_G}\right\}\right].$$

The assumption of the ad hoc function $K(x)$ can be overcome if we solve the coupled DGLAP evolution equations for singlet and gluon structure functions. Following is the procedure by which we solved the coupled DGLAP evolution equations considering the Regge behaviour of structure functions.

Putting equations (3.10) and (3.12) in equations (3.1) and (3.3), we get respectively

$$\frac{\partial F_2^S(x,t)}{\partial t} = \frac{F_2^S(x,t)}{t}f_1(x) + \frac{G(x,t)}{t}f_2(x) \tag{3.34}$$

and

$$\frac{\partial G(x,t)}{\partial t} = \frac{F_2^S(x,t)}{t}f_3(x) + \frac{G(x,t)}{t}f_4(x), \tag{3.35}$$

where

$$f_1(x) = A_f\left[\{3 + 4\ln(1-x)\} + 2\int_x^1 \frac{d\omega}{1-\omega}\left\{(1+\omega^2)\omega^{\lambda_S} - 2\right\}\right],$$

$$f_2(x) = \left[\frac{3}{2}A_f N_f \int_x^1 \{\omega^2 + (1-\omega)^2\}\omega^{\lambda_G}\, d\omega\right],$$

$$f_3(x) = 9A_f \cdot \frac{2}{9}\int_x^1 \left(\frac{1+(1-\omega)^2}{\omega}\right)\omega^{\lambda_S}\, d\omega$$

and

$$f_4(x) = 9A_f \left[ \left\{ \left( \frac{11}{12} - \frac{N_f}{18} \right) + \ln(1-x) \right\} + \int_x^1 d\omega \left\{ \frac{\left(\omega^{\lambda_G+1}-1\right)}{1-\omega} + \left(\omega(1-\omega) + \frac{1-\omega}{\omega}\right)\omega^{\lambda_G} \right\} \right].$$

Let us take, $f_1(x)=P_1$, $f_2(x)=Q_1$, $f_3(x)=R_1$ and $f_4(x)=S_1$. Equations (3.34) and (3.35) results to the simple forms respectively as

$$t.\frac{\partial F_2^S(x,t)}{\partial t} - P_1.F_2^S(x,t) - Q_1.G(x,t) = 0 \tag{3.36}$$

and

$$t.\frac{\partial G(x,t)}{\partial t} - R_1.F_2^S(x,t) - S_1.G(x,t) = 0. \tag{3.37}$$

For a constant value of x, equations (3.36) and (3.37) are simultaneous linear ordinary differential equations in $F_2^S(x,t)$ and $G(x, t)$. We solved these equations by one of the standard methods for solution of ordinary differential equations [Appendix C][69, 70] and the solutions for singlet and gluon structure functions are

$$F_2^S(x,t) = C\left(t^{g_1} + t^{g_2}\right) \tag{3.38}$$

and

$$G(x,t) = C(F_1 t^{g_1} + F_2 t^{g_2}). \tag{3.39}$$

Where $g_1 = \frac{-(U_1-1)+\sqrt{(U_1-1)^2-4V_1}}{2}$, $g_2 = \frac{-(U_1-1)-\sqrt{(U_1-1)^2-4V_1}}{2}$, $U_1=1-P_1-S_1$, $V_1 = S_1.P_1-Q_1.R_1$, $F_1 = (g_1-P_1)/Q_1$, $F_2 = (g_2-P_1)/Q_1$.

Applying initial conditions at $x = x_0$, $F_2^S(x,t)=F_2^S(x_0,t)$ and $G(x,t)=G(x_0,t)$, and at $t = t_0$, $F_2^S(x,t)=F_2^S(x,t_0)$ and $G(x,t)=G(x,t_0)$, the t and x-evolution equations for the singlet and gluon structure functions in LO come as

$$F_2^S(x,t) = F_2^S(x,t_0) \left( \frac{t^{g_1} + t^{g_2}}{t_0^{g_1} + t_0^{g_2}} \right), \tag{3.40}$$

$$F_2^S(x, t) = F_2^S(x_0, t) \left( \frac{t^{g_1} + t^{g_2}}{t^{g_{10}} + t^{g_{20}}} \right), \quad (3.41)$$

$$G(x, t) = G(x, t_0) \left( \frac{F_1 t^{g_1} + F_2 t^{g_2}}{F_1 t_0^{g_1} + F_2 t_0^{g_2}} \right) \quad (3.42)$$

and

$$G(x, t) = G(x_0, t) \left( \frac{F_1 t^{g_1} + F_2 t^{g_2}}{F_{10} t^{g_{10}} + F_{20} t^{g_{20}}} \right), \quad (3.43)$$

where $g_{10}, g_{20}, F_{10}$ and $F_{20}$ are the values of $g_1, g_2, F_1$ and $F_2$ at $x = x_0$. The t and x-evolution equations of deuteron structure function corresponding to equations (3.47) and (3.48) are respectively

$$F_2^d(x, t) = F_2^d(x, t_0) \left( \frac{t^{g_1} + t^{g_2}}{t_0^{g_1} + t_0^{g_2}} \right) \quad (3.44)$$

and

$$F_2^d(x, t) = F_2^d(x_0, t) \left( \frac{t^{g_1} + t^{g_2}}{t^{g_{10}} + t^{g_{20}}} \right). \quad (3.45)$$

## 3.2 Results and Discussion

We have compared our result of deuteron and proton structure functions with the data sets measured by the NMC [71] in muon-deuteron DIS from the merged data sets at incident momenta 90, 120, 200 and 280 GeV² and also with the data sets measured by the Fermilab E665 [72] Collaboration in muon-deuteron DIS at an average beam energy of 470 GeV². For our phenomenological analysis we have considered the data sets of deuteron and proton structure functions in the range what the NMC and E665 collaborations provide at low-x. We considered the QCD cut-off parameter as $\Lambda_{\overline{MS}}$ (N$_f$ = 4) = 323 MeV for $\alpha_s(M_z^2)$ = 0.119± 0.002 [73]. We have compared our results of t and x-evolutions of gluon structure function in LO with MRST 2001, MRST 2004 and GRV1998LO global parameterizations. We have taken the MRST 2001 fit [73] to the CDFIB data [74] for $Q^2 = 20$ GeV$^2$, in which they obtained the optimum global NLO fit with the starting parameterizations of the partons at $Q_0^2=1$ GeV$^2$ given by $xg=123.5x^{1.16}(1-x)^{4.69}(1-3.57x^{0.5}+3.41x)-0.038x^{-0.5}(1-x)^{10}$. The optimum fit corresponds to $\alpha_S(M_Z^2)= 0.119$ i.e. $\Lambda_{\overline{MS}}(N_f=4)=323$ MeV. We have also taken the MRST 2004 fit [75] to the ZEUS [76] and H1 [77] data with $x<0.01$ and $2<Q^2<500$ GeV$^2$ for $Q^2 = 100$ GeV$^2$, in which they have taken parametric form for the starting distribution at $Q_0^2=1$ GeV$^2$ given by

$$xg = A_g x^{-\lambda_g}(1-x)^{3.7}(1+\varepsilon_g\sqrt{x}+\gamma_g x)-A x^{-\delta}(1-x)^{10},$$

where the powers of the (1–x) factors are taken from MRST 2001 fit. The $\lambda_g$, $\varepsilon_g$, A and $\delta$ are taken as free parameters. The value of $\alpha_S(M_Z^2)$ is taken to be the same as in the MRST 2001 fit. We have taken the GRV1998LO parameterization [78] for $10^{-2} \leq x \leq 10^{-5}$ GeV$^2$ and $20 \leq Q^2 \leq 40$ GeV$^2$, where they used H1 [79] and ZEUS [80] high precision data on $G(x, Q^2)$. They have chosen $\alpha_S(M_Z^2)=0.114$ i.e. $\Lambda_{\overline{MS}}(N_f=4)=246$ MeV. The input densities have been fixed using the data sets of HERA [79], SLAC [81], BCDMS [82], NMC [71] and E665 [72]. The resulting input distribution at $Q^2=0.40$ GeV$^2$ is given by $xg=20.80x^{1.6}(1-x)^{4.1}$.

The graphs 'our result' represent the best fit graph of our work with different experimental data sets and parameterization graphs. Data points at lowest-$Q^2$ values are taken as input to test the t-evolution equations and data points at $x<0.1$ is taken as input to test the x-evolution equations. We have compared our results for $K(x) = K$, $ax^b$ and $ce^{dx}$. In our work for deuteron structure function, we found that the value of the structure function remains almost same for a large range of b, $10^{-2}>b>10^{-7}$ and the values remain constant for $b>10^{-2}$. So we choose $b=0.01$ for our calculation. Similarly the value of the structure function remains almost same for a large range of d, $-1>d>-10^{-4}$ and $d>-1$. So we choose $d= -1$ for our calculation. In our work for gluon structure function, we have found the values of the gluon structure function remains almost same for b<0.00001 and for d< 0.0001. So, we have chosen b = 0.00001 and d = 0.0001 and the best fit graphs are observed by changing the values of K, a and c. We are also interested to see the contribution of quark to gluon structure functions at low-x and high-$Q^2$, theoretically which should decrease for x→0, $Q^2$→∞ [41, 83]. According to Regge theory, the high energy (low-x) behaviour of both gluons and sea quarks is controlled by the same singularity factor in the complex angular momentum plane. So, we have taken $\lambda_S = \lambda_{NS} = \lambda_G = \lambda = 0.5$ for our calculation. The values of $\lambda_S$, $\lambda_{NS}$ and $\lambda_G$ should be close to 0.5 in quite a broad range of x [49-52, 75].

In Figure 3.1, we have compared our result of t-evolution of deuteron structure function in LO from equation (3.25) for $\lambda_S$ = 0.5 and K(x)=K with NMC data and the best fit result is found for 2.25≤K≤2.93. Same graphs are found for K(x) =K and $ax^b$ and in this case the best fit results are for 2.33≤ a≤ 3.08 and 2.34≤c≤2.98.

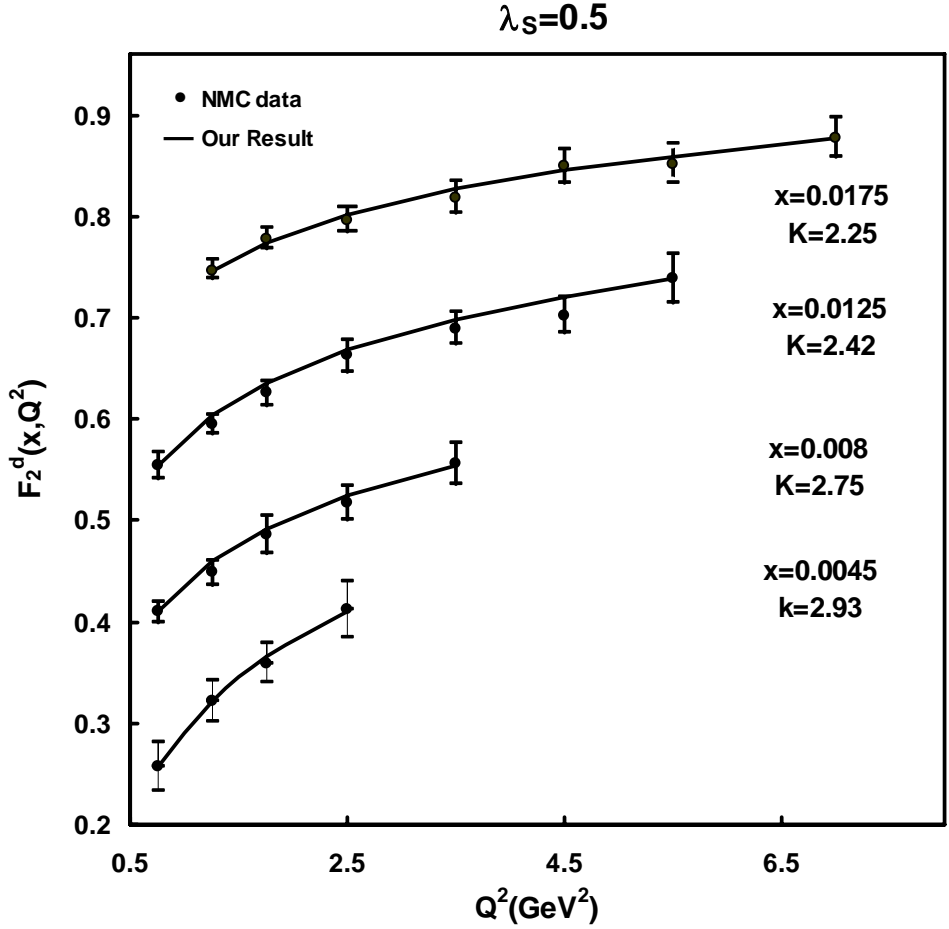

**Figure 3.1**: t-evolution of deuteron structure function in LO at low-x compared with NMC experimental data points.

In Figure 3.2, we find the t-evolution of deuteron structure in LO for representative values of x with K = 2.52 (the average values of K from Figure 3.1) and varying the values of $\lambda_S$. The corresponding values for the best fit results are $0.355 \leq \lambda_S \leq 0.61$. We get the same range of $\lambda_S$ taking average values of a and c as a = 2.63 and c = 2.6 respectively.

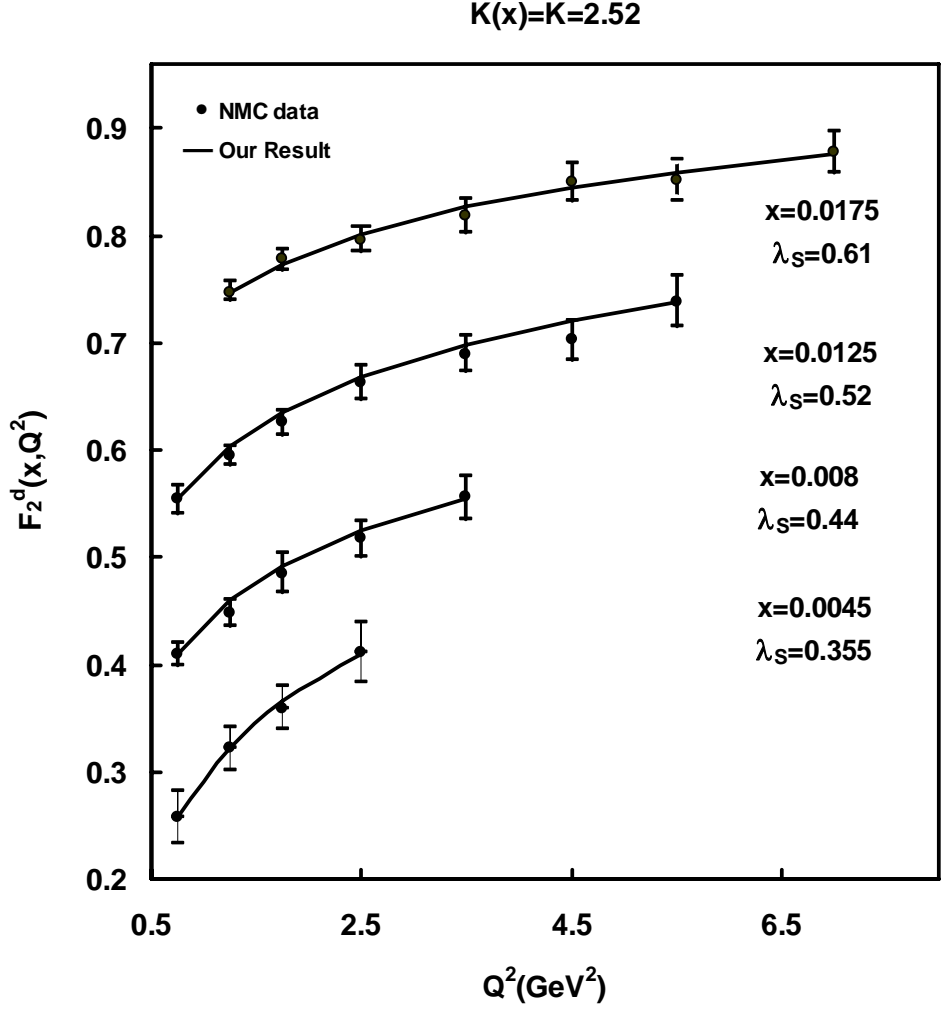

**Figure 3.2**: t-evolution of deuteron structure function in LO at low-x compared with NMC experimental data set. Data are scaled up by +0.15i (i=0, 1, 2, 3) starting from bottom graph.

Figure 3.3 represents our result of x-evolution of deuteron structure function in LO from equation (3.26) for K =0.01 with NMC data. The best fit result is found for $0.09 \leq \lambda_S \leq 0.22$ in the range of $9 \leq Q^2 \leq 20$ GeV$^2$, $0.025 \leq x \leq 0.09$. Here we have kept K fixed at 0.01 since the value of structure function in this case remains almost same for the large range $10^{-6} \leq K \leq 10^{-2}$ and $K > 10^{-2}$.

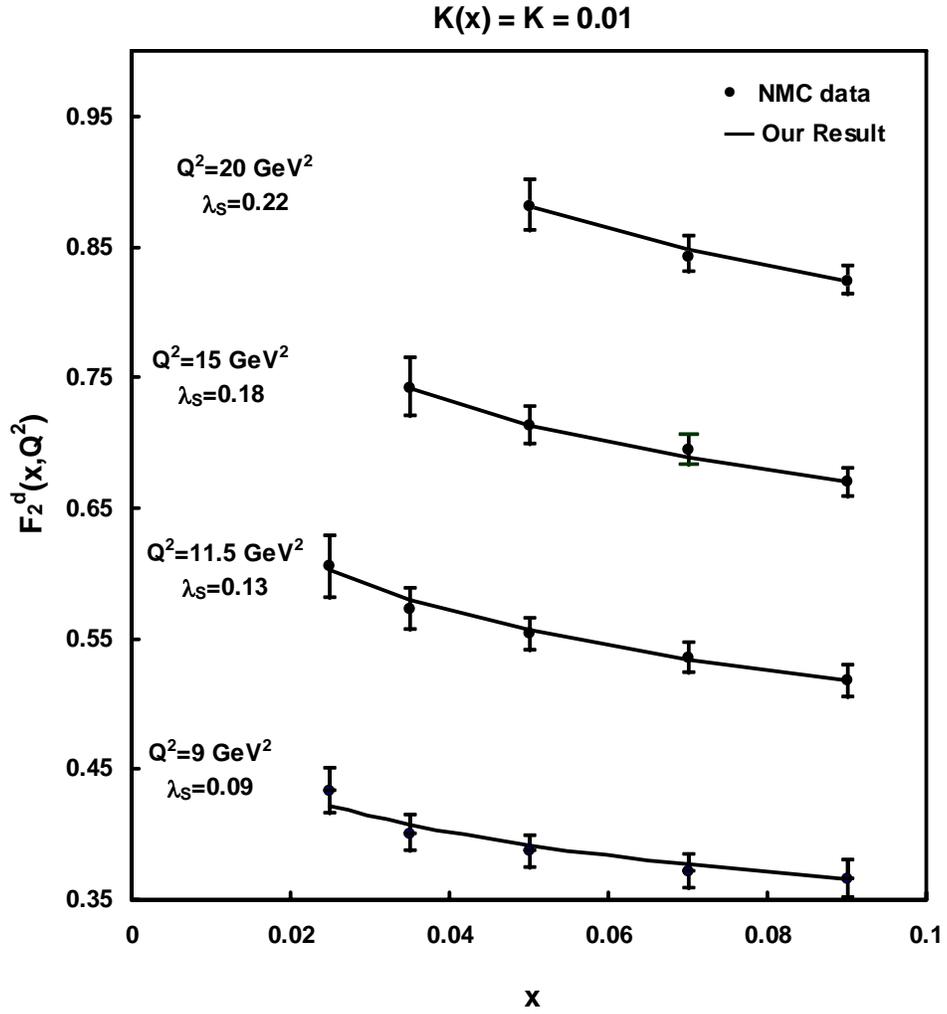

**Figure 3.3**: x-evolutions of deuteron structure function in LO at low-x compared with NMC data set. Data are scaled up by +0.15i (i=0, 1, 2, 3) starting from bottom graph.

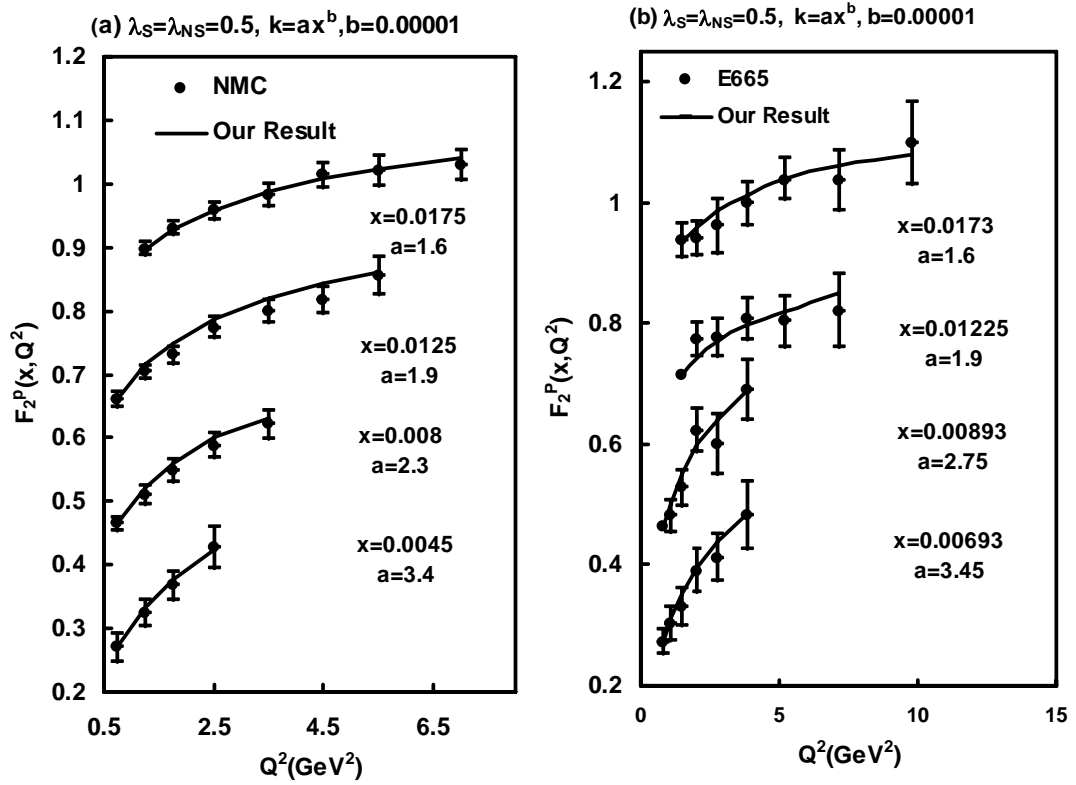

**Figure 3.4**: t-evolution of proton structure function in LO at low-x for $\lambda_S=\lambda_{NS}$ =0.5 and $K(x) = ax^b$ for the representative values of x compared with NMC and E665 data sets. Data are scaled up by +0.2i (i=0, 1, 2, 3) for both NMC and E665 data sets starting from bottom graph.

Figures 3.4(a-b) represents the result of t-evolution of proton structure function from equation (3.27) for $K(x) = ax^b$ with NMC and E665 data sets. Figure 3.4(a) represents the comparison of our result with NMC data. And figure 3.4(b) represents the comparison with E665 data. We get the best fit results for $1.6 \leq a \leq 3.45$. We have also compared our result of t-evolution of proton structure function for $K(x) = K$ and $ce^{-dx}$. Same graphs are found with $1.7 \leq K \leq 3.3$ and $1.6 \leq c \leq 3.4$.

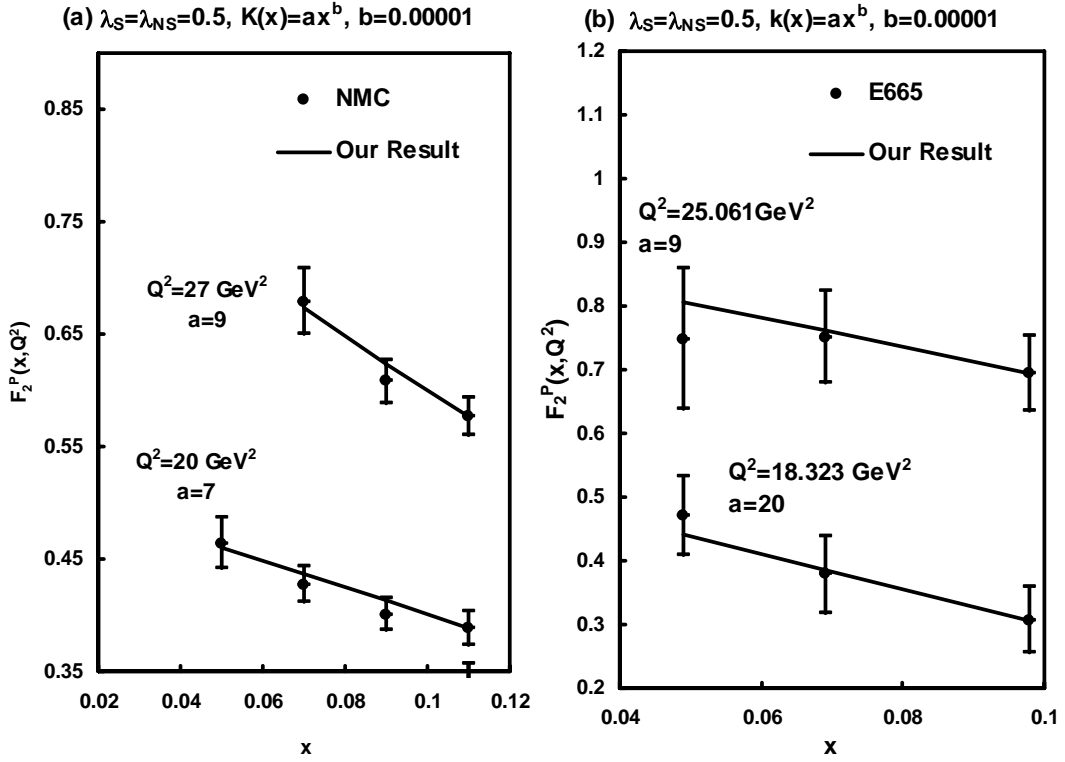

**Figure 3.5**: x -evolution of proton structure function in LO at low-x for $\lambda_S=\lambda_{NS}=0.5$ and $K(x) = ax^b$ for the representative values of x compared with NMC and E665 data sets. Here data are scaled up by +0.3i (i=0, 1) for both NMC and E665 data sets starting from bottom graph.

Figures 3.5(a-b) represent the result of x-evolution of proton structure function from equation (3.28) for K (x) = $ax^b$ with NMC and E665 data sets. Figure 3.5(a) represents the comparison of our result with NMC data. And figure 3.5(b) represents the comparison with E665 data. $9 \leq a \leq 20$ correspond the best fit results. Same graphs are found for K(x) =K and $ce^{-dx}$ where $9 \leq K \leq 20$ and $9 \leq c \leq 20$.

In figures 3.6 and 3.7 we have compared our result of t-evolution of gluon structure function from equation (3.29) with GRV1998LO gluon parameterization at x=$10^{-5}$ and $10^{-4}$ respectively for K(x) = K. At x = $10^{-5}$, we found the best fit result for K = 1.55 and at x = $10^{-4}$ for K = 3. We compared the result for K(x) =$ax^b$ and $ce^{-dx}$ also and found the same graphs as for K (x) = K.

When a = c = 1.55 at x = 10⁻⁵ and a = c = 3 at x = 10⁻⁴, we get the best fit results. The figures show that our results are in good agreement with GRV1998LO parameterization at low-x.

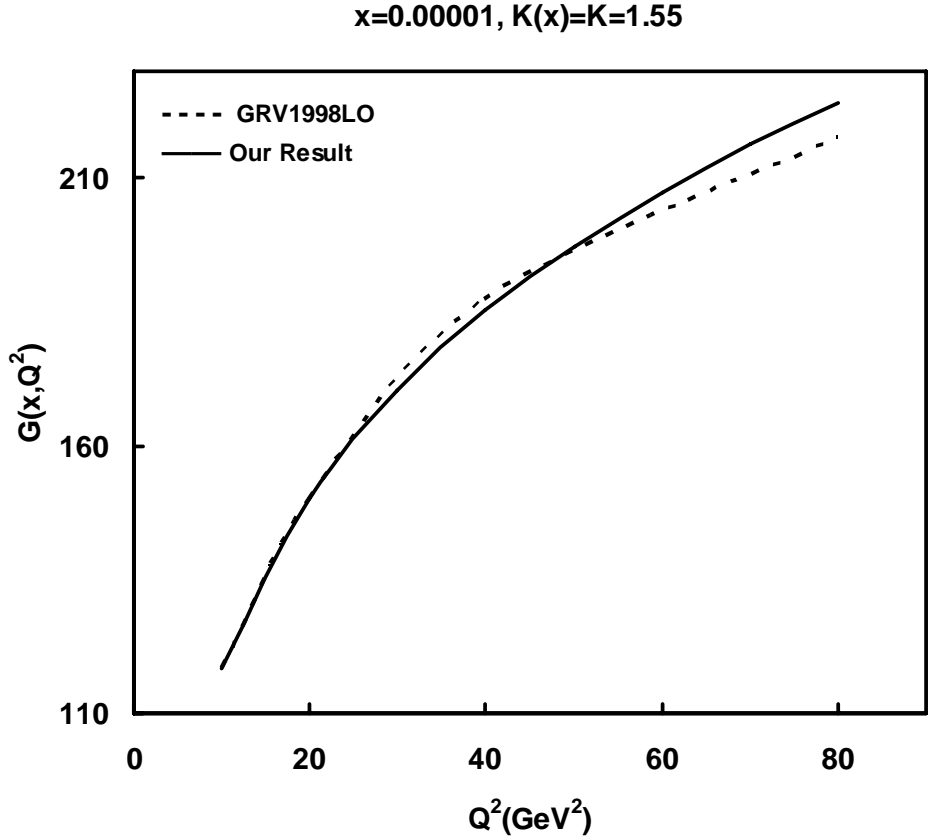

**Figure 3.6**: t -evolution of gluon structure function in LO at low-x for $\lambda_G$= 0.5 and K(x)=K=1.55 for the representative values of x compared with GRV1998LO parameterization graphs.

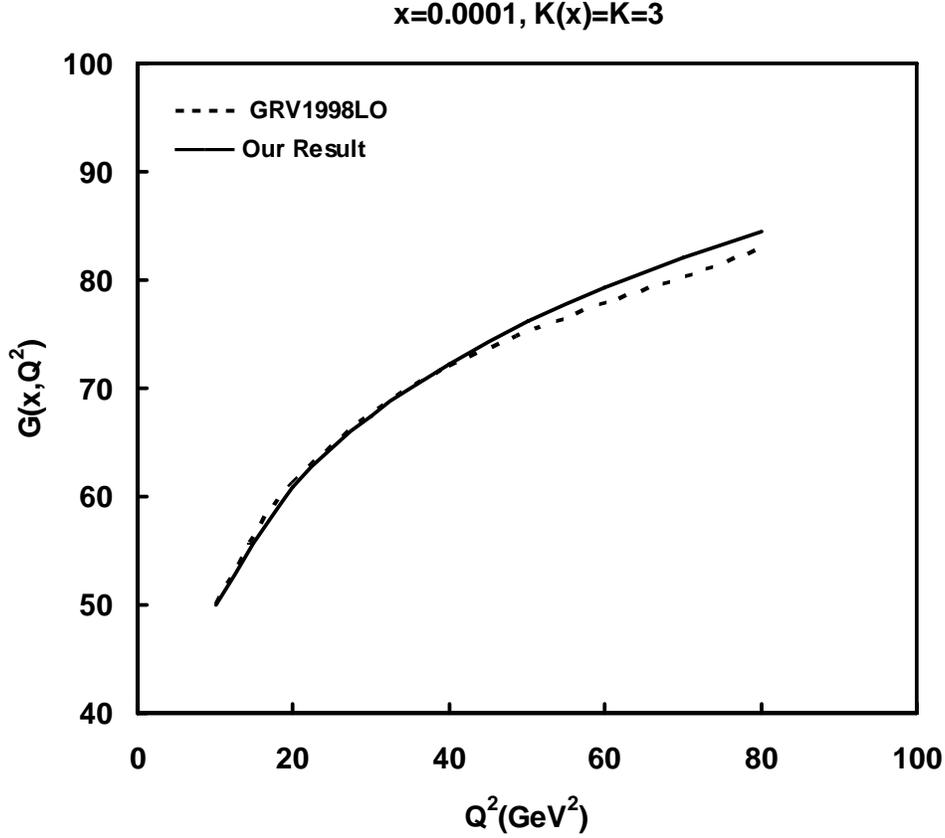

**Figure 3.7**: t -evolution of gluon structure function in LO at low-x for λ$_G$

=0.5 and K(x) = K=3 for the representative values of x compared with GRV1998LO parameterization graphs.

Figures 3.8(a-b) represent our result of x-evolution of gluon structure function from equation (3.30) for K (x) = K with MRST 2001 and MRST 2004 global parameterizations. Figure 3.8(a) represents the comparison of our result with MRST 2001 parameterization at $Q^2$ = 20 GeV². Figure 3.8(b) represents the comparison with MRST 2004 parameterization at $Q^2$ = 100 GeV². We have compared our result for K(x) =ax$^b$ and  ce$^{-dx}$ also and found the same graphs as for K (x) = K. The corresponding values for the best fit results are K=a = c = 0.8 with MRST 2001 parameterization at $Q^2$ = 20 GeV², K=a = c = 0.33 with MRST 2004 parameterization at $Q^2$ = 100 GeV².

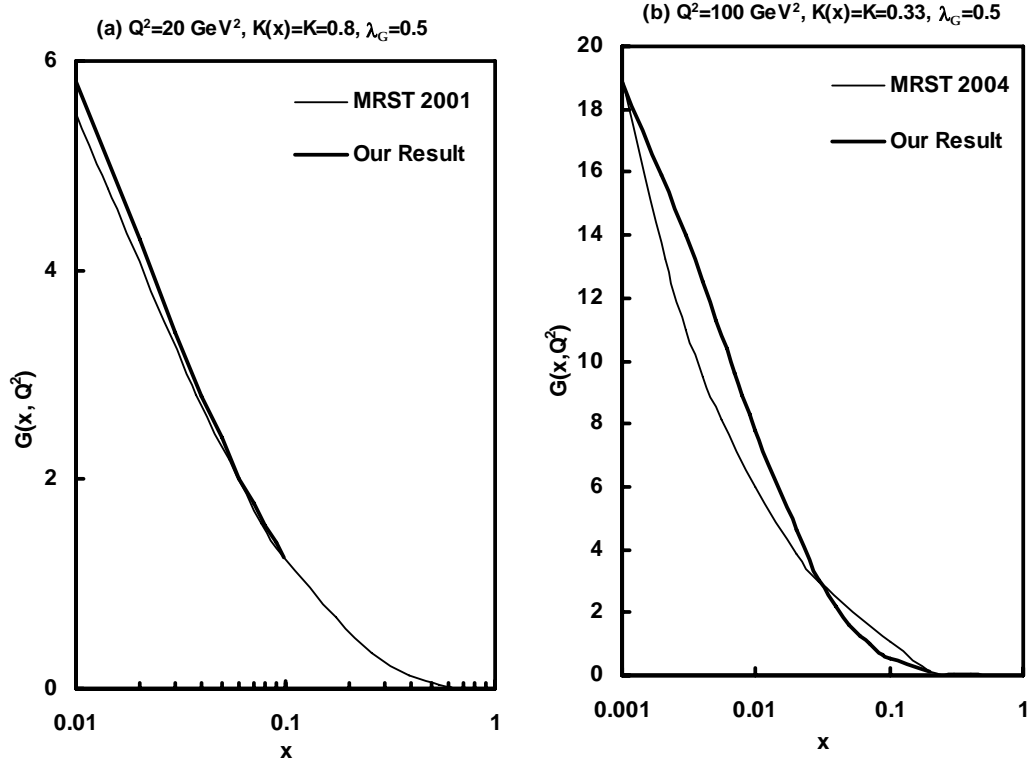

**Figure 3.8**: x-evolution of gluon structure function in LO at low-x for $\lambda_G$ =0.5 and K(x) = K for the representative values of x compared with MRST 2001 and MRST 2004 parameterization graphs

Figures 3.9(a-b) represent the comparison of our result of x-evolution of gluon structure function from equation (3.30) with GRV1998LO parameterization at $Q^2 = 20$ GeV$^2$ and 40 GeV$^2$. K = 0.11 for $Q^2$ =20 GeV$^2$ and K = 0.12 for $Q^2$ =40 GeV$^2$ corresponds the best fit result. In some recent papers [84], Choudhury and Saharia presented a form of gluon structure function at low-x obtained from a unique solution with one single initial condition through the application of the method of characteristics [85]. They have overcome the limitations of non-uniqueness of some of the earlier approaches [65, 86-88]. So, it is theoretically and phenomenologically favoured over the earlier approximations. As we compared both their and our results with GRV1998LO parameterizations we found that our result to GRV1998LO parameterization fits better with decreasing x. We have compared our result for $K(x) = ax^b$ and $ce^{-dx}$

also and found the same graphs as for K (x) = K and a = c = 0.11 at $Q^2$ = 20 GeV$^2$ and a = c = 0.12 at $Q^2$ = 40 GeV$^2$ correspond the best fit results.

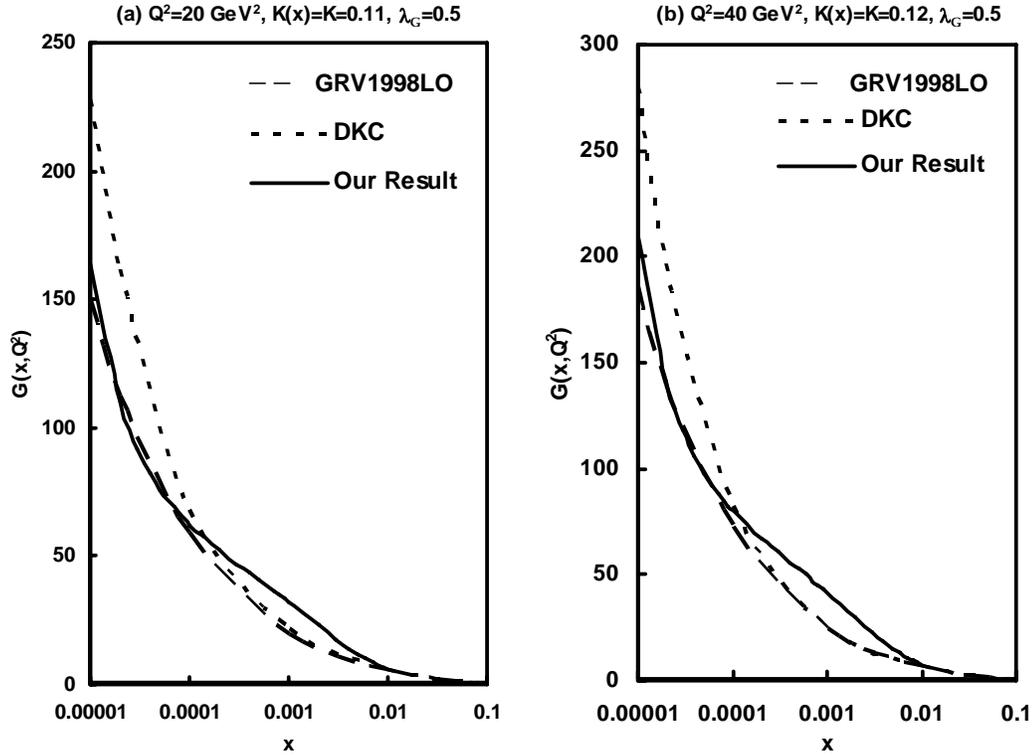

**Figure 3.9**: x-evolution of gluon structure function in LO at low-x for $\lambda_G$ =0.5 and K(x) = K for the representative values of x compared with GRV1998LO parameterization graphs.

Figures 3.10(a-f) represent the sensitivity of the parameters $\lambda$, K, a, b, c and d respectively. Taking the best fit figures to the x-evolution of gluon structure function with MRST 2001 parameterization at $Q^2$ = 20 GeV$^2$, we have given the ranges of the parameters as $0.45 \leq \lambda \leq 0.55$, $0.65 \leq K \leq 0.95$, $0.65 \leq a \leq 0.95$, $0.00001 \leq b \leq 0.05$, $0.65 \leq c \leq 0.95$ and $0.0001 \leq d \leq 0.5$.

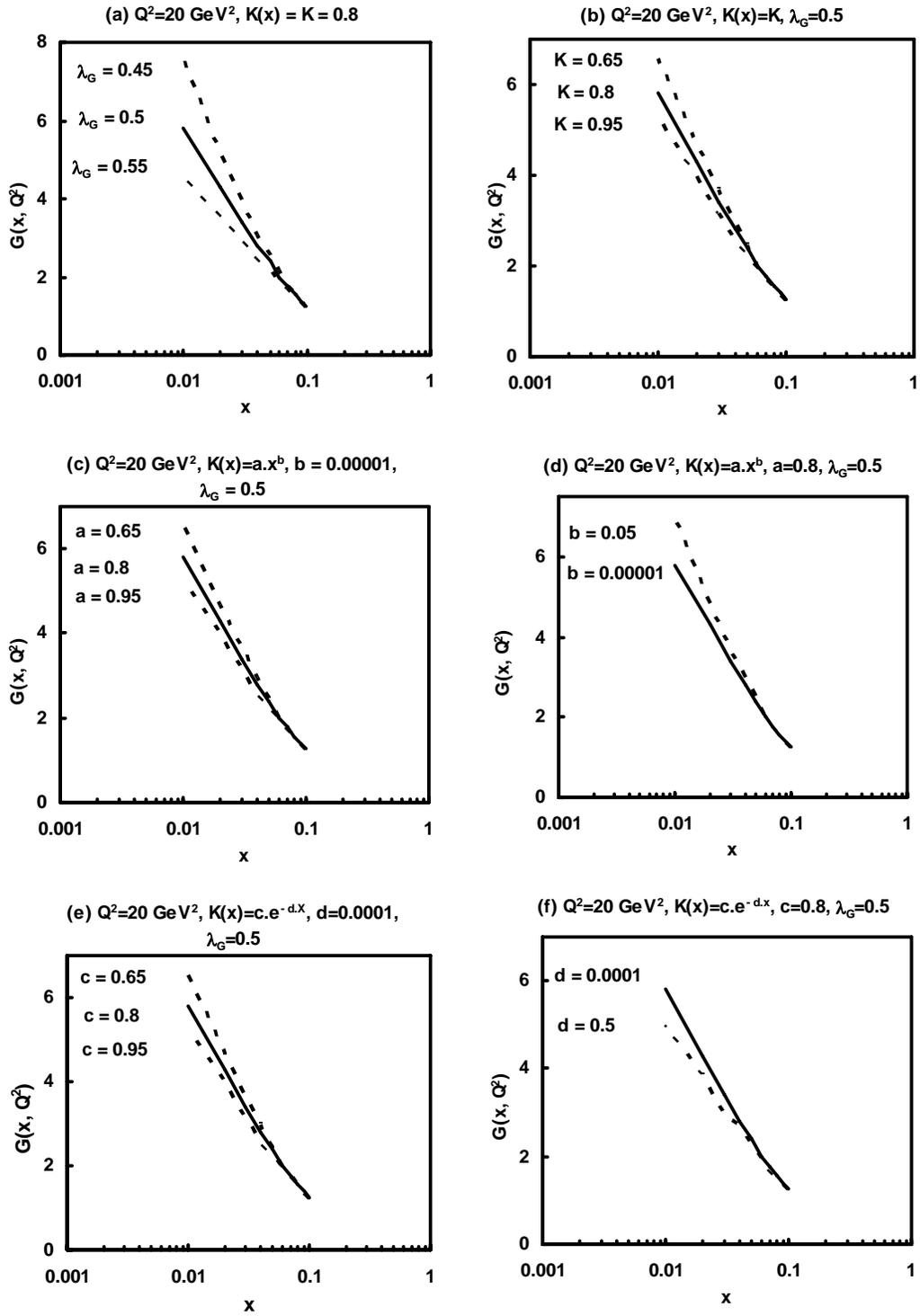

**Figure 3.10:** Fig. 3.10(a-f) show the sensitivity of the parameters $\lambda$, K, a, b, c and d respectively at $Q^2 = 20$ GeV$^2$ with the best fit graphs of our results with MRST 2001 parameterization.

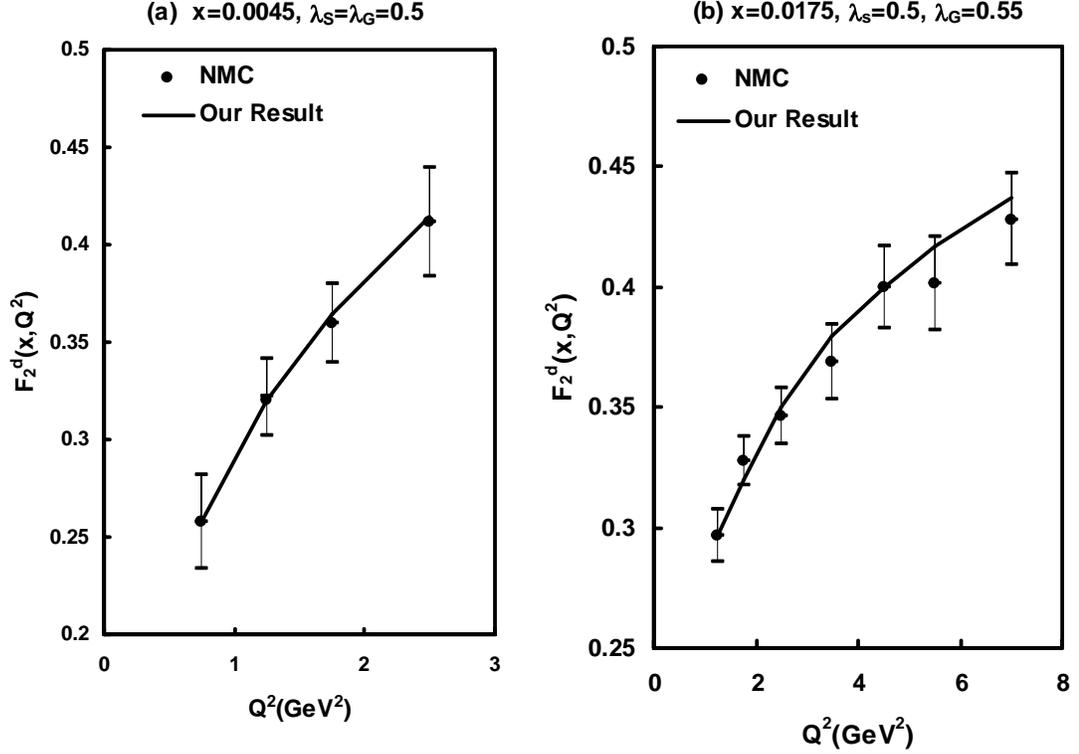

**Figure 3.11**: t evolution of deuteron structure function in LO at low-x obtained by solving coupled equations, compared with NMC data points.

The t and x-evolutions of singlet and gluon structure functions at low-x, obtained by solving coupled DGLAP equations applying Regge behaviour of structure functions are given by equations (3.40) to (3.43) respectively. Figures 3.11(a-b) represent our result of t-evolution of deuteron structure function in LO from equation (3.44) with NMC data set. We get the best fit results with $\lambda_S = \lambda_G = 0.5$ for x=0.0045 and $\lambda_S = 0.5$, $\lambda_G = 0.55$ for x=0.0175.

Figures 3.12(a-b) represent our results for x-evolution of deuteron structure function in LO from equation (3.45) with NMC data set. We find the best fit result corresponds to $\lambda_S = \lambda_G = 0.7$ for both $Q^2=15$ GeV$^2$ and $Q^2=20$ GeV$^2$. Since Regge theory is strictly applicable only for low-x and high-$Q^2$, the best fits of our result for x-evolution of deuteron structure function with NMC data set are not so good and values of $\lambda_S$ and $\lambda_G$ are also not close to 0.5. Due to lack of deuteron data at low-x and high-$Q^2$, we could not check our result for x-evolution of deuteron structure function properly.

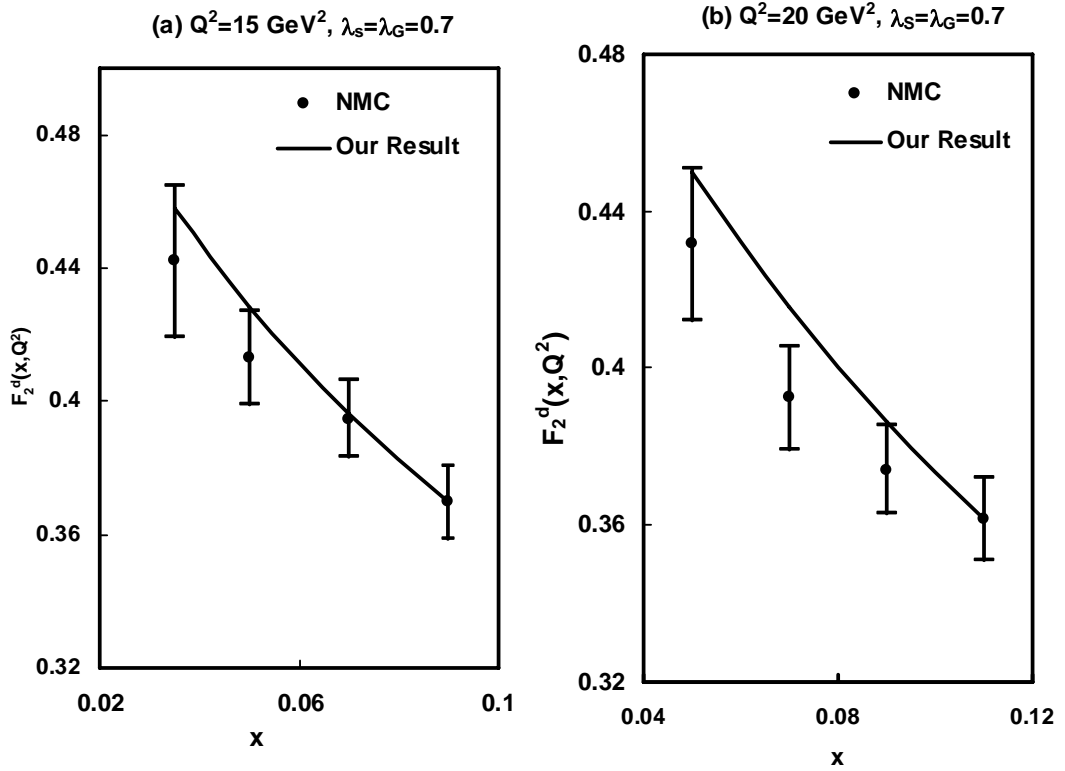

**Figure 3.12**: x-evolution of deuteron structure function in LO at low-x obtained by solving coupled equations, compared with NMC data set.

Figures 3.13(a-b) represent our results of t-evolution of gluon structure function from equation (3.42) with GRV1998LO gluon parameterization at $x=10^{-5}$ and $10^{-4}$ respectively. We get the best fit results with $\lambda_S =0.5$, $\lambda_G= 0.46$ for $x=10^{-5}$ and $\lambda_S =0.5$, $\lambda_G= 0.48$ for $x=10^{-4}$.

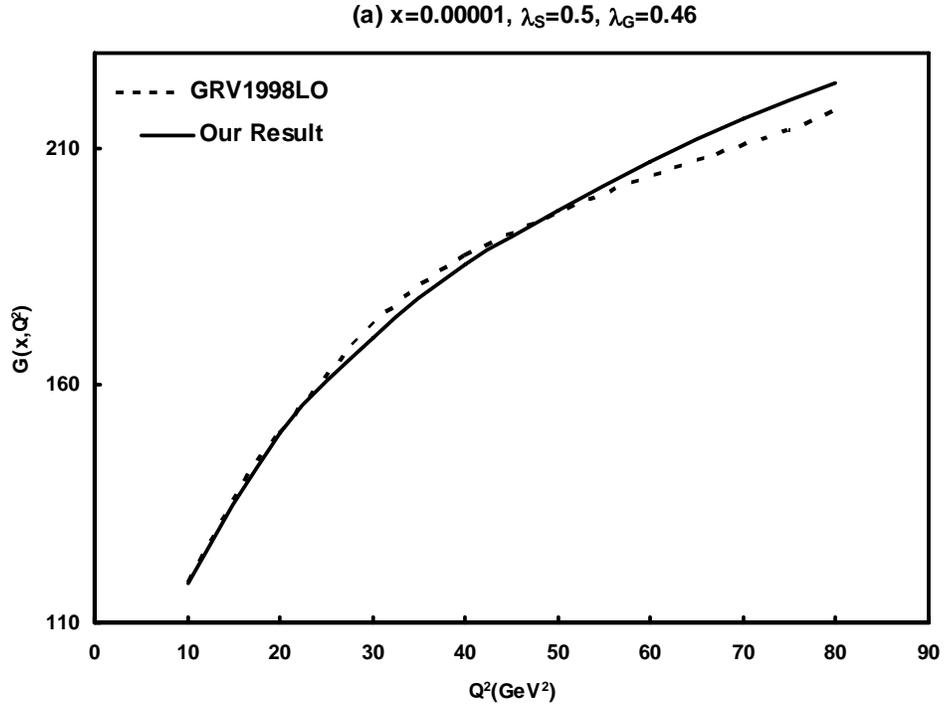

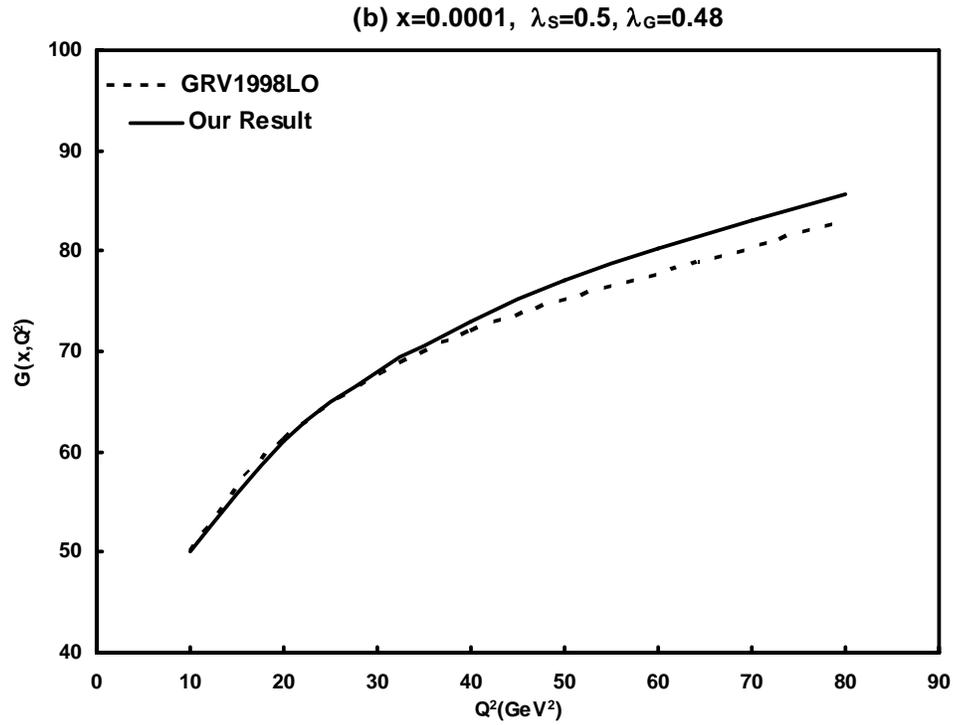

**Figure 3.13**: t -evolution of gluon structure function in LO at low-x which is obtained by solving coupled equations, compared with GRV1998LO parameterization graphs.

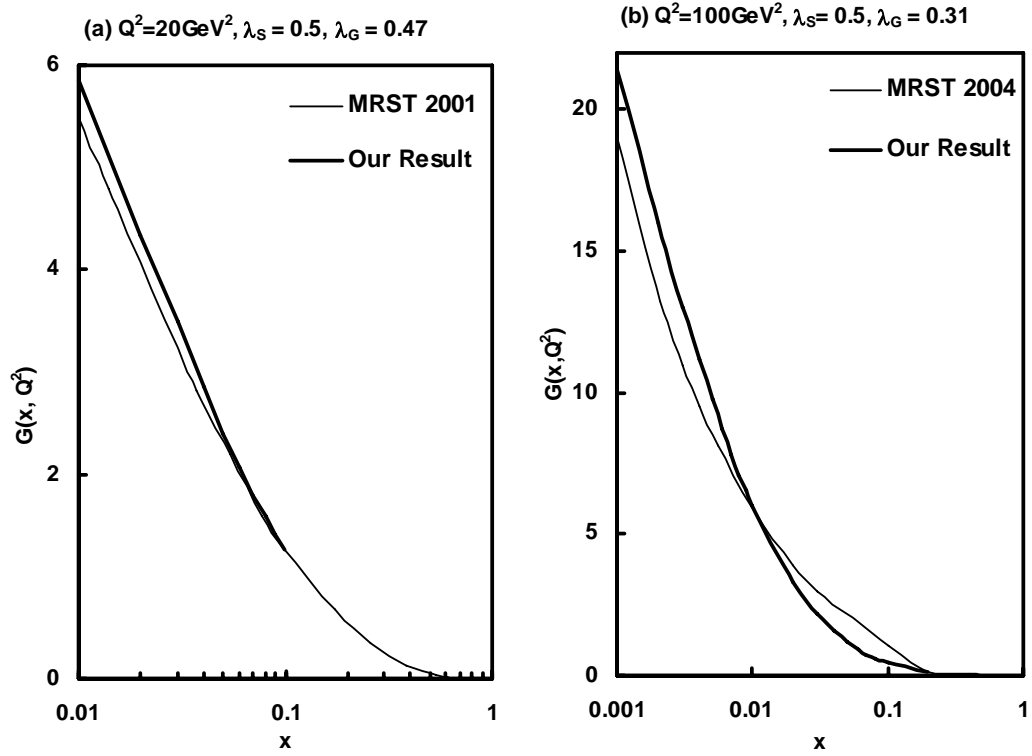

**Figure 3.14**: x-evolution of gluon structure function in LO at low-x obtained by solving coupled equations, compared with MRST 2001 and MRST 2004 parameterization graphs.

Figure 3.14(a) represents our result of x-evolution of gluon structure function from equation (3.43) with MRST 2001 parameterization at $Q^2$ = 20 GeV$^2$. We find the best fit result corresponding to $\lambda_S$ =0.5, $\lambda_G$ = 0.47. Figure 3.14(b) presents our result of x-evolution of gluon structure function from equation (3.43) with MRST 2004 parameterization at $Q^2$ = 100 GeV$^2$. We get the best fit results for $\lambda_S$ =0.5, $\lambda_G$= 0.31.

Figures 3.15 (a-b), represents our result of x-evolution of gluon structure function from equation (3.43) with GRV1998LO parameterization at $Q^2$ = 40 GeV$^2$ and 80 GeV$^2$ respectively. For both the figures the best fit results are for $\lambda_S$ =0.5, $\lambda_G$= 0.3. From the figures it is obvious that our result is best fitted to the GRV1998LO parameterizations for increasing $Q^2$ at low-x.

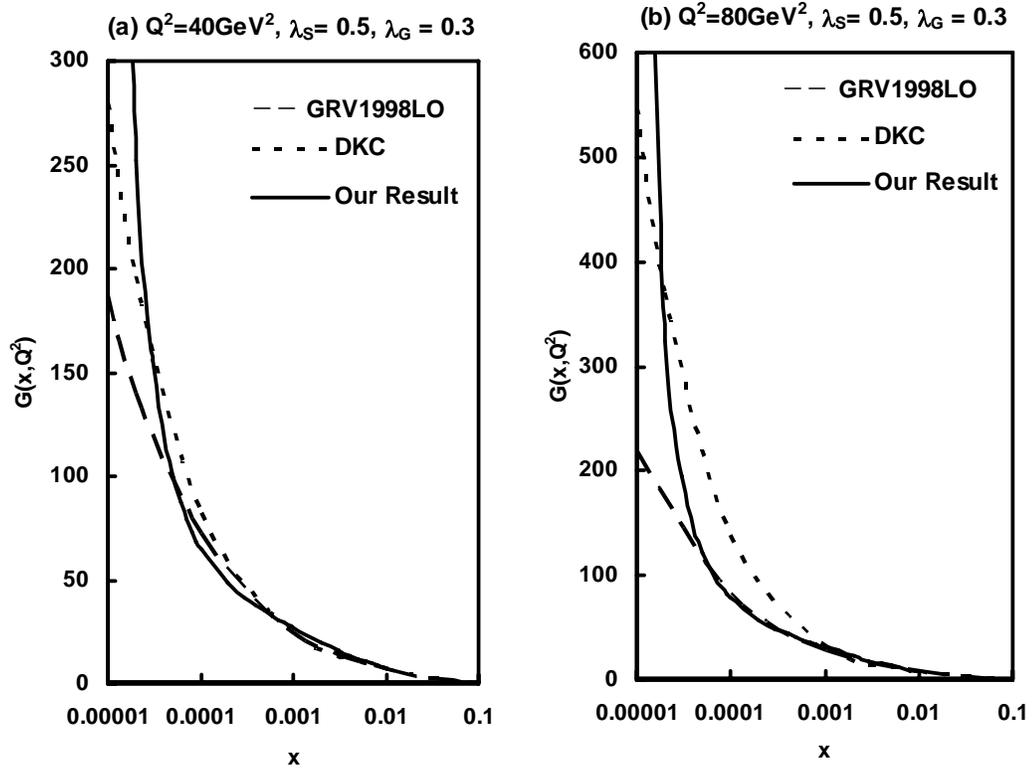

**Fig 3.15**: x-evolution of gluon structure function in LO at low-x which is obtained by solving coupled equations, compared with GRV1998LO parameterization graphs.

Figures 3.16(a-b) represent the sensitivity of the parameters $\lambda_S$ and $\lambda_G$ respectively. Taking the best fit graph of our result to the x-evolution of gluon structure function from equation (3.43) with MRST 2004 parameterization at $Q^2=100$ GeV$^2$, we have given the results for the ranges of the parameters as $0.3 \leq \lambda_S \leq 0.7$ and $0.3 \leq \lambda_G \leq 0.32$. It is observed that $\lambda_G$ is much more sensitive than $\lambda_S$.

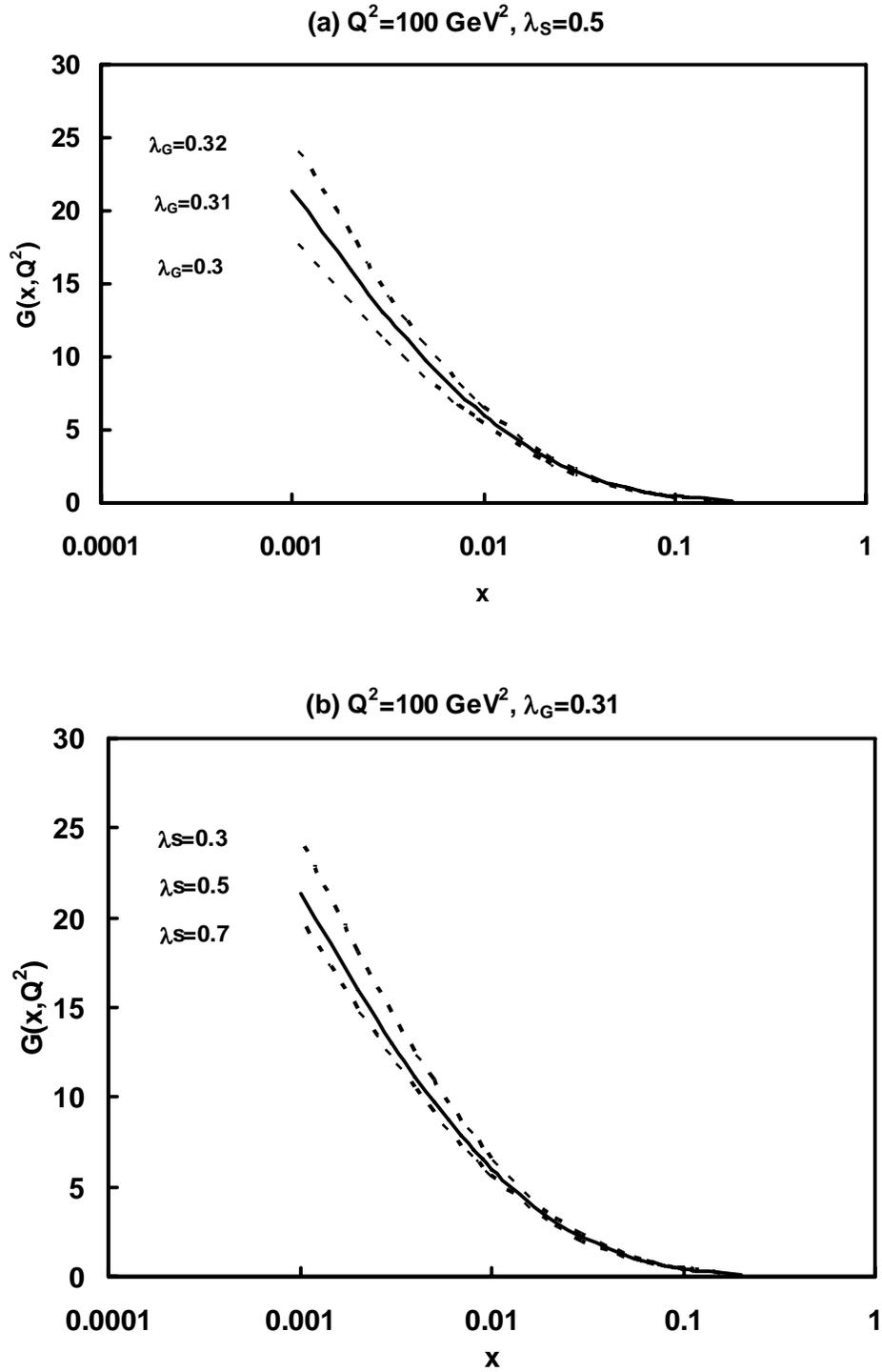

**Fig. 3.16:** Fig. 3.16(a-b) show the sensitivity of the parameters $\lambda_S$ and $\lambda_G$ respectively at $Q^2 = 100$ GeV$^2$ with the best fit graph of our results with GRV1998LO parameterization.

Figures 3.17(a-b) also represent the sensitivity of the parameters $\lambda_S$ and $\lambda_G$ respectively. Considering the best fit graph of our result of t-evolution of deuteron structure function from equation (3.44) with NMC data at x=0.0045, we have given the ranges of the parameters as $0.3 \leq \lambda_S \leq 0.7$ and $0.45 \leq \lambda_G \leq 0.55$. Here also it is observed that $\lambda_G$ is much more sensitive than $\lambda_S$.

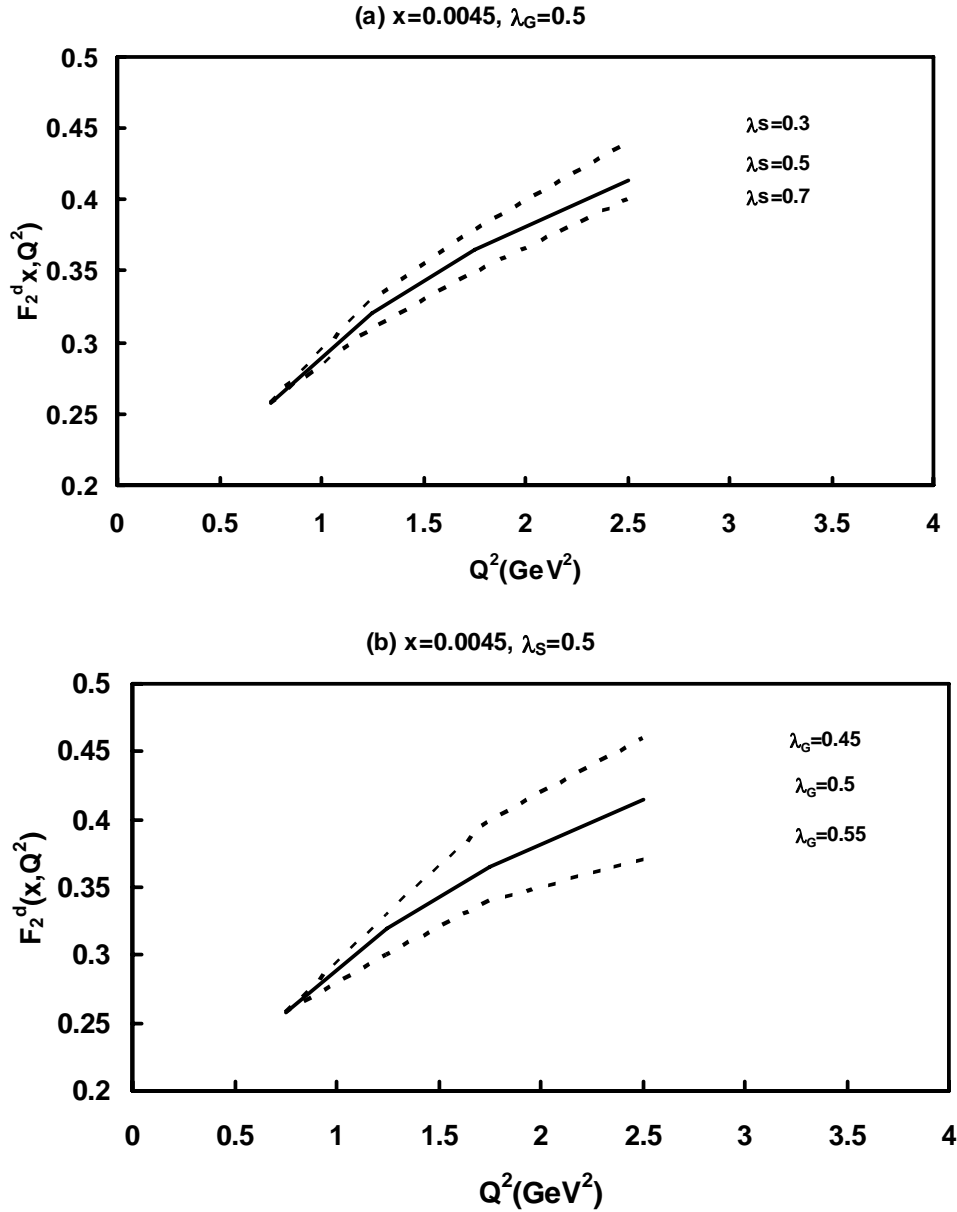

**Figure 3.17:** Fig. 3.17(a) and 3.17(b) show the sensitivity of the parameters $\lambda_S$ and $\lambda_G$ respectively at x = 0.0045 with the best fit graph of our result (solid curve) with NMC data points.

## 3.3 Conclusion

In this chapter by using Regge behaviour of spin-independent structure functions at low-x, we have solved DGLAP evolution equations for singlet, non-singlet and gluon structure functions in LO. Also we have derived the t and x-evolutions of deuteron, proton and gluon structure functions. To overcome the problem of ad hoc assumption of the function K(x), the relation between singlet structure function and gluon structure function, we also solved coupled evolution equations for singlet and gluon structure functions and obtained a new description of t and x-evolutions for both the singlet and gluon structure functions within the Regge limit. We have seen that our results are in good agreement with NMC and E665 experimental data sets for deuteron and proton structure functions, MRST2001, MRST2004 and GRV1998LO global parameterizations for gluon structure function especially at low-x and high-$Q^2$ region. We were also interested to see the contribution of quark to the gluon structure function at different x and $Q^2$ but it has been observed that in our x-$Q^2$ region quark contributes appreciably to gluon structure function. So, we cannot ignore the contribution of quark in that region. We can conclude that Regge behaviour of quark and gluon structure functions is compatible with PQCD at that region assuming that Regge intercepts are almost same for both quark and gluon. We have overcome the limitations that arise from Taylor series expansion method for the analytical solution of the DGLAP evolution equations. In this chapter we solved only leading order evolution equations and seen that next-to-leading order (NLO) and next-next-to-leading order (NNLO) equations are more correct and their solutions give better fit to global data and parameterizations.□

# t and x- Evolutions of Spin-independent DGLAP Evolution Equations in Next-to-Leading Order

In this chapter, we have solved the spin-independent DGLAP evolution equations for singlet, non-singlet and gluon structure functions at low-x in next-to-leading order (NLO) considering Regge behaviour of structure functions and also the coupled equations for singlet and gluon structure functions. The t and x-evolutions of deuteron, proton and gluon structure functions we obtained have been compared with NMC and E665 collaborations data sets and global MRST2001, MRST2004 and GRV1998LO and GRV1998NLO gluon parameterizations respectively.

## 4.1 Theory

The spin-independent DGLAP evolution equations for singlet, non-singlet and gluon structure functions in NLO are given as [52, 65, 89]

$$\frac{\partial F_2^S}{\partial t} - \frac{\alpha_S(t)}{2\pi} I_1^S(x,t) - \left(\frac{\alpha_S(t)}{2\pi}\right)^2 I_2^S(x,t) = 0, \tag{4.1}$$

$$\frac{\partial F_2^{NS}}{\partial t} - \frac{\alpha_S(t)}{2\pi} I_1^{NS}(x,t) - \left(\frac{\alpha_S(t)}{2\pi}\right)^2 I_2^{NS}(x,t) = 0 \tag{4.2}$$

and

$$\frac{\partial G(x,t)}{\partial t} - \frac{\alpha_S(t)}{2\pi} I_1^G(x,t) - \left(\frac{\alpha_S(t)}{2\pi}\right)^2 I_2^G(x,t) = 0. \tag{4.3}$$

The equations upto LO are given in chapter 3 (equations 3.1, 3.2 and 3.3). The NLO contributions are given as

$$I_2^S(x,t) = \left[ (x-1)F_2^S(x,t)\int_0^1 f(\omega)d\omega + \int_x^1 f(\omega)F_2^S\left(\frac{x}{\omega},t\right)d\omega \right.$$

$$\left. + \int_x^1 F_{qq}^S(\omega)F_2^S\left(\frac{x}{\omega},t\right)d\omega + \int_x^1 F_{qg}^S(\omega)G\left(\frac{x}{\omega},t\right)d\omega \right],$$

$$I_2^{NS}(x,t) = (x-1)F_2^{NS}(x,t)\int_0^1 f(\omega)d\omega + \int_x^1 f(\omega)F_2^{NS}\left(\frac{x}{\omega},t\right)d\omega,$$

and

$$I_2^G(x,t) = \int_x^1 \left\{ P_{gg}^2(\omega)G\left(\frac{x}{\omega},t\right) + A(\omega)F_2^S\left(\frac{x}{\omega},t\right) \right\} d\omega.$$

The explicit forms of higher order kernels in NLO are [65, 89, 90]

$$f(\omega) = C_F^2\left[P_F(\omega) - P_A(\omega)\right] + \frac{1}{2}C_F C_A\left[P_G(\omega) + P_A(\omega)\right] + C_F T_R N_f P_{N_f}(\omega),$$

$$F_{qq}^S(\omega) = 2C_F T_R N_f F_{qq}(\omega), \quad F_{qg}^S(\omega) = C_F T_R N_f F_{qg}^1(\omega) + C_G T_R N_f F_{qg}^2(\omega),$$

$$P_F(\omega) = -2\left(\frac{1+\omega^2}{1-\omega}\right)\ln\omega\ln(1-\omega) - \left(\frac{3}{1-\omega} + 2\omega\right)\ln\omega - \frac{1}{2}(1+\omega)\ln^2\omega - 5(1-\omega),$$

$$P_G(\omega) = \left(\frac{1+\omega^2}{1-\omega}\right)\left\{\ln^2\omega + \frac{11}{3}\ln\omega + \frac{67}{9} - \frac{\pi^2}{3}\right\} + 2(1+\omega)\ln\omega + \frac{40}{3}(1-\omega),$$

$$P_{N_F}(\omega) = \frac{2}{3}\left[\left(\frac{1+\omega^2}{1-\omega}\right)\left(-\ln\omega - \frac{5}{3}\right) - 2(1-\omega)\right],$$

$$P_A(\omega) = 2\left(\frac{1+\omega^2}{1+\omega}\right)\int_{\omega/1+\omega}^{1/1+\omega} \frac{dk}{k}\ln\frac{1-k}{k} + 2(1+\omega)\ln\omega + 4(1-\omega),$$

$$F_{qq}(\omega) = \frac{20}{9\omega} - 2 + 6\omega - \frac{56}{9}\omega^2 + \left(1 + 5\omega + \frac{8}{3}\omega^2\right)\ln\omega - (1+\omega)\ln^2\omega,$$

$$F_{qg}^1(\omega) = 4 - 9\omega - (1-4\omega)\ln\omega - (1-2\omega)\ln^2\omega + 4\ln(1-\omega)$$

$$+\left[2\ln^2\left(\frac{1-\omega}{\omega}\right)-4\ln\left(\frac{1-\omega}{\omega}\right)-\frac{2}{3}\pi^2+10\right]P_{qg}^1(\omega),$$

$$F_{qg}^2(\omega)=\frac{182}{9}+\frac{14}{9}\omega+\frac{40}{9\omega}+\left(\frac{136}{3}\omega-\frac{38}{3}\right)\ln\omega-4\ln(1-\omega)-(2+8\omega)\ln^2\omega+$$

$$\left[-\ln^2\omega+\frac{44}{3}\ln\omega-2\ln^2(1-\omega)+2\ln(1-\omega)+\frac{\omega^2}{3}-\frac{218}{9}\right]P_{qg}^1(\omega)+2P_{qg}^1(-\omega)\int_{\omega/1+\omega}^{1/1+\omega}\frac{dz}{z}\ln\frac{1-z}{z},$$

$$P_{qg}^1(\omega)=\omega^2+(1-\omega)^2,$$

$$A(\omega)=C_F^2.A_1(\omega)+C_F.C_G.A_2(\omega)+C_F.T_R.N_F.A_3(\omega),$$

$$A_1(\omega)=-\frac{5}{2}-\frac{7}{2}\omega+\left(2+\frac{7}{2}\omega\right)\ln\omega+\left(-1+\frac{1}{2}\omega\right)\ln^2\omega-2\omega\ln(1-\omega)$$
$$+\{-3\ln(1-\omega)-\ln^2(1-\omega)\}\frac{1+(1-\omega)^2}{\omega},$$

$$A_2(\omega)=\frac{28}{9}+\frac{65}{18}\omega+\frac{44}{9}\omega^2+\left(-12-5\omega-\frac{8}{3}\omega^2\right)\ln\omega+(4+\omega)\ln^2\omega+2\omega\ln(1-\omega)$$
$$+\left(-2\ln\omega\ln(1-\omega)+\frac{1}{2}\ln^2\omega+\frac{11}{3}\ln(1-\omega)+\ln^2(1-\omega)-\frac{1}{6}\pi^2+\frac{1}{2}\right)\frac{1+(1-\omega)^2}{\omega}$$
$$-\frac{1+(1+\omega)^2}{\omega}\int_{\omega/1+\omega}^{1/1+\omega}\frac{dz}{z}\ln\frac{1-z}{z}$$

and

$$A_3(\omega)=-\frac{4}{3}\omega-\left(\frac{20}{9}+\frac{4}{3}\ln(1-\omega)\right)\left(\frac{1+(1-\omega)^2}{\omega}\right).$$

At low-x, for x→0, only gluon splitting function matters [51, 52, 84, 91]. So keeping the full form of other splitting functions, we make some approximation of the splitting function $P^2_{gg}(\omega)$, retaining only its leading term as x→0 [52, 91], i.e., we take $P_{gg}^2(\omega)\cong\frac{52}{3}\frac{1}{\omega}$.

Applying Regge behaviour of structure functions and the relation between singlet and gluon structure functions as given in Chapter 3 (equations (3.7) to (3.9) and equation (3.13)), we get the solution of spin-independent

DGLAP evolution equations for singlet, non-singlet and gluon structure functions in NLO respectively

$$F_2^S(x,t) = C\, t^{H_4(x)}, \qquad (4.4)$$

$$F_2^{NS}(x,t) = C\, t^{H_5(x)} \qquad (4.5)$$

and

$$G(x,t) = C\, t^{H_6(x)}, \qquad (4.6)$$

where C is a constant of integration,

$$H_4(x) = \frac{2}{\beta_0} \cdot f_5(x) + T_0 \cdot \frac{2}{\beta_0} \cdot f_6(x),$$

$$H_5(x) = \frac{2}{\beta_0} \cdot f_7(x) + T_0 \cdot \frac{2}{\beta_0} \cdot f_8(x),$$

$$H_6(x) = \frac{2}{\beta_0} \cdot f_9(x) + T_0 \cdot \frac{2}{\beta_0} \cdot f_{10}(x),$$

$$f_5(x) = \frac{2}{3}\{3 + 4\ln(1-x)\} + \frac{4}{3}\int_x^1 \frac{d\omega}{1-\omega}\{(1+\omega^2)\omega^{\lambda_S} - 2\} + N_f \int_x^1 \left\{\omega^2 + (1-\omega^2)K\left(\frac{x}{\omega}\right)\right\}\omega^{\lambda_S}\, d\omega,$$

$$f_6(x) = (x-1)\int_0^1 f(\omega)\, d\omega + \int_x^1 f(\omega)\omega^{\lambda_S}\, d\omega + \int_x^1 F_{qq}^S(\omega)\omega^{\lambda_S}\, d\omega + \int_x^1 F_{qg}^S(\omega)K\left(\frac{x}{\omega}\right)\omega^{\lambda_S}\, d\omega,$$

$$f_7(x) = \frac{2}{3}\{3 + 4\ln(1-x)\} + \frac{4}{3}\int_x^1 \frac{d\omega}{1-\omega}\{(1+\omega^2)\omega^{\lambda_{NS}} - 2\},$$

$$f_8(x) = (x-1)\int_0^1 f(\omega)d\omega + \int_x^1 f(\omega)\omega^{\lambda_{NS}}\, d\omega,$$

$$f_9(x) = 6\times\left(\frac{11}{12} - \frac{N_f}{18} + \ln(1-x)\right) + 6\times\int_x^1 d\omega\left\{\frac{(\omega^{\lambda_G+1}-1)}{1-\omega} + \left(\omega(1-\omega) + \frac{1-\omega}{\omega}\right)\omega^{\lambda_G}\right.$$

$$\left. + \frac{2}{9}\left(\frac{1+(1-\omega)^2}{\omega}\right)\frac{\omega^{\lambda_G}}{K\left(\frac{x}{\omega}\right)}\right\},$$

and

$$f_{10}(x) = \int_{x}^{1} \left[ \frac{52}{3} \omega^{\lambda_G - 1} + A(\omega) \frac{\omega^{\lambda_G}}{K\left(\frac{x}{\omega}\right)} \right] d\omega.$$

For possible solutions in NLO, we have taken $T(t) = \left(\frac{\alpha_s(t)}{2\pi}\right)$ and the expression for T(t) upto LO correction with the assumption $T^2(t) = T_0 T(t)$ [52, 65, 90], where $T_0$ is a numerical parameter. But $T_0$ is not arbitrary. We choose $T_0$ such that difference between $T^2(t)$ and $T_0 T(t)$ is minimum (Figures 4.1 and 4.2).

The t and x-evolution equations of singlet, non-singlet and gluon structure functions from equations (4.4) to (4.6) are given by

$$F_2^S(x,t) = F_2^S(x,t_0) \left(\frac{t}{t_0}\right)^{H_4(x)}, \tag{4.7}$$

$$F_2^{NS}(x,t) = F_2^{NS}(x,t_0) \left(\frac{t}{t_0}\right)^{H_5(x)}, \tag{4.8}$$

$$F_2^S(x,t) = F_2^S(x_0,t) \, t^{\{H_4(x) - H_4(x_0)\}}, \tag{4.9}$$

$$F_2^{NS}(x,t) = F_2^{NS}(x_0,t) \, t^{\{H_5(x) - H_5(x_0)\}}, \tag{4.10}$$

$$G(x,t) = G(x,t_0) \left(\frac{t}{t_0}\right)^{H_6(x)} \tag{4.11}$$

and

$$G(x,t) = G(x_0,t) \, t^{\{H_6(x) - H_6(x_0)\}}. \tag{4.12}$$

The t and x-evolution equations of deuteron and proton structure functions from equations (4.7) to (4.10) are respectively

$$F_2^d(x,t) = F_2^d(x,t_0) \left(\frac{t}{t_0}\right)^{H_4(x)}, \tag{4.13}$$

$$F_2^d(x,t) = F_2^d(x_0,t) \, t^{\{H_4(x) - H_4(x_0)\}}, \tag{4.14}$$

$$F_2^p(x,t) = F_2^p(x,t_0) \left( \frac{3t^{H_5(x)} + 5t^{H_4(x)}}{3t_0^{H_5(x)} + 5t_0^{H_4(x)}} \right), \tag{4.15}$$

and

$$F_2^p(x,t) = F_2^p(x_0,t) \left( \frac{3t^{H_5(x)} + 5t^{H_4(x)}}{3t^{H_5(x_0)} + 5t^{H_4(x_0)}} \right). \tag{4.16}$$

Here $F_2^d(x,t_0)$, $F_2^p(x,t_0)$, $F_2^d(x_0,t)$ and $F_2^p(x_0,t)$ are the values of the structure functions $F_2^d(x,t)$ and $F_2^p(x,t)$ at $t = t_0$ and $x = x_0$ respectively.

Now ignoring the quark contribution to the gluon structure function we get from the evolution equation (4.3)

$$\frac{\partial G(x,t)}{\partial t} = \frac{\alpha_s(t)}{2\pi} \left\{ 6 \times \left( \frac{11}{12} - \frac{N_f}{18} + \ln(1-x) \right) G(x,t) + 6 \times I_g' \right\}$$

$$+ \left[ \frac{\alpha_s(Q^2)}{2\pi} \right]^2 \int_x^1 \left[ \frac{52}{3} \frac{1}{\omega} G\left( \frac{x}{\omega}, Q^2 \right) \right] d\omega, \tag{4.17}$$

where

$$I_g' = \int_x^1 d\omega \left[ \frac{\omega G\left( \frac{x}{\omega}, t \right) - G(x,t)}{1 - \omega} + \left( \omega(1-\omega) + \frac{1-\omega}{\omega} \right) G\left( \frac{x}{\omega}, t \right) \right].$$

Similarly we get the t and x-evolution equations for the gluon structure function ignoring the quark contribution in NLO and are given by

$$G(x,t) = G(x,t_0) \left( \frac{t}{t_0} \right)^{B_2(x)} \tag{4.18}$$

and

$$G(x,t) = G(x_0,t) \, t^{\{B_2(x) - B_2(x_0)\}}. \tag{4.19}$$

Here

$$B_2(x) = \frac{2}{\beta_0} P_1(x) + T_0 \frac{2}{\beta_0} P_2(x),$$

$$P_1(x) = 6 \times \left( \frac{11}{12} - \frac{N_f}{18} + \ln(1-x) \right) + 6 \times \int_x^1 d\omega \left[ \frac{\left(\omega^{\lambda_G+1} - 1\right)}{1-\omega} + \left( \omega(1-\omega) + \frac{1-\omega}{\omega} \right) \omega^{\lambda_G} \right]$$

and

$$P_2(x) = \int_x^1 \left[ \frac{52}{3} \omega^{\lambda_G - 1} \right] d\omega.$$

We get the solution of spin-independent coupled DGLAP evolution equations for singlet and gluon structure functions in NLO at low-x [Appendix C] respectively as

$$F_2^S(x,t) = C\left( t^{g_3} + t^{g_4} \right) \tag{4.20}$$

and

$$G(x,t) = C(F_3 t^{g_3} + F_4 t^{g_4}), \tag{4.21}$$

Where $g_3 = \dfrac{-(u_2-1) + \sqrt{(u_2-1)^2 - 4v_2}}{2}$, $g_4 = \dfrac{-(u_2-1) - \sqrt{(u_2-1)^2 - 4v_2}}{2}$, $u_2 = 1 - P_2 - S_2$,

$v_2 = S_2.P_2 - Q_2.R_2$, $F_3 = (g_3 - P_2)/Q_2$, $F_4 = (g_4 - P_2)/Q_2$,

$$P_2 = \frac{3}{2} A_f \left[ \frac{2}{3} \{3 + 4\ln(1-x)\} + \frac{4}{3} \int_x^1 \frac{d\omega}{1-\omega} \{(1+\omega^2)\omega^{\lambda_S} - 2\} \right.$$

$$\left. + T_0 \left\{ (x-1) \int_0^1 f(\omega) d\omega + \int_x^1 f(\omega) \omega^{\lambda_S} d\omega + \int_x^1 F_{qq}^S(\omega) \omega^{\lambda_S} d\omega \right\} \right],$$

$$Q_2 = \frac{3}{2} A_f \left[ N_f \int_x^1 \{\omega^2 + (1-\omega)^2\} \omega^{\lambda_G} d\omega + T_0 \int_x^1 F_{qg}^S(\omega) \omega^{\lambda_G} d\omega \right],$$

$$R_2 = \frac{3}{2} A_f \left[ \frac{4}{3} \int_x^1 \left( \frac{1 + (1-\omega)^2}{\omega} \right) \omega^{\lambda_S} d\omega + T_0 \int_x^1 A(\omega) \omega^{\lambda_S} d\omega \right]$$

and

$$S_2 = \frac{3}{2} A_f \left[ 6 \left( \left( \frac{11}{12} - \frac{N_f}{18} \right) + \ln(1-x) \right) + 6 \int_x^1 d\omega \left\{ \frac{(\omega^{\lambda_G+1} - 1)}{1-\omega} + \left( \omega(1-\omega) + \frac{1-\omega}{\omega} \right) \omega^{\lambda_G} \right\} \right.$$
$$\left. + T_0 \int_x^1 \frac{52}{3} \omega^{\lambda_G - 1} d\omega \right]$$

Applying initial conditions, at $x = x_0$, $F_2^S(x,t) = F_2^S(x_0, t)$ and $G(x,t) = G(x_0, t)$, and at $t = t_0$, $F_2^S(x,t) = F_2^S(x, t_0)$ and $G(x,t) = G(x, t_0)$, the t and x-evolution equations for the singlet and gluon structure functions in NLO come as

$$F_2^S(x, t) = F_2^S(x, t_0) \left( \frac{t^{g_3} + t^{g_4}}{t_0^{g_3} + t_0^{g_4}} \right), \tag{4.22}$$

$$F_2^S(x, t) = F_2^S(x_0, t) \left( \frac{t^{g_3} + t^{g_4}}{t^{g_{30}} + t^{g_{40}}} \right), \tag{4.23}$$

$$G(x,t) = G(x, t_0) \left( \frac{F_3 t^{g_3} + F_4 t^{g_4}}{F_3 t_0^{g_3} + F_4 t_0^{g_4}} \right) \tag{4.24}$$

and

$$G(x,t) = G(x_0, t) \left( \frac{F_3 t^{g_3} + F_4 t^{g_4}}{F_{30} t^{g_{30}} + F_{40} t^{g_{40}}} \right), \tag{4.25}$$

where $g_{30}, g_{40}, F_{30}$ and $F_{40}$ are the values of $g_3, g_4, F_3$ and $F_4$ at $x = x_0$. The t and x-evolution equations of deuteron structure function from the equations (4.22) and (4.23) are respectively

$$F_2^d(x, t) = F_2^d(x, t_0) \left( \frac{t^{g_3} + t^{g_4}}{t_0^{g_3} + t_0^{g_4}} \right) \tag{4.26}$$

and

$$F_2^d(x, t) = F_2^d(x_0, t) \left( \frac{t^{g_3} + t^{g_4}}{t^{g_{30}} + t^{g_{40}}} \right). \tag{4.27}$$

## 4.2 Results and Discussion

We have compared our result of deuteron and proton structure functions with the data sets measured by the NMC [71] in muon-deuteron DIS from the merged data sets at incident momenta 90, 120, 200 and 280 GeV² and also with the data sets measured by the Fermilab E665 [72] Collaboration in muon-deuteron DIS at an average beam energy of 470 GeV². For our phenomenological analysis we have considered the data sets of deuteron and proton structure functions in the range what the NMC and E665 collaborations provide at low-x. We considered the QCD cut-off parameter as $\Lambda_{\overline{MS}}(N_f = 4) = 323$ MeV for $\alpha_s(M_z^2) = 0.119 \pm 0.002$ [73]. We have compared our result of t-evolution of gluon structure function in NLO with GRV1998LO [78] and GRV1998NLO [78] global parametrizations and the result of x-evolution with MRST2001 [73], MRST2004 [75], GRV1998LO [78] and GRV1998NLO [78] global parametrizations at medium to high-Q² range. Along with the NLO results, we also presented our LO results from Chapter 3.

The graphs 'our result' represent the best fit graph of our work with different experimental data sets and parameterization graphs. Data points at lowest-Q² values are taken as input to test the t-evolution equations and data points at x < 0.1 is taken as input to test the x-evolution equations. The comparisons of our results with experimental data sets and parameterization graphs are made for $\lambda_S = \lambda_{NS} = \lambda_G$ = constant. Since the value of $\lambda_S$, $\lambda_{NS}$ and $\lambda_G$ should be close to 0.5 in low-x. We have taken $\lambda_S = \lambda_{NS} = \lambda_G = 0.5$.

In Figure 4.1, we plot $T(t)^2$ and $T_0 T(t)$ against Q² in the Q² range $0 \leq Q^2 \leq 30$ GeV² as required by our data used for deuteron and proton structure functions. Here we observe that for $T_0 = 0.108$, errors become minimum in the Q² range $0.75 \leq Q^2 \leq 27$ GeV².

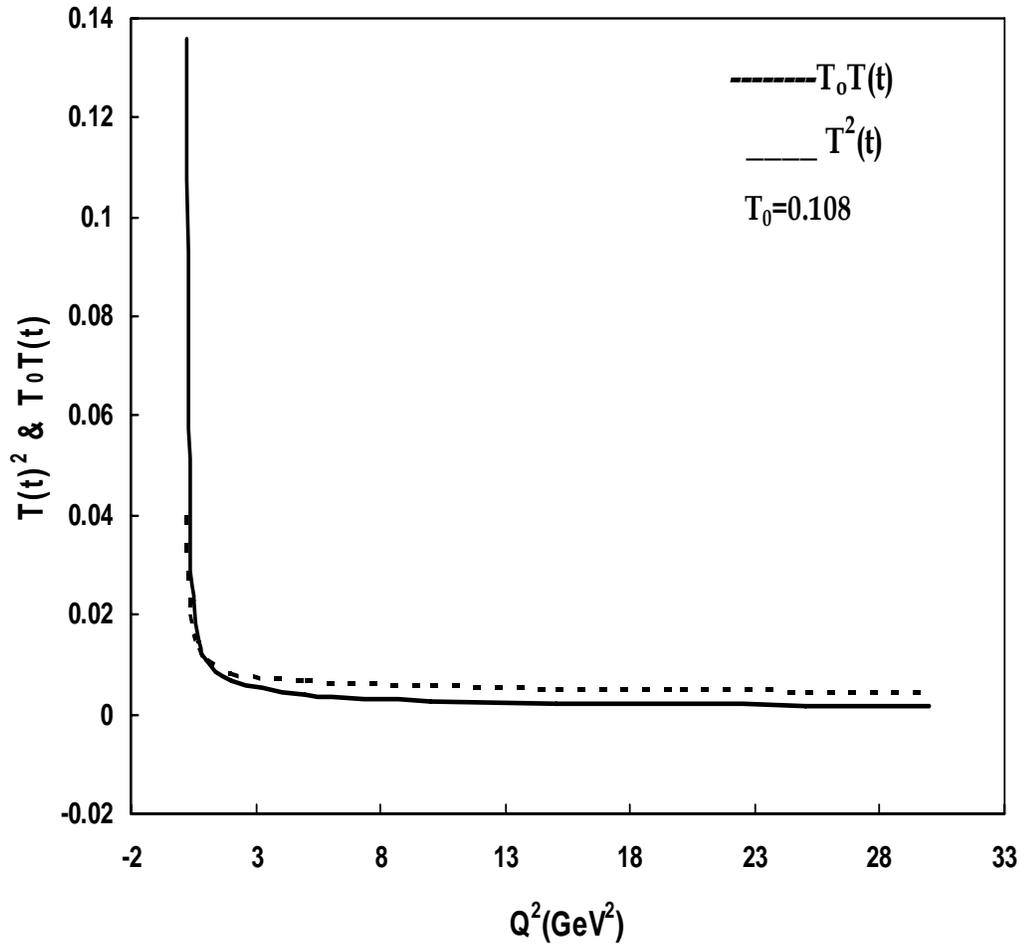

**Figure 4.1:** The variation of $T(t)^2$ and $T_0T(t)$ with $Q^2$.

In Figure 4.2, we plot $T(t)^2$ and $T_0T(t)$ against $Q^2$ in the $Q^2$ range $0 \leq Q^2 \leq 200$ GeV$^2$ as required by our data used for gluon structure function. Here we observe that for $T_0 = 0.05$, errors become minimum in the $Q^2$-range $10 \leq Q^2 \leq 200$ GeV$^2$.

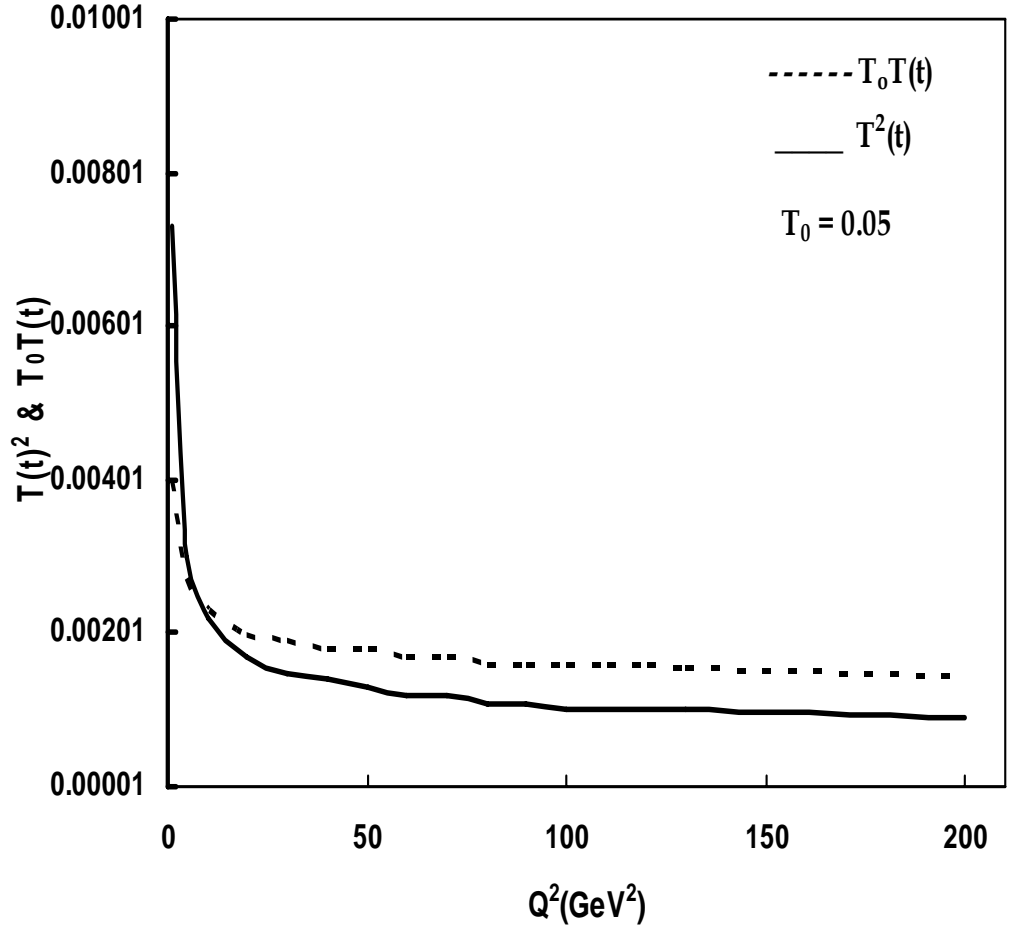

**Figure 4.2:** The variation of T(t)² and $T_0$T(t) with Q².

Figures 4.3(a-b) represent our result of t-evolution of deuteron structure function for the representative values of x in NLO from equation (4.13) compared with NMC and E665 data sets. The best fitted graphs were found for $\lambda_S$ = 0.5, 21≤a≤57 and b=2. We have seen that with increase of x, K(x) decreases. For K(x) =ax$^b$, we get the best fit result of t-evolution of deuteron structure function.

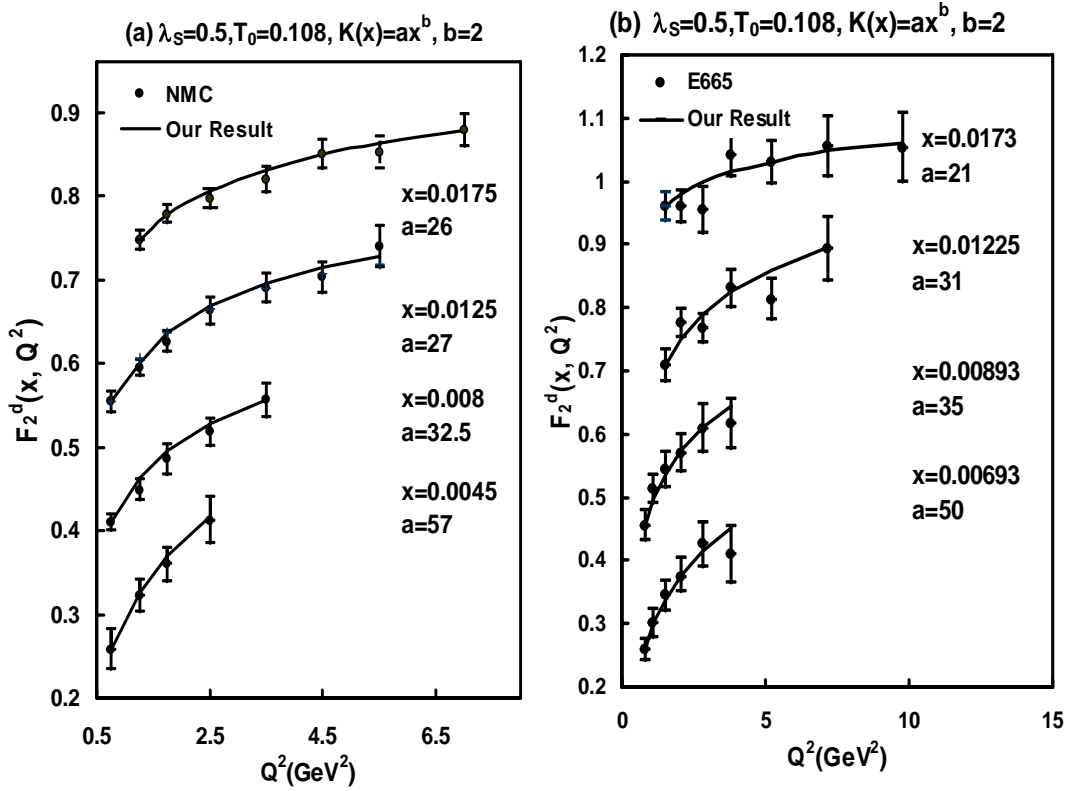

**Figure 4.3:** t-evolution of deuteron structure function in NLO for the representative values of x. Data are scaled up by +0.15i (i=0, 1, 2, 3) for NMC data set and by +0.3i (i=0, 1, 2, 3) for E665 data set starting from bottom graph.

Figures 4.4(a-b) represent our result for x-evolution of deuteron structure function for the representative values of $Q^2$ in NLO from equation (4.14) compared with NMC and E665 data sets. For $K(x) = ax^b$ and $ce^{dx}$, we get the best fit result of x-evolution of deuteron structure function. The best fitted graphs were found for $\lambda_S = 0.5$, $1 \leq a \leq 1.8$ and b=1 in the x-$Q^2$ range of our discussion. Same graphs are observed for $\lambda_S = 0.5$, $0.8 \leq c \leq 1.4$ and d=1.

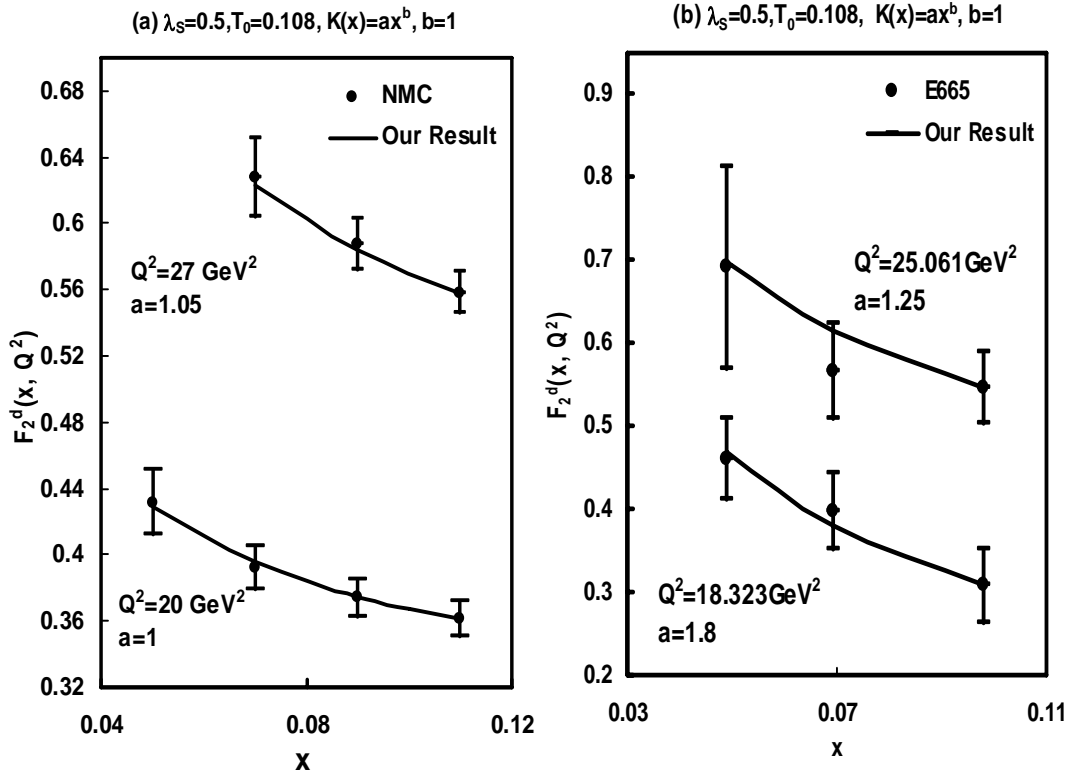

**Figure 4.4**: x -evolution of deuteron structure function in NLO for the representative values of Q². Data are scaled up by +0.3i (i=0, 1) for both NMC and E665 data sets starting from bottom graph.

Figures 4.5(a-b) represent our results for t-evolution of proton structure function for the representative values of x in NLO from equation (4.15) compared with NMC and E665 data sets. Comparison of our result with the data sets for $\lambda_S=\lambda_{NS}$= 0.5, 30≤a≤63 and b=2 gives a good consistency. But the comparison of t-evolution of proton structure function with experimental data sets does not fit well for K(x) = K and $ce^{dx}$. So we present only the results with K(x) =$ax^b$.

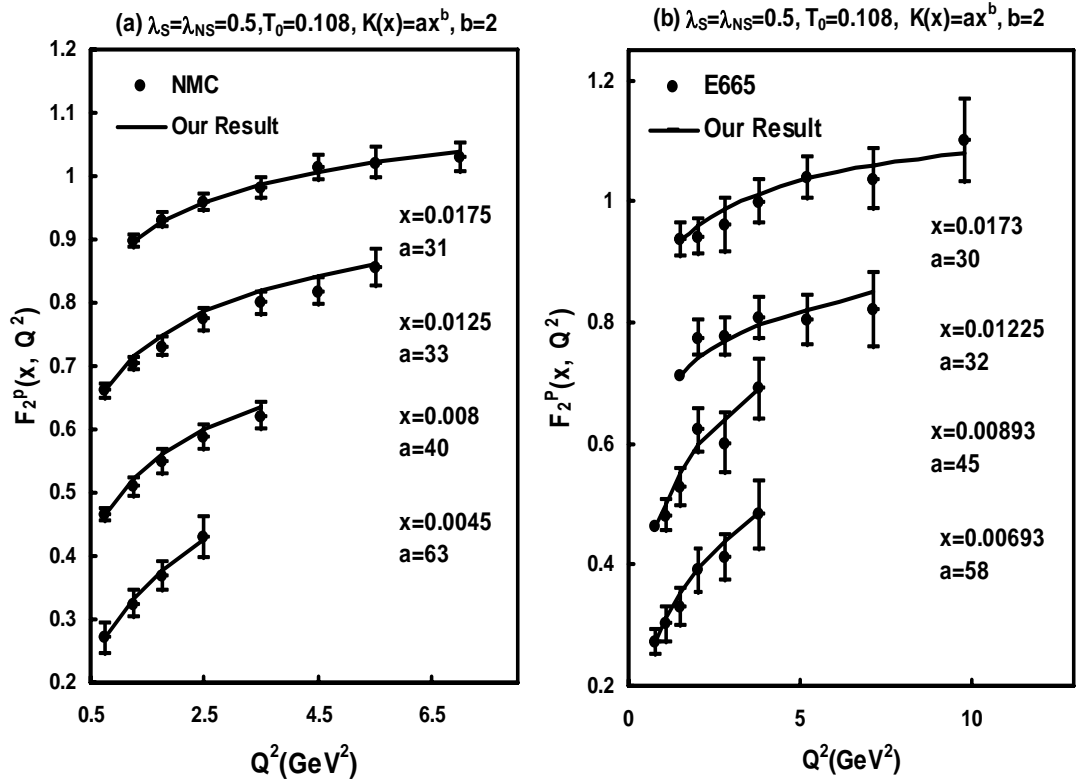

**Figure 4.5:** t-evolution of proton structure function in NLO for the representative values of x. Data are scaled up by +0.2i (i=0, 1, 2, 3) for both NMC and E665 data sets starting from bottom graph.

Figures 4.6(a-b) represent our results for x-evolution of proton structure function for the representative values of $Q^2$ in NLO from equation (4.16) compared with NMC and E665 data sets. Comparison of our result with the data sets for $\lambda_S=\lambda_{NS}= 0.5$, $5 \leq a \leq 15$ and b=2 is good in the x-$Q^2$ range of our discussion. Here also the comparison of x-evolution of proton structure function with experimental data sets does not fit well for $K(x) = K$ and $ce^{dx}$. So we present only the results with $K(x) = ax^b$.

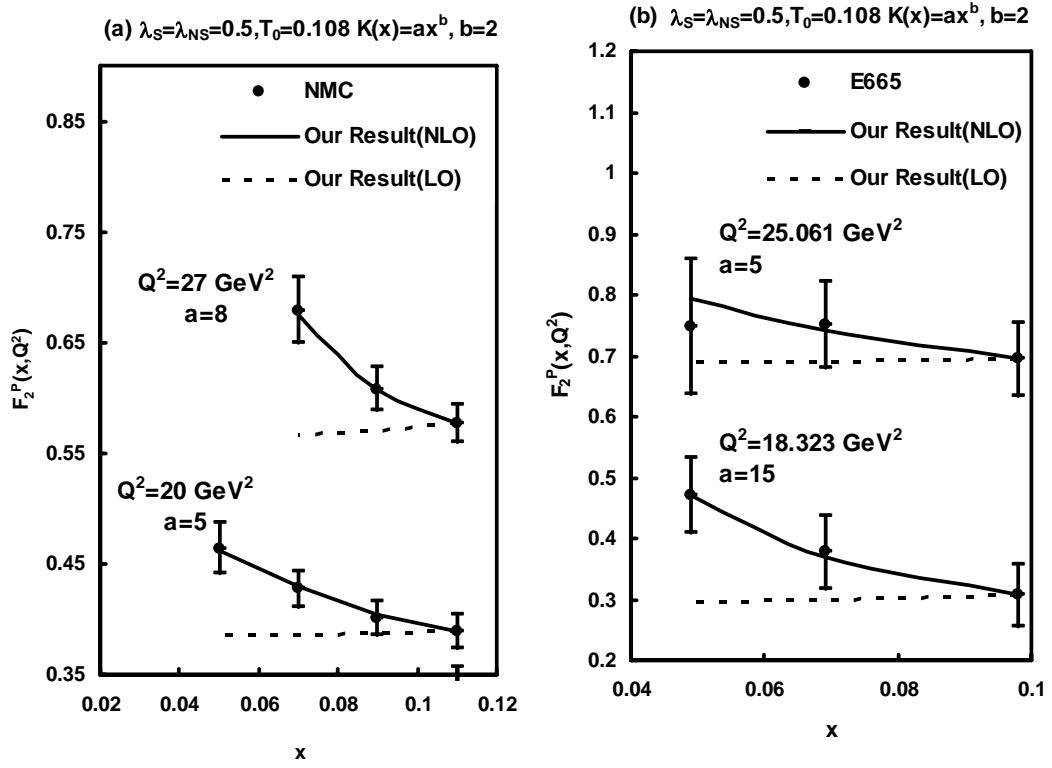

**Figure. 4.6.** x-evolution of proton structure function in NLO for the representative values of $Q^2$. Data are scaled up by +0.3i (i=0, 1) for both NMC and E665 data sets starting from bottom graph.

Figure 4.7 represents our best fit graphs for both LO and NLO results for x-evolution of deuteron structure function from equations (3.26) and (4.14) respectively with NMC data set. In case of LO, the best fitted results are obtained for $\lambda_S=0.5$, K=7, a=7, b=0.001, c=10, d=0.1 at $Q^2=20$ GeV$^2$ and for $\lambda_S=0.5$, K=6.5, a=6.5, b=0.001, c=8.5, d=0.1 at $Q^2=27$ GeV$^2$. In case of NLO, best fitted results are obtained for $\lambda_S=0.5$, $T_0=0.108$, a=1, b=1, c=0.8, d=1 at $Q^2=20$ GeV$^2$ and for $\lambda_S=0.5$, $T_0=0.108$, a=1.05, b=1, c=0.85, d=1 at $Q^2=27$ GeV$^2$. It is seen that NLO plots in the figure have good exponential look than the LO plots.

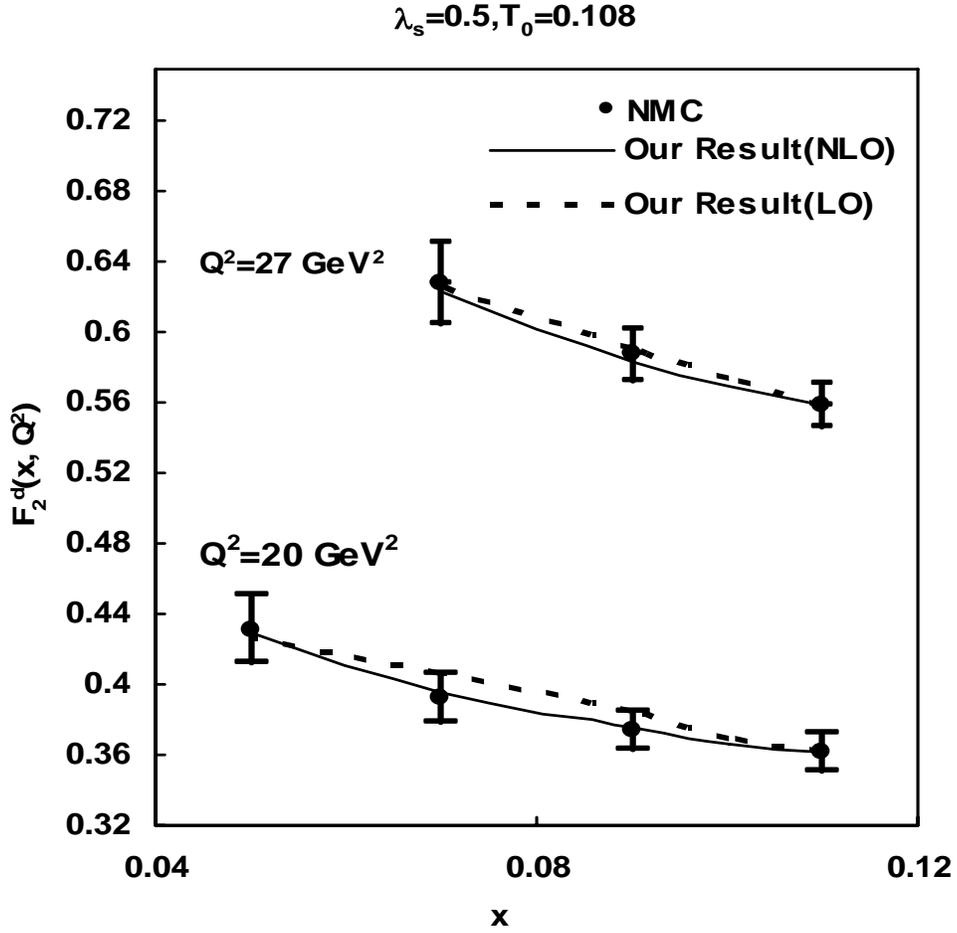

**Figure 4.7:** x-evolution of deuteron structure function from both our LO and NLO results for the representative values of $Q^2$ with NMC data set. Data are scaled up by +0.3i (i=0, 1) starting from bottom graph.

Figures 4.8(a-f) represent the sensitivity of the parameters $T_0$, $\lambda_S$, a, b, c and d in NLO. We have presented the results of sensitivity with our best fitted graph of x-evolution of deuteron structure function from equation (4.14) with the data set of NMC. The curves shift upwards when values of $T_0$, a, c, or d are increased and moves downwards when values of $T_0$, a, c, or d are decreased. If the values of $\lambda_S$ or b increased or decreased the curves goes downwards or upwards respectively. We found the ranges of the parameters as $0.128 \leq T_0 \leq 0.088$, $0.4 \leq \lambda_S \leq 0.6$, $1.1 \leq a \leq 0.9$, $1.15 \leq b \leq 0.85$, $0.87 \leq c \leq 0.73$ and $1.1 \leq d \leq 0.9$ for best fitting with the data.

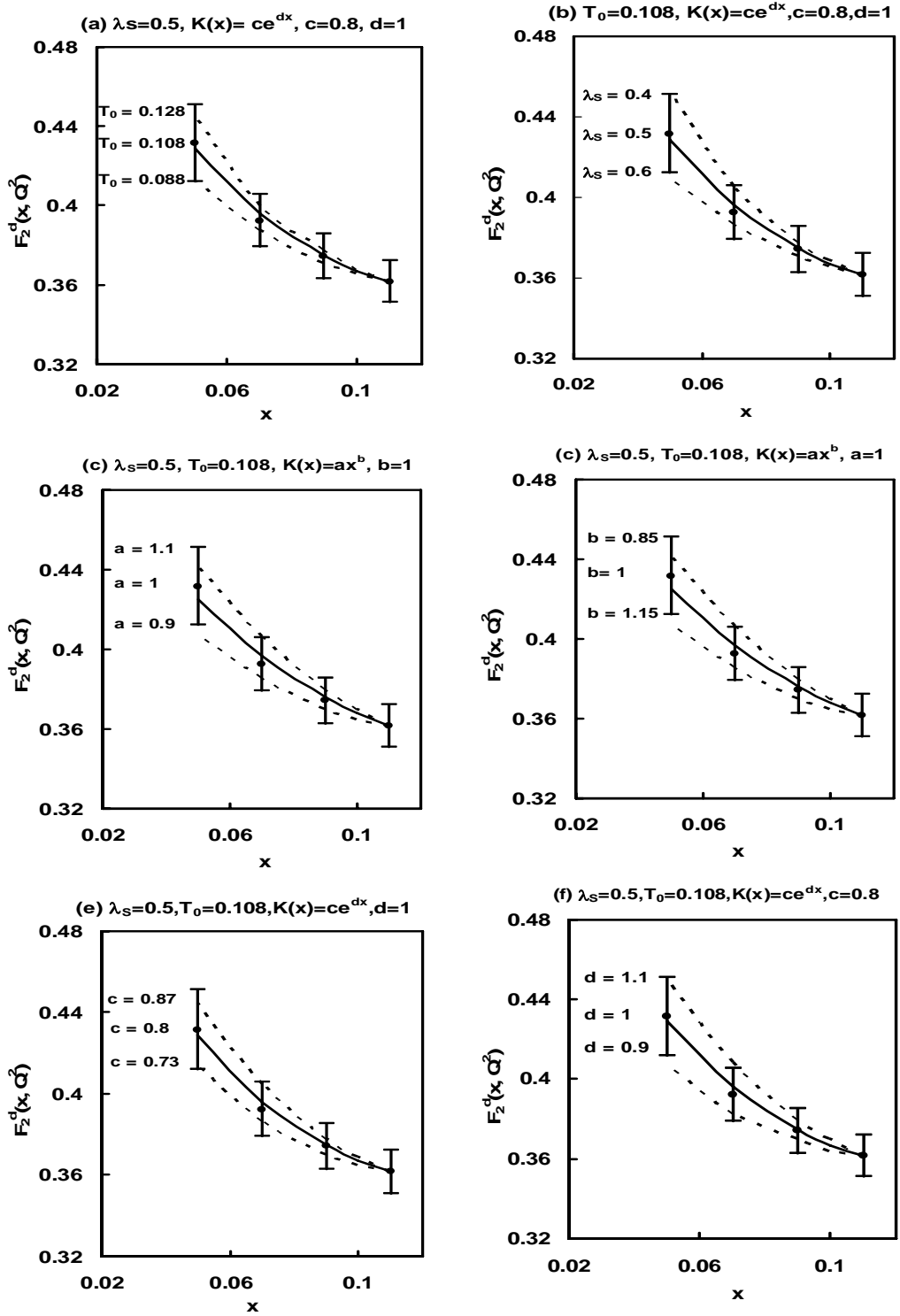

**Figure 4.8**: Sensitivity of the parameters $T_0$, $\lambda_S$, a, b, c and d respectively at $Q^2 = 20$ GeV$^2$ with the best fit graph of our results with NMC data.

We found that the gluon structure function remains almost same for b<0.00001 and d<0.00001. So, we have chosen b = d = 0.00001 for our calculation and the best fit graphs are observed by varying the values of K, a and c. Figure 4.9 represents our result of t-evolution of gluon structure function in NLO from equation (4.11) with GRV1998NLO global parameterization at $x=10^{-4}$ for K(x) = K. We have also compared our results for K(x) = $ax^b$ and $ce^{dx}$. K=a=c=0.4 gives the best fitted graphs.

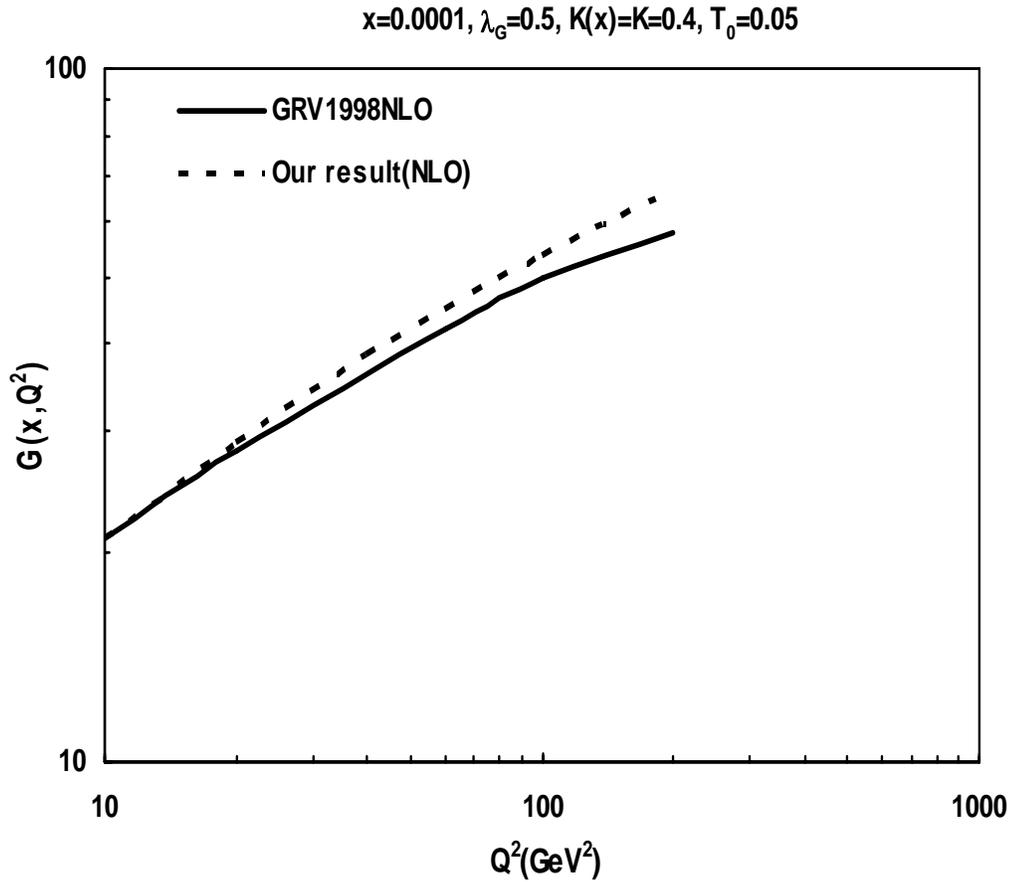

**Figure 4.9:** t-evolution of gluon structure function in NLO for the representative values of x presented with GRV1998NLO global parameterization at $x=10^{-4}$.

Figures 4.10(a-b) represent our result of x-evolution of gluon structure function in NLO from equation (4.12) for K(x) = K with GRV1998NLO global

parameterization at $Q^2$ = 20 and 40 GeV² respectively. For K = - 0.34 at $Q^2$ = 20 GeV² and K = - 0.27 at $Q^2$ =40 GeV², we get the best fitted graphs. When compared our results for $K(x) = ax^b$ and $ce^{dx}$, we found the same graph with a=c= - 0.34 at $Q^2$ = 20 GeV² and a=c= - 0.27 at $Q^2$ =40 GeV².

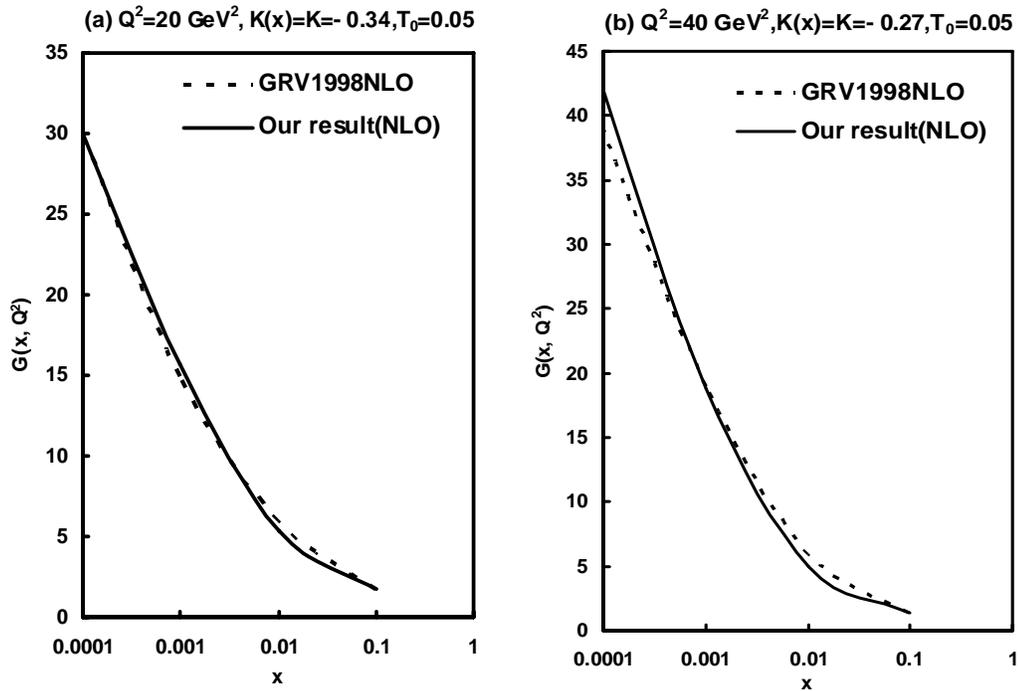

**Figure 4.10:** x-evolution of gluon structure function in NLO with GRV1998NLO global parameterization for the representative values of $Q^2$.

Figures 4.11(a-b) represent our result of x-evolution of gluon structure function in NLO from equation (4.12) for $K(x) = K$ with GRV1998NLO global parameterization at $Q^2$ = 60 and 100 GeV² respectively. For K = - 0.27 at $Q^2$ =60 GeV² and $Q^2$ =100 GeV², we get the best fitted graphs. When compared our results for $K(x) = ax^b$ and $ce^{dx}$, we found the same graph with a=c= - 0.27 at $Q^2$ =60 GeV² and $Q^2$ =100 GeV².

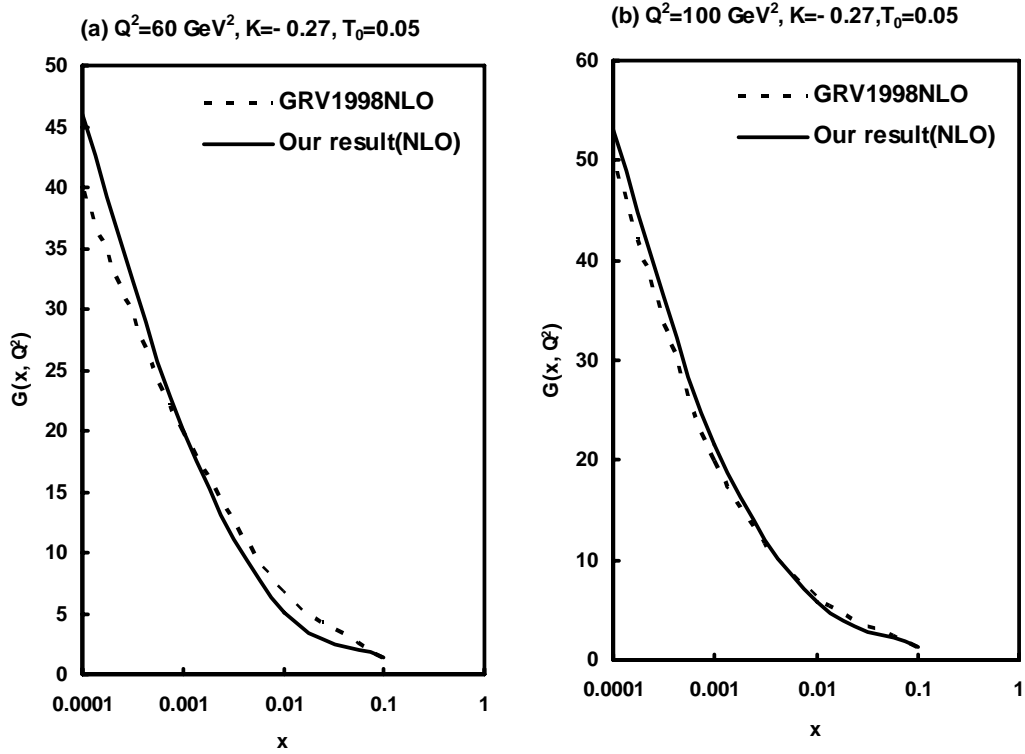

**Figure 4.11:** x-evolution of gluon structure function in NLO with GRV1998NLO global parameterization for the representative values of $Q^2$.

Figures 4.12(a-b) represent our result of x-evolution of gluon structure function in NLO from equation (4.12) for $K(x) = K$ with MRST2004 and GRV1998LO global parameterizations at $Q^2 = 100$ and $20$ GeV$^2$ respectively. Along with NLO results we have also presented our LO results from chapter 3. We get the best fitted graphs for NLO with $K = -0.3$ for $Q^2 = 100$ GeV$^2$ and with $K = -0.19$ for $Q^2 = 20$ GeV$^2$. We also compared our results for $K(x) = ax^b$ and $ce^{dx}$ and found the same graphs as for $K(x) = K$. For $a=c= -0.3$ at $Q^2 = 100$ GeV$^2$ and $a=c= -0.19$ at $Q^2 = 20$ GeV$^2$, we get the best fitted graphs.

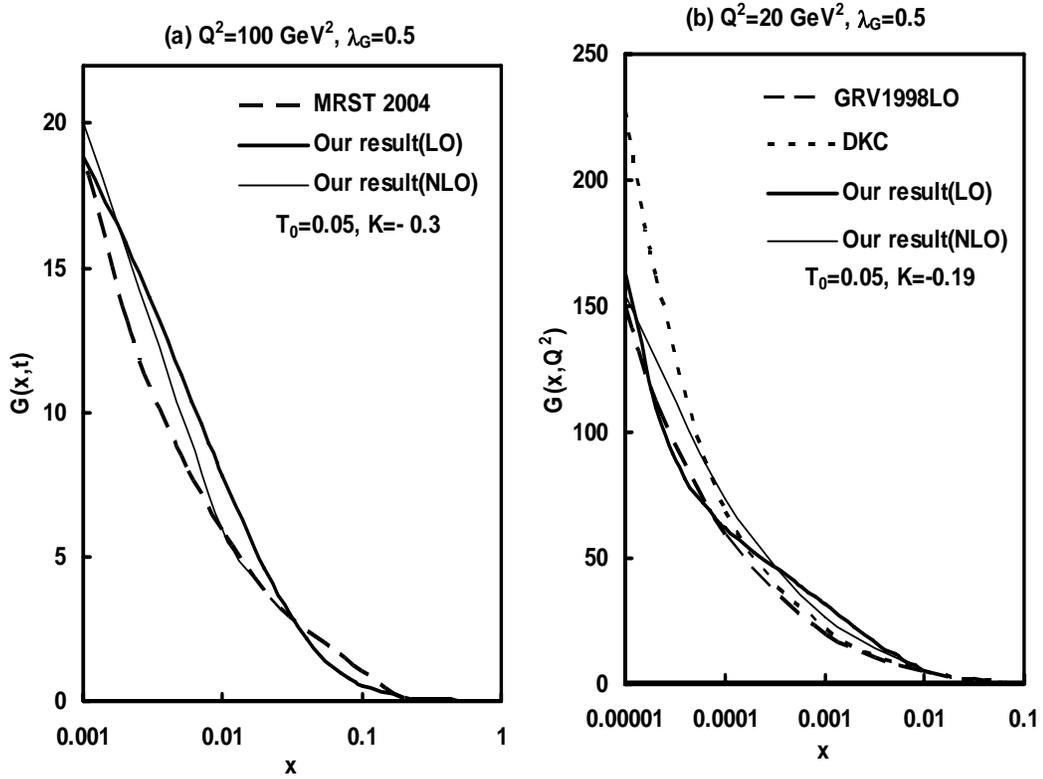

**Figure 4.12:** x-evolution of gluon structure function both in LO and NLO with MRST2004 and GRV1998LO global parameterizations for the representative values of $Q^2$.

Figures 4.13(a-b) represent our result of x-evolution of gluon structure function in NLO from equation (4.12) for $K(x) = K$ with GRV1998LO global parameterization at $Q^2 = 40$ and $80$ GeV$^2$ respectively. $K = -0.19$ corresponds the best fitted graphs for both $Q^2 =40$ GeV$^2$ and $Q^2 =80$ GeV$^2$. We also compared our results for $K(x) = ax^b$ and $ce^{dx}$ and found the same graphs as for $K(x) = K$. In this case for $a=c= -0.19$ at $Q^2 =40$ GeV$^2$ and $Q^2 =80$ GeV$^2$, we get the best fitted graphs. From both the plots of LO and NLO from our result it is seen that the NLO results follows the global parameterization graphs more closely than the LO results.

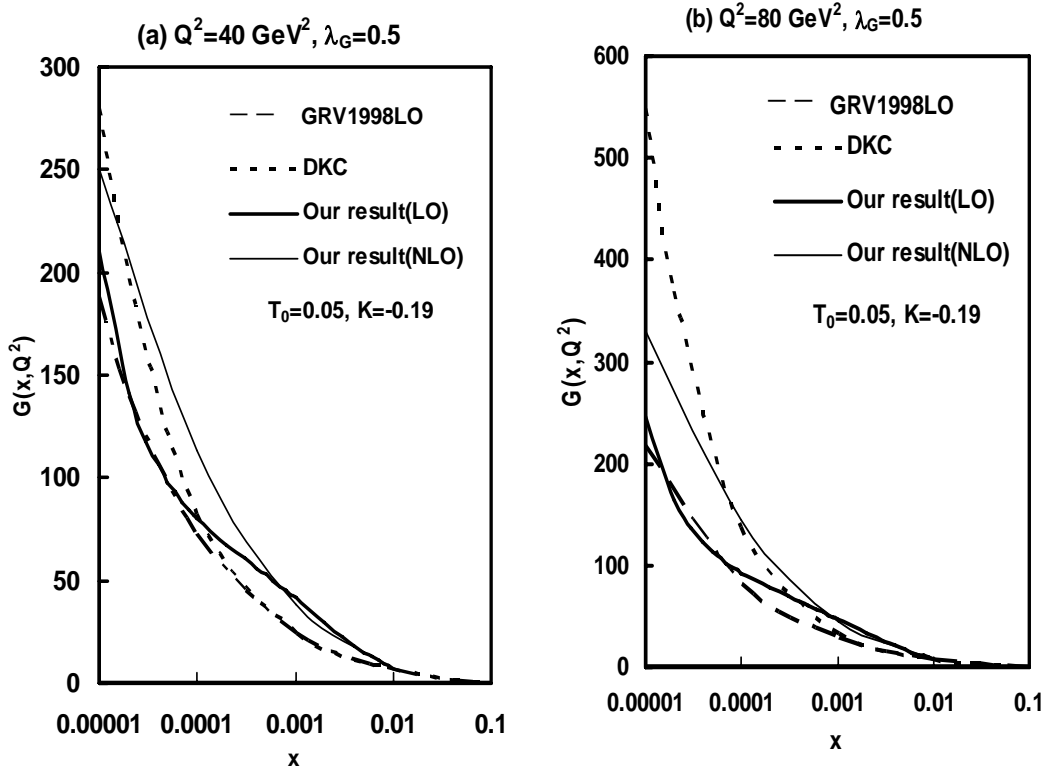

**Figure 4.13:** x-evolution of gluon structure function both in LO and NLO with GRV1998LO global parameterizations for the representative values of $Q^2$.

Figures 4.14(a-d) represent the sensitivity of the parameters $\lambda_G$, K, b and d respectively. Taking the best fit graph of our result of x-evolution of gluon structure function from equation (4.12) in NLO compared with GRV1998NLO global parameterization at $Q^2 = 100$ GeV$^2$, the ranges of the parameters are found as $0.48 \leq \lambda_G \leq 0.52$, $-0.2 \leq K \leq -0.34$, $0.00001 \leq b \leq 0.01$, and $0.00001 \leq d \leq 0.5$. We also checked the ranges of the parameters a and c and found the same graph as that for K. The ranges are found as $-0.2 \leq a \leq -0.34$ and $-0.2 \leq c \leq -0.34$.

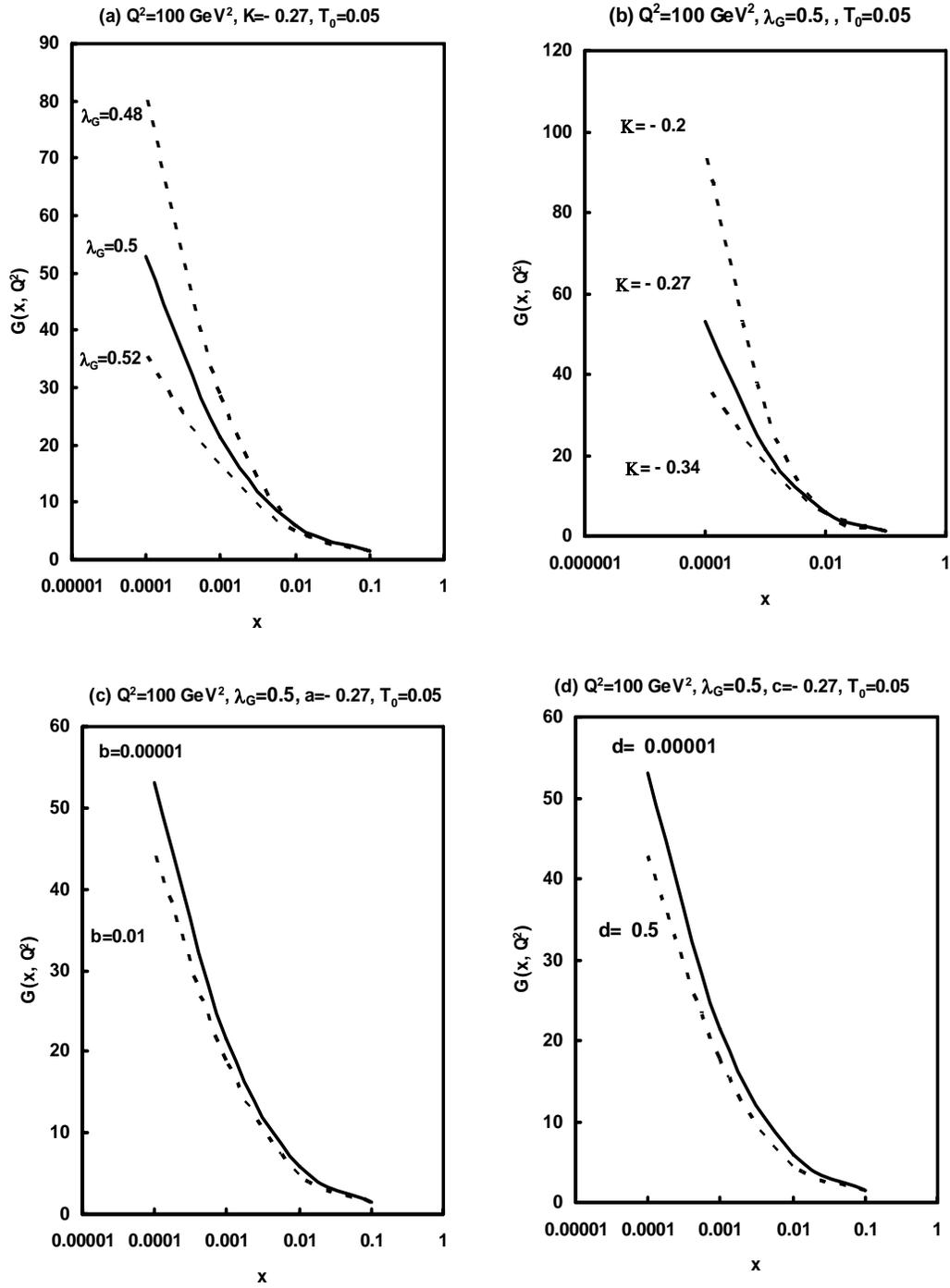

**Figure 4.14:** Sensitivity of the parameters $\lambda_G$, K, b and d respectively at $Q^2$ = 100 GeV$^2$ with the best fit graph of our results for x-evolution of gluon structure function in NLO with GRV1998NLO global parameterization.

The t and x-evolutions of singlet and gluon structure functions at low-x, obtained by solving coupled DGLAP evolution equations in NLO applying Regge behaviour of structure functions are given by equations (4.22) to (4.25) respectively. Along with NLO results we have also presented our LO results from chapter 3.

Figures 4.15(a-b) represent our result of t-evolution of deuteron structure function in NLO from equation (4.26) compared with NMC and E665 data sets respectively. In figure 4.15(a) we compared our result for t-evolution of deuteron structure function in NLO keeping $\lambda_S$ fixed at 0.5 with NMC data set and the best fitted results are observed by varying the value of $\lambda_G$. In the range $0.0045 \leq x \leq 0.0175$, $0.62 \leq \lambda_G \leq 0.85$ correspond to the best fitted graphs. In figure 4.15(b), we compared our result for t-evolution of deuteron structure function in NLO keeping $\lambda_S$ fixed at 0.5 with E665 data set. In the range $0.00693 \leq x \leq 0.0173$, $0.63 \leq \lambda_G \leq 0.9$ correspond the best fitted graphs.

Figures 4.16(a-b) represent our result of x-evolution of deuteron structure function in NLO from equation (4.27) compared with NMC and E665 data sets respectively. In figure 4.16(a), $\lambda_S=0.5$, $\lambda_G=0.7$ and $T_0=0.108$ correspond the best fitted graphs for the range $20 GeV^2 \leq Q^2 \leq 27 GeV^2$. In figure 4.16(b), we compared our result for x-evolution of deuteron structure function in NLO keeping $\lambda_G$ fixed at 0.5 with E665 data set and the best fitted results are observed by varying the value of $\lambda_S$. In the range $18.323 GeV^2 \leq Q^2 \leq 25.061 GeV^2$, $0.3 \leq \lambda_S \leq 0.6$ correspond to the best fitted graphs.

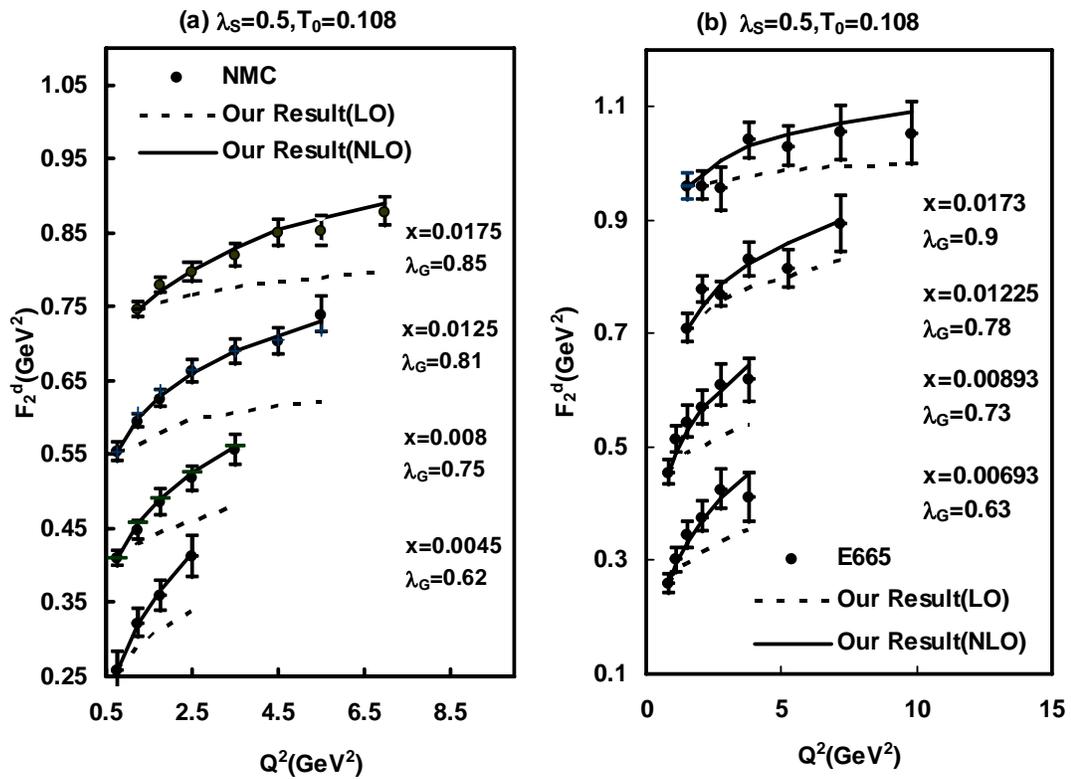

**Figure 4.15**: t-evolution of deuteron structure function both in LO and NLO at low-x which is obtained by solving coupled equations, compared with NMC and E665 data sets for the representative values of x. Data are scaled up by +0.15i (i=0, 1, 2, 3) for NMC data set and by +0.3i (i=0, 1, 2, 3) for E665 data set starting from bottom graph.

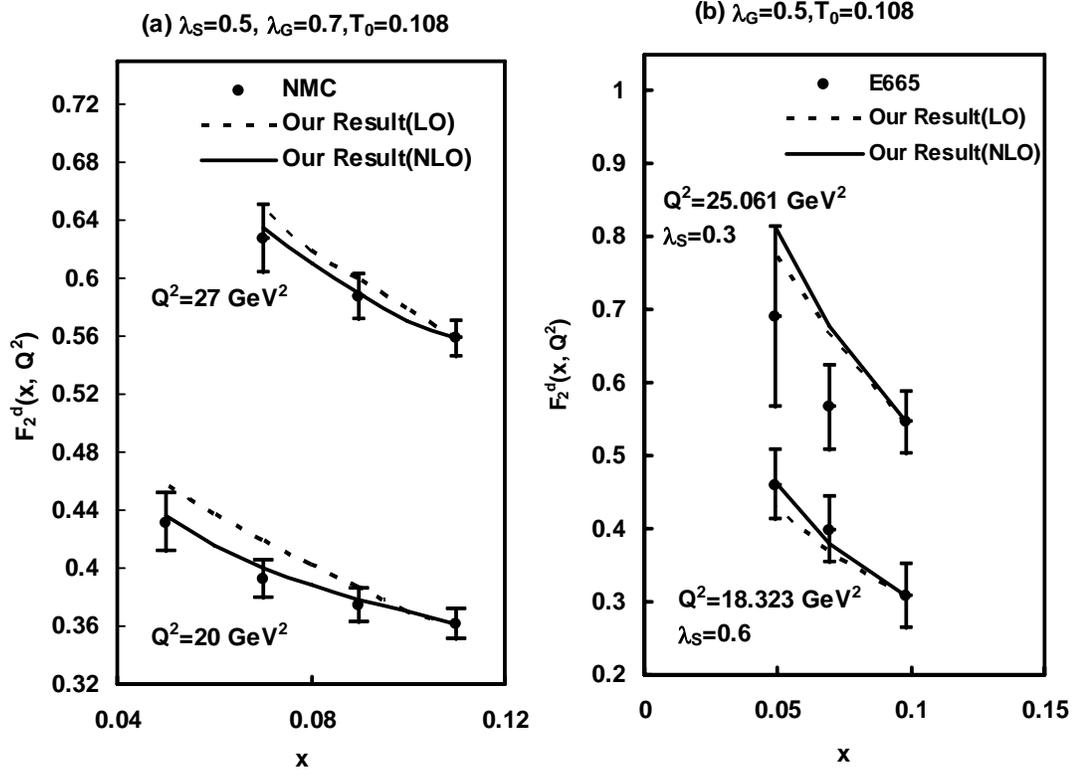

**Figure 4.16**: x -evolution of deuteron structure function in NLO at low-x which is obtained by solving coupled equations, compared with NMC and E665 data sets for the representative values of $Q^2$. Data are scaled up by +0.3i (i=0, 1) for both NMC and E665 data sets starting from bottom graph.

Figure 4.17 represents our result for t-evolution of gluon structure function from equation (4.24) compared with GRV1998NLO global parameterization. We get the best fitted graph for $\lambda_S$ =0.5, $\lambda_G$ =0.45 and $T_0$=0.05 for x=0.0001.

Figures 4.18(a-b) represent our result for x-evolution of gluon structure function in NLO from equation (4.25) compared with GRV1998NLO global parameterization. We get the best fitted graphs for $\lambda_S$=0.5, $\lambda_G$ =0.33 at $Q^2$ =20 GeV$^2$ and $\lambda_S$=0.5, $\lambda_G$ =0.31 at $Q^2$ =40 GeV$^2$.

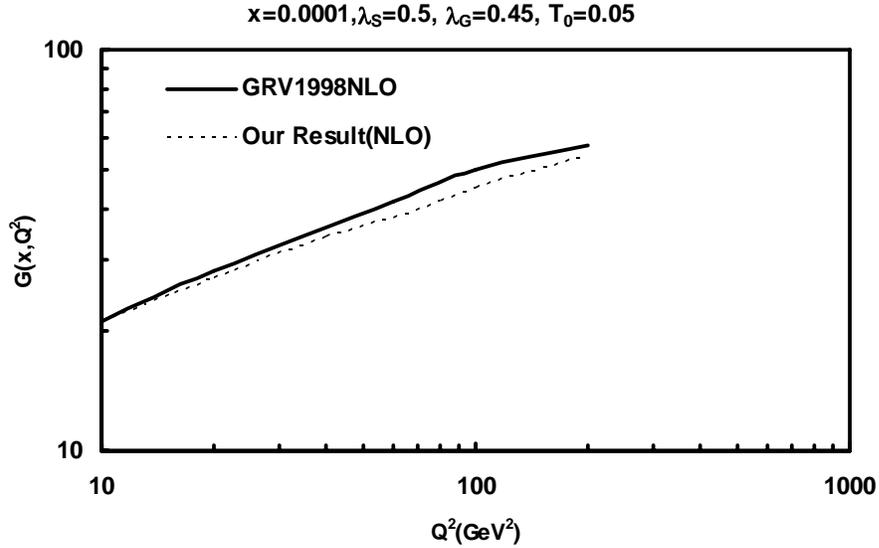

**Figure 4.17**: t-evolution of gluon structure function in NLO obtained by solving coupled equations in NLO, compared with GRV1998NLO global parameterization graph for the representative values of x.

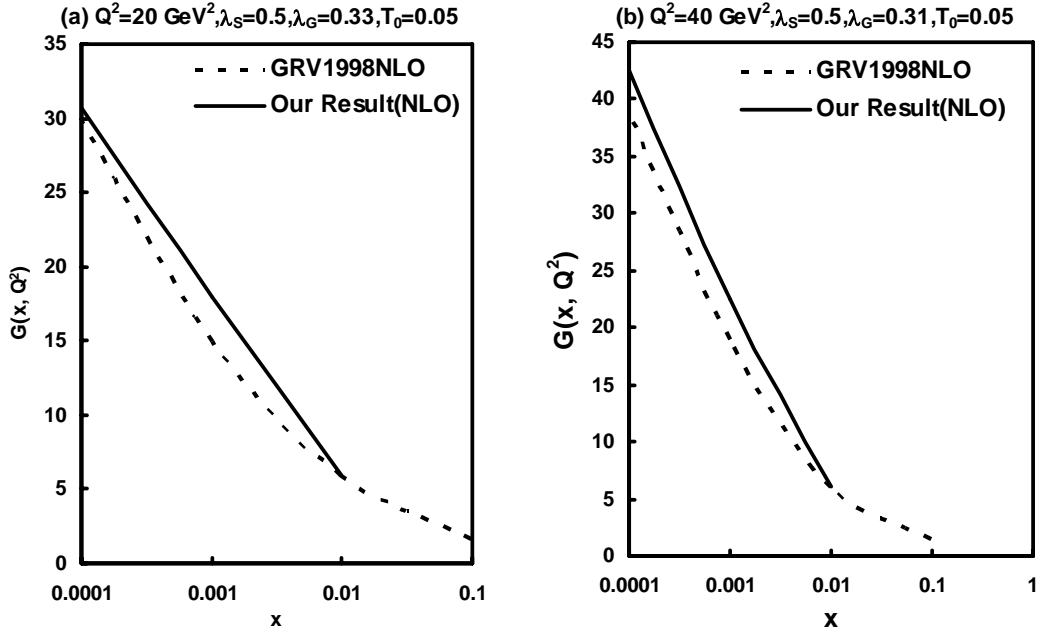

**Figure 4.18**: x -evolution of gluon structure function in NLO obtained by solving coupled equations in NLO compared with GRV1998NLO global parameterization graphs for the representative values of $Q^2$.

Figures 4.19(a-b) represent our result for x-evolution of gluon structure function in NLO from equation (4.25) compared with GRV1998NLO global parameterization. In this case for $\lambda_S=0.5$, $\lambda_G=0.33$ at $Q^2=60$ GeV$^2$ and $\lambda_S=0.5$, $\lambda_G=0.31$ at $Q^2=100$ GeV$^2$ correspond the best fitted graphs.

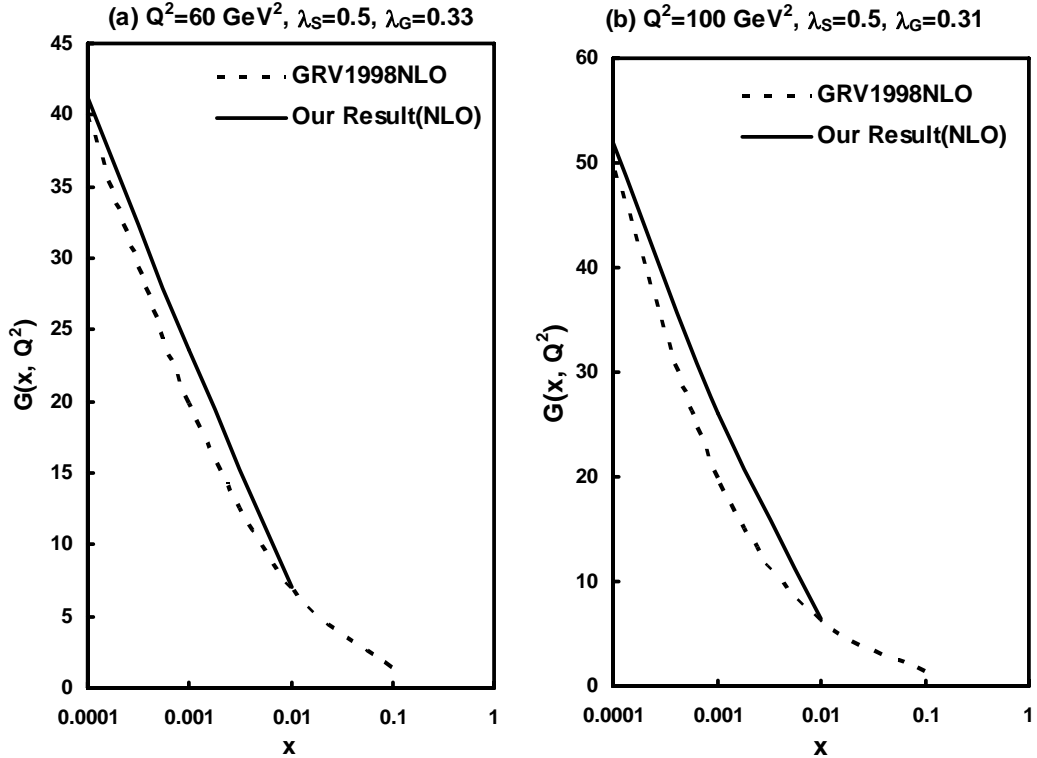

**Figure 4.19**: x -evolution of gluon structure function in NLO obtained by solving coupled equations, compared with GRV1998NLO global parameterization graphs for the representative values of $Q^2$.

Figures 4.20(a-b) represent our result of t-evolution of gluon structure function in NLO from equation (4.24) compared with GRV1998LO global parameterization. $\lambda_S=0.5$, $\lambda_G=0.48$ correspond the best fitted graphs at x=0.00001 and $\lambda_S=0.5$, $\lambda_G=0.5$ correspond the best fitted graphs at x=0.0001. And with the best fitted parameters for NLO, we plotted our LO graphs and from the both LO and NLO graphs it can be easily understood that NLO contribution is appreciable in our region of lower x for $\lambda_S \cong \lambda_G \cong 0.5$.

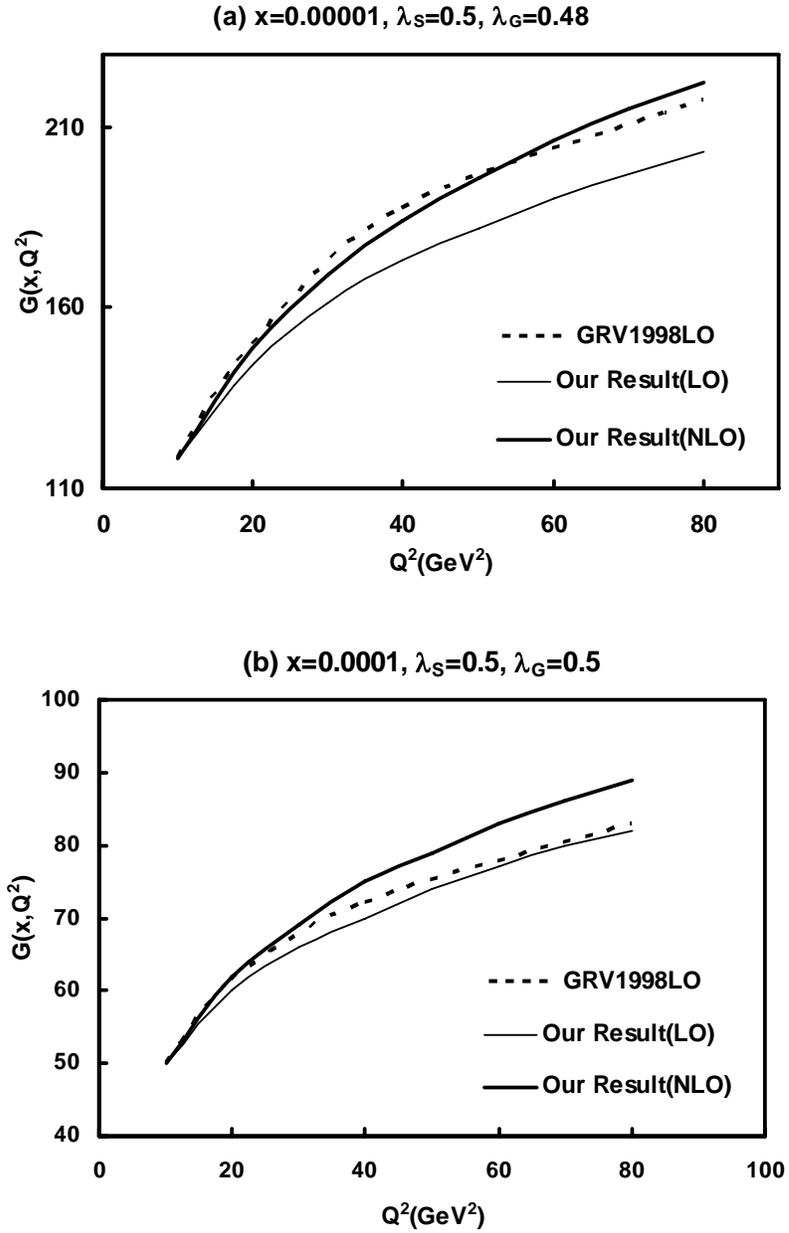

**Figure 4.20**: t -evolution of gluon structure function in both LO and NLO obtained by solving coupled equations, compared with GRV1998LO global parameterization graphs for the representative values of x.

Figures 4.21(a-b) represent our result for x-evolution of gluon structure function from equation (4.25) in compared with MRST2001 and MRST2004

global parameterizations. We get the best fitted graph for $\lambda_S = \lambda_G =0.5$ at $Q^2 =20$ GeV$^2$ of MRST2001 global parameterization and $\lambda_S =0.5$, $\lambda_G = 0.34$ at $Q^2 =100$ GeV$^2$ of MRST2004 global parameterization.

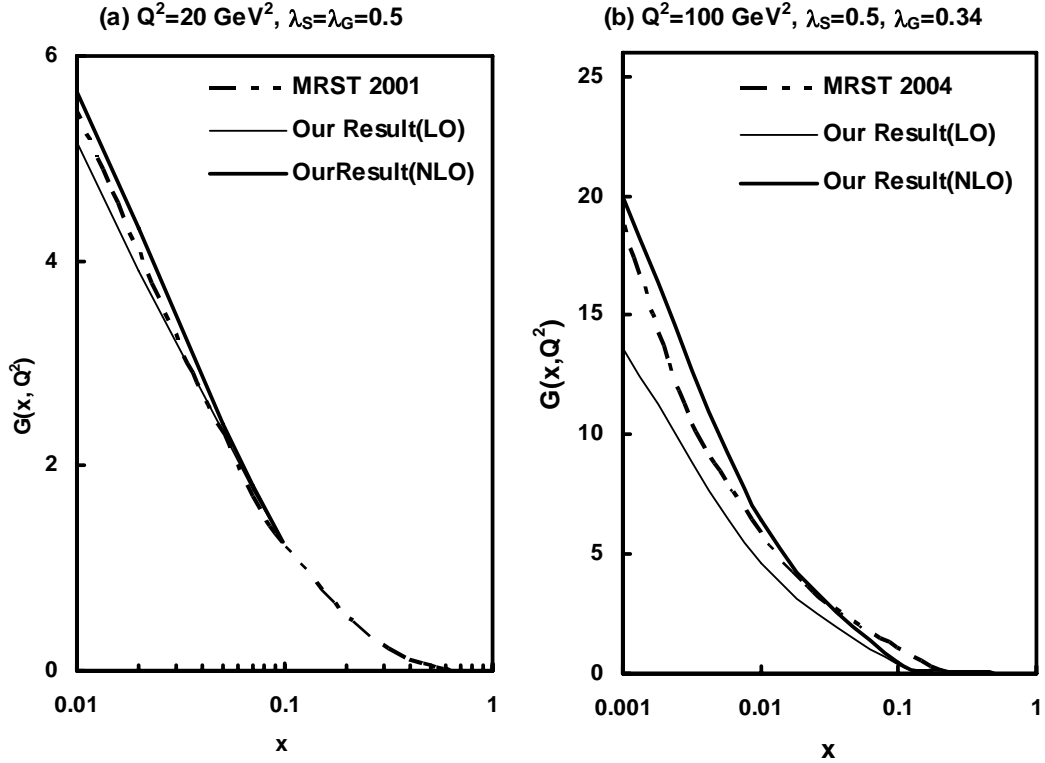

**Figure 4.21**: x -evolution of gluon structure function in NLO obtained by solving coupled equations, compared with MRST2001 and MRST2004 global parameterizations for the representative values of Q².

Figures 4.22(a-b) represent our result for x-evolution of gluon structure function in NLO from equation (4.25) compared with GRV1998LO global parameterization. We get the best fitted graphs for $\lambda_S = 0.5$, $\lambda_G =0.3$ at both $Q^2 =40$ GeV$^2$ and $Q^2 =80$ GeV$^2$ of GRV1998LO global parameterization.

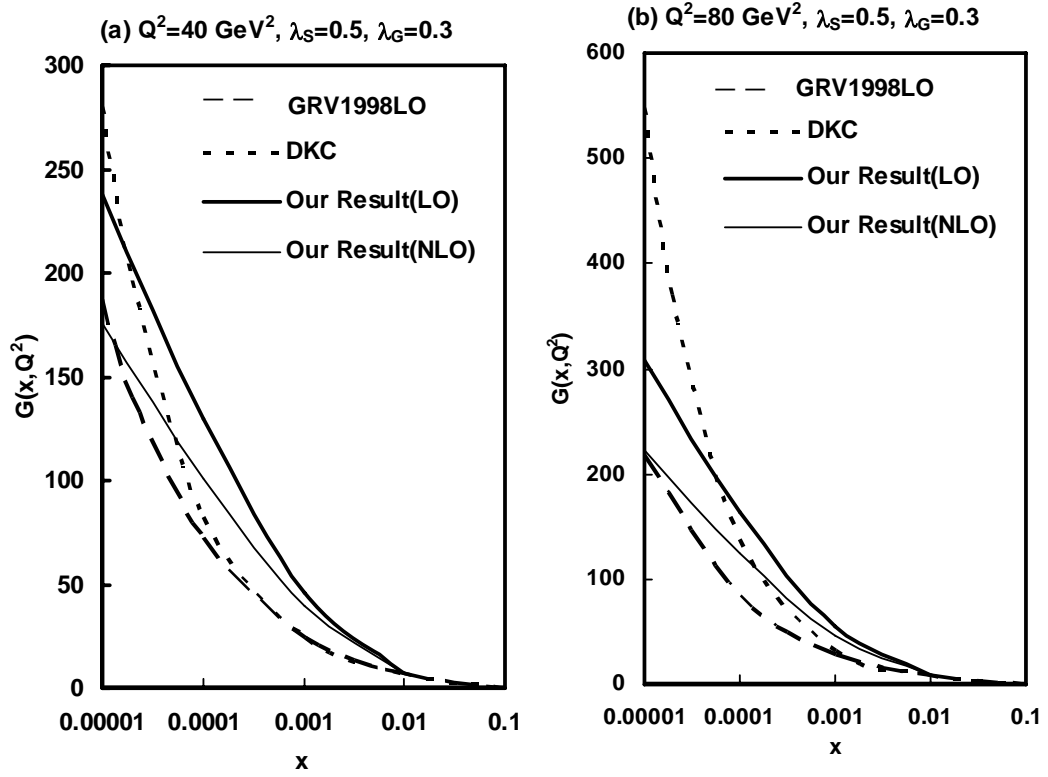

**Figure 4.22**: x-evolution of gluon structure function in NLO obtained by solving coupled equations, compared with GRV1998LO global parameterizations for the representative values of $Q^2$.

Figures 4.23(a-b) represent the sensitivity of the parameters $\lambda_S$ and $\lambda_G$. Here we considered the best fitted x-evolution graph of gluon structure function in NLO from equation (4.25) compared with MRST2004 global parameterizations at $Q^2=100$ GeV$^2$. In figure 4.23(a), keeping $\lambda_G$ fixed at 0.5, we found the range of the parameter $\lambda_S$ as $0.3 \leq \lambda_S \leq 0.7$. In figure 4.23(b), keeping $\lambda_S$ fixed at 0.5, we found the range of the parameter $\lambda_G$ as $0.35 \leq \lambda_G \leq 0.33$.

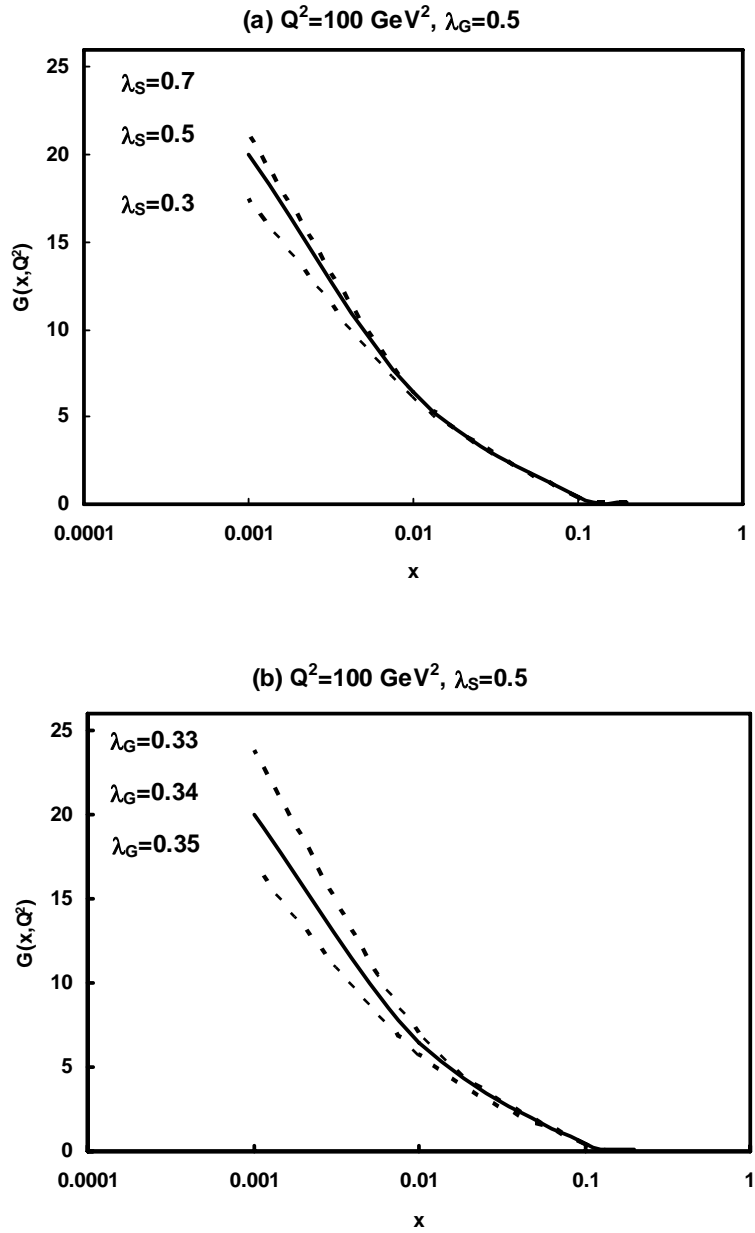

Figure 4.23: Sensitivity of the parameters $\lambda_S$ and $\lambda_G$ at $Q^2 = 100$ GeV$^2$ with the best fit graph of our result of x-evolution of gluon structure function in NLO with MRST2004 global parameterization.

## 4.3 Conclusion

In this chapter we have solved DGLAP evolution equations for singlet, non-singlet and gluon structure functions in NLO using Regge behaviour of spin-independent structure functions. Our results show that deuteron and proton structure functions are in good consistency with NMC and E665 collaborations data sets and also the results of gluon structure function with MRST2001, MRST2004, GRV1998NLO and GRV1998LO global parameterizations. We have compared the x-evolution graphs for deuteron, proton and gluon structure functions from DGLAP evolution equations in both LO and NLO and for all of them NLO shows significantly better fitting to the data sets and parameterizations than that of in LO. So, the higher order terms in NLO has appreciable contribution in the region of low-x to the parton distribution function. The values of $\lambda_S$ and $\lambda_G$ are generally close to 0.5 as predicted by Regge theory.□

# t and x- Evolutions of Spin-independent DGLAP Evolution Equations in Next-Next-to-Leading Order

Here we solved the spin-independent DGLAP evolution equation for singlet and non-singlet structure functions in next-next-to-leading order (NNLO) at low-x. The computation of the three-loop contributions to the anomalous dimensions is needed to complete the NNLO calculations for deep inelastic electron-nucleon scattering. The one and two loop splitting functions have been known for a long time [61, 62, 89] and we presented these splitting functions in previous Chapters 3 and 4. The NNLO corrections should be included in order to arrive at quantitatively reliable predictions for hard processes at present and future high energy colliders. Recently the three loop splitting functions are introduced with a good phenomenological success [92-95]. In this Chapter we present our solutions of spin-independent DGLAP evolution equations for singlet and non-singlet structure functions at low-x in NNLO considering Regge behaviour of structure functions. The t-evolutions of deuteron and proton structure functions thus obtained from singlet and non-singlet structure functions have been compared with NMC and E665 data sets.

## 5.1. Theory

The DGLAP evolution equations for singlet and non-singlet structure functions in NNLO are given as [92-94]

$$\frac{\partial F_2^S}{\partial t} - \frac{\alpha_S(t)}{2\pi} I_1^S(x,t) - \left(\frac{\alpha_S(t)}{2\pi}\right)^2 I_2^S(x,t) - \left(\frac{\alpha_S(t)}{2\pi}\right)^3 I_3^S(x,t) = 0 \tag{5.1}$$

and

$$\frac{\partial F_2^{NS}}{\partial t} - \frac{\alpha_S(t)}{2\pi} I_1^{NS}(x,t) - \left(\frac{\alpha_S(t)}{2\pi}\right)^2 I_2^{NS}(x,t) - \left(\frac{\alpha_S(t)}{2\pi}\right)^3 I_3^{NS}(x,t) = 0, \tag{5.2}$$

where the equations in LO and NLO are as given in Chapters 3 and 4 (equations 3.1, 3.2, 4.1 and 4.2) with their respective kernels. The NNLO contributions are

$$I_3^S(x,t) = \int_x^1 \frac{d\omega}{\omega} \left[ P_{qq}(\omega) F_2^{NS}\left(\frac{x}{\omega}, t\right) + P_{qg}(\omega) G\left(\frac{x}{\omega}, t\right) \right]$$

and

$$I_3^{NS}(x,t) = \int_x^1 \frac{d\omega}{\omega} \left[ P_{NS}^{(2)}(\omega) F_2^{NS}\left(\frac{x}{\omega}, t\right) \right].$$

The explicit forms of higher order kernels in NNLO are

$$P_{qq}(\omega) = P_{NS}^{(2)}(\omega) + P_{PS}^{(2)}(\omega),$$

$$P_{NS}^{(2)}(\omega) = n_f \begin{cases} \begin{cases} L_1(-163.9\,\omega^{-1} - 7.208\,\omega) + 151.49 + 44.51\,\omega \\ -43.12\,\omega^2 + 4.82\,\omega^3 \end{cases}(1-\omega) \\ + L_0 L_1(-173.1 + 46.18 L_0) + 178.04 L_0 + 6.892 L_0^2 \\ + \frac{40}{27}(L_0^4 - 2L_0^3) \end{cases},$$

$$P_{PS}^{(2)}(\omega) \cong \begin{cases} N_f \begin{pmatrix} -5.926 L_1^3 - 9.751 L_1^2 - 72.11 L_1 + 177.4 + 392.9\,\omega \\ -101.4\,\omega^2 - 57.04 L_0 L_1 - 661.6 L_0 + 131.4 L_0^2 \\ -\frac{400}{9} L_0^3 + \frac{160}{27} L_0^4 - 506.0\,\omega^{-1} - \frac{3584}{27}\,\omega^{-1} L_0 \end{pmatrix} \\ + N_f^2 \begin{pmatrix} 1.778 L_1^2 + 5.944 L_1 + 100.1 - 125.2\,\omega \\ + 49.26\,\omega^2 - 12.59\,\omega^3 - 1.889 L_0 L_1 + 61.75 L_0 \\ + 17.89 L_0^2 + \frac{32}{27} L_0^3 + \frac{256}{81}\,\omega^{-1} \end{pmatrix} \end{cases}(1-\omega)$$

and

$$P_{qg}(\omega) \cong N_f \begin{pmatrix} \dfrac{100}{27}L_1^4 - \dfrac{70}{9}L_1^3 - 120.5L_1^2 + 104.42L_1 + 2522 \\ -3316\omega + 2126\omega^2 + L_0L_1(1823 - 25.22L_0) - 252.5\omega L_0^3 \\ + 424.9L_0 + 881.5L_0^2 - \dfrac{44}{3}L_0^3 + \dfrac{536}{27}L_0^4 - 1268.3\omega^{-1} \\ -\dfrac{896}{3}\omega^{-1}L_0 \end{pmatrix}$$

$$+ N_f^2 \begin{pmatrix} \dfrac{20}{27}L_1^3 + \dfrac{200}{27}L_1^2 - 5.496L_1 - 252.0 + 158.0\omega + 145.4\omega^2 \\ -139.28\omega^3 - 98.07\omega L_0^3 + 11.70\omega L_0^3 \\ -L_0L_1(53.09 + 80.616L_0) - 254.0L_0 - 90.80L_0^2 \\ -\dfrac{376}{27}L_0^3 - \dfrac{16}{9}L_0^4 + \dfrac{1112}{243}\omega^{-1} \end{pmatrix},$$

with $L_0 = \ln(\omega)$ and $L_1 = \ln(1-\omega)$. Here results are from direct x-space evolution and $P_{NS}^{(2)}(\omega), P_{PS}^{(2)}(\omega)$ and $P_{qg}(\omega)$ are calculated using FORTRAN package [96]. Except for x-values very close to zero of $P_{NS}^{(2)}(\omega)$, this parameterizations deviate from the exact expressions by less than one part in thousand, which can be consider as sufficiently accurate. For a maximal accuracy for the convolutions with quark densities, slight adjustment should do using low integer moments [93].

Applying Regge behaviour of structure functions as given by equations (3.7) and (3.8) and the relation between singlet and gluon structure functions by equation (3.13), the solutions of spin-independent DGLAP evolution equations for singlet and non-singlet structure functions in NNLO come out as

$$F_2^S(x,t) = C t^{H_7(x)} \tag{5.3}$$

and

$$F_2^{NS}(x,t) = C t^{H_8(x)}, \tag{5.4}$$

Where C is an arbitrary constant,

$$H_7(x) = \dfrac{3}{2} A_f \left[ \left\{ \dfrac{2}{3}\{3 + 4\ln(1-x)\} + f_{11}(x) \right\} + T_0 f_{12}(x) + T_1 f_{13}(x) \right],$$

$$H_8(x) = \frac{3}{2} A_f \left[ \left\{ \frac{2}{3} \{3 + 4 \ln(1-x)\} + f_{14}(x) \right\} + T_0 f_{15}(x) + T_1 f_{16}(x) \right],$$

$$f_{11}(x) = \frac{4}{3} \int_x^1 \frac{d\omega}{1-\omega} \left[(1+\omega^2)\omega^{\lambda_S} - 2\right] + N_f \int_x^1 \{\omega^2 + (1-\omega)^2\} K\left(\frac{x}{\omega}\right) \omega^{\lambda_S} d\omega,$$

$$f_{12}(x) = \left[ (x-1)\int_0^1 f(\omega) d\omega + \int_x^1 f(\omega)\omega^{\lambda_S} d\omega + \int_x^1 F_{qq}^S(\omega) \omega^{\lambda_S} d\omega + \int_x^1 F_{qg}^S(\omega) K\left(\frac{x}{\omega}\right) \omega^{\lambda_S} d\omega \right],$$

$$f_{13}(x) = \int_x^1 \frac{d\omega}{\omega} \left[ P_{qq}(\omega) \omega^{\lambda_S} + P_{qg}(\omega) K\left(\frac{x}{\omega}\right) \omega^{\lambda_S} \right],$$

$$f_{14}(x) = \frac{4}{3} \int_x^1 \frac{d\omega}{1-\omega} \left[(1+\omega^2)\omega^{\lambda_{NS}} - 2\right],$$

$$f_{15}(x) = \left[ (x-1)\int_0^1 f(\omega) d\omega + \int_x^1 f(\omega)\omega^{\lambda_{NS}} d\omega \right]$$

and

$$f_{16}(x) = \int_x^1 \frac{d\omega}{\omega} \left[ P_{NS}^{(2)}(\omega) \omega^{\lambda_{NS}} \right].$$

Similarly as in chapter 4, for possible solutions in NNLO, we have taken $T(t) = \left(\frac{\alpha_S(t)}{2\pi}\right)$ and the expression for T(t) upto LO correction with the assumptions $T(t)^2 = T_0 T(t)$ and $T(t)^3 = T_0(T(t))^2 = T_1 T(t)$ [52, 65, 90], where $T_0$ and $T_1$ are numerical parameters. But $T_0$ and $T_1$ are not arbitrary. We choose $T_0$, $T_1$ such that difference between $T^2(t)$, $T_0 T(t)$ and $T^3(t)$, $T_1 T(t)$ are minimum in the region of our discussion (Figure 5.1).

Applying initial conditions at $t = t_0$, $F_2^S(x,t) = F_2^S(x,t_0)$ and $F_2^{NS}(x,t) = F_2^{NS}(x,t_0)$, and at $x = x_0$, $F_2^S(x,t) = F_2^S(x_0,t)$ and $F_2^{NS}(x,t) = F_2^{NS}(x_0,t)$, we find the t an x-evolutions for the singlet and non-singlet structure functions from equations (5.3) and (5.4) in NNLO respectively as

$$F_2^S(x,t) = F_2^S(x,t_0)\left(\frac{t}{t_0}\right)^{H_7(x)}, \tag{5.5}$$

$$F_2^S(x,t) = F_2^S(x_0,t)\, t^{\{H_7(x)-H_7(x_0)\}}, \tag{5.6}$$

$$F_2^{NS}(x,t) = F_2^{NS}(x,t_0)\left(\frac{t}{t_0}\right)^{H_8(x)} \tag{5.7}$$

and

$$F_2^{NS}(x,t) = F_2^{NS}(x_0,t)\, t^{\{H_8(x)-H_8(x_0)\}}. \tag{5.8}$$

The t and x-evolutions of deuteron and proton structure functions corresponding to equations (5.5) to (5.8) are respectively

$$F_2^d(x,t) = F_2^d(x,t_0)\left(\frac{t}{t_0}\right)^{H_7(x)}, \tag{5.9}$$

$$F_2^d(x,t) = F_2^d(x_0,t)\, t^{\{H_7(x)-H_7(x_0)\}}, \tag{5.10}$$

$$F_2^P(x,t) = F_2^P(x,t_0)\left(\frac{3t^{H_8(x)} + 5t^{H_7(x)}}{3t_0^{H_8(x)} + 5t_0^{H_7(x)}}\right) \tag{5.11}$$

and

$$F_2^P(x,t) = F_2^P(x_0,t)\left(\frac{3t^{H_8(x)} + 5t^{H_7(x)}}{3t^{H_8(x_0)} + 5t^{H_7(x_0)}}\right). \tag{5.12}$$

Here $F_2^d(x,t_0)$, $F_2^P(x,t_0)$, $F_2^d(x_0,t)$ and $F_2^P(x_0,t)$ are the values of the structure functions $F_2^d(x,t)$ and $F_2^P(x,t)$ at $t=t_0$ and $x=x_0$ respectively.

## 5.2 Results and Discussion

We have compared our result of deuteron and proton structure functions with the data sets measured by the NMC [71] and E665 [72] Collaborations. We choose the QCD cut-off parameter as $\Lambda_{\overline{MS}}$ ($N_f$ = 4) = 323 MeV for $\alpha_s(M_z^2)$ = 0.119± 0.002 [73] for the phenomenological analysis of our result. Along with the NNLO results we also presented our NLO and LO results from Chapters 3 and 4.

The comparisons of our results with experimental data sets are made for $\lambda_S=\lambda_{NS}=0.5$. We compared our results for K(x) = K, $ax^b$ and $ce^{dx}$, where K, a, b, c and d are constants. Data points at lowest-$Q^2$ values and at highest-x values (for x<0.1) are taken as input to test the evolution equations. In our work, we found the value of the deuteron and proton structure functions remain almost same for b<0.00001 and for d<0.00001. So, we have chosen b = d = 0.00001 for our analysis and the best fit graphs are observed by varying the values of K, a and c.

In Figure 5.1, we have plotted both $T(t)^2$, $T_0T(t)$ and $T(t)^3$, $T_1T(t)$ against $Q^2$ where $Q^2$ range is 0.5 ≤ $Q^2$≤ 30 $GeV^2$. From the plots it is obvious that errors become minimum for $T_0$ = 0.05 and $T_1$ = 0.0028.

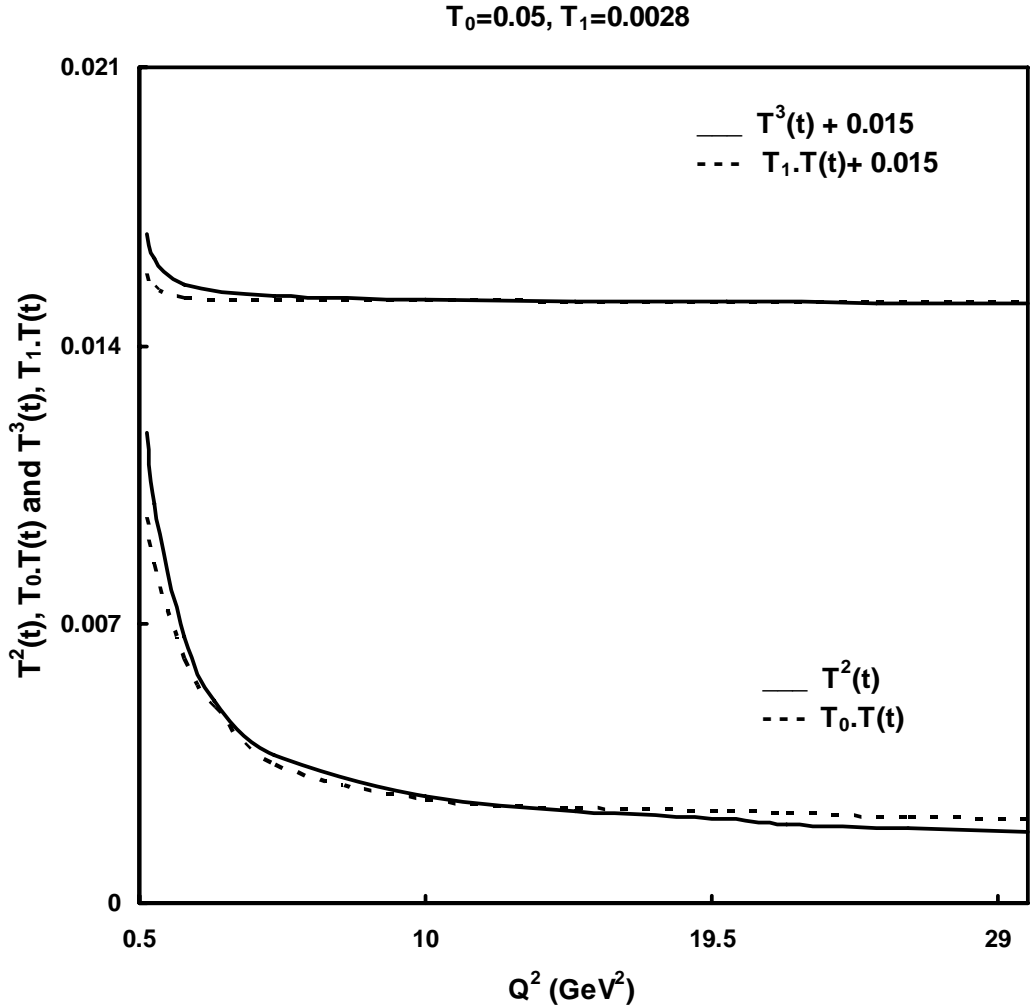

**Figure 5.1:** The variation of $T^2(t)$, $T_0.T(t)$ and $T^3(t)$, $T_1.T(t)$ with $Q^2$.

Figure 5.2 represents our result of t-evolution of deuteron structure function from equation (5.9) for $K(x) = ce^{dx}$ with NMC data set. For $-0.778 \leq c \leq -0.6145$ we get the best fit results. Same graphs are found for $K(x) = K$ and $ax^b$ and in this case the best fit results are found for $-0.778 \leq K \leq -0.6145$ and $-0.778 \leq a \leq -0.6145$.

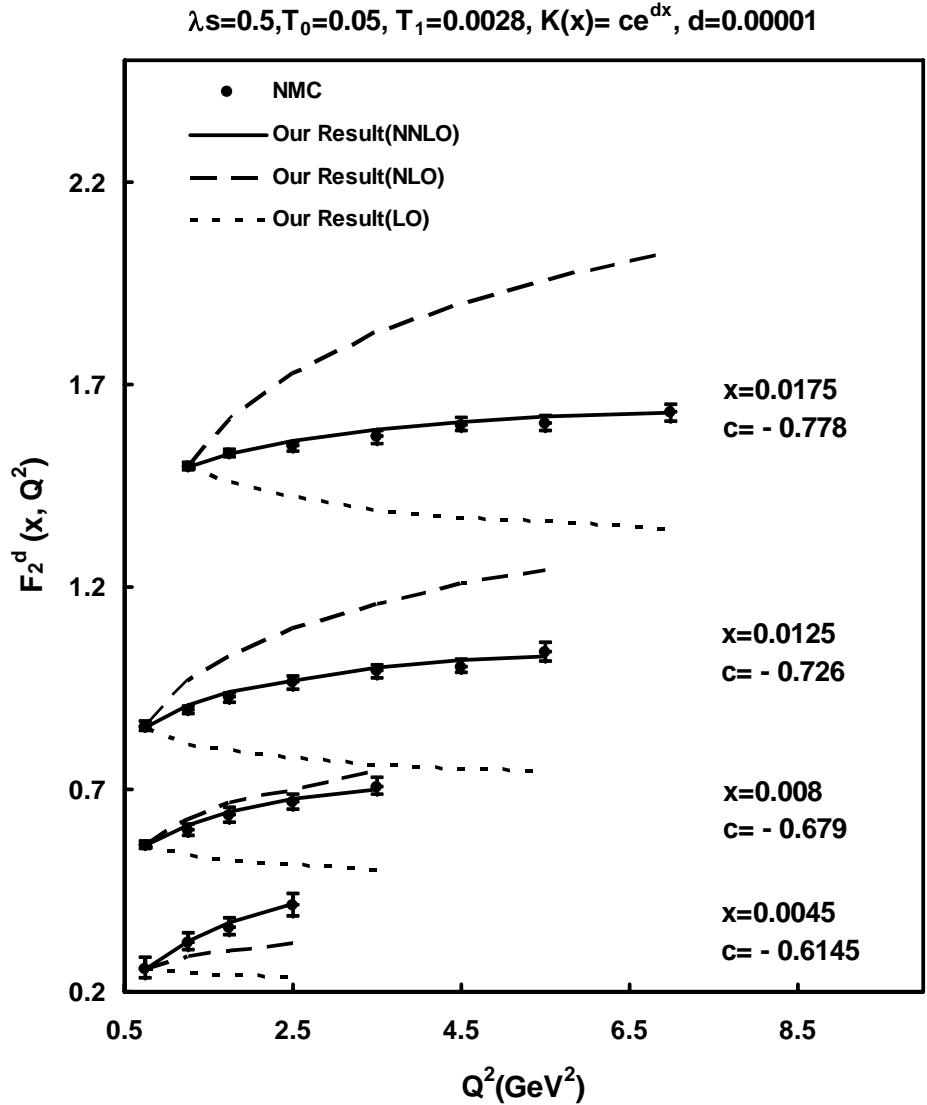

**Figure 5.2**: t-evolution of deuteron structure function in LO, NLO and NNLO compared with NMC data set for the representative values of x. Data are scaled up by +0.3i (i=0, 1, 2, 4) starting from bottom graph.

Figure 5.3 represents our result of t-evolution of deuteron structure function from equation (5.9) for $K(x) = ce^{dx}$ with E665 data set. We get the best fit results for $-0.775 \leq c \leq -0.634$. Same graphs are found for $K(x) = K$ and $ax^b$ with $-0.775 \leq K \leq -0.634$ and $-0.775 \leq a \leq -0.634$.

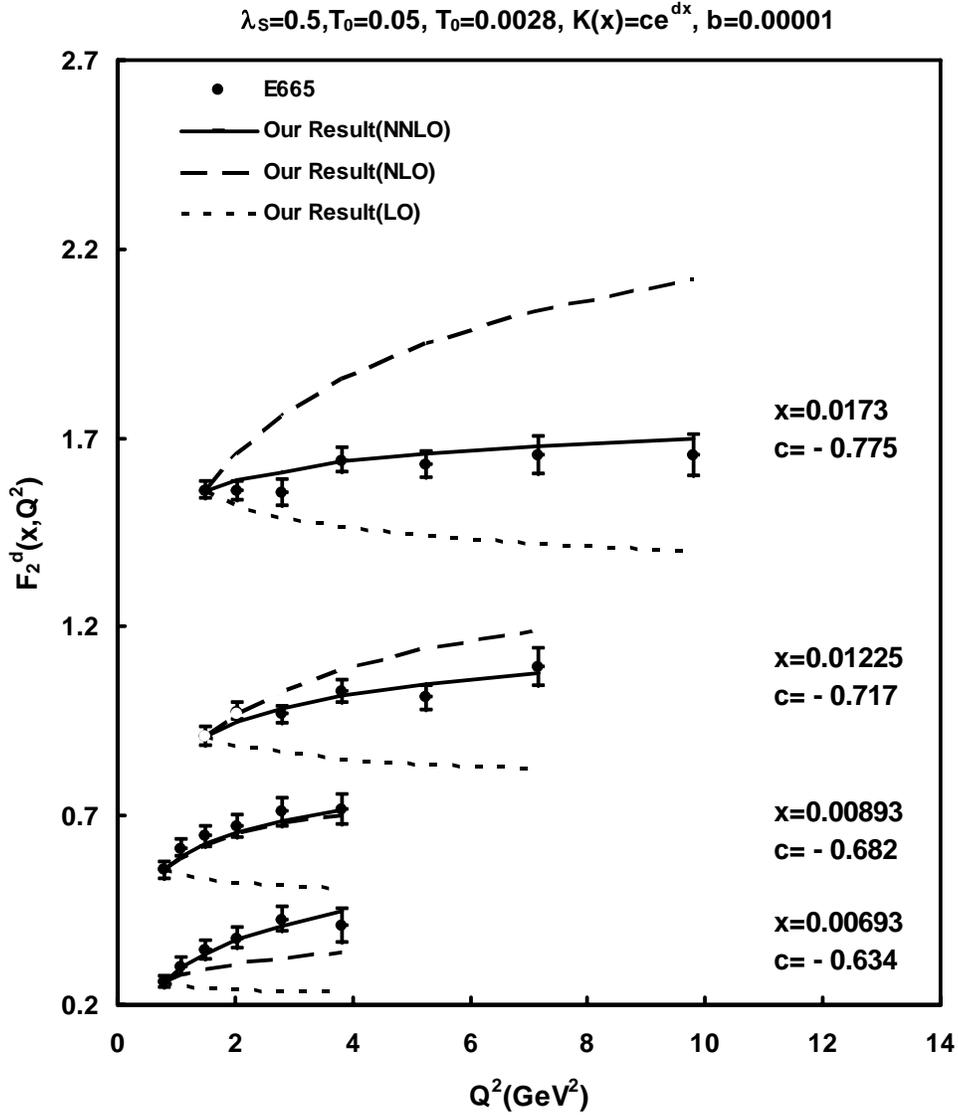

**Figure 5.3**: t-evolution of deuteron structure function in LO, NLO and NNLO compared with E665 data set for the representative values of x. Data are scaled up by +0.3i (i=0, 1, 2, 4) starting from bottom graph.

Figure 5.4 represents our result of t-evolution of proton structure function from equation (5.11) for K(x) = $ce^{dx}$ with NMC data set. We get the best fit results for - 0.755≤c≤ - 0.6085. Same graphs are found for K(x) =K and $ax^b$ with - 0.755≤K≤ - 0.6085 and - 0.755≤a≤ - 0.6085.

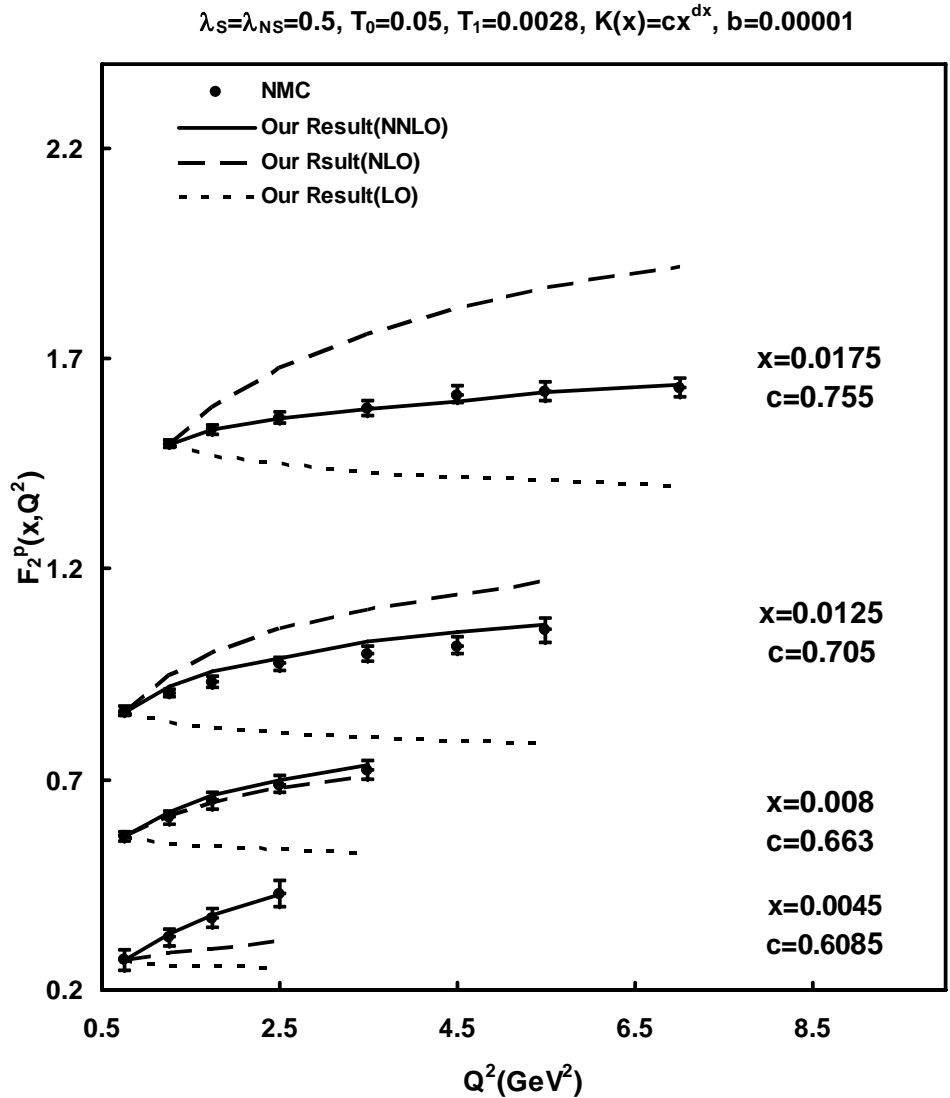

**Figure 5.4**: t-evolution of proton structure function in LO, NLO and NNLO compared with NMC data set for the representative values of x. Data are scaled up by +0.3i (i=0, 1, 2, 4) starting from bottom graph.

Figure 5.5 represents our result of t-evolution of proton structure function from equation (5.11) for K (x) = $ce^{dx}$ with E665 data set. For - 0.755≤c≤ - 0.617 we get the best fit results. Same graphs are found for K(x) =K and $ax^b$ with - 0.755≤K≤ - 0.617 and - 0.755≤a≤ - 0.617.

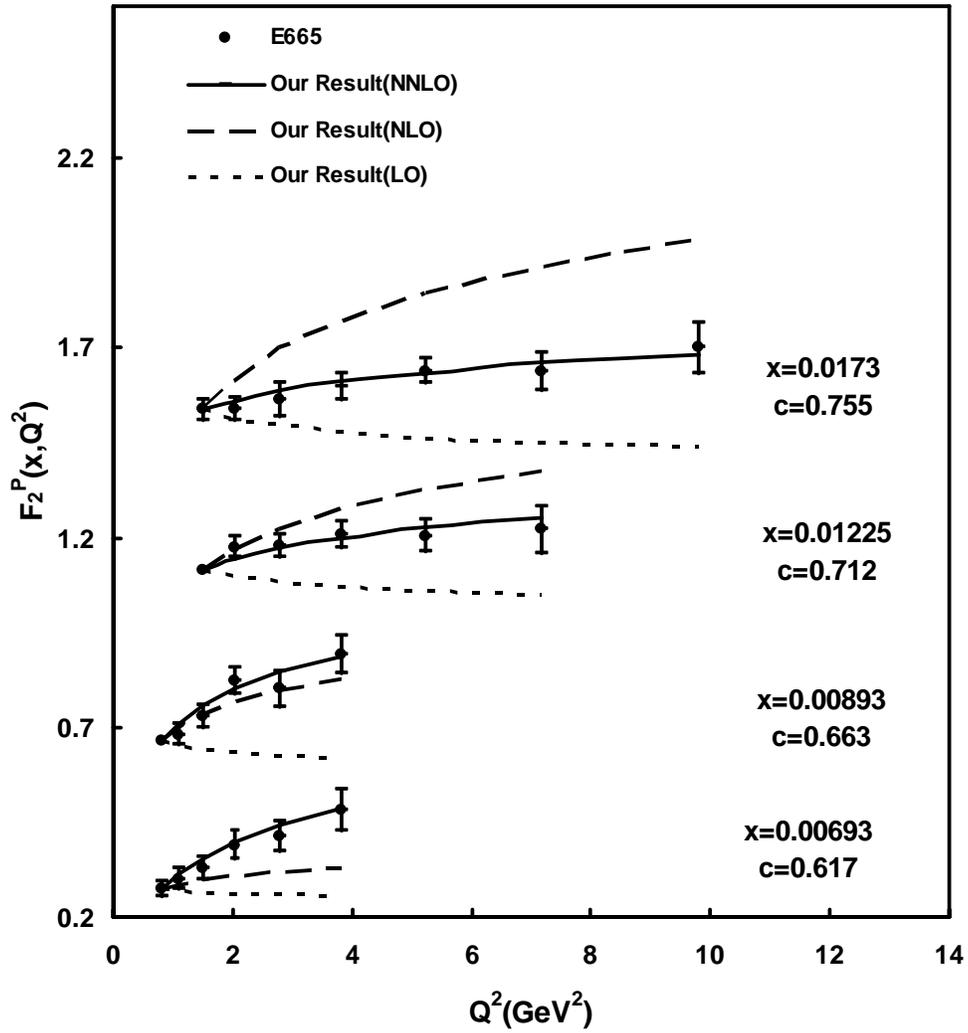

**Figure 5.5**: t -evolution of proton structure function in LO, NLO and NNLO compared with E665 data set for the representative values of x. Here data are scaled up by +0.4i (i=0, 1, 2, 3) starting from bottom graph.

## 5.3 Conclusion

In this chapter, we have considered the Regge behaviour of spin-independent singlet and non-singlet structure functions in NNLO to solve DGLAP evolution equations and presented approximate analytical solutions of these equations. We have seen that the results of t-evolutions of deuteron and proton structure functions are in good consistency with the NMC and E665 collaboration data sets. Since $T_1$ is much smaller than $T_0$, as from equations (5.9) to (5.12), we expect NNLO contributions to the DGLAP evolution equations should be small. But from the comparison of t-evolution graphs of our results of LO, NLO and NNLO, it is seen that NNLO corrections have significant effect. Though we have simplified our solution through numerical variables $T_0$ and $T_1$, but they are not chosen arbitrarily.□

*Part II*

# Spin-dependent DGLAP evolution equations at low-x

<div style="text-align: right;">**Chapter 6**</div>

# t and x-Evolutions of Spin-dependent DGLAP Evolution Equations in Leading Order

DIS of polarized electrons and muons off polarized targets has been used to study the internal spin structure of the nucleon. The most abundant and accurate experimental information we have so far comes from the so called longitudinal spin-dependent structure function $g_1$ which is obtained with longitudinally polarized leptons on longitudinally polarized protons, deuterons, and $^3$He targets and it allows separate determination of spin-dependent deuteron, proton and neutron structure functions [97-104].

Here we have presented our solutions of spin-dependent DGLAP evolution equations for singlet, non-singlet and gluon structure functions at low-x in LO considering Regge behaviour of spin-dependent structure functions at this limit. We solved each equation for singlet, non-singlet and gluon structure functions and also the coupled equations for singlet and gluon structure functions. The evolutions of deuteron, proton and neutron structure functions thus obtained have been compared with SLAC-E-154, SLAC-E-143 and SMC collaborations data sets. And the evolution of gluon structure function has been compared with the result obtained by numerical method.

# 6.1 Theory

The spin-dependent DGLAP evolution equations for singlet, non-singlet and gluon structure functions [60, 104, 105] in LO are respectively

$$\frac{\partial g_1^S(x,t)}{\partial t} - \frac{\alpha_S(t)}{2\pi} J_1^S(x,t) = 0, \qquad (6.1)$$

$$\frac{\partial g_1^{NS}(x,t)}{\partial t} - \frac{\alpha_S(t)}{2\pi} J_1^{NS}(x,t) = 0 \qquad (6.2)$$

and

$$\frac{\partial \Delta G(x,t)}{\partial t} - \frac{\alpha_S(t)}{2\pi} J_1^G(x,t) = 0, \qquad (6.3)$$

where $g_1^S, g_1^{NS}$ and $\Delta G$ are the spin-dependent singlet, non-singlet and gluon structure functions respectively,

$$J_1^S(x,t) = \frac{2}{3}\{3 + 4\ln(1-x)\} g_1^S(x,t) + \frac{4}{3}\int_x^1 \frac{d\omega}{1-\omega}\left\{(1+\omega^2) g_1^S\left(\frac{x}{\omega},t\right) - 2 g_1^S(x,t)\right\}$$

$$+ N_f \int_x^1 \{\omega^2 - (1-\omega)^2\} \Delta G\left(\frac{x}{\omega},t\right) d\omega,$$

$$J_1^{NS}(x,t) = \frac{2}{3}\{3 + 4\ln(1-x)\} g_1^{NS}(x,t) + \frac{4}{3}\int_x^1 \frac{d\omega}{1-\omega}\left\{(1+\omega^2) g_1^{NS}\left(\frac{x}{\omega},t\right) - 2 g_1^{NS}(x,t)\right\},$$

$$J_1^G(x,t) = \left\{6\left(\frac{11}{12} - \frac{N_f}{18} + \ln(1-x)\right) \Delta G(x,t) + 6 I_G\right\}$$

and

$$I_G = \int_x^1 d\omega \left[\frac{1}{2}\left\{\frac{(1+\omega^4) \Delta G\left(\frac{x}{\omega},t\right) - 2\Delta G(x,t)}{1-\omega} + \left(\frac{1}{\omega} + \omega^3 - \frac{(1-\omega)^3}{\omega}\right) \Delta G\left(\frac{x}{\omega},t\right)\right\}\right.$$

$$\left. + \frac{2}{9}\left(\frac{1-(1-\omega)^2}{\omega}\right) g_1^S\left(\frac{x}{\omega},t\right)\right].$$

Spin-dependent deuteron, proton and neutron structure functions interms of spin-dependent singlet, non-singlet and gluon structure functions [60] can be written as

$$g_1^d(x,t) = \frac{5}{9} g_1^S(x,t),  \qquad (6.4)$$

$$g_1^p(x,t) = \frac{3}{18} g_1^{NS}(x,t) + \frac{5}{18} g_1^S(x,t) \qquad (6.5)$$

and

$$g_1^n(x,t) = \frac{5}{18} g_1^S(x,t) - \frac{3}{18} g_1^{NS}(x,t). \qquad (6.6)$$

The low-x behaviour of spin-dependent structure functions for fixed-$Q^2$ is the Regge limit of the spin-dependent DIS where the Regge pole exchange model should be applicable [49, 106-112]. The Regge behaviour for spin-dependent singlet, non-singlet and gluon structure functions has the general form $A_i(x,t) = T_i(t) x^{-\beta_i}$ [18, 60, 113], where $A_i(x, t)$ are the structure functions, $T_i(t)$ are some functions of t and $\beta_i$ are the respective Regge intercepts of the trajectory. Let us take $\beta_i$'s as $\beta_S$, $\beta_{NS}$ and $\beta_G$ for the spin-dependent singlet, non-singlet and gluon structure functions respectively. Hence

$$g_1^S(x,t) = T_4(t) x^{-\beta_S}, \qquad (6.7)$$

$$g_1^{NS}(x,t) = T_5(t) x^{-\beta_{NS}} \qquad (6.8)$$

and

$$\Delta G(x,t) = T_6(t) x^{-\beta_G}. \qquad (6.9)$$

Therefore,

$$g_1^S\left(\frac{x}{\omega},t\right) = T_4(t) \omega^{\beta_S} x^{-\beta_S} = g_1^S(x,t) \omega^{\beta_S}, \qquad (6.10)$$

$$g_1^{NS}\left(\frac{x}{\omega},t\right) = T_5(t) \omega^{\beta_{NS}} x^{-\beta_{NS}} = g_1^{NS}(x,t) \omega^{\beta_{NS}} \qquad (6.11)$$

and

$$\Delta G\left(\frac{x}{\omega},t\right)=T_6(t)\,\omega^{\beta_G}\,x^{-\beta_G}=\Delta G(x,t)\,\omega^{\beta_G}, \tag{6.12}$$

where $T_4(t)$, $T_5(t)$ and $T_6(t)$ are functions of t, and $\beta_S$, $\beta_{NS}$ and $\beta_G$ are the Regge intercepts for spin-dependent singlet, non-singlet and gluon structure functions respectively. This form of Regge behaviour is well supported by the work carried out by namely Ziaja [111], Soffer and Teryaev [58] and Badelek and Kwiecinski [112].

The DGLAP evolution equations of spin-dependent singlet, non-singlet and gluon structure functions given by equations (6.1) to (6.3) are in the same forms of derivative with respect to t, so the relation between singlet and gluon structure functions will come out in terms of x at fixed-$Q^2$. So, similarly as the spin-independent cases, we can consider for spin-dependent case also the ansatz [113]

$$\Delta G(x,t)=K(x)g_1^S(x,t) \tag{6.13}$$

for simplicity, where $K(x)$ is a parameter to be determined from phenomenological analysis and we assume $K(x) = K$, $ax^b$ or $ce^{dx}$, where K, a, b, c and d are constants.

Putting equations (6.7), (6.10) and 6.13) in equation (6.1) we arrive at

$$\frac{\partial g_1^S(x,t)}{\partial t}-\frac{g_1^S(x,t)}{t}J_1(x)=0, \tag{6.14}$$

where

$$J_1(x)=A_f\left[\{3+4\ln(1-x)\}+2\int_x^1\frac{d\omega}{1-\omega}\left\{(1+\omega^2)\omega^{\beta_S}-2\right\}\right.$$
$$\left.+\frac{3}{2}N_f\int_x^1\{\omega^2-(1-\omega)^2\}K\left(\frac{x}{\omega}\right)\omega^{\beta_S}\,d\omega\right],$$

Integrating equation (6.14) we get

$$g_1^S(x,t)=C\,t^{J_1(x)}, \tag{6.15}$$

where C is a constant of integration and $A_f = 4/(33 - 2N_f)$. At $t = t_0$, equation (6.15) gives

$$g_1^S(x, t_0) = C \, t_0^{J_1(x)} . \tag{6.16}$$

From equations (6.15) and (6.16) we get

$$g_1^S(x, t) = g_1^S(x, t_0) \left(\frac{t}{t_0}\right)^{J_1(x)}, \tag{6.17}$$

which gives the t-evolution of spin-dependent singlet structure function in LO. Again at $x = x_0$, equation (6.15) gives

$$g_1^S(x_0, t) = C \, t^{J_1(x_0)}. \tag{6.18}$$

From equations (6.15) and (6.18), we get

$$g_1^S(x, t) = g_1^S(x_0, t) \, t^{\{J_1(x) - J_1(x_0)\}}, \tag{6.19}$$

which gives the x-evolution of spin-dependent singlet structure function at LO.

Similarly we get the solution of spin-dependent DGLAP evolution equations for non-singlet and gluon structure functions in LO at low-x from the standard DGLAP evolution equations (6.2) and (6.3) respectively as

$$g_1^{NS}(x, t) = C \, t^{J_2(x)} \tag{6.20}$$

and

$$\Delta G(x, t) = C \, t^{J_3(x)}, \tag{6.21}$$

where

$$J_2(x) = A_f \left[ \{3 + 4 \ln(1-x)\} + 2 \int_x^1 \frac{d\omega}{1-\omega} \{(1 + \omega^2) \omega^{\beta_{NS}} - 2\} \right],$$

and

$$J_3(x) = 9A_f \left[ \left( \frac{11}{12} - \frac{N_f}{18} + \ln(1-x) \right) + \frac{1}{2} \int_x^1 d\omega \left\{ \frac{(1+\omega^4)\omega^{\beta_G} - 2}{1-\omega} + \left( \frac{1}{\omega} + \omega^3 - \frac{(1-\omega)^3}{\omega} \right) \omega^{\beta_G} \right\} \right.$$
$$\left. + \frac{2}{9} \int_x^1 d\omega \left( \frac{1-(1-\omega)^2}{\omega} \right) \frac{\omega^{\beta_G}}{K\left(\frac{x}{\omega}\right)} \right].$$

The t and x-evolutions of spin-dependent non-singlet structure function in LO at low-x are given as

$$g_1^{NS}(x,t) = g_1^{NS}(x,t_0) \left( \frac{t}{t_0} \right)^{J_2(x)} \tag{6.22}$$

and

$$g_1^{NS}(x,t) = g_1^{NS}(x_0,t) \, t^{\{J_2(x) - J_2(x_0)\}}. \tag{6.23}$$

The t and x-evolution equations of spin-dependent deuteron, proton and neutron structure functions from equations (6.17), (6.19), (6.22), and (6.23) are respectively

$$g_1^d(x,t) = g_1^d(x,t_0) \left( \frac{t}{t_0} \right)^{J_1(x)}, \tag{6.24}$$

$$g_1^d(x,t) = g_1^d(x_0,t) \, t^{\{J_1(x) - J_1(x_0)\}}, \tag{6.25}$$

$$g_1^P(x,t) = g_1^P(x,t_0) \left( \frac{3t^{J_2(x)} + 5t^{J_1(x)}}{3t_0^{J_2(x)} + 5t_0^{J_1(x)}} \right), \tag{6.26}$$

$$g_1^P(x,t) = g_1^P(x_0,t) \left( \frac{3t^{J_2(x)} + 5t^{J_1(x)}}{3t^{J_2(x_0)} + 5t^{J_1(x_0)}} \right), \tag{6.27}$$

$$g_1^n(x,t) = g_1^n(x,t_0) \left( \frac{5t^{J_1(x)} - 3t^{J_2(x)}}{5t_0^{J_1(x)} - 3t_0^{J_2(x)}} \right) \tag{6.28}$$

and

$$g_1^n(x,t) = g_1^n(x_0,t) \left( \frac{5t^{J_1(x)} - 3t^{J_2(x)}}{5t^{J_1(x_0)} - 3t^{J_2(x_0)}} \right). \tag{6.29}$$

Here $g_1^d(x,t_0)$, $g_1^P(x,t_0)$, $g_1^n(x,t_0)$, $g_1^d(x_0,t)$, $g_1^P(x_0,t)$ and $g_1^n(x_0,t)$ are the values of the spin-dependent structure functions $g_1^d(x,t)$, $g_1^P(x,t)$ and $g_1^n(x,t)$ at $t = t_0$ and $x = x_0$ respectively.

Similarly the t and x-evolutions of spin-dependent gluon structure functions in LO at low-x are given as

$$\Delta G(x,t) = \Delta G(x,t_0) \left(\frac{t}{t_0}\right)^{J_3(x)} \tag{6.30}$$

and

$$\Delta G(x,t) = \Delta G(x_0,t) \, t^{\{J_3(x) - J_3(x_0)\}}. \tag{6.31}$$

Now ignoring the quark contribution to the gluon structure function we get from the standard DGLAP evolution equation (6.3)

$$\frac{\partial \Delta G(x,t)}{\partial t} - \frac{\alpha_s(t)}{2\pi} J_4^G(x,t) = 0, \tag{6.32}$$

$$J_4^G(x,t) = \left\{ 6\left(\frac{11}{12} - \frac{N_f}{18} + \ln(1-x)\right) \Delta G(x,t) + 6 I'_G \right\}$$

and

$$I'_G = \int_x^1 d\omega \left[ \frac{1}{2} \left\{ \frac{(1+\omega^4)\Delta G\left(\frac{x}{\omega},t\right) - 2\Delta G(x,t)}{1-\omega} + \left(\frac{1}{\omega} + \omega^3 - \frac{(1-\omega)^3}{\omega}\right) \Delta G\left(\frac{x}{\omega},t\right) \right\} \right]$$

By the same procedure as above, we get the t and x-evolution equations for the spin-dependent gluon structure function ignoring the quark contribution in LO at low-x respectively as

$$\Delta G(x,t) = G(x,t_0) \left(\frac{t}{t_0}\right)^{B_3(x)} \tag{6.33}$$

and

$$\Delta G(x,t) = G(x_0,t) \, t^{\{B_3(x) - B_3(x_0)\}}, \tag{6.34}$$

where

$$B_3(x)=9A_f\left[\left(\frac{11}{12}-\frac{N_f}{18}+\ln(1-x)\right)+\frac{1}{2}\int_x^1 d\omega\left\{\frac{(1+\omega^4)\omega^{\beta_G}-2}{1-\omega}+\left(\frac{1}{\omega}+\omega^3-\frac{(1-\omega)^3}{\omega}\right)\omega^{\beta_G}\right\}\right]$$

Same as in the spin-independent case, here also the actual functional form of K(x) can be determined by simultaneous solutions of coupled equations of spin-dependent singlet and gluon structure functions. So to overcome the assumption of the ad hoc function K(x), we have to derive the solution of coupled DGLAP evolution equations for spin-dependent singlet and gluon structure functions at low-x in LO considering Regge behaviour of structure functions. We get the solution of coupled DGLAP evolution equations for spin-dependent singlet and gluon structure functions in LO at low-x [Appendix C], respectively as

$$g_1^S(x,t)=C\left(t^{g_5}+t^{g_6}\right) \tag{6.35}$$

and

$$\Delta G(x,t)=C(F_5 t^{g_5}+F_6 t^{g_6}), \tag{6.36}$$

Where

$$g_5=\frac{-(U_3-1)+\sqrt{(U_3-1)^2-4V_3}}{2}, \quad g_6=\frac{-(U_3-1)-\sqrt{(U_3-1)^2-4V_3}}{2}, \quad U_3=1-P_3-S_3,$$

$V_3 = S_3.P_3-Q_3.R_3$, $F_5 = (g_5-P_3)/Q_3$, $F_6 = (g_6-P_3)/Q_3$,

$$P_3=A_f\{3+4\ln(1-x)\}+2\int_x^1\frac{d\omega}{1-\omega}\{(1+\omega^2)\omega^{\beta_S}-2\},$$

$$Q_3=\frac{3}{2}A_f N_f\int_x^1\{\omega^2-(1-\omega)^2\}\omega^{\beta_G}\,d\omega,$$

$$R_3=9A_f\frac{2}{9}\int_x^1\left(\frac{1-(1-\omega)^2}{\omega}\right)\omega^{\beta_S}d\omega$$

And

$$S_3=9A_f\left[\left(\left(\frac{11}{12}-\frac{N_f}{18}\right)+\ln(1-x)\right)+\frac{1}{2}\int_x^1 d\omega\left\{\frac{(1+\omega^4)\omega^{\beta_G}-2}{1-\omega}+\left(\frac{1}{\omega}+\omega^3-\frac{(1-\omega)^3}{\omega}\right)\omega^{\beta_G}\right\}\right].$$

Then we find the t and x-evolution equations for the spin-dependent singlet and gluon structure functions in LO respectively as

$$g_1^S(x,t) = g_1^S(x,t_0) \left( \frac{t^{g_5} + t^{g_6}}{t_0^{g_5} + t_0^{g_6}} \right), \tag{6.37}$$

$$g_1^S(x,t) = g_1^S(x_0,t) \left( \frac{t^{g_5} + t^{g_6}}{t^{g_{50}} + t^{g_{60}}} \right), \tag{6.38}$$

$$\Delta G(x,t) = \Delta G(x,t_0) \left( \frac{F_5 t^{g_5} + F_6 t^{g_6}}{F_5 t_0^{g_5} + F_6 t_0^{g_6}} \right) \tag{6.39}$$

and

$$\Delta G(x,t) = \Delta G(x_0,t) \left( \frac{F_5 t^{g_5} + F_6 t^{g_6}}{F_{50} t^{g_{50}} + F_{60} t^{g_{60}}} \right), \tag{6.40}$$

where $g_{50}, g_{60}, F_{50}$ and $F_{60}$ are the values of $g_5, g_6, F_5$ and $F_6$ at $x = x_0$. The t and x-evolution equations of spin-dependent deuteron structure function from equations (6.37) and (6.38) are respectively

$$g_1^d(x,t) = g_1^d(x,t_0) \left( \frac{t^{g_5} + t^{g_6}}{t_0^{g_5} + t_0^{g_6}} \right), \tag{6.41}$$

$$g_1^d(x,t) = g_1^d(x_0,t) \left( \frac{t^{g_5} + t^{g_6}}{t^{g_{50}} + t^{g_{60}}} \right). \tag{6.42}$$

## 6.2 Results and Discussion

In this chapter, we have compared the results of t and x-evolutions of spin-dependent deuteron, proton and neutron structure functions in LO with different experimental data sets measured by the SLAC-E-143 [114], SLAC-E-154 [115] and SMC [116] collaborations and the result of x-evolution of spin-dependent gluon structure function in LO with the graph obtained by numerical method [111]. The SLAC-E-143 collaborations data sets give the measurement of the spin-dependent structure function of deuteron, proton and neutron in deep inelastic scattering of spin-dependent electrons at incident energies of 9.7, 16.2 and 29.1 GeV on a spin-dependent Ammonia target. Data cover the kinematical x range 0.024 to 0.75 and $Q^2$-range from 0.5 to 10 $GeV^2$. The SMC collaborations data sets give the final results of measurements of the virtual photon asymmetry of deuteron and proton and the spin-dependent deuteron and proton structure functions in DIS of 100 GeV and 190 GeV spin-dependent muons on spin-dependent deuterons and protons. The data cover the kinematic range of x from 0.0008 to 0.7 and $Q^2$-range from 0.2 to 100 $GeV^2$. The SLAC-E-154 collaborations data set give the measurement of the spin-dependent structure function of the neutron in DIS spin-dependent electrons at an incident energy of 48.3 GeV on a spin-dependent Helium-3 target. Data cover the kinematical x-range 0.014 to 0.7 and mean $Q^2$-range from 1 to 17 $GeV^2$.

The graphs 'our result' represent the best fit graph of our work with different experimental data sets and numerical result. Data points at lowest-$Q^2$ values are taken as input to test the t-evolution equations and data points at x<0.1 is taken as input to test the x-evolution equations. We have compared our results for $K(x) = K$, $ax^b$ and $ce^{dx}$. Since $K(x)$ is a function of x only, in our analysis the t-evolution of deuteron, proton and neutron structure functions do not show significant change with the variation of the form of $K(x)$. Here we have presented the result of t-evolution of spin-dependent deuteron, proton and neutron structure functions for $K(x) = ax^b$. For x-evolution of spin-dependent deuteron and gluon structure functions, we found that $K(x) = ax^b$ correspond the best fit graphs whereas for x-evolution of spin-dependent proton and neutron structure functions $K(x)=ce^{dx}$ correspond the best fit graphs. The values of $\beta_S$, $\beta_{NS}$ and $\beta_G$ should be close to 0.5 in quite a broad range of x [41]. We have taken $\beta_S = \beta_{NS} = \beta_G = \beta = 0.5$ in our analysis.

Figure 6.1, represents our result of t-evolution of spin-dependent deuteron structure function in LO from equation (6.24) for $\beta_S = 0.5$ compared with SLAC-E-143 data set for representative values of x and the best fit result is found for a=1 and b= - 0.75.

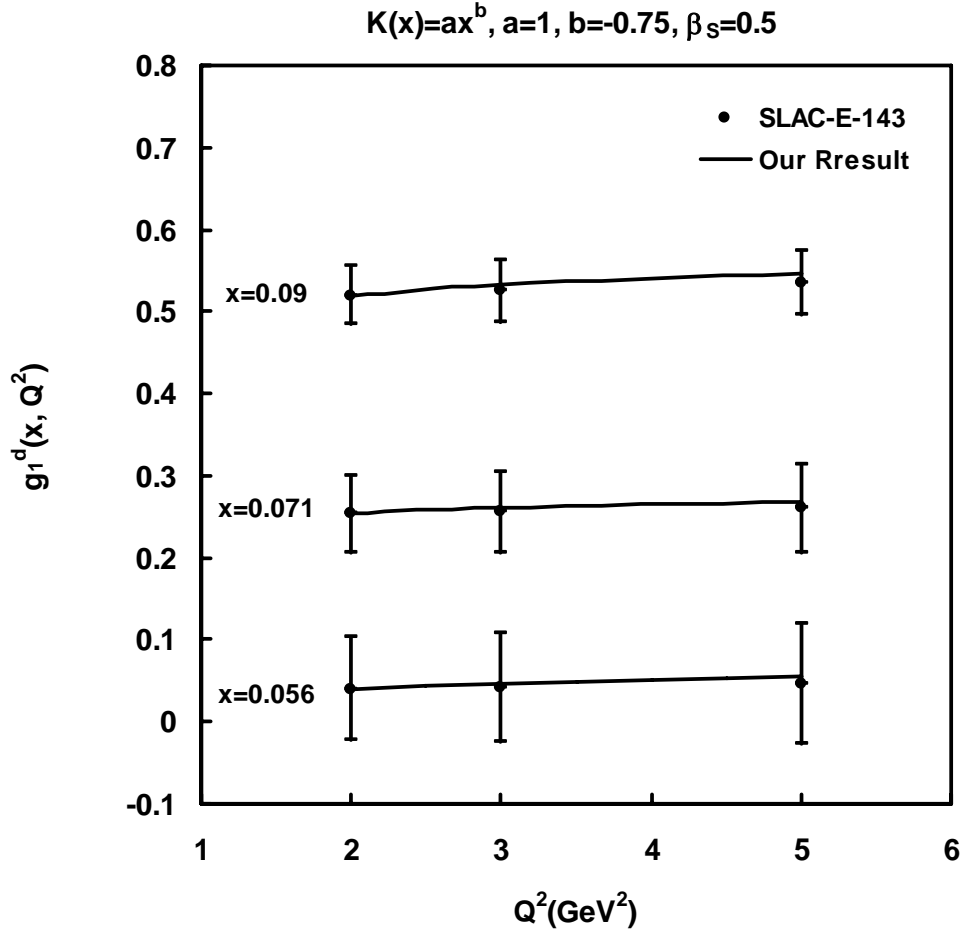

**Figure 6.1**: t -evolution of spin-dependent deuteron structure function in LO at low-x compared with SLAC-E-143 experimental data points. Data are scaled up by +0.2i (i=0, 1, 2) starting from bottom graph.

Figure 6.2(a-b) represent our result of x-evolution of spin-dependent deuteron structure function in LO from equation (6.25) compared with SLAC-E-143 and SMC collaborations data sets for representative values of $Q^2$.

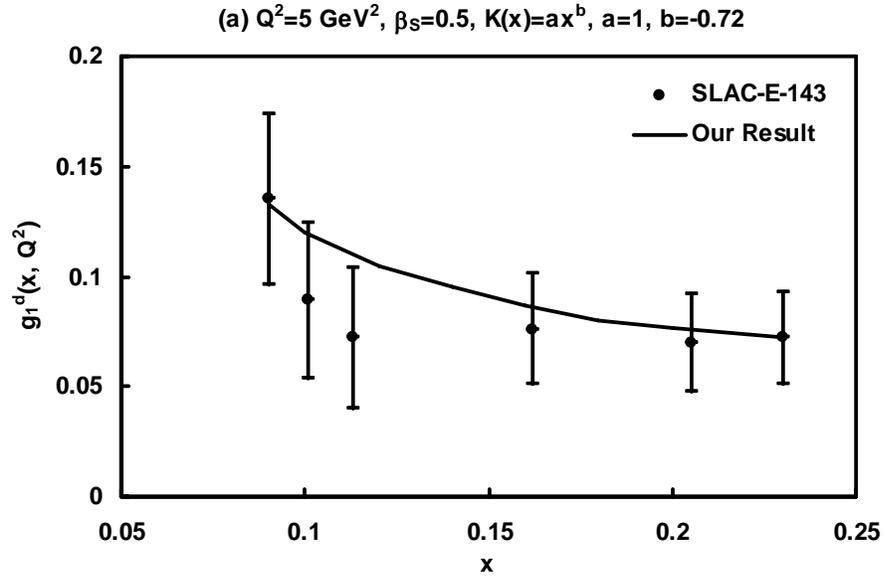

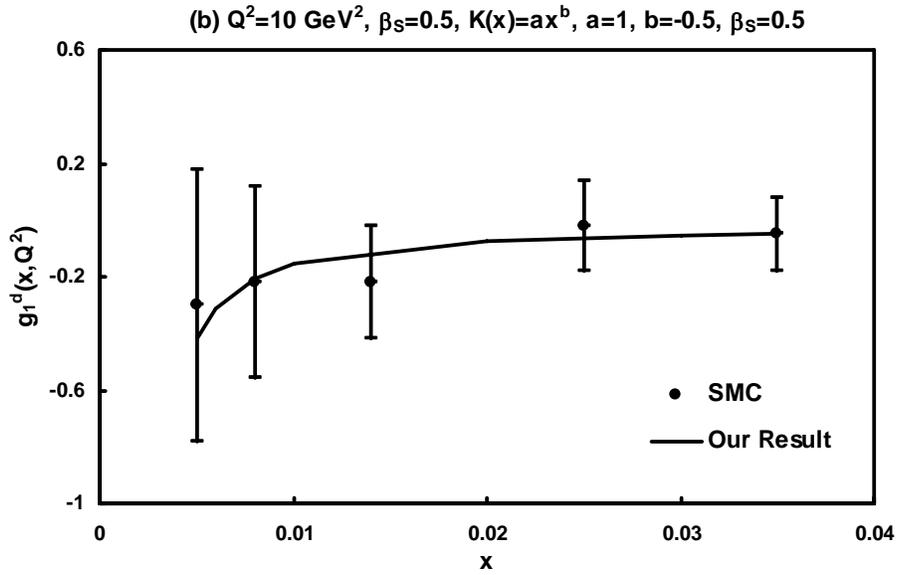

**Figure 6.2**: x -evolution of spin-dependent deuteron structure function in LO at low-x for the representative values of $Q^2$ compared with SLAC-E-143 and SMC collaborations data sets.

Figure 6.2(a) represents the comparison of the result with SLAC-E-143 collaborations data set for $Q^2$= 5 GeV². In this case for a=1 and b= - 0.72, we get the best fit result. Figure 6.2(b) represents the comparison with SMC collaborations data set for $Q^2$= 10 GeV². In this case for a=1 and b = - 0.5, we get the best fit result.

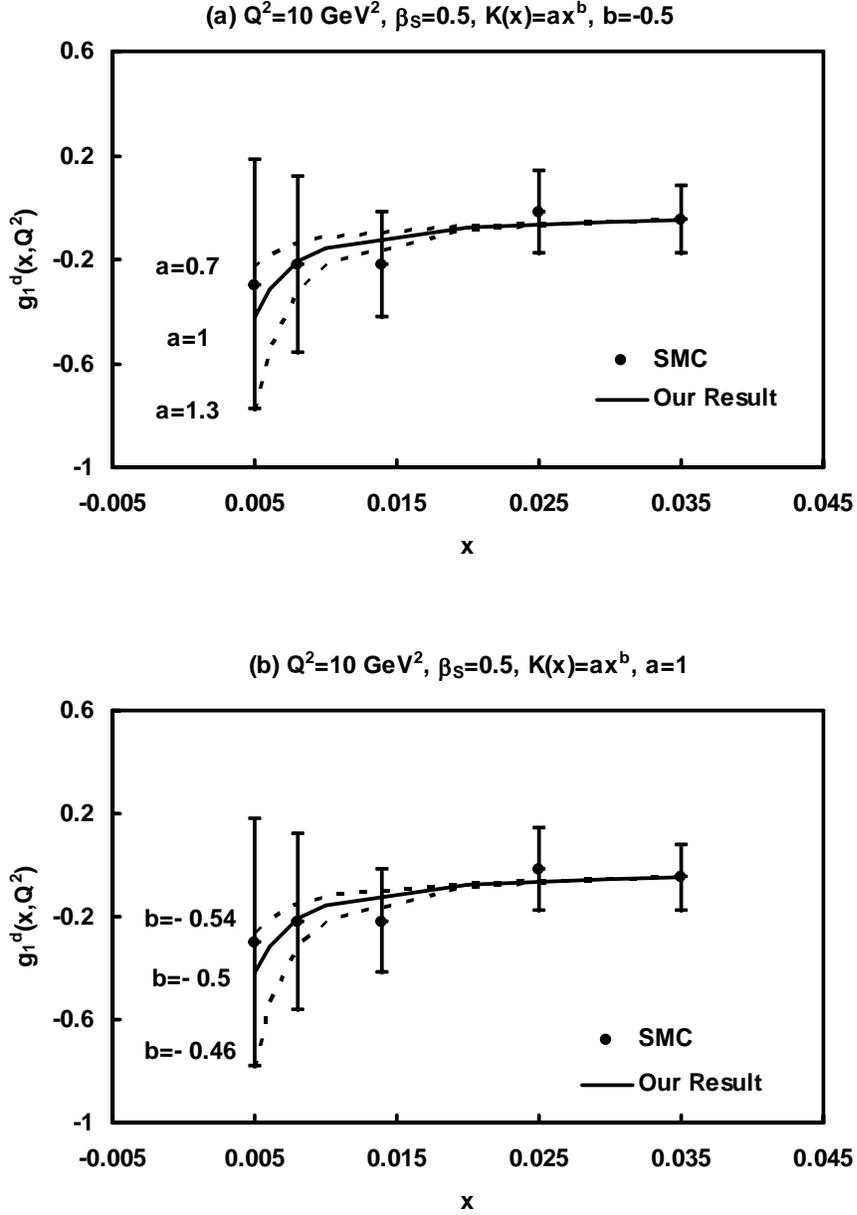

**Figure 6.3:** Sensitivity of the parameters a and b respectively at $Q^2$ = 10 GeV$^2$ with the best fit graphs of the results with SMC collaborations data set.

Figures 6.3(a-b) represents the sensitivity of the parameters a and b respectively. Taking the best fit figures to the x-evolution of spin-dependent deuteron structure function from equation (6.25) compared with SMC collaborations data set at $Q^2$ = 10 GeV$^2$, we have given the ranges of the parameters as $0.7 \leq a \leq 1.3$ and $-0.54 \leq b \leq -0.46$.

Figure 6.4 represents the result of t-evolution of spin-dependent proton structure function in LO from equation (6.26) for $\beta_S = \beta_{NS} = 0.5$ with SLAC-E-143 data set for representative values of x and the best fit result is found for a=1 and b= - 0.55.

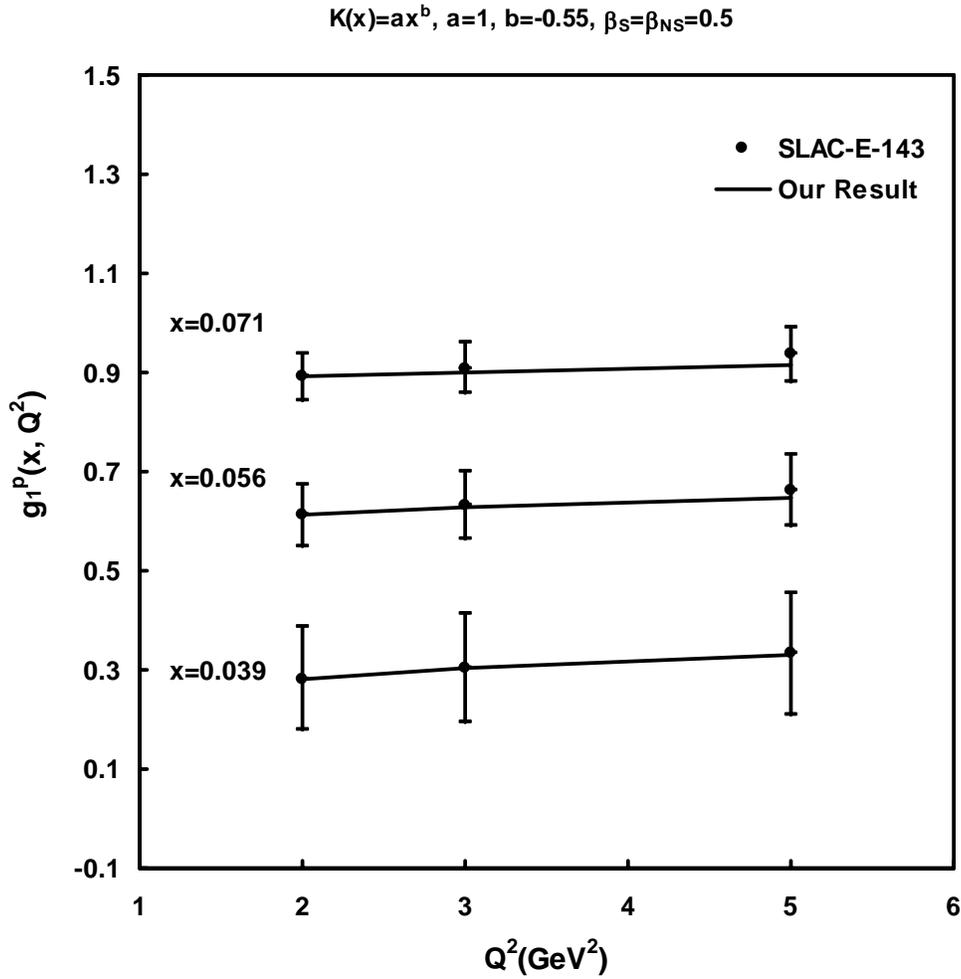

**Figure 6.4**: t -evolution of spin-dependent proton structure function in LO at low-x compared with SLAC-E-143 experimental data points. Data are scaled up by +0.3i (i=0, 1, 2) starting from bottom graph.

Figure 6.5(a-b) represents the result of x-evolution of spin-dependent proton structure function in LO from equation (6.27) for $K(x) = ce^{dx}$ with SLAC-E-143 and SMC collaborations data sets.

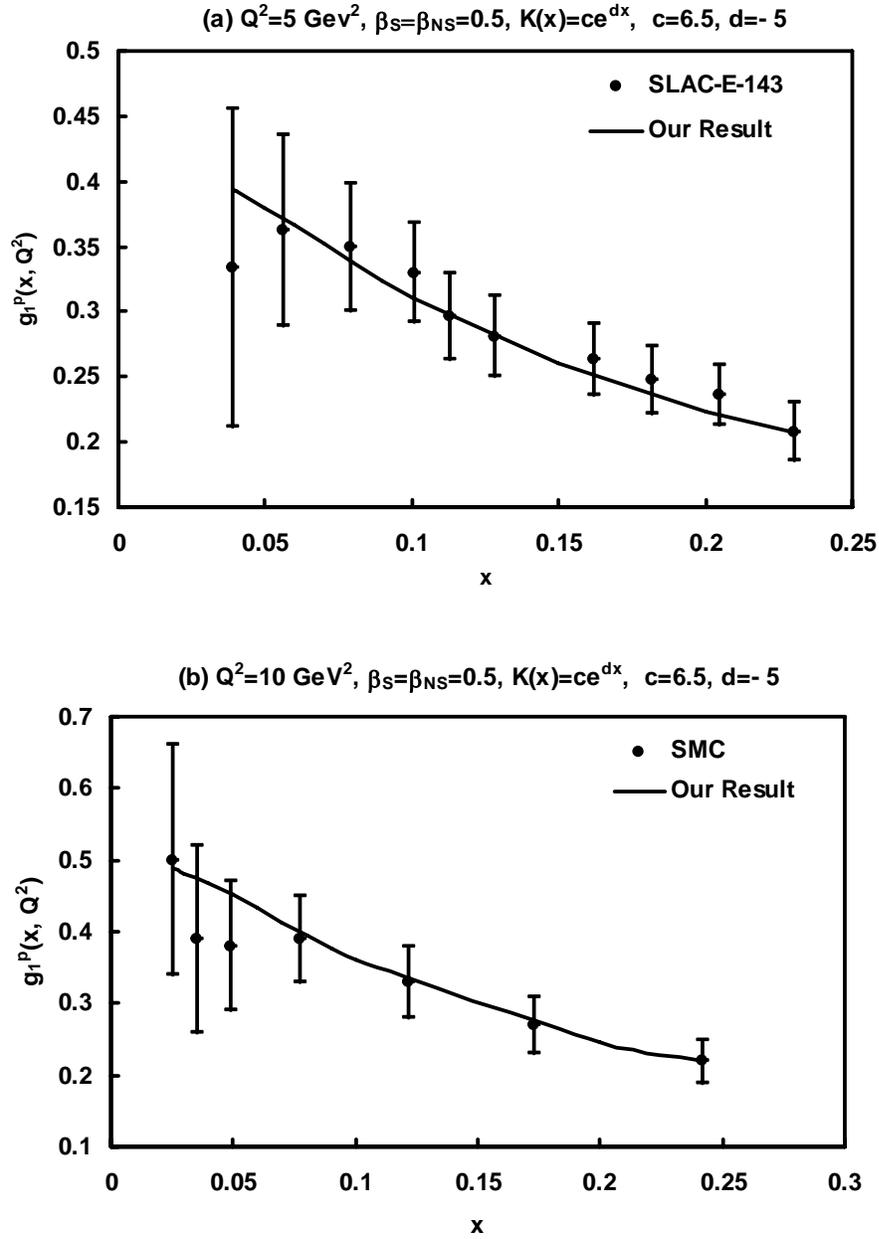

**Figure 6.5**: x -evolution of spin-dependent proton structure function in LO at low-x for the representative values of $Q^2$ compared with SLAC-E-143 and SMC collaborations data sets.

Figure 6.5(a) represent the comparison of the result with SLAC-E-143 collaborations data set for $Q^2$= 5 GeV². And figure 6.5(b) shows the comparison with SMC collaborations data set for $Q^2$= 10 GeV². For c=6.5 and d= - 5 we get the best fit results for both the cases.

In Figure 6.6, we have compared the result of t-evolution of spin-dependent neutron structure function in LO from equation (6.28) for $\beta_S = \beta_{NS} = 0.5$ with SLAC-E-143 data set for $K(x) = ax^b$ for representative values of x and the best fit result is found for a=1 and b= - 0.42.

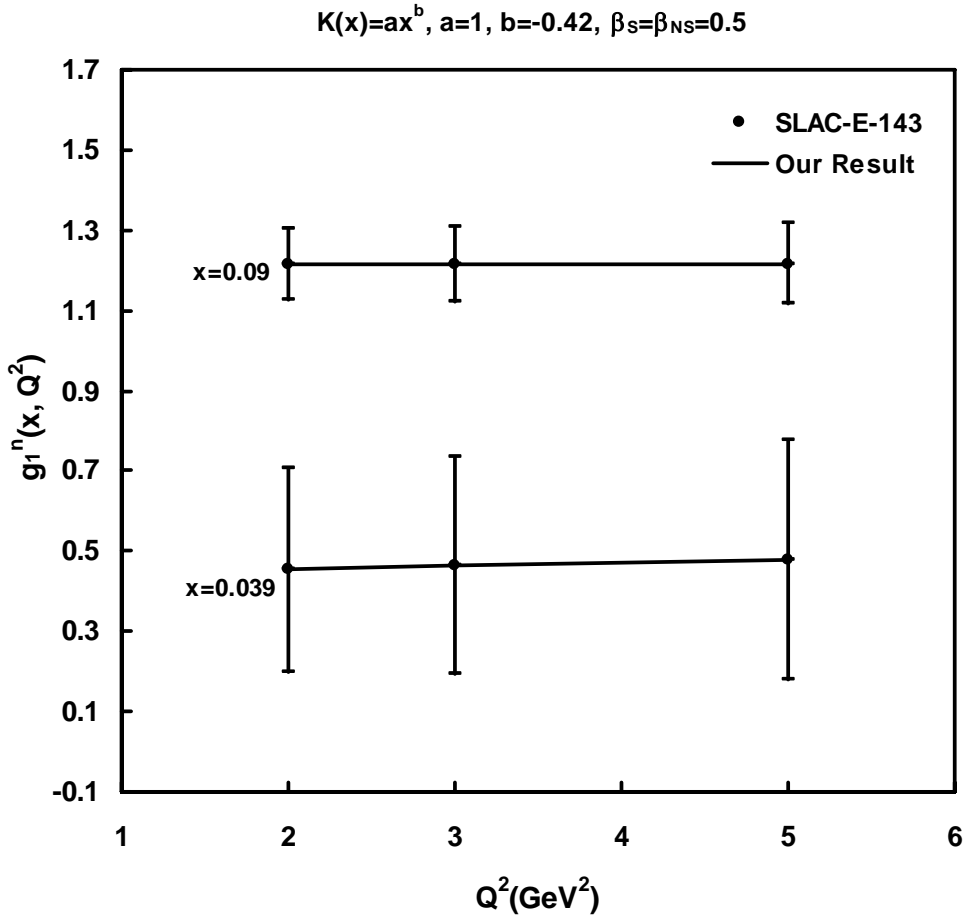

**Figure 6.6**: t -evolution of spin-dependent neutron structure function in LO at low-x compared with SLAC-E-143 experimental data points. Data are scaled up by +0.3i (i=1, 4) starting from bottom graph.

Figure 6.7(a-b) represent the result of x-evolution of spin-dependent neutron structure function in LO from equation (6.29) compared with SLAC-E-154 and SLAC-E-143 collaborations data sets and in both the cases, for c=15 and d= - 5 we get the best fit results.

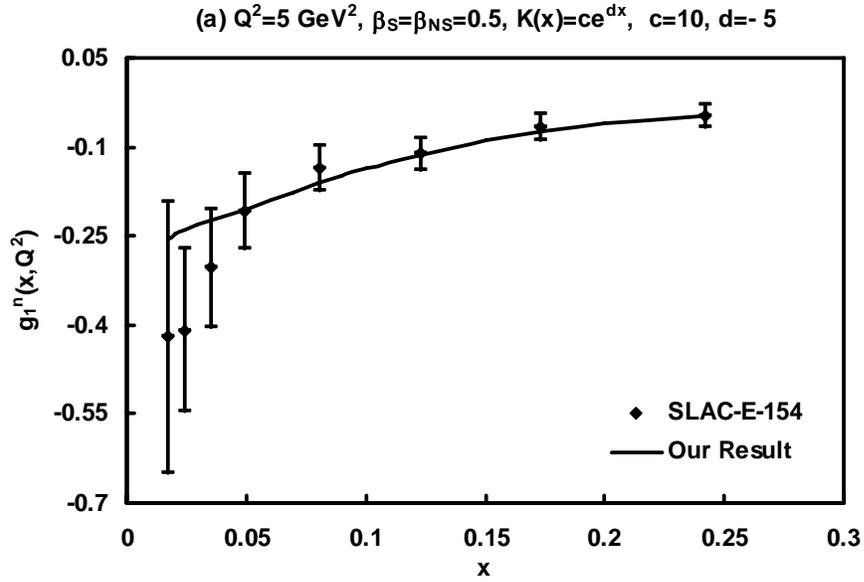

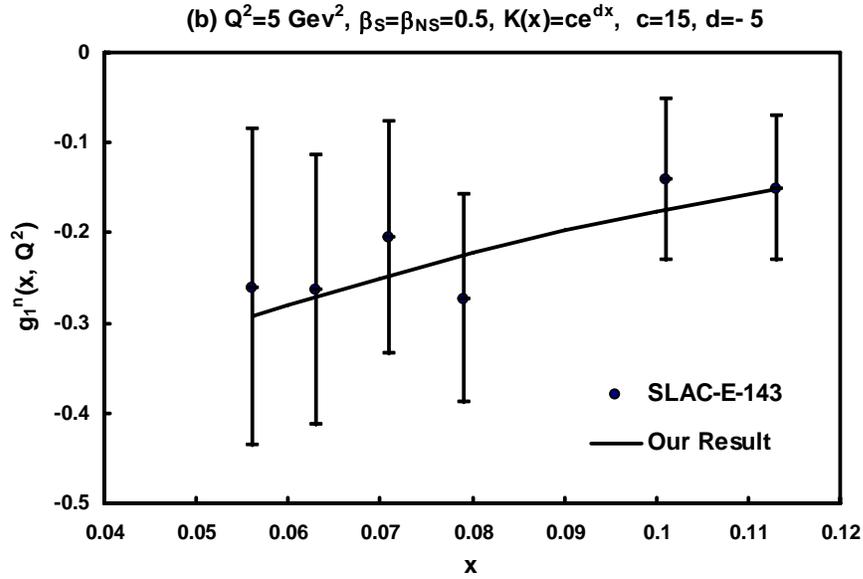

**Figure 6.7**: x -evolution of spin-dependent neutron structure function in LO at low-x for the representative values of x compared with SLAC-E-154 and SLAC-E-143 collaborations data sets.

Figures 6.8(a-b) represent the sensitivity of the parameters c and d respectively. Taking the best fit graphs of the x-evolution of spin-dependent neutron structure function from equation (6.29) compared with SLAC-E-143 collaborations data set at $Q^2 = 5$ GeV$^2$, we observe the ranges of the parameters as $5 \leq c \leq 25$ and $-9 \leq d \leq -1$.

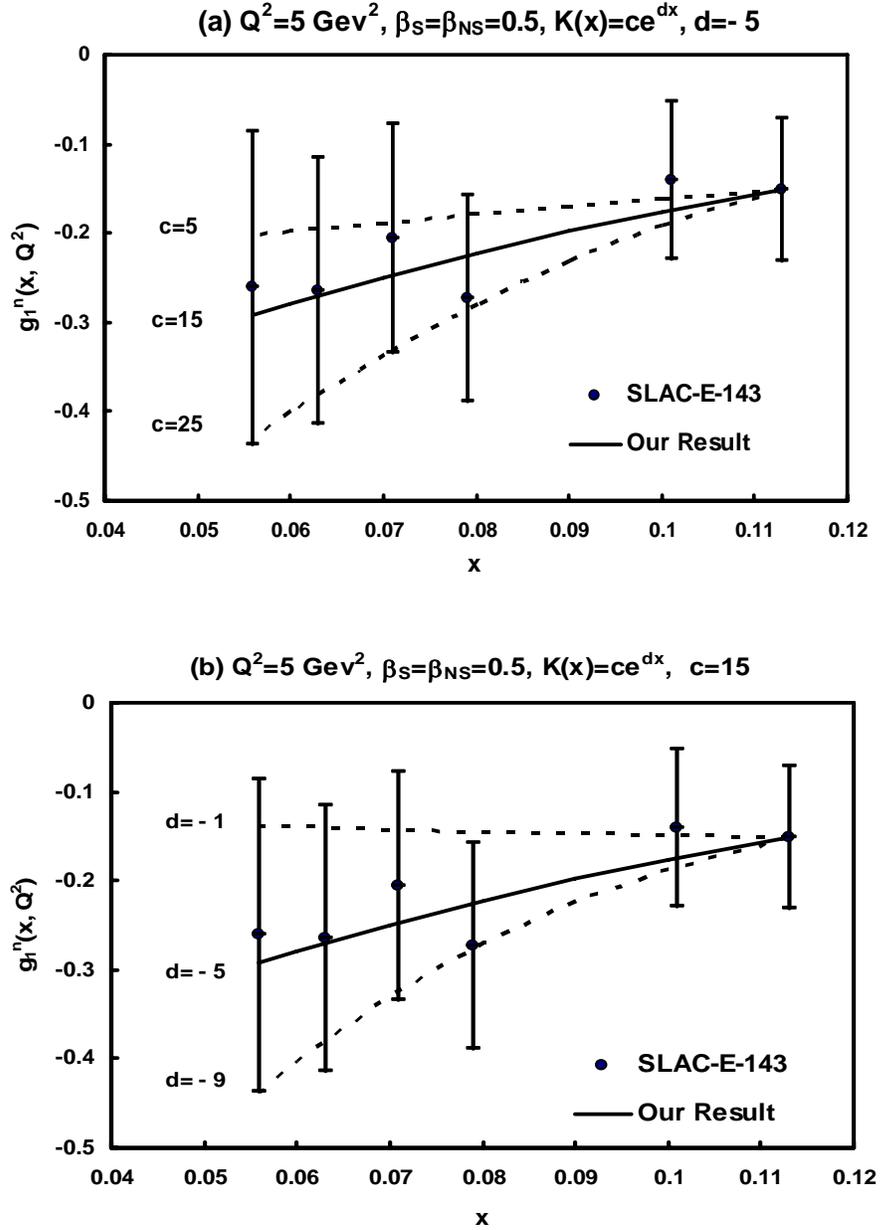

**Figure 6.8:** Sensitivity of the parameters c and d respectively at $Q^2 = 5$ GeV$^2$ with the best fit graphs of the results with SLAC-E-143 collaborations data set.

Figure 6.9 represent our result of x-evolution of spin-dependent gluon structure function in LO from equation (6.31) with the graph obtained by solving unified evolution equation by numerical method. For a=0.01 and b= 0.01 we get the best fit result.

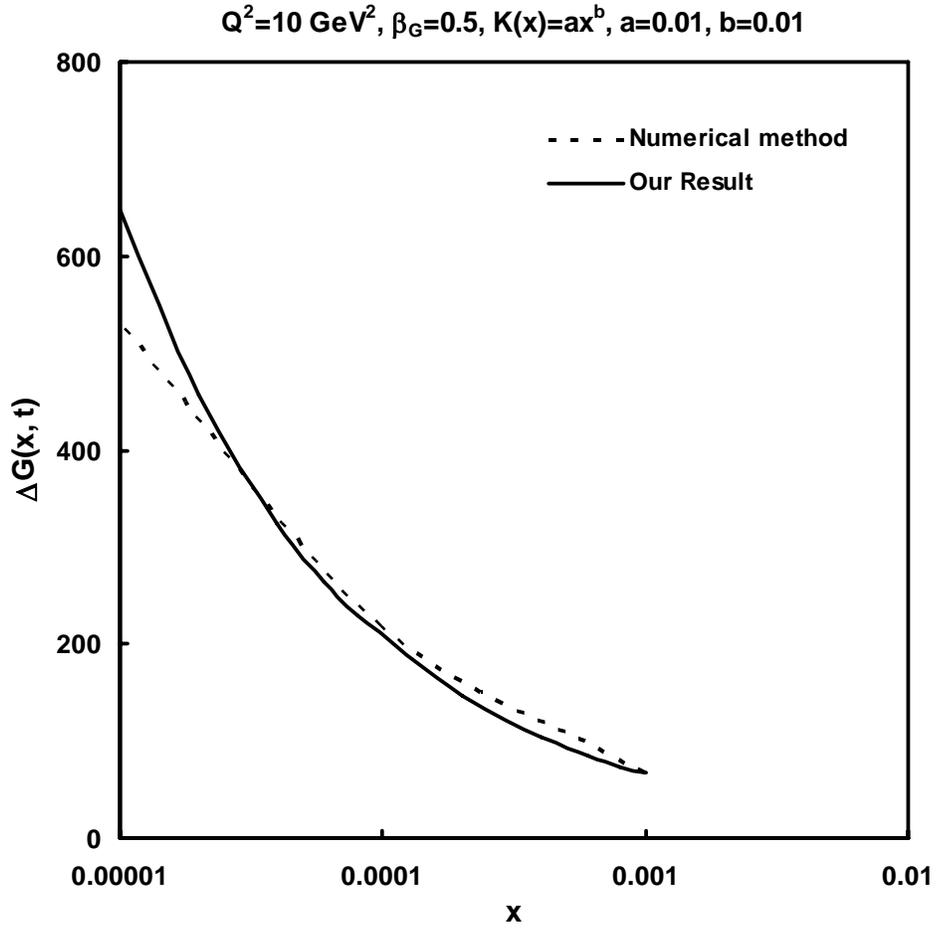

**Figure 6.9**: x-evolution of spin-dependent gluon structure function in LO at low-x for the representative values of $Q^2$ compared with the graph obtained by solving unified evolution equation by numerical method.

Figure 6.10 represents the result of t-evolution of deuteron structure function in LO with SLAC-E-143 collaborations data sets from equation (6.41).

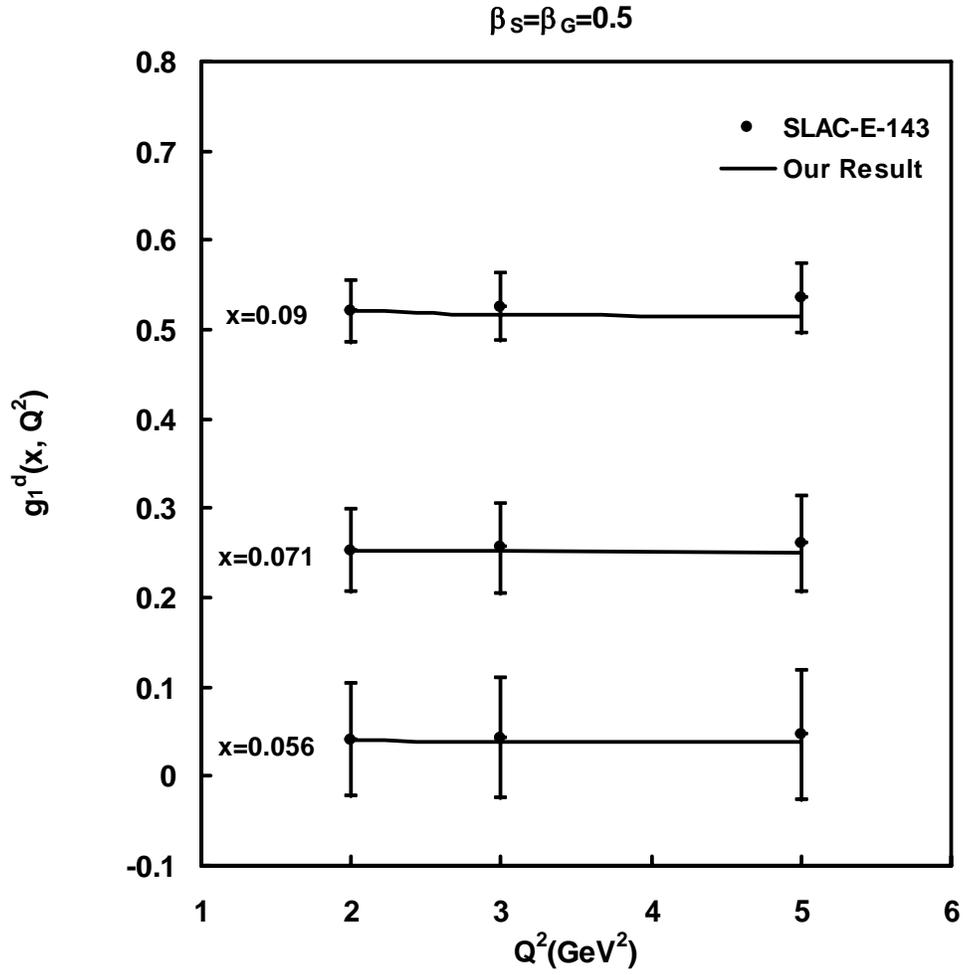

**Figure 6.10**: t evolution of spin-dependent deuteron structure function in LO at low-x obtained by solving coupled equations, compared with SLAC-E-143 collaborations data set.

## 6.3 Conclusion

In this chapter by using Regge behaviour of spin-dependent structure functions at low-x, we have solved DGLAP evolution equations for singlet, non-singlet and gluon structure functions in LO and derived the t and x-evolutions of spin-dependent deuteron, proton, neutron and gluon structure functions. To overcome the problem of ad hoc assumption of the function K(x) we solved coupled evolution equations for singlet and gluon structure functions. The results are in good consistency with SLAC-E-154, SLAC-E-143, SMC collaborations data sets for spin-dependent deuteron, proton and neutron structure functions and the result from numerical method for spin-dependent gluon structure function at low-x and high-$Q^2$ region. We can conclude that Regge behaviour of spin-dependent quark and gluon structure functions are compatible with PQCD at that region assuming the Regge intercept almost same for both quark and gluon. The values of $\beta_S$ and $\beta_G$ are generally close to 0.5 as predicted by Regge theory. □

# Chapter 7

# t and x- Evolutions of Spin-dependent DGLAP Evolution Equations in Next-to-Leading Order

Here we present our solutions of spin-dependent DGLAP evolution equations for singlet, non-singlet and gluon structure functions at low-x in NLO considering Regge behaviour of spin-dependent structure functions. We solved each equation for singlet, non-singlet and gluon structure functions and also the coupled equations for singlet and gluon structure functions. The evolutions of deuteron, proton and neutron structure functions thus obtained have been compared with SLAC-E-154, SLAC-E-143 and SMC collaborations data sets. And the evolution of gluon structure function has been compared with the result obtained by numerical method.

## 7.1 Theory

The spin-dependent DGLAP evolution equations for singlet, non-singlet and gluon structure functions [104, 105] in NLO are respectively

$$\frac{\partial g_1^S(x,t)}{\partial t} - \frac{\alpha_S(t)}{2\pi} J_1^S(x,t) - \left(\frac{\alpha_S(t)}{2\pi}\right)^2 J_2^S(x,t) = 0, \tag{7.1}$$

$$\frac{\partial g_1^{NS}(x,t)}{\partial t} - \frac{\alpha_S(t)}{2\pi} J_1^{NS}(x,t) - \left(\frac{\alpha_S(t)}{2\pi}\right)^2 J_2^{NS}(x,t) = 0 \tag{7.2}$$

and

$$\frac{\partial \Delta G(x,t)}{\partial t} - \frac{\alpha_S(t)}{2\pi} J_1^G(x,t) - \left(\frac{\alpha_S(t)}{2\pi}\right)^2 J_2^G(x,t) = 0, \qquad (7.3)$$

where the equations in LO are as given in chapter 6 (equations 6.1, 6.2 and 6.3) and the NLO contributions are given as [117-119, 105]

$$J_2^S(x,t) = (x-1) g_1^S(x,t) \int_0^1 f(\omega) d\omega + \int_0^1 f(\omega) g_1^S\left(\frac{x}{\omega},t\right) d\omega$$

$$+ \int_x^1 \Delta P_{qq}^S(\omega) g_1^S\left(\frac{x}{\omega},t\right) d\omega + \int_x^1 \Delta P_{qg}^S(\omega) \Delta G\left(\frac{x}{\omega},t\right) d\omega,$$

$$J_2^{NS}(x,t) = (x-1) g_1^{NS}(x,t) \int_0^1 f(\omega) d\omega + \int_x^1 f(\omega) g_1^{NS}\left(\frac{x}{\omega},t\right) d\omega + \int_x^1 \Delta P_{qq}^S(\omega) g_1^{NS}\left(\frac{x}{\omega},t\right) d\omega,$$

and

$$J_2^G(x,t) = (x-1)\Delta G(x,t) \int_0^1 \Delta P_{gg}^S(\omega) d\omega + \int_x^1 \Delta P_{gg}^S(\omega) \Delta G\left(\frac{x}{\omega},t\right) d\omega + \int_x^1 \Delta P_{gq}^S(\omega) g_1^S\left(\frac{x}{\omega},t\right) d\omega,$$

where $f(\omega)$ comes from the unpolarized NLO splitting functions (as given in Chapter 4),

$$\Delta P_{qq}^S(\omega) = 2 C_F T_R N_f \Delta P_{qq}(\omega),$$

$$\Delta P_{qq}(\omega) = (1-\omega) - (1-3\omega)\ln\omega - (1+\omega)\ln^2\omega,$$

$$\Delta P_{qg}^S(\omega) = C_F T_R N_f \Delta P_{qg}^1(\omega) + C_G T_R N_f \Delta P_{qg}^2(\omega),$$

$$\Delta P_{qg}^1(\omega) = -22 + 27\omega - 9\ln\omega + 8(1-\omega)\ln(1-\omega)$$
$$+ (2\omega - 1)\left[2\ln^2(1-\omega) - 4\ln(1-\omega)\ln\omega + \ln^2\omega - \frac{2\pi^2}{3}\right],$$

$$\Delta P_{qg}^2(\omega) = 2(12 - 11\omega) - 8(1-\omega)\ln(1-\omega) + 2(1+8\omega)\ln\omega$$
$$- 2\left\{\ln^2(1-\omega) - \frac{\pi^2}{6}\right\}(2\omega - 1) - \left\{2\int_{\omega/1+\omega}^{1/1+\omega} \frac{dz}{z} \ln\frac{1-z}{z} - 3\ln^2\omega\right\}(-2\omega - 1),$$

$$I_g' = \int_x^1 d\omega \left[\frac{1}{2}\left\{\frac{(1+\omega^4)\Delta G\left(\frac{x}{\omega},t\right) - 2\Delta G(x,t)}{1-\omega} + \left(\frac{1}{\omega} + \omega^3 - \frac{(1-\omega)^3}{\omega}\right)\Delta G\left(\frac{x}{\omega},t\right)\right\}\right]$$

$$+\frac{2}{9}\left(\frac{1-(1-\omega)^2}{\omega}\right)g_1^S\left(\frac{x}{\omega},t\right)\bigg],$$

$$\Delta P_{gg}^S(\omega)=-C_G T_R N_f \Delta P_{gg}^1(\omega)-C_F T_R N_f \Delta P_{gg}^2(\omega)+C_G^2 \Delta P_{gg}^3(\omega),$$

$$\Delta P_{gg}^1(\omega)=4(1-\omega)+\frac{4}{3}(1+\omega)\ln\omega+\frac{20}{9}\left(\frac{1}{(1-\omega)}-2\omega+1\right),$$

$$\Delta P_{gg}^2(\omega)=10(1-\omega)+2(5-\omega)\ln\omega+2(1+\omega)\ln^2\omega,$$

$$\Delta P_{gg}^3(\omega)=\frac{1}{3}(29-67\omega)\ln\omega-\frac{19}{2}(1-\omega)+4(1+\omega)\ln^2\omega$$

$$-2\int_{\omega/1+\omega}^{1/1+\omega}\frac{dz}{z}\ln\frac{1-z}{z}\left(\frac{1}{(1+\omega)}+2\omega+1\right)$$

$$+\left\{\frac{67}{9}-4\ln(1-\omega)\ln\omega+\ln^2\omega-\frac{\pi^2}{3}\right\}\left(\frac{1}{(1-\omega)}-2\omega+1\right),$$

$$\Delta P_{gq}^S(\omega)=C_F T_R N_f \Delta P_{gq}^1(\omega)+C_F^2 \Delta P_{gq}^2(\omega)+C_F C_G \Delta P_{gq}^3(\omega),$$

$$\Delta P_{gq}^1(\omega)=-\frac{4}{9}(\omega+4)-\frac{4}{3}(2-\omega)\ln(1-\omega),$$

$$\Delta P_{gq}^2(\omega)=-\frac{1}{2}-\frac{1}{2}(4-\omega)\ln\omega-(2+\omega)\ln(1-\omega)+\left\{-4-\ln^2(1-\omega)+\frac{1}{2}\ln^2\omega\right\}(2-\omega)$$

and

$$\Delta P_{gq}^3(\omega)=(4-13\omega)\ln\omega+\frac{1}{3}(10+\omega)\ln(1-\omega)+\frac{1}{9}(41+35\omega)$$

$$+\frac{1}{2}\left\{-2\int_{\omega/1+\omega}^{1/1+\omega}\frac{dz}{z}\ln\frac{1-z}{z}+3\ln^2\omega\right\}(2+\omega)$$

$$+\left\{\ln^2(1-\omega)-2\ln(1-\omega)\ln\omega-\frac{\pi^2}{6}\right\}(2-\omega).$$

Applying Regge behaviour of spin-dependent structure functions and the relation between spin-dependent singlet and gluon structure functions as given in Chapter 6 (equations (6.7) to (6.9) and equation (6.13)), we get the solution of DGLAP evolution equations for spin-dependent singlet, non-singlet and gluon structure functions in NLO respectively as

$$g_1^S(x,t)=C\,t^{J_4(x)}, \tag{7.4}$$

$$g_1^{NS}(x,t) = C\, t^{J_5(x)} \tag{7.5}$$

and

$$\Delta G(x,t) = C\, t^{J_6(x)}, \tag{7.6}$$

where C is a constant of integration,

$$J_4(x) = \frac{2}{\beta_0} f_{17}(x) + T_0 \frac{2}{\beta_0} f_{18}(x),$$

$$J_5(x) = \frac{2}{\beta_0} f_{19}(x) + T_0 \frac{2}{\beta_0} f_{20}(x),$$

$$J_6(x) = \frac{2}{\beta_0} f_{21}(x) + T_0 \frac{2}{\beta_0} f_{22}(x),$$

$$f_{17}(x) = \frac{2}{3}\{3 + 4\ln(1-x)\} + \frac{4}{3}\int_x^1 \frac{d\omega}{1-\omega}\{(1+\omega^2)\omega^{\beta_S} - 2\} + N_f \int_x^1 \{\omega^2 - ((1-\omega)^2\}K\left(\frac{x}{\omega}\right)\omega^{\beta_S}\, d\omega,$$

$$f_{18}(x) = (x-1)\int_0^1 f(\omega)\, d\omega + \int_x^1 f(\omega)\omega^{\beta_S}\, d\omega$$

$$+ \int_x^1 \Delta P_{qq}^S(\omega)\omega^{\beta_S}\, d\omega + \int_x^1 \Delta P_{qg}^S(\omega)\omega^{\beta_S} K\left(\frac{x}{\omega}\right)\, d\omega,$$

$$f_{19}(x) = \frac{2}{3}\{3 + 4\ln(1-x)\} + \frac{4}{3}\int_x^1 \frac{d\omega}{1-\omega}\{(1+\omega^2)\omega^{\beta_{NS}} - 2\},$$

$$f_{20}(x) = (x-1)\int_0^1 f(\omega)\, d\omega + \int_x^1 f(\omega)\omega^{\beta_{NS}}\, d\omega + \int_x^1 \Delta P_{qq}^S(\omega)\omega^{\beta_{NS}}\, d\omega,$$

$$f_{21}(x) = 6\left(\frac{11}{12} - \frac{N_f}{18} + \ln(1-x)\right) + 3\int_x^1 d\omega \left\{\frac{(1+\omega^4)\omega^{\beta_G} - 2}{1-\omega} + \left(\frac{1}{\omega} + \omega^3 - \frac{(1-\omega)^3}{\omega}\right)\omega^{\beta_G}\right\}$$

$$+ \frac{4}{3}\int_x^1 d\omega \left(\frac{1-(1-\omega)^2}{\omega}\right)\frac{\omega^{\beta_G}}{K\left(\frac{x}{\omega}\right)}$$

and

$$f_{22}(x) = (x-1)\int_0^1 \Delta P_{gg}^S(\omega)\, d\omega + \int_x^1 \Delta P_{gg}^S(\omega)\omega^{\beta_G}\, d\omega + \int_x^1 \Delta P_{gq}^S(\omega)\frac{\omega^{\beta_G}}{K(x/\omega)}\, d\omega\ .$$

Similarly as in the spin-independent case, here also for possible solutions in NLO, For possible solutions in NLO, we have taken $T(t) = \left(\dfrac{\alpha_s(t)}{2\pi}\right)$ and the expression for T(t) upto LO correction with the assumption $T^2(t) = T_0 T(t)$ [52, 65, 90], where $T_0$ is a numerical parameter. But $T_0$ is not arbitrary. We choose $T_0$ such that difference between $T^2(t)$ and $T_0 T(t)$ is minimum (Figure 7.1).

Applying initial conditions at $t = t_0$, $g_1^S(x,t) = g_1^S(x,t_0)$ and $g_1^{NS}(x,t) = g_1^{NS}(x,t_0)$, and at $x = x_0$, $g_1^S(x,t) = g_1^S(x_0,t)$ and $g_1^{NS}(x,t) = g_1^{NS}(x_0,t)$, The t and x-evolution equations of spin-dependent singlet and non-singlet structure functions corresponding to equations (7.4) to (7.5) are respectively

$$g_1^S(x,t) = g_1^S(x,t_0) \left(\dfrac{t}{t_0}\right)^{J_4(x)}, \tag{7.7}$$

$$g_1^S(x,t) = g_1^S(x_0,t)\, t^{\{J_4(x) - J_4(x_0)\}}, \tag{7.8}$$

$$g_1^{NS}(x,t) = g_1^{NS}(x,t_0) \left(\dfrac{t}{t_0}\right)^{J_5(x)} \tag{7.9}$$

and

$$g_1^{NS}(x,t) = g_1^{NS}(x_0,t)\, t^{\{J_5(x) - J_5(x_0)\}}. \tag{7.10}$$

Using equations (7.7), (7.8), (7.9), and (7.10) as before we get the t and x-evolution equations of spin-dependent deuteron, proton and neutron structure functions in NLO as

$$g_1^d(x,t) = g_1^d(x,t_0) \left(\dfrac{t}{t_0}\right)^{J_4(x)}, \tag{7.11}$$

$$g_1^d(x,t) = g_1^d(x_0,t)\, t^{\{J_4(x) - J_4(x_0)\}}, \tag{7.12}$$

$$g_1^p(x,t) = g_1^p(x,t_0) \left(\dfrac{3t^{J_5(x)} + 5t^{J_4(x)}}{3t_0^{J_5(x)} + 5t_0^{J_4(x)}}\right), \tag{7.13}$$

$$g_1^p(x,t) = g_1^p(x_0,t) \left( \frac{3t^{J_5(x)} + 5t^{J_4(x)}}{3t^{J_5(x_0)} + 5t^{J_4(x_0)}} \right), \quad (7.14)$$

$$g_1^n(x,t) = g_1^n(x,t_0) \left( \frac{5t^{J_4(x)} - 3t^{J_5(x)}}{5t_0^{J_4(x)} - 3t_0^{J_5(x)}} \right) \quad (7.15)$$

and

$$g_1^n(x,t) = g_1^n(x_0,t) \left( \frac{5t^{J_4(x)} - 3t^{J_5(x)}}{5t^{J_4(x_0)} - 3t^{J_5(x_0)}} \right). \quad (7.16)$$

Here $g_1^d(x,t_0), g_1^p(x,t_0), g_1^n(x,t_0), g_1^d(x_0,t), g_1^p(x_0,t)$ and $g_1^n(x_0,t)$ are the values of the spin-dependent structure functions $g_1^d(x,t), g_1^p(x,t)$ and $g_1^n(x,t)$ at $t = t_0$ and $x = x_0$ respectively.

Similarly the t and x-evolutions of spin-dependent gluon structure functions in NLO at low-x are given as

$$\Delta G(x,t) = \Delta G(x,t_0) \left( \frac{t}{t_0} \right)^{J_6(x)} \quad (7.17)$$

and

$$\Delta G(x,t) = \Delta G(x_0,t) t^{\{J_6(x) - J_6(x_0)\}}. \quad (7.18)$$

Now ignoring the quark contribution to the spin-dependent gluon structure function and pursuing the same procedure as in Chapter 6, we get the t and x-evolution equations for the spin-dependent gluon structure function ignoring the quark contribution in NLO at low-x respectively as

$$G(x,t) = G(x,t_0) \left( \frac{t}{t_0} \right)^{B_4(x)} \quad (7.19)$$

and

$$G(x,t) = G(x_0,t) t^{\{B_4(x) - B_4(x_0)\}}, \quad (7.20)$$

where

$$B_4(x) = \frac{2}{\beta_0} \cdot f_{23}(x) + T_0 \cdot \frac{2}{\beta_0} \cdot f_{24}(x),$$

$$f_{23}(x) = 6\left(\frac{11}{12} - \frac{N_f}{18} + \ln(1-x)\right) + 3\int_x^1 d\omega \left[\frac{(1+\omega^4)\omega^{\beta_G} - 2}{1-\omega} + \left(\frac{1}{\omega} + \omega^3 - \frac{(1-\omega)^3}{\omega}\right)\omega^{\beta_G}\right],$$

and

$$f_{24}(x) = (x-1)\int_0^1 \Delta P_{gg}^S(\omega) d\omega + \int_x^1 \Delta P_{gg}^S(\omega) \omega^{\beta_G} d\omega.$$

We get the solution of the coupled DGLAP evolution equations for spin-dependent singlet and gluon structure functions in NLO at low-x, respectively as [Appendix C]

$$g_1^S(x,t) = C\left(t^{g_7} + t^{g_8}\right) \tag{7.21}$$

and

$$\Delta G(x,t) = C(F_7 t^{g_7} + F_8 t^{g_8}), \tag{7.22}$$

where $g_7 = \dfrac{-(u_4-1) + \sqrt{(u_4-1)^2 - 4v_4}}{2}$, $g_8 = \dfrac{-(u_4-1) - \sqrt{(u_4-1)^2 - 4v_4}}{2}$, $u_4 = 1 - P_4 - S_4$,

$v_4 = S_4 \cdot P_4 - Q_4 \cdot R_4$, $F_7 = (g_7 - P_4)/Q_4$, $F_8 = (g_8 - P_4)/Q_4$,

$$P_4 = \frac{3}{2} A_f \left[\frac{2}{3}\{3 + 4\ln(1-x)\} + \frac{4}{3}\int_x^1 \frac{d\omega}{1-\omega}\{(1+\omega^2)\omega^{\beta_S} - 2\}\right.$$

$$\left. + T_0 \left\{(x-1)\int_0^1 f(\omega) d\omega + \int_x^1 f(\omega)\omega^{\beta_S} d\omega \int_x^1 \Delta P_{qq}^S(\omega) \omega^{\beta_S} d\omega\right\}\right],$$

$$Q_4 = \frac{3}{2} A_f \left[N_f \int_x^1 \{\omega^2 - (1-\omega)^2\}\omega^{\beta_G} d\omega + T_0 \int_x^1 \Delta P_{qg}^S(\omega)\omega^{\beta_G} d\omega\right],$$

$$R_4 = \frac{3}{2} A_f \left[\frac{4}{3}\int_x^1 \left(\frac{1-(1-\omega)^2}{\omega}\right)\omega^{\beta_S} d\omega + T_0 \int_x^1 \Delta P_{gq}^S(\omega)\omega^{\beta_S} d\omega\right]$$

and

$$S_4 = \frac{3}{2} A_f \left[6\left\{\left(\frac{11}{12} - \frac{N_f}{18}\right) + \ln(1-x)\right\} + 3\int_x^1 d\omega \left\{\frac{(1+\omega^4)\omega^{\beta_G} - 2}{1-\omega} + \left(\frac{1}{\omega} + \omega^3 - \frac{(1-\omega)^3}{\omega}\right)\omega^{\beta_G}\right\}\right]$$

$$+ T_0 \left\{ (x-1) \int_0^1 \Delta P_{gg}^S(\omega) \, d\omega + \int_x^1 \Delta P_{gg}^S(\omega) \, \omega^{\beta_G} \, d\omega \right\} \right] .$$

Then we find the t and x-evolution equations for the spin-dependent singlet and gluon structure functions in NLO respectively as

$$g_1^S(x,t) = g_1^S(x,t_0) \left( \frac{t^{g_7} + t^{g_8}}{t_0^{g_7} + t_0^{g_8}} \right), \qquad (7.23)$$

$$g_1^S(x,t) = g_1^S(x_0,t) \left( \frac{t^{g_7} + t^{g_8}}{t^{g_{70}} + t^{g_{80}}} \right), \qquad (7.24)$$

$$\Delta G(x,t) = \Delta G(x,t_0) \left( \frac{F_7 t^{g_7} + F_8 t^{g_8}}{F_7 t_0^{g_7} + F_8 t_0^{g_8}} \right) \qquad (7.25)$$

and

$$\Delta G(x,t) = \Delta G(x_0,t) \left( \frac{F_7 t^{g_7} + F_8 t^{g_8}}{F_{70} t^{g_{70}} + F_{80} t^{g_{80}}} \right), \qquad (7.26)$$

where $g_{70}$, $g_{80}$, $F_{70}$ and $F_{80}$ are the values of $g_7$, $g_8$, $F_7$ and $F_8$ at $x = x_0$. The t and x-evolution equations of spin-dependent deuteron structure function from equations (7.23) and (7.24) are respectively

$$g_1^d(x,t) = g_1^d(x,t_0) \left( \frac{t^{g_7} + t^{g_8}}{t_0^{g_7} + t_0^{g_8}} \right), \qquad (7.27)$$

and

$$g_1^d(x,t) = g_1^d(x_0,t) \left( \frac{t^{g_7} + t^{g_8}}{t^{g_{70}} + t^{g_{80}}} \right). \qquad (7.28)$$

## 7.2  Results and Discussion

In this chapter, we have compared the results of t and x-evolutions of spin-dependent deuteron, proton and neutron structure functions in NLO with different experimental data sets measured by the SLAC-E-143 [114], SLAC-E-154 [115] and SMC [116] collaborations and the result of x-evolution of spin-dependent gluon structure function in NLO with the graph obtained by numerical method [111].

Each graph representing 'our result' is the best fit graph of our work with the experimental data sets and numerical method. Data points from experimental data sets, at lowest-$Q^2$ values are taken as input to test the t-evolution equations of the results and at x <0.1 is taken as input to test the x-evolution equations of the results. Similarly as given in Chapter 6, we compared the results for $K(x) = ax^b$ and $ce^{dx}$, where a, b, c and d are constants. As $K(x)$ is a function of x only, in our work the t-evolution of deuteron, proton and neutron structure functions do not show significant change with the variation of the form of $K(x)$. Here we have presented the result of t-evolution of spin-dependent deuteron, proton and neutron structure functions for $K(x) = ax^b$. For x-evolution of spin-dependent deuteron and gluon structure functions, we found the best fitted graphs correspond to $K(x) = ax^b$ whereas in our work for x-evolution of spin-dependent proton and neutron structure functions, we found the best fitted graphs correspond to $K(x)=ce^{dx}$. We have taken $\beta_S = \beta_{NS} = \beta_G = \beta = 0.5$ in our calculation. Along with the NLO results we have also presented our LO results from Chapter 6.

In Figure 7.1 we have plotted $T(t)^2$ and $T_0T(t)$ against $Q^2$ in the $Q^2$-range $0 \leq Q^2 \leq 30$ GeV². Here we observe that for $T_0 = 0.108$, errors become minimum in the range $0 \leq Q^2 \leq 30$ GeV².

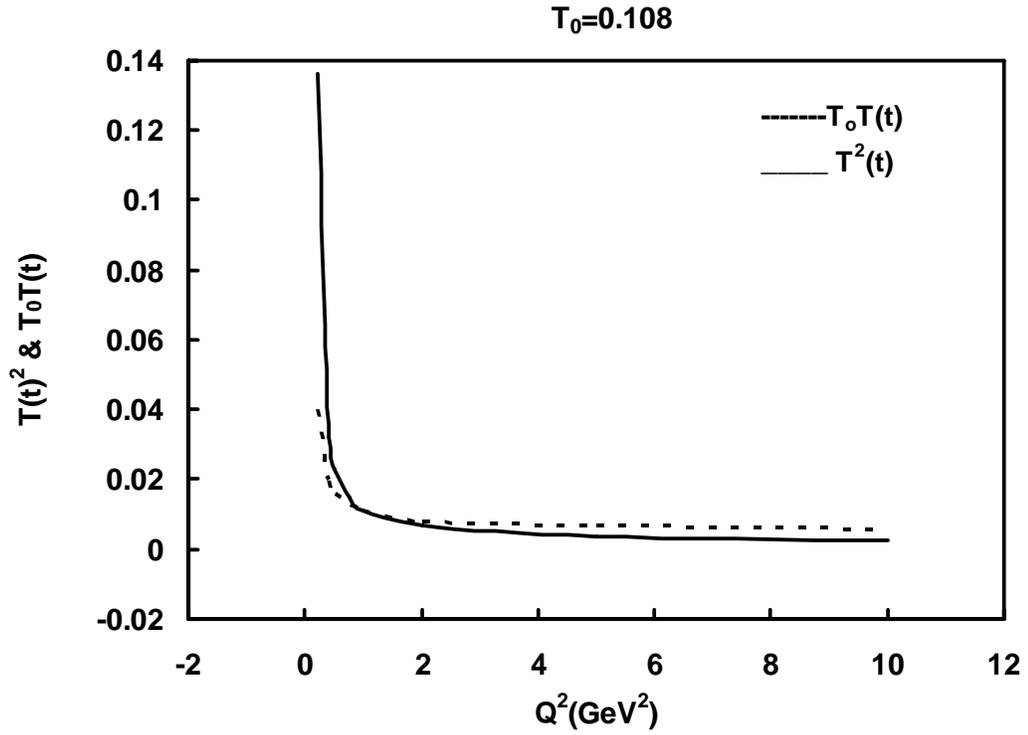

**Figure 7.1:** The variation of $T(t)^2$ and $T_0T(t)$ with $Q^2$.

Figure 7.2 represents the result of t-evolution of spin-dependent deuteron structure function in NLO from equation (7.11) with SLAC-E-143 data set. The best fit result is found for a=1 and b= - 0.75.

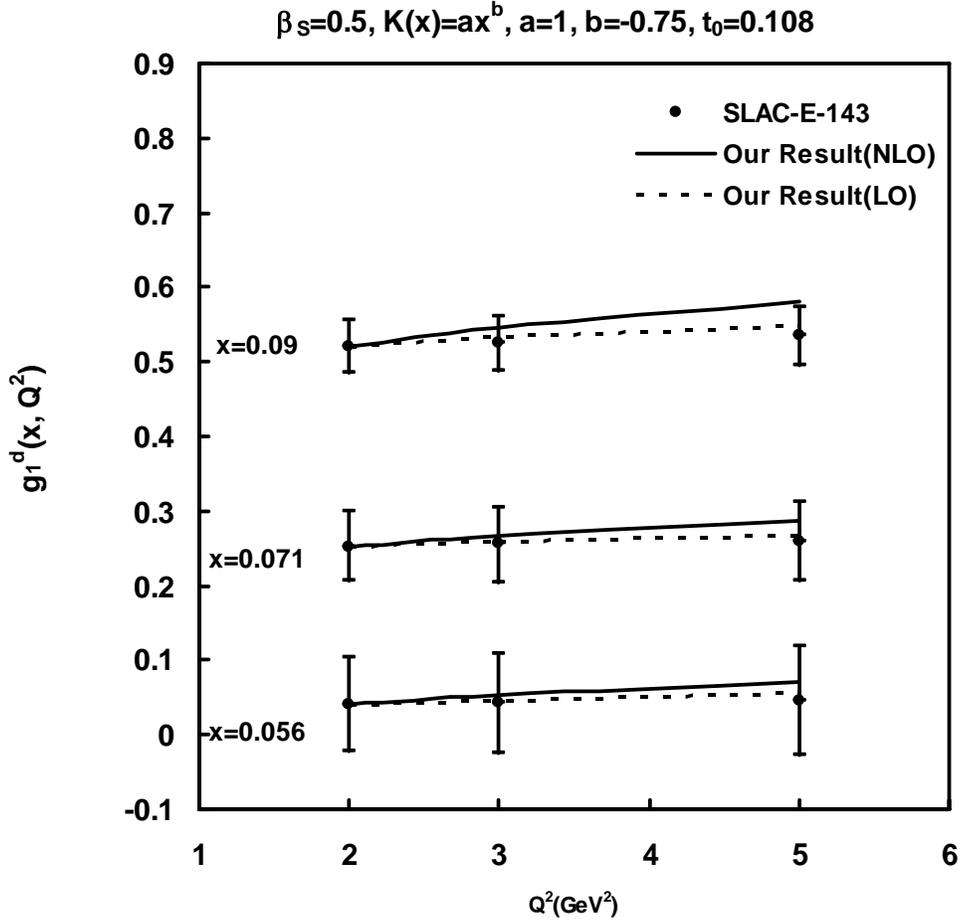

**Figure 7.2**: t-evolution of spin-dependent deuteron structure function in NLO at low-x compared with SLAC-E-143 data set. Data are scaled up by +0.2i (i=0, 1, 2) starting from bottom graph.

Figure 7.3(a-b) represent the result of x-evolution of spin-dependent deuteron structure function in NLO from equation (7.12) with SLAC-E-143 and SMC collaborations data sets for representative values of $Q^2$. Figure 7.3(a) represents the comparison of the result with SLAC-E-143 collaborations data set for $Q^2$= 5 GeV$^2$ and a=1 and b= - 0.72 correspond the best fit result. Figure 7.3(b) represent the comparison with SMC collaborations data set for $Q^2$= 10 GeV$^2$ and a=1 and b= - 0.4 correspond the best fit result.

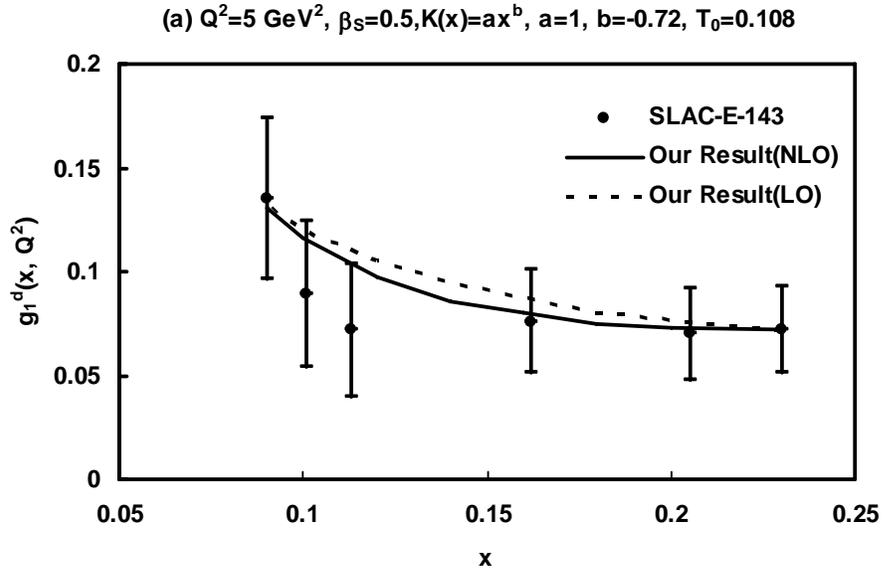

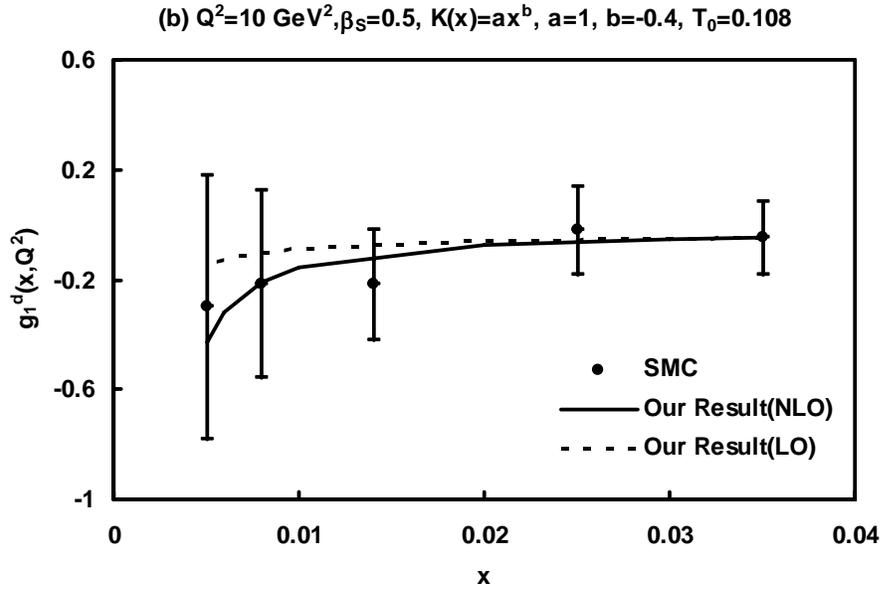

**Figure 7.3**: x-evolution of spin-dependent deuteron structure function in NLO at low-x for the representative values of $Q^2$ compared with SLAC-E-143 and SMC collaborations data sets.

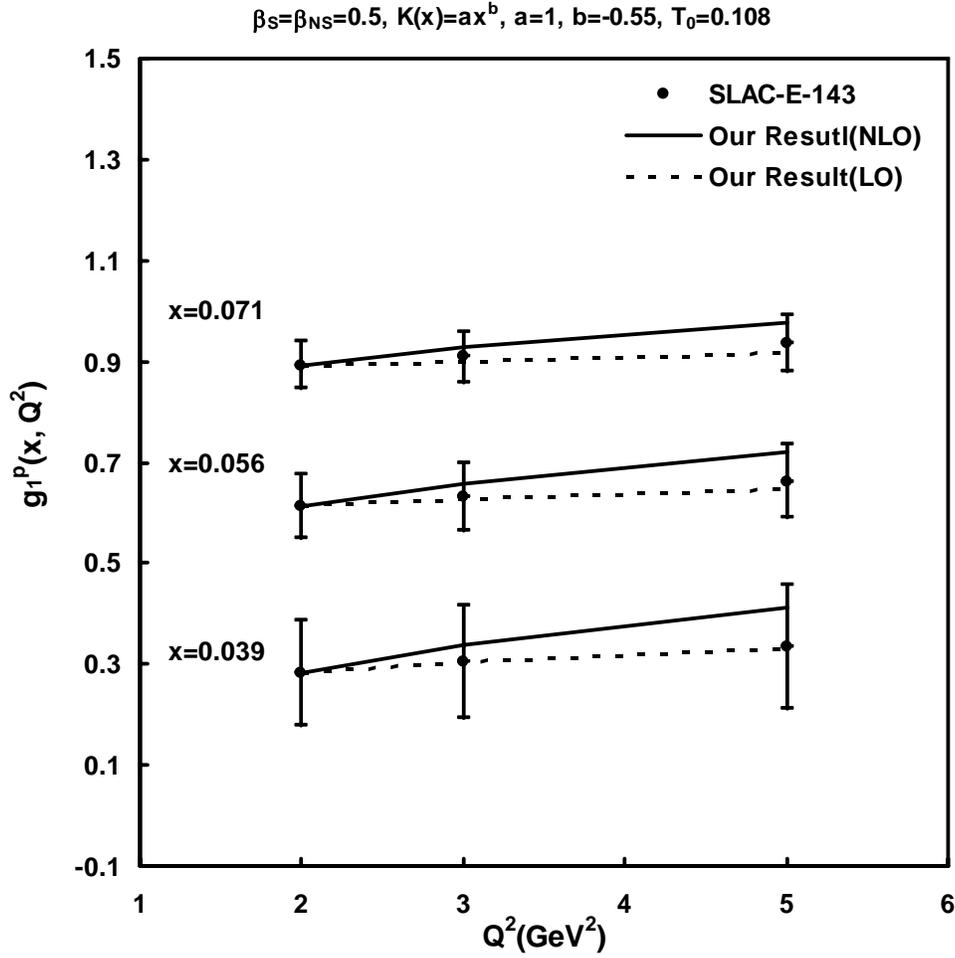

**Figure 7.4**: t-evolution of spin-dependent proton structure function in NLO at low-x compared with SLAC-E-143 experimental data points. Data are scaled up by +0.3i (i=0, 1, 2) starting from bottom graph.

Figure 7.4 represents the result of t-evolution of spin-dependent proton structure function in NLO from equation (7.13) with SLAC-E-143 data set and the best fit result is found for a=1 and b= - 0.55.

Figure 7.5(a-b) represent the result of x-evolution of spin-dependent proton structure function in NLO from equation (7.14) with SLAC-E-143 and SMC collaborations data sets. Figure 7.5(a) represents the comparison of our result with SLAC-E-143 collaborations data set for $Q^2$= 5 GeV$^2$. Figure 7.5(b)

shows the comparison with SMC collaborations data set for $Q^2 = 10$ GeV$^2$. In both the cases, for c=6.5 and d= - 5 we get the best fit results.

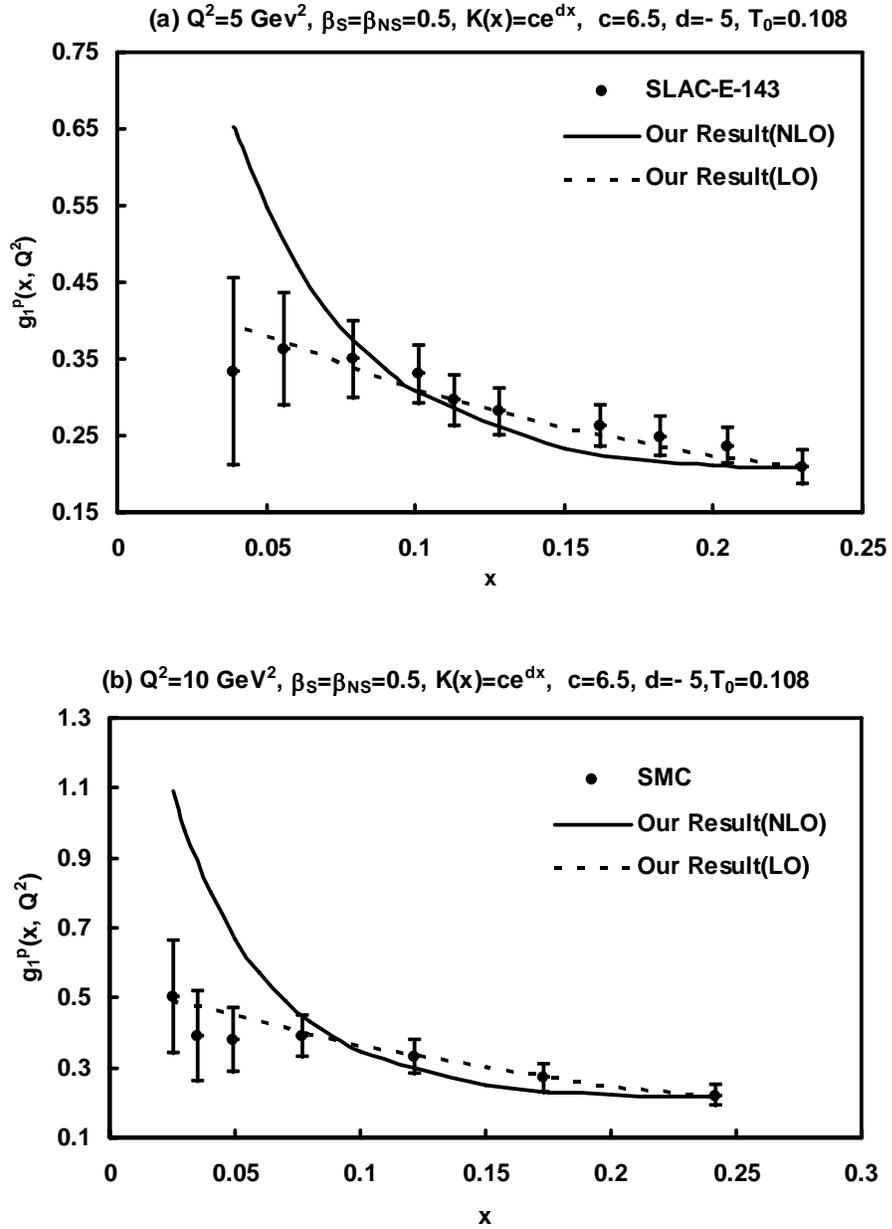

**Figure 7.5**: x-evolution of spin-dependent proton structure function in NLO at low-x for the representative values of $Q^2$, compared with SLAC-E-143 and SMC collaborations data sets.

Figure 7.6 represents the result of t-evolution of spin-dependent neutron structure function in NLO from equation (7.15) with SLAC-E-143 data set and the best fit result is found for a=1 and b= - 0.42.

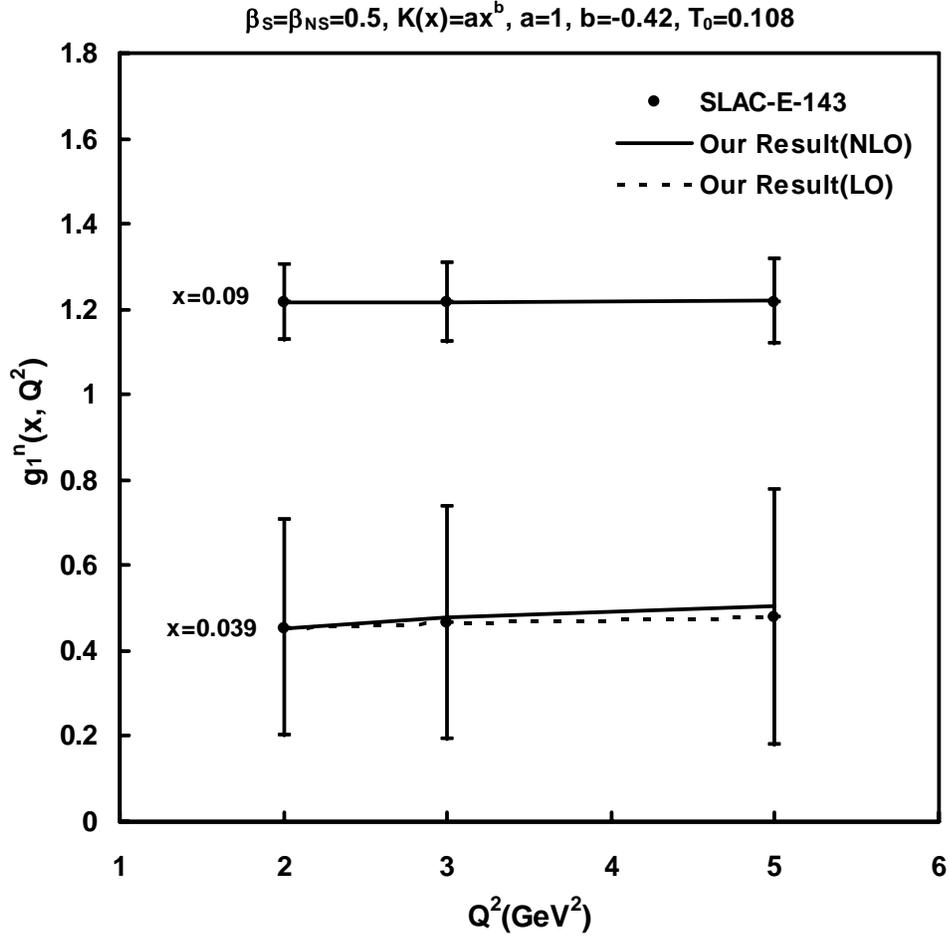

**Figure 7.6**: t-evolution of spin-dependent neutron structure function in NLO at low-x compared with SLAC-E-143 experimental data points. Data are scaled up by +0.3i (i=1, 4) starting from bottom graph.

Figure 7.7 (a-b) represent the result of x-evolution of spin-dependent neutron structure function in NLO from equation (7.16) with SLAC-E-154 and SLAC-E-143 collaborations data sets. Figure 7.7(a) represents the comparison of our result with SLAC-E-154 collaborations data set for $Q^2$= 5 GeV$^2$. Figure 7.7(b) represents the comparison with SLAC-E-143 collaborations data set for $Q^2 = 5$ GeV$^2$. In both the cases, c=5 and d= -5 correspond the best fit results.

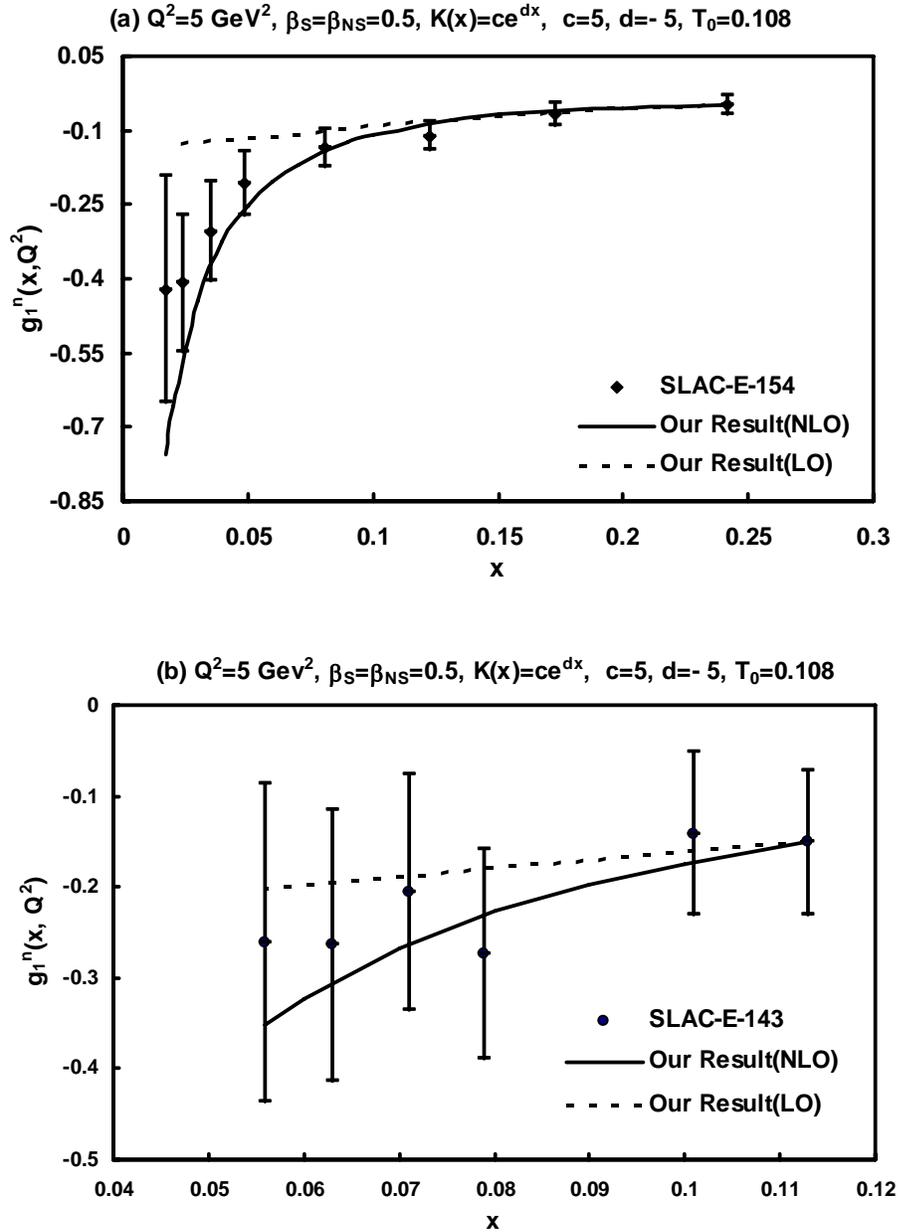

**Figure 7.7**: x-evolution of spin-dependent neutron structure function in NLO at low-x for the representative values of $Q^2$ compared with SLAC-E-154 and SLAC-E-143 collaborations data sets.

Figure 7.8 represents the result of x-evolution of spin-dependent gluon structure function in NLO from equation (7.18) with the graph obtained by numerical method and a=0.01 and b= 0.003 correspond the best fit result.

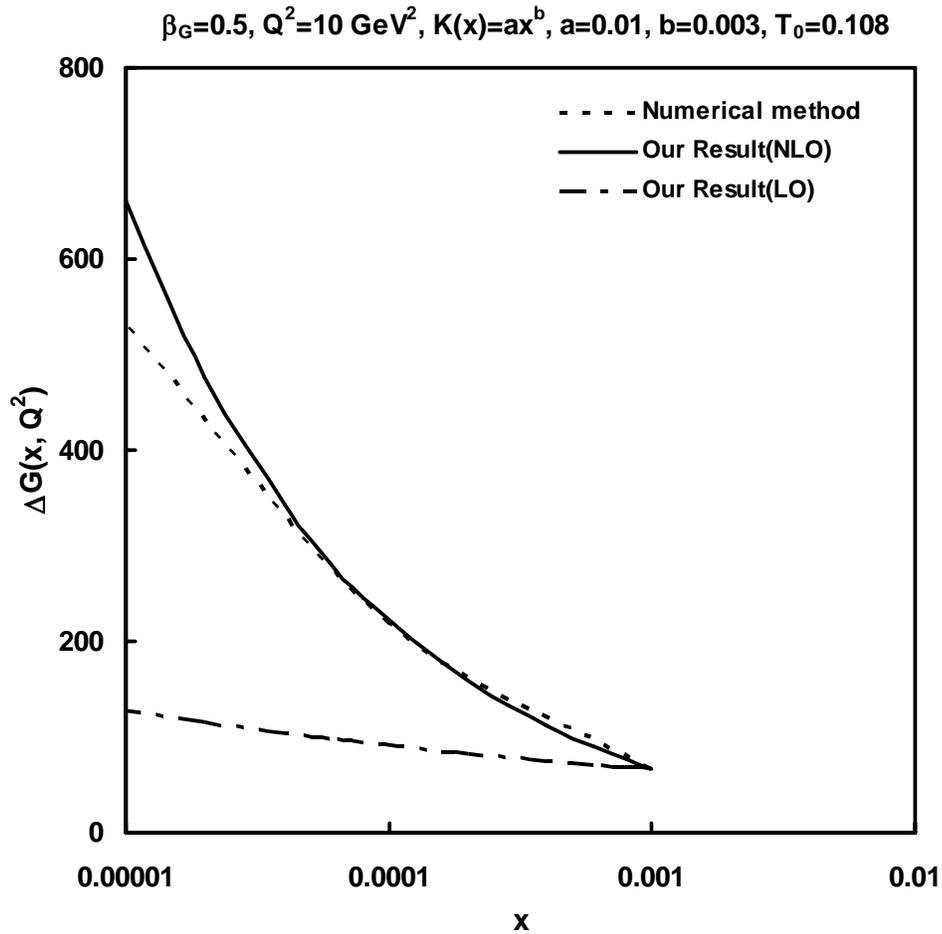

**Figure 7.8**: x-evolution of spin-dependent gluon structure function in NLO at low-x for the representative values of $Q^2$ compared with the graph obtained by numerical method.

Figure 7.9 represents the result of t-evolution of deuteron structure function in NLO from equation (7.27) with SLAC-E-143 collaborations data set. The best fit graph of the result in NLO overlaps with the graph of the result in LO (from Chapter 6) for $\beta_S =\beta_G =0.5$ and $T_0=0.108$.

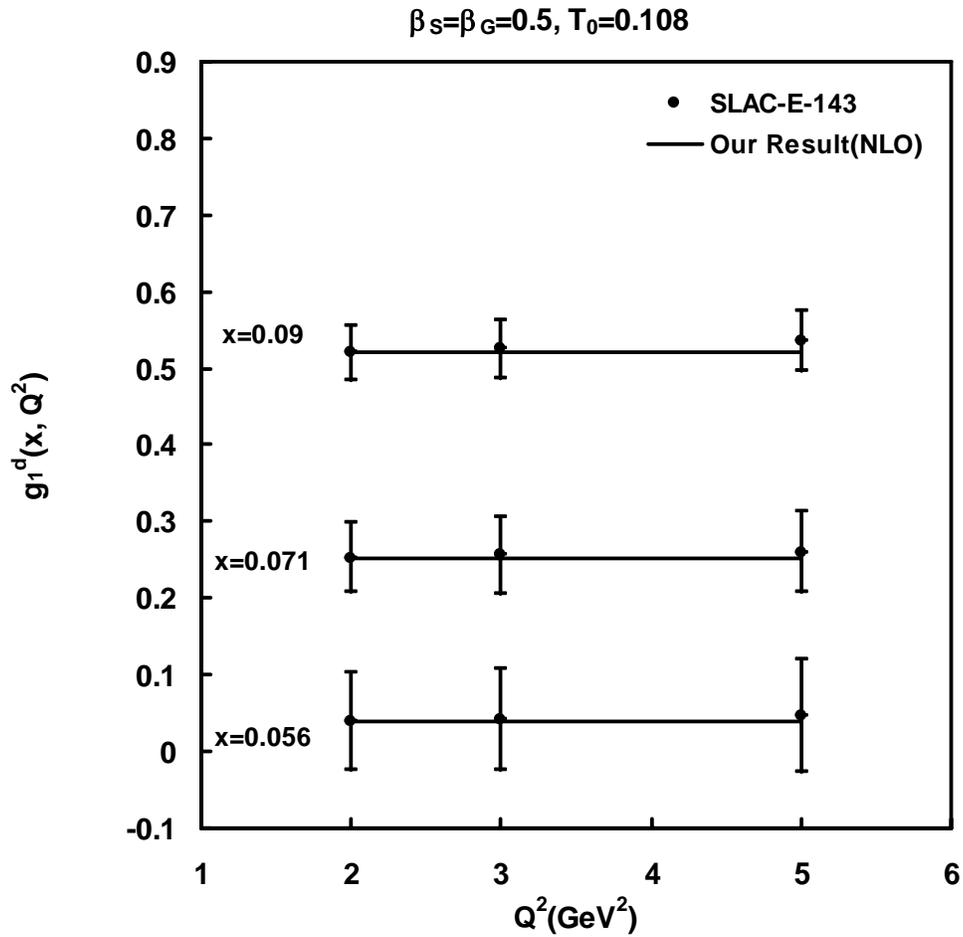

**Figure 7.9**: t evolution of spin-dependent deuteron structure function in NLO at low-x compared with SLAC-E-143 collaborations data set.

## 7.3 Conclusion

In this chapter we have solved DGLAP evolution equations for singlet, non-singlet and gluon structure functions in NLO using Regge behaviour of spin-dependent structure functions. The results are in good consistency with SLAC-E-154, SLAC-E-143, SMC collaborations data sets and with the result obtained by solving unified evolution equation by numerical method especially at low-x and high-$Q^2$ region. The x-evolution graphs for deuteron, proton and gluon structure functions in both LO and NLO are compared and for all of them NLO shows significantly better fitting to the data sets than that of in LO. So, NNLO corrections have significant effect and we cannot ignore the contribution of NNLO terms in our region of work i.e. in low-x and high $Q^2$ region. Whereas the t-evolution graphs for deuteron and proton structure functions for both LO and NLO are almost same. □

Chapter 8

# Conclusion

In this thesis we solved spin-independent DGLAP evolution equations upto next-next-to-leading order (NNLO) and spin-dependent DGLAP evolution equations upto next-to-leading order (NLO) by applying Regge behaviour of structure functions and obtained approximate solutions for both spin-independent and spin-dependent singlet, non-singlet and gluon structure functions. We derived t and x-evolutions of deuteron, proton, neutron and gluon structure functions and compared them with experimental data sets, parameterizations and result from numerical method with satisfactory phenomenological success. In all the results, from experiments, global fits or numerical method it is seen that, all the mentioned structure functions increase when x decreases for fixed values of $Q^2$ and when $Q^2$ increases for fixed values of x.

We have seen that our results of spin-independent deuteron and proton structure functions are in good agreement with NMC and E665 collaborations data sets and the results of spin-independent gluon structure function are in good agreement with MRST2001, MRST2004, GRV1998NLO and GRV1998LO global parameterizations. Results of spin-dependent structure functions are also in good agreement with SLAC-E-154, SLAC-E-143, SMC collaborations data sets and with the curve obtained by solving unified evolution equation by numerical method especially at low-x and high-$Q^2$ region. Again from our results of the best fitted graphs, it is clear that the deuteron, proton, neutron and gluon structure functions in NLO show significantly better fitting to the parameterizations than that of in LO. So, the higher order term in NLO has appreciable contribution in the low-x region to the parton distribution functions. In Chapter 5, we have seen that $T_1$ is much smaller than $T_0$, so, we expect NNLO contributions to the DGLAP evolution equations should be small.

But from the comparison of t-evolution graphs of our results for LO, NLO and NNLO, it is seen that NNLO corrections have significant effect. This indicates that we should not ignore the contribution of NNLO terms in our region of work i.e. in low-x and high $Q^2$ region. Contributions from higher and higher orders to the DGLAP evolution equations will be smaller and smaller and ultimately leading to insignificant effect to the structure functions. It has been observed that in our x-$Q^2$ region of discussion, quark contributes appreciably to gluon structure function. But we have not given the comparison here between the results with quarks and without quarks since the results without quarks are far from the data sets as well as the parametrizations graphs. Though we have derived t and x-evolutions for deuteron, proton, neutron and gluon structure functions in LO and NLO for both spin-independent and spin-dependent cases and in NNLO spin-independent case, we can not establish completely unique solutions because of the assumptions K(x), $T_0$ and $T_1$. We can get more accurate solutions of DGLAP evolution equations for both spin-independent and spin-dependent cases if we solve the coupled equations for singlet and gluon structure functions since in that case we need not to consider the ad-hoc function K(x). By using Regge behaviour we tried and succeeded to solve coupled DGLAP evolution equations and finally overcome the problem of considering the function K(x). For simplified solutions of DGLAP evolution equations we have considered numerical variables $T_0$ and $T_1$, not arbitrarily. The values are chosen such that differences between $T^2(t)$ and $T_0\, T(t)$ and $T^3(t)$ and $T_1\, T(t)$ are negligible.

   In this thesis we have overcome the limitations that arise from Taylor series expansion method. Generally, the input x-distribution functions are taken arbitrarily whereas our analytical method of solution gives the x-distribution directly from the solutions. The number of arbitrary parameters used for our method of solution is also less and the ranges of those values are also very narrow. Number of parameters is much less when we solved coupled evolution equations. So, we can conclude that Regge behaviour of structure functions can give the solution for DGLAP evolution equations in a simple manner which can be taken as simple alternative to all other methods.□

# REFERENCES

[1]. D Griffiths, '*Introduction to Elementary Particles*', John Wiley & Sons, New York (1987)

[2]. J. S. Lilley, '*Nuclear Physics: Principles and Applications*', John Wiley & Sons, New York (2002)

[3]. Gordon L. Kane '*Modern Elementary Particle Physics*', West View Press (1993)

[4]. Kenneth S. Krane, '*Introductory Nuclear Physics*', John Wiley & Sons, New York (1988)

[5]. G I Kopylov, '*Elementary Kinematics of Elementary Particles*', Mir Publishers, Moscow (1983)

[6]. B. Lal and N. Subrahmanyam, '*Atomic and Nuclear Physics*', S. Chand & Company (1986)

[7]. F Halzen and A D Martin, '*Quarks and Lepton, An Introductory Course in Modern Particle Physics*', John Wiley and Sons, New York (1990)

[8]. P D B Collins, A D Martin and R H Dalitz, '*Hadron Interactions*', Adam Hilger, Bristol (1984)

[9]. C. Sutton, '*Elementary Particles*', Encyclopedia Britannica, 1994-2001(2001).

[10]. '*Elementary Particles*', Encyclopedia, Microsoft Encarta, 1993-2001(2001).

[11]. W. Greiner, S. Schramm and E, Stein, '*Quantum Chromodynamics*', Springer-Verlag, New York (1994)

[12]. E Reya, Phys. Rep. **69** (1981) 195

[13]. B Foster, International Journal of Modern Physics **A 13** (1998) 1543

[14]. G Wolf, '*First Results from HERA*', DESY (1992) 92

[15]. R. K. Ellis, W. J. Stirling and B. R. Webber, '*QCD and Collider Physics*', Cambridge University Press, Cambridge (1996)

[16]. J. F. Donoghue, E. Golowich and B. R. Holstein, '*Dynamics of the Standard Model*', Cambridge university Press, Cambridge (1996)

[17]. G Jariskog and D Rein, eds, Proc. '*Large Hardon Collider Workshop*', CERN report, CERN **90-10** (1990) 685

[18]. E. Leader, '*Spin in Particle Physics*', Cambridge University Press (2001)

[19]. *E155*, P. L. Anthony et al., *SLAC-PUB* (2002) 8813

[20]. G. Igo, Phys. Rev. D **63** (2001) 057501

[21]. *EMC*, J. Ashman et al., Phys. Lett. B **206** (1988) 364

[22]. *SMC*, B. Adeva et al., Phys. Rev. D **54** (1996) 6620; E143, abe et al., Phys. Rev. D **58** (1998) 112003 ; E154, abe et al., Phys. Lett. B **405** (1997) 180 ; *E155*, P. L. Anthony et al., Phys. Lett. B **493** (2000) 19

[23]. *COMPASS* proposal, *CERN/SPSLC* 96-14, *SPSLC/P* 297,1996

[24]. *HERMES*, A. Airapetian et al., Phys. Lett. B **442** (1998) 484

[25]. R. Basu, Pramana, Journal of Physics, **51** (1998) 205

[26]. B. Badelek et. al., DESY (1991) 91 ; arXiv:hep-ph/9612274 (1996)

[27]. G. Soyez, Phys. Rev. D **69** (2004) 096005; hep-ph/0306113 (2003) ; hep-ph/0401177 (2004)

[28]. A. Donnachie and P. V. Landshoff, Phys.Lett. B **533** (2002) 277; Phys. Lett. B **550** (2002) 160

[29]. Ya Ya Balitski and L N Lipatov, Sov. J. Nucl. Phys. **28** (1978) 822

[30]. E A Kuraev, L N Lipatov and V S Fadin, Sov. Phys. JETP **44** (1976) 443

[31]. E A Kuraev, L N Lipatov and V S Fadin, Sov. Phys. JETP **45** (1977) 199

[32]. L V Gribov, E M Levin and M G Ryskin, Phys. Rep.**1009** (1983) 1

[33]. L V Gribov, E M Levin and M G Ryskin, Nucl. Phys. **B188** (1981) 555

[34]. L V Gribov, E M Levin and M G Ryskin, Zh. Eksp. Theo. Fiz. **80** (1981) 2132; Sov. Phys. JETP **53** (1981) 1113.

[35]. M Ciafaloni, Nucl. Phys. **B 296** (1988) 49

[36]. S Catani, F Fiorani and G Marchesini, Phys. Lett. **B 234** (1990) 339

[37]. S Catani, F Fiorani and G Marchesini, Nucl. Phys. **B 336** (1990) 18

[38]. G Marchesini, Nucl. Phys. **B 445** (1995) 49

[39]. A. D. Martin and T. D. Spearman, '*Elementary particle theory*', North-Holland (1970)

[40]. R. J. Eden, P. V. Landshoff, D. I. olive and J. C. Polkinghorne, '*The analytic S-matrix'*, Cambridge University Press (1966)

[41]. P. D. B. Collins, '*An introduction to Regge theory and high energy Physics*', Cambridge University Press (1977)

[42]. T. Regge, Nuovo Cimento, **14** (1959) 951

[43]. T. Regge, Nuovo Cimento, **18** (1960) 947

[44]. M. Froissart, Phys. Rev. **123** (1961) 1053

[45]. V. N. Gribov, JETP **41,** (1961) 667

[46]. A. Martin, Nuovo Cimento, **42** (1966) 930

[47]. E. C. Titchmarsh, '*The theory of functions*', Oxford University Press (1939)

[48]. V.S. Fadin, R. Fiore, M.G. Kozlov, A.V. Reznichenko, arXiv:hep-ph/0602006 (2006)

[49]. J. K. Sarma, G. K. Medhi, Eur. Phys. J. C **16** (2000) 481

[50]. U. Jamil and J.K. Sarma, IJAMES **2** (2008) 69

[51]. U. Jamil and J.K. Sarma, Pramana, Journal of Physics, **69** (2007) 167

[52]. U. Jamil and J.K. Sarma, accepted in Pramana, Journal of Physics.

[53]. A. Saikia, Pramana, Journal of Physics, **48** (1997) 1

[54]. V. kotikov and G. Parente, Phys. Lett. B **379** (1996) 195

[55]. A. Donnachie and P. V. Landshoff, Phys. Lett. B **437** (1998) 408

[56]. J. Gayler, arXiv:hep-ph/0206062 (2002)

[57]. Badelek, Acta Phys.Polon. B **34** (2003) 2943

[58]. J. Soffer and O. V. Teryaev, PhysRevD.**56** (1997) 1549

[59]. P. Desgrolard, A. Lengyel, E. Martynov, arXiv:hep-ph/0110149 (2001)

[60]. J. Kwiecinski and B. Ziaja, Phys. Rev. D **60** (1999) 054004.

[61]. G. Altarelli and G. Parisi, Nucl. Phys. *B* **126,** (1997) 298

[62]. L. F. Abbott, W. B. Atwood and R. M. Barnett, Phys. Rev. D **22** (1980) 582

[63]. A. Cafarella et al. Nucl. Phys. B **748** (2006) 253; hep-ph/ 0512358

[64]. J. K. Sarma and B. Das, Phys. Lett. B **304** (1993) 323

[65]. R. Rajkhowa and J. K. Sarma, Indian J. Phys. **79 (1)** (2005) 55; Indian J. Phys. **78** (9) (2004) 979

[66]. R. Baishya and J. K. Sarma, PhysRevD.**74** (2006) 107702, arXiv: hep-ph/0707.0914

[67]. S. Grandshteyn and I. M. Ryzhik, *'Tables of Integrals, Series and Products'*, edited by A. Jeffrey, (Academic Press, New York 1965)

[68]. J. B. Scarborough, *'Numerical Mathematical Analysis'*, John Hopkins Press, Baltimore (1996)

[69]. J. N. Sharma and Dr. R. K. Gupta, *'Differential Equations'*, Krishna Prakashan Mandir, Meerut (1990)

[70]. Sneddon, *'Elements of Partial Differential Equations'*, McGraw-Hill, New York (1957)

[71]. M. Arneodo et al., *NMC Collaboration*, Nucl. Phys. *B* **483** (1997) 3

[72]. Adams et al. Phys.Rev.D**54**, (1996) 3006

[73]. A. D. Martin, R. G. Roberts, W. J. Stirling and R. S. Throne, Eur. Phys. J. C **23** (2002) 73

[74]. T. Affolder et al., [*CDF Collaboration*], Phys. Rev. D**64** (2001) 032001

[75]. A. D. Martin, M. G. Ryskin and G. Watt, arXiv:hep-ph/0406225 (2004)

[76]. S. Chekanov et al. [*ZEUS Collaboration*], Eur. Phys. J. C**21** (2001) 443

[77]. Adloff et al. [*H1 Collaboration*], Eur. Phys. J. C**21** (2001) 33

[78]. M. Glück, E. Reya and A. Vogt, Z. Phys. C**67** (1995) 433; Eur. Phys. J. C**5** (1998) 461

[79]. S. Aid et al. [*H1 Collaboration*], Nucl. Phys. B**470** (1996) 3

[80]. M. Derrick et al. [*ZEUS Collaboration*], Z. Phys. C**69** (1996) 607

[81]. L. W. Whitlow et al., Phys. Lett. B**282** (1992) 475; L. W. Whitlow, SLAC-report (1990) 357

[82]. A. C. Benvenuti et al., [*BCDMS Collaboration*], Phys. Lett. B **223** (1989) 485

[83]. Vernon D. Barger, Roger J. N. Phillips, *'Collider Physics'* (West View Press 1996)

[84]. K. Choudhury and P. K. Sahariah, Pramana-J. Phys. **58** (2002) 599; **60** (2003) 563

[85]. S. J. Farlow, '*Partial differential equations for scientists and engineers*' (John Willey, New York 1982)

[86]. K. Choudhury and A. Saikia, Pramana-J. Phys. **29** (1987) 385; **33** (1989) 359; 34, 85 (1990); **38** (1992) 313

[87]. D. K. Choudhury and J. K. Sarma, Pramana-J. Phys. **38** (1992) 481; **39** (1992) 273

[88]. J. K. Sarma, D. K. Choudhury, G. K. Medhi, Phys. Lett. B **403** (1997) 139

[89]. W Furmanski and R Petronzio, Phys. Lett. B 97, 437 (1980); Ref. TH-2933-*CERN* (KEK scanned version); Ref. TH-2815-*CERN* (KEK scanned version)

[90]. A Deshamukhya and D K Choudhury, Proc. 2nd Regional Conf. Phys. Research in North-East, Guwahati, India, October, (2001) 34

[91]. D. K. Choudhury and P. K. Sahariah, Pramana-J. Phys. **65** (2005) 193

[92]. W. L. van Neerven and A. Vogt, Nucl. Phys. B **588** (2000) 345; Nucl. Phys. B **568** (2000) 263; Phys. Lett. B **490** (2000) 111

[93]. S. Moch, J. A. M. Vermaseren and A. Vogt, Nucl. Phys. Proc. Suppl. **116** (2003) 100; Nucl. Phys. B **688** (2004) 101

[94]. S. Moch, J. A. M. Vermaseren, Nucl. Phys. B **573** (2000) 853

[95]. A. Cafarella, C. Coriano and m. Guzzi, arXiv: hep-ph/0803.0462 (2008)

[96]. T. Gehrmann and E. Remiddi, Comput. Phys. Commun. **141** (2001) 296

[97]. P. Chiappetta and J. Soffer, Phys. Rev. D **31** (1985) 1019

[98]. Bourrely and J. Soffer, Phys. Rev. D **53** (1996) 4067

[99]. X. Jiang, G. A. Navarro, R. Sassot, Eur. Phys. J. C **47** (2006) 81

[100]. Ziaja, Phys. Rev. D **66** (2002) 114017

[101]. K. Abe et. al., Phys. Rev. Lett. **78** (1997) 815

[102]. Preparata, P. G. Ratchiffe and J. Soffer, Phys. Rev. D **42** (1990) 930

[103]. Adams et. al., Phys. Rev. D **56** (1997) 5330

[104]. V. N. Gribov and L. N. Lipatov, Sov. J. Nucl. Phys. **15** (1972) 438; Yu. L. Dokshitzer, Sov. Phys. JETP **46** (1977) 641.

[105]. M. Hirai, S. Kumano and Myiama, Comput. Phys. Commun. **108** (1998) 38.

[106]. N. I. Kochelev et. al., Phys. Rev. D **67** (2003) 074014

[107]. S. D. Bass and M. M. Brisudova, Eur. Phys. J. A **4** (1999) 251.

[108]. N. Bianchi and E. Thomas, Phys. Lett. B **450** (1999) 439

[109]. The *COMPASS* Collaboration, E. S. Ageev et. al., arXiv:hep-ex/0701014 (2007)

[110]. Gaby Rädel, Acta Physica Polonica B **29** (1998) 1295

[111]. B. Ziaja, Eur. Phys. J. C **28** (2003) 475; arXiv:hep-ph/0304268 (2003)

[112]. B. Badelek and J. Kwiecinski, arXiv:hep-ph/9812297 (1998)

[113]. Nazir Hussain Shah and J K Sarma, Phys. Rev. D **77** (2008) 074023

[114]. Abe et al., *SLAC-PUB*-7753 (1998); Phys. Rev. D **58** (1998) 112003

[115]. Abe et al., Phys.Lett.B **405** (1997) 180

[116]. B. Adeva et al., Phys.Rev.D **58** (1998) 112001

[117]. J. Kodaira et. al., Phys.Rev.D **20** (1979) 627; Nucl. Phys. B **165** (1980) 129

[118]. R. Mertig and W. L. Van Neerven, Z. Phys. C **70** (1996) 637

[119]. W. Vogelsang, Nucl. Phys. B **475** (1996) 47

# APPENDICES

## A. Optical theorem

The scattering amplitudes are not directly measurable. What are actually determined in a scattering experiment are the momenta, energies and spin polarizations of all the n-particles which are produced in a given two particle collision 1+2→n. The scattering cross-section, $\sigma_{12}^{tot}$, for this process is defined as the total transfer rate per unit incident flux [41]. The total DIS cross-section for particles 1 and 2 is obtained over all possible final states containing different numbers of particles which is given as

$$\sigma_{12}^{tot} = \sum_{n=2}^{\infty} \sigma_{12 \to n}.$$

And $\sigma_{12}^{tot}$ satisfies a remarkable unitarity relation which is known as the Optical theorem. For a elastic scattering 1+2→1+2, this relation can be shown as

$$\sigma_{12}^{tot} = \frac{1}{2q_{s12}\sqrt{s}} \operatorname{Im}\{\langle i|A|i\rangle\},$$

where $q_{S12}$ is the three momentum, equal but opposite for the two particles.

## B. Carlson's theorem

More precisely Carlson's theorem states that: if f(l) is regular and of the form $O(e^{k|l|})$, where k<π, for Re{l}>N and f(l)=0 for an infinite sequence of integers, l=n, n+1, n+2,....., then f(l) =0 identically. Thus if we were to write

$$A_l^3(t) = A_l^{FG}(t) + f(l,t),$$

where $A_l^{FG}(t)$ is obtained from the Froissart-Gribov projection and f(l, t)=0 for integer l, the theorem tells us that either $A_l^3(t) \to 0$ as $|l| \to \infty$ or f(l, t) vanishes everywhere [41].

## C. Solution of simultaneous linear ordinary differential equations

Let us consider the partial differential equations of A(x, t) and B(x, t) given as [69]

$$t \cdot \frac{\partial A(x,t)}{\partial t} - P.A(x,t) - Q.B(x,t) = 0 \tag{1}$$

and

$$t \cdot \frac{\partial B(x,t)}{\partial t} - R.A(x,t) - S.B(x,t) = 0, \tag{2}$$

where A(x, t) and B(x, t) are functions of x and t. For a constant value of x, the equations are simultaneous linear ordinary differential equations in A(x, t) and B(x, t). Differentiating equation (1) with respect to t we get

$$t \frac{d^2 A(x,t)}{dt^2} + \frac{dA(x,t)}{dt} - P\frac{dA(x,t)}{dt} - Q\frac{dB(x,t)}{dt} = 0.$$

Multiplying this equation by t we get

$$t^2 \frac{d^2 A(x,t)}{dt^2} + t\frac{dA(x,t)}{dt} - P.t\frac{dA(x,t)}{dt} - Q.t\frac{dB(x,t)}{dt} = 0. \tag{3}$$

Putting the value of $t\frac{dB(x,t)}{dt}$ from equation (2) in equation (3) we get

$$t^2 \frac{d^2 A(x,t)}{dt^2} + t\frac{dA(x,t)}{dt} - P.t\frac{dA(x,t)}{dt} - Q.(RA(x,t) + S.B(x,t)) = 0. \tag{4}$$

Equation (1) gives $B(x,t) = \frac{1}{Q}\left(t\frac{dA(x,t)}{dt} - P.A(x,t)\right).$ \hfill (5)

From equations (4) and (5) we have

$$t^2 \frac{d^2 A(x,t)}{dt^2} + (1-P).t\frac{dA(x,t)}{dt} - Q.R.A(x,t) - S.t.\frac{dA(x,t)}{dt} - S.P.A(x,t) = 0.$$

Therefore,

$$t^2 \frac{d^2 A(x,t)}{dt^2} + (1-P-S).t\frac{dA(x,t)}{dt} + (S.P - Q.R).A(x,t) = 0,$$

or

$$t^2 \frac{d^2 A(x,t)}{dt^2} + U.t \frac{dA(x,t)}{dt} + V.A(x,t) = 0, \qquad (6)$$

where U=1-P-S and V=S.P-Q.R. Putting $t=e^Z$,

$$\frac{dA(x,t)}{dt} = \frac{dA(x,t)}{dZ} \frac{dZ}{dt} = \frac{dA(x,t)}{dZ} \frac{1}{\frac{dt}{dZ}} = \frac{dA(x,t)}{dZ} \frac{1}{e^Z} = \frac{dA(x,t)}{dZ} \frac{1}{t},$$

or

$$t.\frac{dA(x,t)}{dt} = \frac{dA(x,t)}{dZ} \quad \text{or} \quad t.\frac{d}{dt} \equiv \frac{d}{dZ}.$$

Now,

$$t.\frac{d}{dt}\left(t.\frac{dA(x,t)}{dt}\right) = t^2 \frac{d^2 A(x,t)}{dt^2} + t \frac{dA(x,t)}{dt}.$$

Therefore,

$$t^2.\frac{d^2 A(x,t)}{dt^2} = t.\frac{d}{dt}\left(t\frac{dA(x,t)}{dt}\right) - t.\frac{dA(x,t)}{dt} = \left(t.\frac{d}{dt} - 1\right)t.\frac{dA(x,t)}{dt} = \left(\frac{d}{dZ} - 1\right)\frac{dA(x,t)}{dZ}.$$

Let us replace d/dZ by D so that

$$t^2 \frac{d^2 A(x,t)}{dt^2} = (D-1).D.A(x,t).$$

Therefore, equation (6) becomes

$$\left(D^2 + (U-1).D + V\right)A(x,t) = 0. \qquad (7)$$

Let, $A(x,t) = e^{mZ}$ be the solution of the equation (7). So we get

$$m^2 + (U-1)m + V = 0,$$

which gives

$$m = \frac{-(U-1) \pm \sqrt{(U-1)^2 - 4V}}{2}.$$

Let us take

$$\frac{-(U-1) + \sqrt{(U-1)^2 - 4V}}{2} = g_1 \quad \text{and} \quad \frac{-(U-1) - \sqrt{(U-1)^2 - 4V}}{2} = g_2.$$

So, the solution of the equation (7) is

$A(x,t) = C_1 e^{g_1 Z} + C_2 e^{g_2 Z}$, where $C_1$ and $C_2$ are arbitrary constants,

$\Rightarrow A(x,t) = C_1 t^{g_1} + C_2 t^{g_2}$. (8)

Now,

$\dfrac{dA(x,t)}{dt} = C_1 \cdot g_1 \cdot t^{g_1-1} + C_2 \cdot g_2 \cdot t^{g_2-1}$.

Putting the values of $\dfrac{dA(x,t)}{dt}$ and $A(x,t)$ in equation (1) we get

$t \cdot (C_1 \cdot g_1 \cdot t^{g_1-1} + C_2 \cdot g_2 \cdot t^{g_2-1}) - P \cdot (C_1 t^{g_1} + C_2 t^{g_2}) - Q \cdot B(x,t) = 0$.

Solving for B(x, t) we get

$B(x,t) = C_1 F_1 t^{g_1} + C_2 F_2 t^{g_2}$, (9)

where $F_1 = (g_1-P)/Q$ and $F_2 = (g_2-P)/Q$. As $C_1$ and $C_2$ are only arbitrary constants, we can take $C_1 = C_2 = C$. Hence the forms of $A(x,t)$ and $B(x,t)$ become

$A(x,t) = C(t^{g_1} + t^{g_2})$ (10)

and

$B(x,t) = C(F_1 t^{g_1} + F_2 t^{g_2})$. (11)

Equations (10) and (11) give the solution of simultaneous linear ordinary differential equations (1) and (2). □